\documentclass[aps,amsmath,amssymb, notitlepage,
twocolumn,
nofootinbib,
superscriptaddress
]{revtex4-1}

\makeatletter

\newcommand{\Rmnum}[1]{\expandafter\@slowromancap\romannumeral #1@}
\makeatother

\usepackage{color}
\definecolor{darkblue}{rgb}{0.,0.,0.4}
\definecolor{darkred}{rgb}{0.5,0.,0.}
\usepackage[pdftex,colorlinks=true,linkcolor=darkblue,citecolor=blue,urlcolor=darkred]{hyperref}

\usepackage{amsmath}
\usepackage{tabularx,graphicx}
\usepackage{epstopdf}
\usepackage{graphicx}
\usepackage{latexsym}
\usepackage{amssymb}
\usepackage{amsmath}
\usepackage{color}
\usepackage{psfrag}
\usepackage{bbm}
\usepackage{bm}
\usepackage{titlesec}
\usepackage{dsfont}
\usepackage{feynmp}
\usepackage{slashed}
\usepackage{mathtools}

\newcommand{\mc}[1]{\mathcal{#1}}

\newcommand{\beq}{\begin{eqnarray}}
\newcommand{\eeq}{\end{eqnarray}}

\newcommand{\la}{\langle}
\newcommand{\ra}{\rangle}

\newcommand{\bsp}{\begin{split}}
\newcommand{\esp}{\end{split}}

\newcommand{\eff}{{\rm eff}}

\newcommand{\half}{\frac{1}{2}}

\begin{document}
\title{Symmetry enriched U(1) quantum spin liquids}
\author{Liujun Zou}
\affiliation{Department of Physics, Harvard University, Cambridge, MA 02138, USA}
\affiliation{Department of Physics, Massachusetts Institute of Technology, Cambridge, MA 02139, USA}
\author{Chong Wang}
\affiliation{Department of Physics, Harvard University, Cambridge, MA 02138, USA}
\author{T. Senthil}
\affiliation{Department of Physics, Massachusetts Institute of Technology, Cambridge, MA 02139, USA}
\date{\today}

\begin{abstract}
We classify and characterize three dimensional $U(1)$ quantum spin liquids (deconfined $U(1)$ gauge theories) with global symmetries.
These spin liquids have an emergent gapless photon and emergent electric/magnetic excitations (which we assume are gapped).
We first discuss in great detail the case with time reversal and $SO(3)$ spin rotational symmetries. We find there are 15 distinct such quantum spin liquids based on the properties of bulk excitations. We show how to interpret them as gauged symmetry-protected topological states (SPTs). Some of these states possess fractional response to an external $SO(3)$ gauge field, due to  which we dub them ``fractional topological paramagnets".   We identify 11 other anomalous states that can be grouped into 3 anomaly classes. 
The classification is further refined  by weakly coupling these quantum spin liquids to bosonic Symmetry Protected Topological (SPT) phases  with the same symmetry.  This refinement does not modify the bulk excitation structure but modifies universal surface properties. Taking this refinement into account, we find there are 168 distinct such $U(1)$ quantum spin liquids.  After this warm-up we provide a general framework to classify symmetry enriched $U(1)$ quantum spin liquids for a large class of symmetries. As a more complex example, we discuss $U(1)$ quantum spin liquids with time reversal and $Z_2$ symmetries in detail. Based on the properties of the bulk excitations, we find there are 38 distinct such spin liquids that are anomaly-free. There are also 37 anomalous $U(1)$ quantum spin liquids with this symmetry. Finally, we briefly discuss the classification of $U(1)$ quantum spin liquids enriched by some other symmetries.
\end{abstract}

\maketitle

\tableofcontents

\section{Introduction}

Symmetry and entanglement both play important roles in understanding quantum phases of matter.   It is by now well known that  the ground states of quantum many-body systems may be in phases characterized by long-range entanglement between local degrees of freedom.  Global symmetry may be realized in interesting ways in such long-range entangled phases.    The simplest (and best understood)  cases  are gapped topologically ordered quantum phases as exemplified by the fractional quantum Hall states. The long-range entanglement in the fractional quantum Hall ground state wavefunctions enables gapped quasiparticle excitations showing fractional statistics and fractional charge.   The fractional statistics is a fundamental  phenomenon that follows from the topological order, while the fractional charge describes the implementation of the global $U(1)$ charge conservation symmetry in this state.

Another prototypical class of states that possess long-range entanglement  are quantum spin liquid phases of insulating magnets.  A wide variety of quantum spin liquids have been described theoretically.
Their universal low energy physics is (in most known examples) described by a deconfined emergent gauge theory coupled to matter fields.  In the presence of global symmetries it is necessary to also specify the symmetry implementation in this low energy theory.   Indeed two phases with the same structure of long-range entanglement (eg, same low energy gauge theory) can still be sharply distinguished by their symmetry implementations.  This leads to a symmetry protected distinction between symmetry unbroken phases of matter (as is familiar from the theory of topological band insulators).

It is useful to distinguish two very broad classes of spin liquids. The simplest and best understood are ones in which all excitations are gapped. These gapped spin liquids are topologically ordered - they have well defined quasiparticle excitations with non-local `statistical' interactions,   ground state degeneracies on topologically non-trivial manifolds, etc. Global symmetries can be implemented non-trivially in topologically ordered phases. For instance a symmetry may be fractionalized.  Topological phases in the presence of global symmetries have been dubbed ``Symmetry Enriched Topological" (SET) matter. Thus symmetry protected distinctions between
 different SET phases may be much more striking than in topological band insulators.  Though much of the early work on spin liquids dealt with SET phases, it is only in
 the last few years that there has been tremendous and systematic progress in understanding their full structure and classification in  two dimensional systems\cite{Levin2012,NeupertSantosRyuEtAl2011,SantosNeupertRyuEtAl2011, EssinHermele2012,MesarosRan2012,HungWan2013,LuVishwanath2013,Wang2013,Barkeshli2014, Tarantino2015,Lan2016}.  Some limited progress has been made for three dimensional  SET phases as well\cite{Xu2013a,ChenHermele2016,Ning2016}. A different broad class of spin liquid phases have gapless excitations. These are much less understood theoretically though they have tremendous experimental relevance.

In this paper, we study a particularly simple class of quantum spin liquids in three spatial dimensions (3D) with an emergent gapless photon excitation. Their low energy dynamics  is described by a deconfined $U(1)$ gauge theory. Microscopic models for such phases were described in Refs \onlinecite{Motrunich2002,HermeleFisherBalents2004,Motrunich2005,Levin2005, SavaryBalents2011,SavaryBalents2016,RoechnerBalentsSchmidt2016}.   The emergence of the photon is necessarily accompanied by the emergence of quasiparticles carrying electric and/or magnetic charges
that couple to the photon. We will restrict attention to phases where these `charged' matter fields are all gapped\footnote{The problem of gapless matter fields coupled to a (compact) $U(1)$ gauge field is an interesting and extensively studied problem. For some representative papers from the condensed matter literature see Refs. \onlinecite{Affleck1988,LeeNagaosa1992,WenLee1996,Hermele2004,LeeLee2005,Lee2008,Senthil2008a,Zou2016}.  A full classification of such phases with gapless matter fields is beyond the reach of currently available theoretical tools.}. One of our main focuses is on the realization of such $U(1)$ quantum spin liquids in $3D$ magnets with spin $SO(3)$ and time reversal ${\cal T}$ symmetries. After warming up with this example, we will describe a general framework to classify symmetry enriched $U(1)$ quantum spin liquids with a large class of symmetries. Then we will apply this framework to the more complicated case where the symmetry is $Z_2\times\mc{T}$. We will also briefly discuss such $U(1)$ quantum spin liquids enriched by some other symmetries. In previous work by two of us\cite{Wang2016} (see also Ref. \onlinecite{Wang2013}) we described the various such phases when time reversal is the only global internal symmetry.  The extension to $SO(3) \times {\cal T}$, $Z_2\times\mc{T}$ and other symmetries is non-trivial and requires some conceptual and technical advances which we describe in detail in this paper.

For the case with $SO(3)\times\mc{T}$ symmetry, we find that there are 15 families of such $U(1)$ quantum spin liquids which may be distinguished by the symmetry realizations on the gapped electric/magnetic excitations.  We describe the physical properties of these states.  We will show that there are two such quantum spin liquids which have a ``fractional" response to a background external $SO(3)$ gauge field. For this reason we dub them ``Fractional Topological Paramagnets".
They are closely analogous to the fractional topological insulators discussed theoretically.

Each of the 15 families is further refined when the quantum spin liquid phase is combined with a Symmetry Protected Topological (SPT) phase of the underlying spin system protected by the same $SO(3) \times {\cal T}$ symmetry.  This does not change the bulk excitation spectrum but manifests itself in different boundary properties. As described in our previous work\cite{Wang2016} this refinement can be non-trivial: some but not all SPT phases can be ``absorbed" by the spin liquid and not lead to a new state of matter. Including this refinement we find a total of 168 different such $U(1)$ quantum spin liquids with $SO(3) \times {\cal T}$ symmetry.

For the case with $Z_2\times\mc{T}$ symmetry, we find there are 38 distinct of such $U(1)$ quantum spin liquids based on the properties of the bulk fractional excitations.  We also obtain the classification for such spin liquids with some other symmetries.

Studying symmetry enriched $U(1)$ quantum spin liquids is of conceptual and practical importance not only for quantum magnetism, but has far reaching connections to many other topics in modern theoretical physics.
First as emphasized in previous papers\cite{Wang2016}, there is a very useful connection to the theory of Symmetry Protected Topological (SPTs) insulators of bosons/fermions.
It is very helpful to view these $U(1)$ quantum spin liquids as the gauged version of some SPTs with a $U(1)$ symmetry, i.e. these quantum spin liquids can be obtained by coupling the relevant SPTs to a dynamical $U(1)$ gauge field.  There  are actually two distinct  ways in which the same $U(1)$ QSL can be viewed as a gauged $U(1)$ SPT - either as a gauged SPT of the electric charge or a gauged SPT of the magnetic monopole. This leads to a generalization of the standard electric-magnetic duality of three dimensional Maxwell theory which  incorporates the realization of global symmetry \cite{Wang2016,Metlitski2016,Metlitski2015,qeddual}.   In the presence of a boundary, this $3+1$-dimensional ``symmetry-enriched" electric-magnetic duality  implies interesting and non-trivial dualities between  $2+1$-dimensional quantum field theories\cite{Wang2015,Metlitski2015,dualdrMAM,qeddual}. This line of thinking has proven to be very powerful in studying difficult problems in  strongly-correlation physics in two space dimensions. Examples include   quantum hall systems, especially the half-filled Landau level\cite{sonphcfl,Metlitski2016,Wang2016a,Geraedts197,WangSenthil2016,WangCooperHalperinStern}, interacting topological insulator surfaces\cite{Wang2015,Metlitski2015,dualdrMAM}, quantum electrodynamics in $2+1$ dimensions\cite{qeddual,seiberg2} and a class of Landau-forbidden quantum phase transitions known as deconfined quantum criticality\cite{dqcpdual}. The lower dimensional dualities are also interesting on their own as nontrivial results in $2+1$ dimensional quantum field theory\cite{Seiberg2016,karchtong,murugan,kachrubosonization}. Therefore, we discuss in detail the relation between different symmetry enriched $U(1)$ quantum spin liquids and various SPTs.

The rest of the paper is organized as follows. In Sec. \ref{sec:classification}, we enumerate all possible $SO(3)\times\mc{T}$ symmetric $U(1)$ quantum spin liquid states based on the properties of their bulk fractional excitations. However,  we will find that   11 of them are anomalous in the sense that these states cannot be realized in any three dimensional spin system with time reversal and $SO(3)$ spin rotational symmetries. We will present various ways of understanding the 15 non-anomalous families. In particular, we describe their physical properties and their construction as gauged SPTs. In Sec. \ref{sec: anomalies}, we explain why the other 11 states are anomalous. In Sec.~\ref{sec: FTP} we discuss the topological response of the $U(1)$ spin liquids to an $SO(3)$ probe gauge field, which leads to the notion of ``fractional topological paramagnets". In Sec. \ref{sec:QSL+SPT}, we combine the non-anomalous $U(1)$ quantum spin liquids with 3D bosonic SPTs with the same symmetry, and discuss how the presence of the SPTs further enriches the classification of the quantum spin liquids. After warming up with the example of $SO(3)\times\mc{T}$ symmetric $U(1)$ quantum spin liquids, in Sec. \ref{sec: general framework} we describe a general framework to classify symmetry enriched $U(1)$ quantum spin liquids for a large class of symmetries. We will apply the general framework to classify $Z_2\times\mc{T}$ symmetric $U(1)$ quantum spin liquids in Sec. \ref{sec: Z2Z2T}, and to classify $U(1)$ quantum spin liquids with some other symmetries in Sec. \ref{sec: other symmeties}. Finally, we conclude in Sec. \ref{sec:discussion}. The appendices contain some supplementary details, and some contents there are interesting and important, albeit rather technical.


\section{$U(1)$ quantum spin liquids enriched by time reversal and $SO(3)$ spin rotational symmetries} \label{sec:classification}

We will start by considering systems of interacting spins on a lattice with $SO(3) \times {\cal T}$ symmetries. The microscopic Hilbert space thus has a tensor product structure.  Further all local operators in this Hilbert space will transform under linear representations of the $SO(3) \times {\cal T}$ symmetry ({\em i.e} integer spin and Kramers singlet). Also, these local operators can only create bosonic excitations.
Our goal is to classify and characterize U(1) quantum spin liquids  that can emerge in such systems with the simplifying assumption that only the emergent photon is gapless.

A first cut understanding of the different possible such $U(1)$ spin liquids is obtained by focusing on the properties of the gapped matter excitations,  such as their statistics and their quantum numbers under the relevant symmetries\cite{Wang2016}.  In three dimensions, the statistics of particles can be either bosonic or fermionic. Under time reversal symmetry, they can be Kramers doublets or non-Kramers. Under $SO(3)$, they can either be in a linear representation (spin-1) or its projective representation (spin-1/2). Note that any excitation with integer spin can be reduced to spin-$1$ by binding local excitations ({\em i.e} excitations created by local operators), and half-integer spin excitations can be similarly reduced to ones with minimal spin-$1/2$.  Thus the only distinction is between linear and projective realizations of the global symmetry.

\begin{figure}
  \centering
  \includegraphics[width=0.3\textwidth]{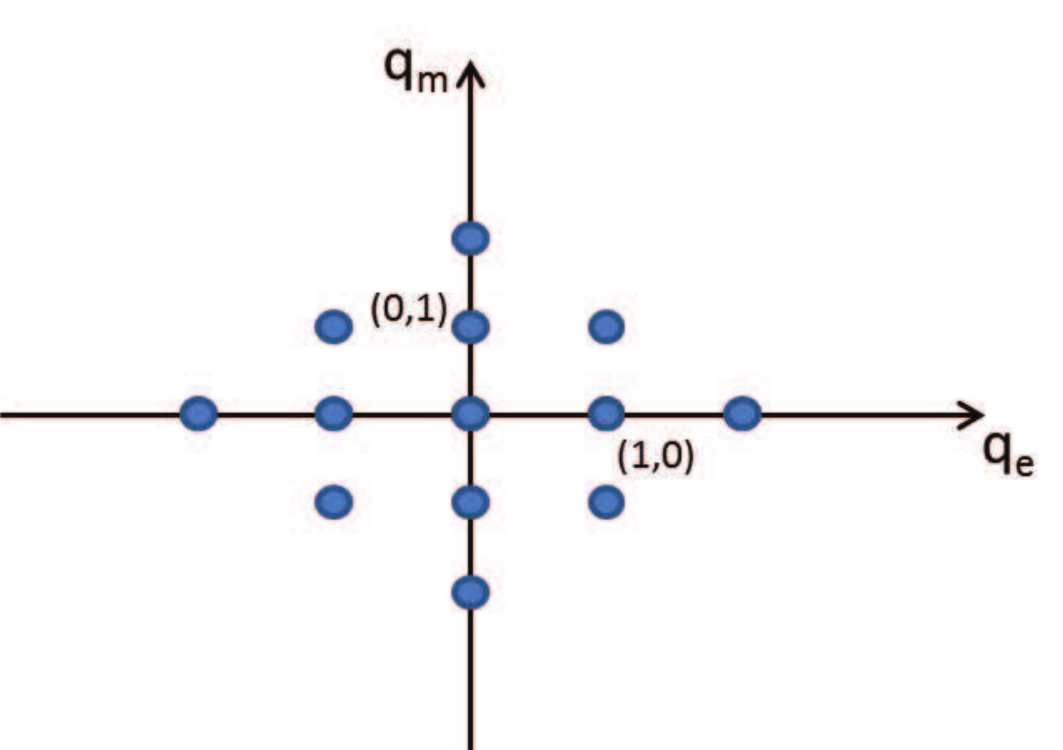}
  \caption{Charge-monopole lattice at $\theta=0\ ({\rm mod}\ 2\pi)$.}\label{fig:charge-monopole-lattice1}
  \includegraphics[width=0.3\textwidth]{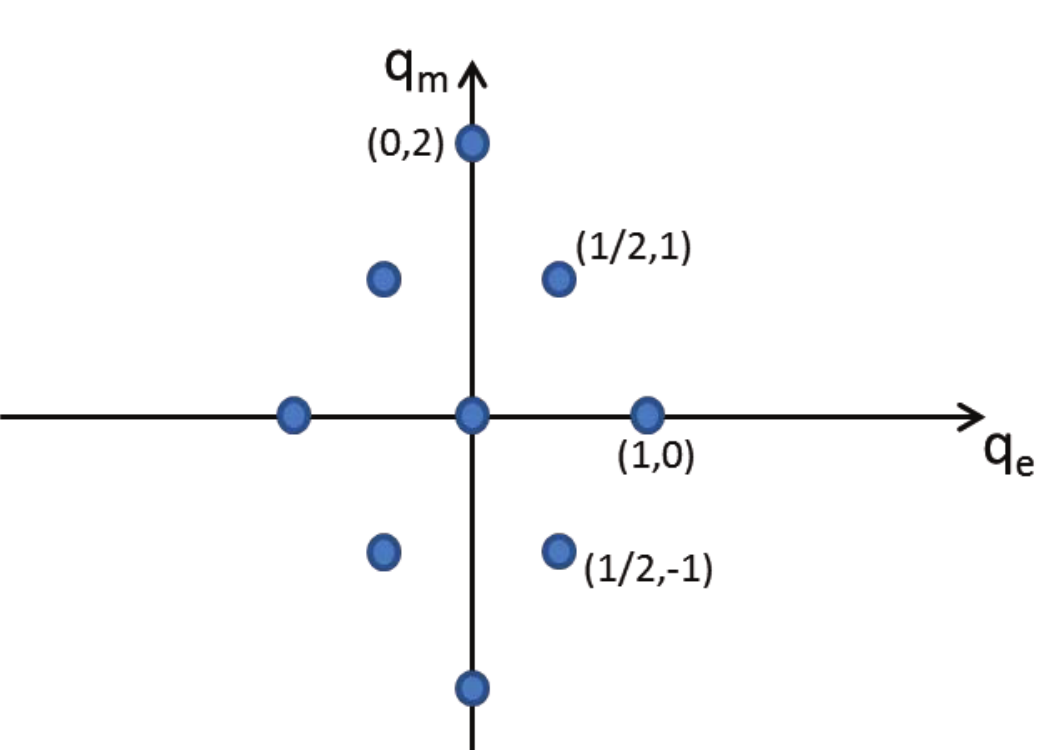}\\
  \caption{Charge-monopole lattice at $\theta=\pi\ ({\rm mod}\ 2\pi)$.}\label{fig:charge-monopole-lattice2}
\end{figure}

In the presence of time reversal symmetry, it is helpful to integrate out the gapped matter fields and consider the effective theory of the photon field. In general this effective theory has the form
\beq \label{eq: gauge theory}
\mc{L}_\eff=\mc{L}_{{\rm Maxwell}}+\frac{\theta}{4\pi^2}\vec E\cdot\vec B
\eeq
where $\mc{L}_{{\rm Maxwell}}$ represents the usual Maxwell Lagrangian, and the second term is of topological character. It is customary to define time-reversal transform such that the electric charge is invariant, namely under time reversal $\vec E\rightarrow\vec E$ and $\vec B\rightarrow-\vec B$. This definition also implies that $\theta\rightarrow-\theta$ under time-reversal. It is also known that $\theta$ is periodic with a period $4\pi$ if the elementary electric charge is a boson, or a period $2\pi$ if the elementary electric charge is a fermion (see Refs. \onlinecite{VishwanathSenthil2013,MetlitskiKaneFisher2013}  for arguments from a condensed matter perspective).  In all cases, the possible electric and magnetic charges of excitations form a two-dimensional lattice, and there are only two distinct configurations of this charge-monopole lattice, i.e. $\theta=0\ ({\rm mod}\ 2\pi)$ and $\theta=\pi\ ({\rm mod}\ 2\pi)$, as shown in Fig. \ref{fig:charge-monopole-lattice1} and Fig. \ref{fig:charge-monopole-lattice2}, respectively. For notational simplicity, we will denote these two cases by $\theta=0$ and $\theta=\pi$, respectively. Notice we take the normalization that the elementary electric charge is $1$, and the minimal magnetic charge is such that it emits $2\pi$ flux seen by the elementary charge.

It is natural to ask whether time-reversal can act on the charge-monopole lattice in more complicated ways. Some examples were discussed in Ref.~\onlinecite{Seiberg2016}, in which the charge-monopole lattices undergoes a rotation (also known as $S$-duality transform) under time-reversal. However, in those examples the theories can be redefined, through appropriate electric-magnetic duality transforms, into the conventional form with the canonical time-reversal transform ($\vec E\rightarrow\vec E$ and $\vec B\rightarrow-\vec B$). In general, such a redefinition should always be possible if the theory, while preserving time-reversal symmetry, has a weakly coupled limit (with gauge coupling $e^2\ll1$).

Denote an excitation with electric charge $q_e$ and magnetic charge $q_m$ by $(q_e,q_m)$. When $\theta=0$, the lattice of charge-monopole excitations is generated by the two particles $(1,0)$ which we denote $E$ and $(0,1)$ which we denote $M$.  Then  the distinct possibilities for the statistics and quantum numbers of $E$ and $M$ will correspond to distinct $U(1)$ quantum spin liquids.  Under time reversal an excitation with nonzero magnetic charge is transformed to another excitation that differs from the original one by a nonlocal operation. It is then meaningless to discuss whether these excitations are Kramers doublet or not, because $T^2$ is not a gauge invariant quantity for them\cite{Wang2013,Wang2014}. On the other hand, all the pure electric charges should have well-defined $T^2$, and they are either Kramers singlet ($T^2=1$) or Kramers doublet ($T^2=-1$). More details can be found in Appendix \ref{app: remarks on TR}.

For the case with $\theta=\pi$, time reversal interchanges $(\half,1)$ and $(\half,-1)$, which generate the entire charge-monopole lattice. In this case, we will still denote $E$ as the $(1,0)$ excitation, but we denote $M$ as the $(0,2)$ excitation.

\subsection{Quantum spin liquids with $\theta=0$}
\label{qsltheta0}

\begin{figure}[h!]
  \centering
  \includegraphics[width=0.45\textwidth]{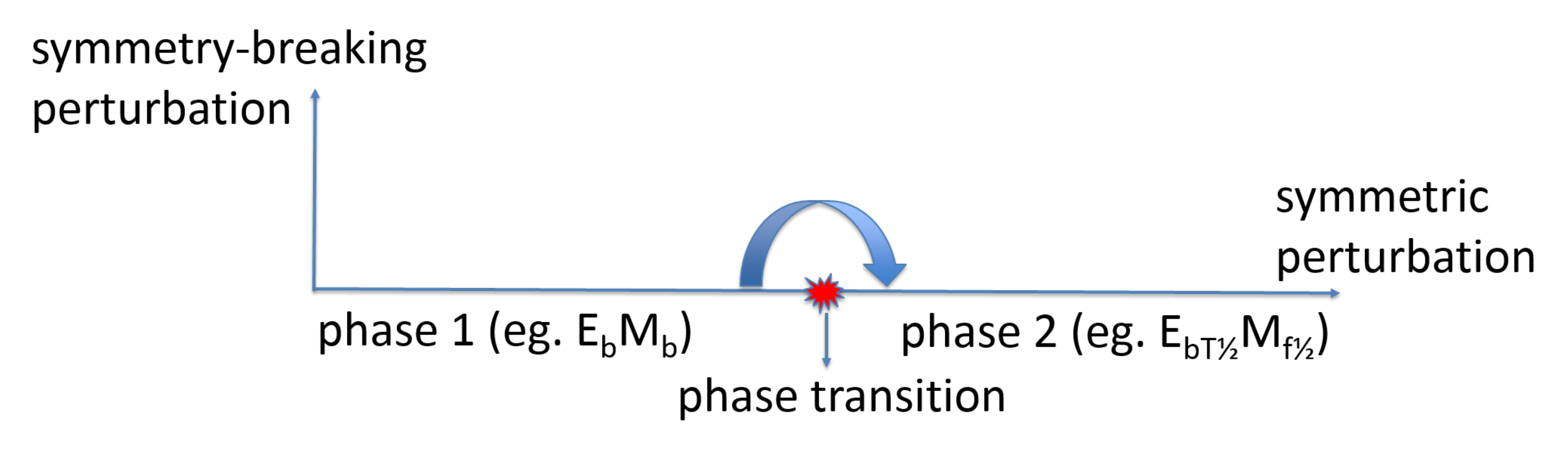}\\
  \caption{Symmetry protected distinctions among symmetry enriched $U(1)$ quantum spin liquids. For example, with $SO(3)\times\mc{T}$ symmetries, two phases, $E_bM_b$ and $E_{bT\half}M_{f\half}$ in this example, cannot be connected without crossing a phase transition. When the symmetry is broken, they can be connected without crossing a phase transition.}\label{fig: symmetry-protected-distinction}
\end{figure}

We start with phases where $\theta=0$. Let us consider the distinct possibilities for the $E$ and $M$ particles.  Note that  U(1) quantum spin liquids with both $E$ and $M$ fermionic are anomalous, i.e., they cannot be realized in a strictly three dimensional bosonic system but they can be realized as the surface of some four dimensional bosonic systems.\cite{Wang2014,Kravec2015,Thorngren2014} We will therefore restrict to situations in which at most one of $E$ and $M$ is a  fermion. Consider the case where $E$ is a boson. Naively then $E$ may have $SO(3)$ spin $S = 0$ or $S = 1/2$, and may be Kramers singlet or doublet, while $M$ may be either a boson or fermion, and may have $S = 0$ or $1/2$. This gives 16 distinct possibilities.  If instead $E$ is a fermion, it may again have $S = 0$ or $1/2$, and $T^2 = \pm 1$ while $M$ must be a boson but may have $S = 0$ or $1/2$, corresponding to 8 distinct possibilites. In total this gives 24 distinct possibilities for the $E$ and $M$ particles which each correspond to a distinct symmetry enriched $U(1)$ QSL (see Figure \ref{fig: symmetry-protected-distinction}).
However we will argue below that of these 10 are anomalous ({\em i.e} the symmetry implementation is inconsistent in a strictly $3+1$-D  system and is only consistent at the boundary of a $4+1$-dimensional SPT phase). We will discard these so that there are only 14 distinct possibilities for the $E$ and $M$ particles at $\theta = 0$. These will describe 14 distinct families of $U(1)$ QSLs.

In Table \ref{table:theta=0}, we list these distinct possible families, and introduce labels for them that we will use in the rest of the paper.
The rest of this subsection will explain how to obtain these 14 spin liquids and Sec. \ref{sec: anomalies} will show that the other 10 spin liquids are anomalous.

\begin{table}
\begin{tabular}{|c|c|c|c|c|}
\hline
&$T^2_E$&$S_E$&$S_M$&comments\\
\hline
$E_bM_b$&1&1&1&E: trivial, M: trivial\\
\hline
$E_bM_f$&1&1&1&E: $eCmC$, M: trivial\\
\hline
$E_bM_{b\half}$&1&1&$\half$&E: $eCm\half$, M: trivial\\
\hline
$E_bM_{f\half}$&1&1&$\half$&E: $eCmC\half$, M: trivial\\
\hline
$E_{bT}M_b$&-1&1&1&E: trivial, M: $eCmT$\\
\hline
$E_{bT}M_f$&-1&1&1&E: $(eCmC)_T\epsilon$, M: n=2 TSC\\
\hline
$E_{b\half}M_b$&1&$\half$&1&E: trivial, M: $eCm\half$\\
\hline
$E_{bT\half}M_b$&-1&$\half$&1&E:trivial, M: $eCmT\half$\\
\hline
$E_{bT\half}M_{f\half}$&-1&$\half$&$\half$&E: $\theta=2\pi$, M: n=2 TSC\\
\hline
$E_fM_b$&1&1&1&E: trivial, M: $eCmC$\\
\hline
$E_{fT}M_b$&-1&1&1&E: trivial, M: $eCmCT$\\
\hline
$E_{f\half}M_b$&1&$\half$&1&E: trivial, M: $eCmC\half$\\
\hline
$E_{f\half}M_{b\half}$&1&$\half$&$\half$&E: n=2 TI, M: $\theta=2\pi$\\
\hline
$E_{fT\half}M_b$&-1&$\half$&1&E: trivial, M: $eCmCT\half$\\
\hline
\end{tabular}
\begin{tabular}{|c|c|c|c|c|}
\hline
&$T^2_E$&$S_E$&$S_M$&comments\\
\hline
$E_{b\half}M_{b\half}$&1&$\half$&$\half$&anomalous (class I)\\
\hline
$E_fM_{b\half}$&1&1&$\half$&anomalous (class I)\\
\hline
$E_{fT}M_{b\half}$&-1&1&$\half$&anomalous (class II)\\
\hline
$E_{b\half}M_f$&1&$\half$&1&anomalous (class II)\\
\hline
$E_{bT\half}M_f$&-1&$\half$&1&anomalous (class II)\\
\hline
$E_{bT\half}M_{b\half}$&-1&$\half$&$\half$&anomalous (class II)\\
\hline
$E_{bT}M_{f\half}$&-1&1&$\half$&anomalous (class III)\\
\hline
$E_{b\half}M_{f\half}$&1&$\half$&$\half$&anomalous (class III)\\
\hline
$E_{bT}M_{b\half}$&-1&1&$\half$&anomalous (class III)\\
\hline
$E_{fT\half}M_{b\half}$&-1&$\half$&$\half$&anomalous (class III)\\
\hline
\end{tabular}
\caption{List of U(1) quantum spin liquids at $\theta=0$. The subscripts ``b" and ``f" refer to bosonic or fermionic statistics of the associated particle, respectively. $T^2_E=1$ ($T^2_E=-1$) means the electric charge is a Kramers singlet (doublet). $S_E$ and $S_M$ refer to the spin of the corresponding particle under $SO(3)$. In this table the spin liquids with both $E$ and $M$ fermions are not listed, because they are known to be anomalous. We identified ten more anomalous spin liquids, and they are divided into three classes. More details can be found in the main texts.} \label{table:theta=0}
\end{table}

Among the 14 quantum spin liquids, the 6 of them in which none of $E$ or $M$ carries spin-1/2 have been described in detail previously\cite{Wang2016} {\footnote{Here we just note that from the point of view of $E$, $E_{bT}M_f$ can be viewed as a bosonic SPT with $\left((U(1)\rtimes\mc{T})/Z_2\right)\times SO(3)$ symmetry. On its surface there can be a symmetric $Z_2$ topological order, where both $e$ and $m$ carry charge-1/2 under $U(1)$ and spin-1 under $SO(3)$. Interestingly, time reversal exchanges $e$ and $m$, while their neutral bound state $\epsilon$ is a Kramers singlet. This surface state is labelled as $(eCmC)_T\epsilon$.}. Below we demonstrate how the other 8 can be constructed. Many of these spin liquids can be obtained simply. Specifically if either $E$ or $M$ is a trivial boson ({\em i.e} has $S = 0$ and (for $E$ particles) $T_E^2 = 1$), then the corresponding spin liquid is obtained by gauging a trivial insulator of the other particle. For instance, to obtain $E_{b\frac{1}{2}}M_b$, start with a trivial insulator formed by bosons with $S = 1/2$ and a conserved $U(1)$ charge that is even under time reversal. Coupling this charge  to a dynamical $U(1)$ gauge field produces a quantum spin liquid which is precisely $E_{b\frac{1}{2}}M_b$.  If instead we wanted to obtain  $E_{b}M_{b\frac{1}{2}}$, we begin with a trivial insulator of a boson with $S = 1/2$ and a conserved $U(1)$ charge that is odd under time reversal.  Gauging this insulator produces  $E_{b}M_{b\frac{1}{2}}$. This kind of  construction clearly works for 6 of the 8 phases where one of $E$ or $M$ is a trivial boson while the other has $S = 1/2$.  It is instructive to also understand these phases from a different `dual' perspective where we will need to gauge the $U(1)$ symmetry of some SPTs with symmetries that contain a $U(1)$ subgroup.  We explain this first below. This will also set the stage to understand the two interesting  remaining cases
where neither $E$ nor $M$ is a trivial boson (these are $E_{bT\half}M_{f\half}$ and $E_{f\half}M_{b\half}$).

\begin{enumerate}
\item $E_{b}M_{b\frac{1}{2}}$

From the point of view of $M$ (that is, viewing $M$ as the gauge charge), $E_{b}M_{b\frac{1}{2}}$ can be viewed as a gauged trivial bosonic insulator with symmetry $\left((U(1)\times SU(2))/Z_2\right)\times\mc{T}$. From the point of view of $E$, it can be viewed as a gauged SPT with symmetry $(U(1)\rtimes\mc{T})\times SO(3)$, where the microscopic boson is a Kramers singlet. This SPT is denoted by $eCm\half$, which means that it can have a surface topologically ordered (STO) state with $Z_2$ topological order, where the topological sectors, $(1,e,m,\epsilon)$, have $e$ carrying charge-1/2 under the $U(1)$ symmetry and $m$ carrying spin-1/2 under the $SO(3)$ symmetry.
This SPT is discussed in more details in Appendix \ref{app: eCmhalf}.

\item $E_bM_{f\half}$

From the point of view of $M$, $E_bM_{f\half}$ can be viewed as a gauged trivial fermionic insulator with symmetry $\left((U(1)\times SU(2))/Z_2\right)\times\mc{T}$. From the point of view of $E$, it can be viewed as a gauged bosonic SPT with symmetry $(U(1)\rtimes\mc{T})\times SO(3)$, where the microscopic boson is a Kramers singlet. We denote this SPT by $eCmC\half$, which can be viewed as a combination of $eCm\half$ and $eCmC$, a well-known SPT with symmetry $U(1)\times\mc{T}$ or $U(1)\rtimes\mc{T}$.\cite{VishwanathSenthil2013,Wang2013,MetlitskiKaneFisher2013} In fact, $eCmC$ is still a nontrivial SPT even if there is additional $SO(3)$ symmetry that commutes with $U(1)\times\mc{T}$ or $U(1)\rtimes\mc{T}$.

\item $E_{b\half}M_b$

From the point of view of $E$, $E_{b\half}M_b$ can be viewed as a gauged trivial bosonic insulator with $((U(1)\rtimes\mc{T})\times SU(2))/Z_2$, where the microscopic bosons are Kramers singlets. From the point of view of $M$, it can be viewed as the gauged $eCm\half$, but with symmetry $U(1)\times\mc{T}\times SO(3)$.

\item $E_{bT\half}M_b$

From the point of view of $E$, $E_{bT\half}M_b$ can be viewed as a gauged trivial bosonic insulator with $((U(1)\rtimes\mc{T})\times SU(2))/Z_2^2$, where the microscopic bosons are Kramers doublets. From the point of view of $M$, it can be viewed as a gauged SPT $eCmT\half$ under symmetry $U(1)\times\mc{T}\times SO(3)$. This SPT can be viewed as a combination of $eCm\half$ and $eCmT$, another well-known SPT with symmetry $U(1)\times\mc{T}$.\cite{VishwanathSenthil2013,Wang2013} It can be shown that this is still a nontrivial SPT even if there is additional $SO(3)$ symmetry that commutes with $U(1)\times\mc{T}$.

\item $E_{f\half}M_b$

From the point of view of $E$, $E_{f\half}M_b$ can be viewed as a gauged trivial fermionic insulator with $((U(1)\rtimes\mc{T})\times SU(2))/Z_2$, where the microscopic fermions are Kramers singlet. From the point of view of $M$, it can be viewed as a gauged $eCmC\half$ with symmetry $U(1)\times\mc{T}\times SO(3)$.

\item $E_{fT\half}M_b$

From the point of view of $E$, $E_{fT\half}M_b$ can be viewed as a gauged trivial fermionic insulator with $((U(1)\rtimes\mc{T})\times SU(2))/Z_2^2$, where the microscopic fermions are Kramers doublets. From the point of view of $M$, it can be viewed as a gauged $eCmCT\half$ under symmetry $U(1)\times\mc{T}\times SO(3)$. This SPT can be viewed as a combination of $eCm\half$, $eCmC$ and $eCmT$.

We now turn to the last 2 cases  $E_{bT\half}M_{\half}$ and $E_{f\half}M_{b\half}$. As both the $E$ and $M$ are non-trivial in these spin liquids, in both the electric and magnetic pictures they should be viewed as gauged SPTs. We state their construction here and describe their properties in greater detail later. We will see that they should be viewed as ``Fractional Topological Paramagnets".

\item $E_{bT\half}M_{f\half}$

From the point of view of $M$, $E_{bT\half}M_{f\half}$ can be viewed as a gauged $n=2$ topological superconductor with symmetry $\left((U(1)\times SU(2))/Z_2\right)\times\mc{T}$. From the point of view of $E$, it can be viewed as a gauged bosonic $\theta=2\pi$ SPT under symmetry $((U(1)\rtimes\mc{T})\times SU(2))/Z_2^2$, where the microscopic bosons are Kramers doublets.

\item $E_{f\half}M_{b\half}$

From the point of view of $E$, $E_{f\half}M_{b\half}$ can be viewed as a gauged $n=2$ topological insulator of fermions with symmetry $((U(1)\rtimes\mc{T})\times SU(2))/Z_2$, where the microscopic fermions are Kramers singlets. From the point of view of $M$, it can be viewed as a gauged bosonic $\theta=2\pi$ SPT with symmetry $\left((U(1)\times SU(2))/Z_2\right)\times\mc{T}$. The properties of this SPT is described in Ref. \onlinecite{WangSenthil2016}.

\end{enumerate}

\subsection{Quantum spin liquids with $\theta=\pi$}

Now we turn to $U(1)$ quantum spin liquids with $\theta=\pi$. At $\theta=\pi$, the charge-monopole lattice is shown in Fig. \ref{fig:charge-monopole-lattice2}. Because time reversal symmetry exchanges $(\half,1)$ and $(\half,-1)$, they should have the same statistics and quantum numbers. Further, they have $\pi$ mutual braiding statistics. This implies that $E$, the bound state of $(\half,1)$ and $(\half,-1)$, has to be a Kramers doublet spin-1 fermion.\cite{Wang2014} Also, because $(-\half,1)$ dyon is the antiparticle of $(\half,-1)$ dyon, it has the same properties as $(\half,1)$. Due to the mutual $\pi$ braiding between $(\half,1)$ and $(-\half,1)$, their bound state, $M$, is also a fermion that carries spin-1 and is non-Kramers. Similar thoughts imply that the statistics and quantum numbers of $(\half,1)$ will determine the statistics and quantum numbers of all gapped excitations. So the classification of $U(1)$ spin liquids with $\theta=\pi$ is equivalent to the classification of the statistics and quantum numbers of the $(\half,1)$ dyon.

It is known that $(\half,1)$ must be a boson. \cite{Wang2014,Kravec2015,Wang2016} Under time reversal symmetry, $T^2$ is not a gauge invariant quantity for $(\half,1)$, so it is non-Kramers. Under $SO(3)$, it can carry either spin-1 or spin-1/2. We will denote the former by $(E_{fT}M_f)_\theta$ and the latter by $(E_{fT}M_f)_{\theta\half}$. These states are summarized in Table \ref{table: theta=pi}.

$(E_{fT}M_f)_\theta$ has been described in detail in Ref. \onlinecite{Wang2016}. From the point of view of $E$, it can be viewed as a gauged free fermion topological insulator with symmetry $\left(((U(1)\rtimes\mc{T}))/Z_2\right)\times SO(3)$, where the microscopic fermions are Kramers doublets. From the point of view of $M$, it can be viewed as a gauged $n=1$ free fermion topological superconductor with symmetry $U(1)\times\mc{T}\times SO(3)$.

In Sec. \ref{sec: anomalies}, we will show that $(E_{fT}M_f)_{\theta\half}$ is anomalous.

\begin{table} \label{table: theta=pi}
\begin{tabular}{|c|c|c|}
\hline
&$S_D$&comments\\
\hline
$(E_{fT}M_f)_\theta$&1&E: TI, M: n=1 TSC\\
\hline
$(E_{fT}M_f)_{\theta\half}$&$\half$&anomalous, class II\\
\hline
\end{tabular}
\caption{List of U(1) quantum spin liquids at $\theta=\pi$. $S_D=1$ ($S_D=\half$) represents the case where the $(\half,1)$ dyon carries spin-1 (spin-1/2).} \label{table:theta=pi}
\end{table}

\section{Anomalous quantum spin liquids with $SO(3)\times\mc{T}$ symmetry} \label{sec: anomalies}

In the enumeration in Sec. \ref{sec:classification}, 11 states are claimed to be anomalous, where 10 of them have $\theta=0$ and 1 has $\theta=\pi$. In this section we will provide arguments to demonstrate these anomalies. We start with the 10 with $\theta=0$.

\subsection{Anomalous states with $\theta=0$}
\label{anomalyargument}

The 10 anomalous quantum spin liquid states at $\theta=0$ are grouped into three classes, such that within each class any one of them can be obtained by coupling another in the same class and some non-anomalous quantum spin liquids. For illustration, let us demonstrate how to obtain $E_fM_{b\half}$ by coupling $E_{b\half}M_{b\half}$ and $E_{f\half}M_{b\half}$, a non-anomalous quantum spin liquid. To do this, one can couple $E_{b\half}M_{b\half}$ and $E_{f\half}M_{b\half}$, and condense the bound state of the monopole of $E_{b\half}M_{b\half}$ and the anti-monopole of $E_{f\half}M_{b\half}$. This bound state is a trivial boson, so condensing it will not break any symmetry. After this condensation, the electric charge of $E_{b\half}M_{b\half}$ and that of $E_{f\half}M_{b\half}$ will be confined together, and the resulting bound state is a fermion that carries no nontrivial quantum number. This is precisely $E_fM_{b\half}$.

The above example shows the relation between the two anomalous quantum spin liquids in class I. We denote this relation by
\beq
E_{b\half}M_{b\half}
\xleftrightarrow{E_{f\half}M_{b\half}}
E_fM_{b\half}
\eeq
The relations among the two other classes are listed here:
\begin{itemize}

\item[]class II:
\beq
\begin{split}
&E_{b\half}M_f
\xleftrightarrow{E_{bT}M_f}
E_{bT\half}M_f\\
&\xleftrightarrow{E_{bT\half}M_{f\half}}
E_{bT\half}M_{b\half}
\xleftrightarrow{E_{f\half}M_{b\half}}
E_{fT}M_{b\half}
\end{split}
\eeq

\item[]class III:
\beq
\begin{split}
&E_{b\half}M_{f\half}
\xleftrightarrow{E_{bT\half}M_{f\half}}
E_{bT}M_{f\frac{1}{2}}\\
&\xleftrightarrow{E_{bT}M_f}
E_{bT}M_{b\half}
\xleftrightarrow{E_{f\half}M_{b\half}}
E_{fT\half}M_{b\half}
\end{split}
\eeq

\end{itemize}

Because of these relations, given that the other 14 quantum spin liquids can be realized in strictly three dimensional bosonic systems, showing that any one of the states of a certain class is anomalous is sufficient to show the entire class is anomalous. Below we will show that $E_{b\half}M_{b\half}$, $E_{bT\half}M_{b\half}$ and $E_{bT}M_{b\half}$ are anomalous.

States of matter that realize a global symmetry non-anomalously allow a consistent coupling of background gauge fields. In our context a non-anomalous realization of $SO(3)$ symmetry thus implies that we can consistently couple background $SO(3)$ gauge fields. Conversely anomalous states can  be detected by finding inconsistencies when such background gauge fields are turned on.

Let us therefore  couple our spin liquids to a  probe $SO(3)$ gauge field. Because $\pi_1(SO(3))=Z_2$, there are monopole configurations of this $SO(3)$ gauge field that are classified by $Z_2$. \cite{Wu1975}

One explicit expression of a nontrivial $SO(3)$ monopole configuration is:
\beq \label{eq:SO(3) monopole}
A^1_\mu=A^2_\mu=0, A^3_\mu=A_{U(1),\mu}
\eeq
where $\mathcal{A}_\mu=\sum A^a_\mu T^a$ is the Lie algebra valued $SO(3)$ gauge field with $T^a$ the generators. $A_{U(1),\mu}$ is the $U(1)$ gauge field configuration of a $U(1)$ monopole.\cite{Wu1975} One of the physical consequences of this $SO(3)$ monopole is a Berry phase factor of an excitation going around a closed loop around it:
\beq \label{eq:Berry phase}
\exp{\left(i\frac{\Omega}{2}S^z\right)}
\eeq
where $\Omega$ is the solid angle of the closed loop with respect to the monopole and $S^z$ is the representation of one of the generators of $SO(3)$. For spin-1 particles, $S^z$ can be taken to be $S^z_{S=1}={\rm diag}(1,0,-1)$. For spin-1/2 particles, $S^z$ can be taken to be $S^z_{S=1/2}={\rm diag}(1/2,-1/2)$. This formula can be easily obtained by borrowing the well-known result of the Berry phase factor of a $U(1)$ charge moving around a $U(1)$ monopole and using (\ref{eq:SO(3) monopole}).

Now consider a Dirac string that ends at this monopole. According to (\ref{eq:Berry phase}), moving around an infinitesimal loop around the Dirac string, a spin-1 particle will get a unit phase factor, which seems normal. But a spin-1/2 particle will see a phase factor of $-1$, which is unphysical. To cancel this phase factor, another defect that also gives a $-1$ phase factor to spin-1/2 particles around the Dirac string needs to be trapped at the $SO(3)$ monopole. We will denote such an $SO(3)$ monopole (with this defect included) by $\mc{M}_{SO(3)}$.

Next we argue that the defect trapped at another $SO(3)$ monopole with
\beq
A^1_\mu=A^2_\mu=0, A^3_\mu=-A_{U(1),\mu}
\eeq
can be essentially the same as the one trapped at the previous $SO(3)$ monopole, $\mc{M}_{SO(3)}$. This is because this new $SO(3)$ monopole can be obtained by performing on $\mc{M}_{SO(3)}$ a $\pi$-rotation around any axis on the $xy$-plane. In the presence of $SO(3)$ symmetry, the defect trapped by it should be the same as that trapped in $\mc{M}_{SO(3)}$ up to a spin rotation. We denote this $SO(3)$ (anti)monopole (with the same defect included) by $\mc{M}_{SO(3)}'$. Notice if an $\mc{M}_{SO(3)}$ and an $\mc{M}_{SO(3)}'$ are fused together, the $SO(3)$ gauge field background will be cancelled, and what remains will be an excitation of the original system without any $SO(3)$ gauge field. These imply that the defect trapped in these $SO(3)$ monopoles can be viewed as ``half" of an excitation of the quantum spin liquid.

For the quantum spin liquid states $E_{b\half}M_{b\half}$ and $E_{bT\half}M_{b\half}$, the Dirac string of a bare $SO(3)$ monopole will give any excitation with spin-1/2 a $-1$ phase factor. These excitations all satisfy $\Delta q=q_e-q_m$ is odd. For these excitations, a $(Q_e,Q_m)$ dyon with odd $Q_e$ and $Q_m$ will give rise to a phase $2\pi(Q_eq_m-Q_mq_e)$ around an infinitesimal loop around the Dirac string, so ``half" of such a $(Q_e,Q_m)$ dyon will give a phase, which is an odd multiple of $\pi$, that exactly cancels the $-1$ phase factor due to the bare $SO(3)$ monopole. One can also check this $-1$ phase factor cannot be cancelled by ``half" of any other type of excitations, where at least one of $Q_e$ and $Q_m$ is even.

According to the argument above, fusing a $\mc{M}_{SO(3)}$ with $\mc{M}_{SO(3)}'$ here should give rise to an $(Q_e,Q_m)$ dyon, with both $Q_e$ and $Q_m$ odd integers. That is,
\beq \label{eq: wrong-fusion-1}
\mc{M}_{SO(3)}\times\mc{M}_{SO(3)}'\sim D_{(Q_e,Q_m)}
\eeq
However, the $(Q_e,Q_m)$ dyon is a fermion as long as both $Q_e$ and $Q_m$ are odd,\cite{Goldhaber1976} and this is inconsistent: $\mc{M}'$ and $\mc{M}$ cannot have any nontrivial mutual Berry phase since they differ merely by a continuous gauge rotation, so the bound state of the two cannot be a fermion. Therefore, the above fusion rule is physically impossible. This shows that all states in class I and class II are anomalous.

The anomalies in the states discussed above do not involve time reversal symmetry in an essential way, but this is not the case for $E_{bT}M_{b\frac{1}{2}}$. For $E_{bT}M_{b\frac{1}{2}}$, the analogous fusion rule we will obtain is
\beq \label{eq: wrong-fusion-2}
\mc{M}_{SO(3)}\times\mc{M}_{SO(3)}'\sim D_{(Q_e,Q_m)}
\eeq
with an odd $Q_e$ and an even $Q_m$.  As $Q_m$ is even, we can always bind $-Q_m/2$ $U(1)$ monopoles to $\mc{M}_{SO(3)}$ and $\mc{M}_{SO(3)}'$ to cancel their magnetic charges.  Thus $Q_m$ can be taken to be zero in the above fusion rule. In this case, the time reversal partners of the $\mc{M}_{SO(3)}$ (and $\mc{M}_{SO(3)}'$) will differ from itself only by a local operator. This implies that  they have a well-defined value for $T^2$. However this is also seen to be impossible: first note that a $(Q_e,0)$ dyon with odd $Q_e$ is a Kramers doublet in this case and all microscopic degrees of freedom are Kramers singlet.  Suppose the fusion rule in Eqn. \ref{eq: wrong-fusion-2} is possible, then $\mc{M}_{SO(3)}$ and $\mc{M}_{SO(3)}'$ should satisfy $T^4=-1$. The argument in Appendix \ref{app: remarks on TR} shows this is impossible unless there are microscopic Kramers doublets, which is absent by assumption. Therefore, $E_{bT}M_{b\half}$ and hence all states in class III are anomalous.

\subsection{Anomalous state with $\theta=\pi$}

Now we show $(E_{fT}M_f)_{\theta\half}$ is also anomalous. The simplest way to see this is to first ignore time reversal symmetry, then from the point of view of $(\half,1)$ and $(\half,-1)$ dyons, this spin liquid is just $E_{b\half}M_{b\half}$. We have shown $E_{b\half}M_{b\half}$ is anomalous with $SO(3)$ symmetry alone even without using time reversal symmetry, this implies $(E_{fT}M_f)_{\theta\half}$ must be anomalous. Another way to see the anomaly is to notice the relation
\beq
(E_{fT}M_f)_{\theta\half}\xleftrightarrow{(E_{fT}M_f)_{\theta}} E_{fT}M_{b\half}
\eeq
This also shows $(E_{fT}M_f)_{\theta\half}$ is anomalous, and in the presence of time reversal symmetry its anomaly belongs to class II.

A more direct argument similar to the ones used above goes as follows. In this case, all $(q_e,q_m)$ dyons with $q_e$ an half-odd-integer and $q_m$ an odd integer carry spin-1/2. This implies the following fusion rule
\beq
\mc{M}_{SO(3)}\times\mc{M}_{SO(3)}'\sim D_{(Q_e,Q_m)}
\eeq
with $Q_e=2n$ and $Q_m=4m+2$, or $Q_e=2n+1$ and $Q_m=4m$, where $n$ and $m$ are integers. One can check this dyon must be a fermion, which in turn shows that this spin liquid is anomalous.

\subsection{Some comments}

The above arguments show that the 11 quantum spin liquids cannot be realized in strictly three dimensions made of bosons if the symmetry $SO(3)\times\mc{T}$ is present. Careful readers may have noticed that the descendent states of these anomalous states will still be anomalous if the symmetry is broken down to $(U(1)\rtimes Z_2)\times\mc{T}\cong O(2)\times\mc{T}$,
where $U(1)$ is the spin rotation around one axis, say, the $z$ axis, and $Z_2$ is a discrete $\pi$-spin rotation around an axis perpendicular to the $z$ axis. In this case, we can couple the system to a $U(1)$ gauge field corresponding to the spin rotational symmetry around the $z$ axis, then $\mc{M}_{SO(3)}$ and $\mc{M}_{SO(3)}'$ become the monopoles of this $U(1)$ gauge field, and the analogous equations of (\ref{eq: wrong-fusion-1}) and (\ref{eq: wrong-fusion-2}) still hold. These two monopoles are mapped into each other by the $Z_2$ transformation. Because this unitary $Z_2$ transformation flips both $S_z$, the spin component along the $z$ direction, and the field value of the $U(1)$ gauge field corresponding to $S_z$ rotational symmetry, there is no mutual statistics between these two monopoles. Therefore, all previous arguments still apply.} In fact, we conjecture even if the symmetry is broken down to $Z_2\times Z_2\times\mc{T}$, the descendant states of these anomalous states will still be anomalous{\footnote{However, if the symmetry is broken down to $U(1)\times\mc{T}$, the descendants of all the anomalous states will become non-anomalous (see Sec. \ref{subsec: SO(N)Z2T}).}}. In Appendix \ref{app: 4d surface} we will show they can be realized as the surface of some four dimensional short-range entangled bosonic systems. In particular, four dimensional bosonic SPT states with only $SO(3)$ symmetry were discussed in Ref.~\onlinecite{Chen2013} using group cohomology, where the SPT states have a $\mathbb{Z}_2$ classification. This is consistent with our result: the only anomalous $U(1)$ spin liquid with $SO(3)$ symmetry is $E_{b\frac{1}{2}}M_{b\frac{1}{2}}$.

If these states were not anomalous, they could also be viewed as some gauged SPTs. So their anomalies imply the impossibilities of some SPTs, which is discussed in more general terms in Sec. \ref{subsec: general-anomaly-detection}. One such example is given in Appendix \ref{app: eCmhalf}.

We would also like to mention that, although the anomalies of these states are shown by examining the $SO(3)$ monopoles, an alternative argument independent of the $SO(3)$ monopoles is sketched in Sec. \ref{sec: general framework}.

In passing, we notice that the fact that ``half" of a dyon is confined by itself does not invalidate our arguments. In fact, the phenomenon where a defect is unphysical unless it traps a confined object is familiar. The most familiar example may be that in a conventional two dimensional superconductor obtained by condensing charge-$1$ bosonic chargons from a spin-charge separated described by a $Z_2$ gauge theory, a $\pi$-flux always appears with a vison, which is confined by itself in the superconducting phase.\cite{Senthil2000}

\section{Fractional topological paramagnets} \label{sec: FTP}

In this section we study the topological response of the spin liquid phases to an external $SO(3)$ gauge field that couples with the $SO(3)$ spin degrees of freedom. In particular we show that the two phases $E_{bT\frac{1}{2}}M_{f\frac{1}{2}}$ and $E_{f\frac{1}{2}}M_{b\frac{1}{2}}$ exhibit nontrivial fractionalized topological response, due to which we dub them ``fractional topological paramagnets".

We start with non-fractionalized (short-range entangled) bosonic phases with $SO(3)\times\mathcal{T}$ symmetry, coupled with a background $SO(3)$ gauge field $\mathcal{A}_{\mu}$. Since the bulk dynamics is trivial by assumption, we can integrate out all the bulk degrees of freedom and ask about the effective response theory for the $SO(3)$ gauge field $\mathcal{A}$. The simplest topological response is a theta-term:
\beq
S_{\Theta}=\frac{\Theta}{16\pi^2}\int{\rm tr}_{SO(3)}F\wedge F,
\eeq
where $F$ is the $SO(3)$ field strength. The normalization is chosen so that if the $SO(3)$ symmetry is broken down to $U(1)\sim SO(2)$, the term becomes a theta-term for the $U(1)$ gauge field with familiar normalization.

It is important to realize that the period of $\Theta$ is $4\pi$ for purely bosonic systems, in contrary to fermionic systems where the period is $2\pi$. In fact a bosonic short-range entangled phase with $\Theta=2\pi$ is a nontrivial SPT state protected by $SO(3)\times\mathcal{T}$. The physics behind is what is known as the ``statistical Witten effect"\cite{MetlitskiKaneFisher2013}: consider inserting a monopole configuration of $\mathcal{A}$ of the form of Eq.~\eqref{eq:SO(3) monopole}, we can ask about the $SO(3)$ charge carried by this monopole. But since the monopole configuration already breaks the symmetry down to $SO(2)\sim U(1)$, we can only ask about the $U(1)$ charge it carries. The standard Witten effect implies that the monopole carries $U(1)$ charge $q_s=\Theta/2\pi=1$. We can bind a gauge charge to the monopole to neutralize the gauge charge, but this converts the monopole to a fermion\cite{Goldhaber1976}.

The above argument also shows that for short-range entangled bosonic phases with $SO(3)\times\mathcal{T}$, the minimal nontrivial $\Theta$-angle is $2\pi$ since under time-reversal $\Theta\to-\Theta$. However, it is also known that for long-range entangled (fractionalized) phases, time-reversal symmetry could be compatible with smaller $\Theta$-angles\cite{fracTI1,fracTI2}. This is because the effective period of $\Theta$ is reduced due to the presence of fractionalized excitations. More formally, in the presence of emergent dynamical gauge fields, it is more appropriate to integrate out only the gapped matter fields and keep the low energy dynamics of the gauge field explicit. The response theory is then correctly captured by a $\Theta$-term and a dynamical term
\beq
\tilde{S}_{\Theta}=\frac{\Theta}{16\pi^2}\int{\rm tr}_{SO(3)}F\wedge F+S'_{\Theta}[a_{\mu},\mathcal{A}_{\mu}],
\eeq
where the second term involves the dynamical gauge field $a_{\mu}$. It is this $\tilde{S}_{\Theta}$ that has a reduced period of $\Theta$. We will explain this point in more concrete examples at the end of Sec.~\ref{ftppiover2}. However, to understand the physics, it suffices to simply study the properties of an $SO(3)$ magnetic monopole (the Witten effect) carefully -- we will mainly focus on this approach here.

We argue below, in the context of $U(1)$ spin liquids, that the effective period of $\Theta$ is reduced to $\pi$ when spin-$1/2$ excitations are allowed in the bulk. This allows, in principle, time-reversal symmetric phases with $\Theta=\pi/2$ (mod $\pi$).

\subsection{Triviality of $\Theta=\pi$}
\label{trivialpi}

First, we need to show that $\Theta=\pi$ is in some sense trivial if (and only if) there are spin-$1/2$ excitations (either $E$ or $M$ particle). Our argument proceeds by carefully studying the Witten effect. Consider again a monopole of $\mathcal{A}$ of the form of Eq.~\eqref{eq:SO(3) monopole}, denoted by $\mathcal{M}_{SO(3)}$. In general it could carry both the $SO(2)$ charge $q_s=\Theta/2\pi=1/2$, and the electric-magnetic charge of the dynamical $U(1)$ gauge field $(q_e,q_m)$. We denote this object as $\mathcal{M}_{SO(3)}=(q_e,q_m,q_s,q_{\mathcal{M}})=(q_e,q_m,1/2,1)$. Time-reversal symmetry implies that the object $\tilde{\mathcal{M}}_{SO(3)}=(q_e,-q_m,-q_s,q_{\mathcal{M}})=(q_e,-q_m,-1/2,1)$ must also exist in the spectrum, and it must have the same statistics with $\mathcal{M}_{SO(3)}$. One can think of $\tilde{\mathcal{M}}_{SO(3)}$ as $\mathcal{M}_{SO(3)}$ attached with a $(0,2q_m,1,0)$ particle (which implies that this particle should exist in the excitation spectrum). Notice that if $q_m=0$, this attachment will change the statistics of $\mathcal{M}_{SO(3)}$ from boson to fermion (or vice versa), and $\mathcal{M}_{SO(3)}$ cannot have the same statistics with $\tilde{\mathcal{M}}_{SO(3)}$ -- this is precisely why in the absence of fractionalization, $\Theta=\pi$ cannot be time-reversal symmetric for a bosonic system. Now with nonzero $(q_m,q_e)$, the issue can be cured by another statistical transmutation if
\beq
\label{cond1}
2q_eq_m+\mathcal{S}_{(0,2q_m,1,0)}=1 \hspace{5pt} ({\rm mod} \hspace{2pt} 2),
\eeq
where $\mathcal{S}_{(0,2q_m,1,0)}=0$ if the $(0,2q_m,1,0)$ particle is a boson, and $\mathcal{S}_{(0,2q_m,1,0)}=1$ if it is a fermion.

Furthermore, any excitations in the (ungauged) $U(1)$ spin liquid $(q_e',q_m',q_s',0)$ should satisfy the general Dirac quantization condition with respect to $\mathcal{M}_{SO(3)}$:
\beq
\label{cond2}
q_eq_m'-q_mq_e'-q_s'=0\hspace{5pt}({\rm mod}\hspace{2pt} 1).
\eeq
The two conditions Eq.~\eqref{cond1} and \eqref{cond2}, together with the existence of $(0,2,1,0)$ in the spectrum, strongly constrains the allowed values of $(q_e,q_m)$ for $\mathcal{M}_{SO(3)}$ and the allowed spectra of the $U(1)$ spin liquids. For example, the $(E_{fT}M_f)_{\theta}$ spin liquid could satisfy Eq.~\eqref{cond1} with $q_e=0,q_m=1$, but this choice inevitably violates Eq.~\eqref{cond2} with $q_e'=1/2,q_m'=1,q_s'=0$. A related phase $(E_{fT}M_f)_{\theta\frac{1}{2}}$ could satisfy all conditions since the test particle for Eq.~\eqref{cond2} should have $q_e'=1/2,q_m'=1,q_s'=1/2$ -- the problem is that this phase is anomalous and cannot be realized in three dimensions on its own. It can be seen after some careful examination, that among the anomaly-free $U(1)$ spin liquids, only those with either $E$ or $M$ particle (but not both) carrying spin-$1/2$ are allowed for $\Theta=\pi$. These include $(E_bM_{b\frac{1}{2}}, E_bM_{f\frac{1}{2}}, E_{b\frac{1}{2}}M_b, E_{bT\frac{1}{2}}M_b, E_{f\frac{1}{2}}M_b, E_{fT\frac{1}{2}}M_b)$. The values of $q_e$ and $q_m$ for $\mathcal{M}_{SO(3)}$ are chosen in the following way: if $E$ particle carries spin-$1/2$, then $q_e=1$ (mod $2$) and $q_m=1/2$ (mod $1$); if $M$ carries spin-$1/2$, then $q_e=1/2$ (mod $1$) and $q_m=1$ (mod $2$). This choice is needed to satisfy the Dirac quantization condition Eq.~\eqref{cond2}. It is also easy to check that Eq.~\eqref{cond1} is satisfied (for those states without anomaly).

It is now easy to see why $\Theta=\pi$ should be considered trivial. In those spin liquids where $E$ particles carry spin-$1/2$, we can bind an $E$ particle to $\mathcal{M}_{SO(3)}$. This gives another, equally legitimate, $SO(3)$ monopole with $q_s=0$ and $q_e=0$. We still have $q_m=1/2$ for the monopole, but this is simply a consequence of the spin-$1/2$ carried by $E$ particle which should be true regardless of what value $\Theta$ takes. Therefore one can equivalently view this phase as having $\Theta=0$ (mod $2\pi$) (notice that both $q_s=0$ and $q_e=0$ for the redefined monopole are important to draw this conclusion). The argument is identical if $M$ particles carry spin-$1/2$ instead. There is still the ambiguity whether the redefined $q_s=0$ monopole is a boson or a fermion, but this is simply about whether $\Theta=0$ or $\Theta=2\pi$ (mod $4\pi$) -- or whether a boson SPT state has been stacked on top of the $U(1)$ spin liquid. We therefore conclude that for a $U(1)$ spin liquid, $\Theta=\pi$ is trivial.

\subsection{$\Theta=\pi/2$: Fractional Topological Paramagnets}
\label{ftppiover2}

We now argue that the two $U(1)$ spin liquid phases $E_{bT\frac{1}{2}}M_{f\frac{1}{2}}$ and $E_{f\frac{1}{2}}M_{b\frac{1}{2}}$ effectively have $\Theta=\pi/2$, and hence can be called ``Fractional Topological Paramagnets".

Again we consider a monopole $\mathcal{M}_{SO(3)}$. In general it could carry both an $SO(2)$ charge $q_s$, and the electric-magnetic charge of the dynamical $U(1)$ gauge field $(q_e,q_m)$. Since both the fundamental electric and magnetic excitations ($E$ and $M$) of the two spin liquids carry spin-$1/2$, according to the argument in Sec.~\ref{anomalyargument} we require $(q_e,q_m)=(1/2,1/2)$ for $\mathcal{M}_{SO(3)}$, up to integer shifts. We denote this object as $\mathcal{M}_{SO(3)}=(q_e,q_m,q_s,q_{\mathcal{M}})=(1/2,1/2,q_s,1)$. Time-reversal symmetry implies that the object $\tilde{\mathcal{M}}_{SO(3)}=(q_e,-q_m,-q_s,q_{\mathcal{M}})=(1/2,-1/2,-q_s,1)$ must also exist in the spectrum. Now take the anti-particle of $\tilde{\mathcal{M}}_{SO(3)}$ and bind it together with $\mathcal{M}_{SO(3)}$, we get an object $(0,1,2q_s,0)$. Since this object does not carry magnetic charge of the $SO(3)$ gauge field, it must exist in the $U(1)$ spin liquid phase before coupling to $\mathcal{A}^{SO(3)}$. But in the spin liquid phase any particle with $q_m=1$ and $q_e=0$ must carry spin-$1/2$. Therefore $2q_s=1/2$ (mod $1$) and $q_s=1/4$ (mod $1/2$). This implies an effective $\Theta=\pi/2$ (mod $\pi$).

One can also ask whether $E_{bT\frac{1}{2}}M_{f\frac{1}{2}}$ and $E_{f\frac{1}{2}}M_{b\frac{1}{2}}$ are the only two ($\mathcal{T}$-invariant) $U(1)$ spin liquids with $\Theta=\pi/2$. An argument similar to that in Sec.~\ref{trivialpi} for $\Theta=\pi$ shows that these two are indeed the only $U(1)$ spin liquids with $\Theta=\pi/2$.

The fractional value of $\Theta$ for the two spin liquids can be understood quite easily if they are viewed as some gauge SPT states (as discussed in Sec.~\ref{qsltheta0}). For concreteness we take $E_{f\frac{1}{2}}M_{b\frac{1}{2}}$ as example (the logic will be parallel for the other state). This state can be obtained from $E_{f\frac{1}{2}}M_b$ by putting the fermionic $E$ particles into a topological band. The corresponding surface state for $E$ will have two Dirac cones -- one for each spin. It is well known\cite{SchnyderRyuLudwig} that this state, when coupled to an $SU(2)$ gauge field, induces a theta-term for the $SU(2)$ gauge field at $\Theta_{SU(2)}=\pi$. This implies $\Theta=\pi/2$ for the $SO(3)$ gauge field.

The Witten effect is also easy to study in this picture: an $SO(3)$ monopole $\mathcal{M}_{SO(3)}$ is viewed by the spin-$1/2$ $E$ particles as a half-monopole. Therefore it should bind with a magnetic charge $q_m=1/2$ (mod $1$). Let's choose $q_m=1/2$. The monopole is then viewed as a $q_m=1$ monopole by the spin-up fermion $f_{\uparrow}$, and a $q_m=0$ object viewed by the spin-down fermion $f_{\downarrow}$. Since each fermion (up or down) has one Dirac cone on the surface, similar to the usual topological insulator, the $\mathcal{M}_{SO(3)}$ monopole will trap half of the charge of an $f_{\uparrow}$ fermion, which gives $q_e=1/2$ and $q_s=1/4$, in agreement with what was obtained earlier using a direct argument.

Alternatively, one can obtain the $E_{f\frac{1}{2}}M_{b\frac{1}{2}}$ state from $E_bM_{b\frac{1}{2}}$ by putting the spin-$1/2$ boson $M$ into a bosonic topological insulating state. The result should be identical, even though the bosonic state is harder to picture due to the lack of non-interacting limit.

We can make the picture slightly more precise by writing down the response theory. We first consider the electric picture, viewing the state as a gauged fermion SPT. This is the more convenient choice if the gauge coupling for the Maxwell term $e^2$ is weak. Integrating out the fermion matter field gives (on a general oriented manifold $Y^4$)
\begin{widetext}
\beq
\label{Lthetapiover2}
\tilde{S}_{\Theta=\pi/2}=\frac{\pi}{2}\left(\frac{1}{16\pi^2}\int{\rm tr}_{SO(3)}F\wedge F+\frac{1}{2\pi^2}\int f\wedge f+\frac{1}{2\cdot24\pi^2}\int{\rm tr}R\wedge R  \right),
\eeq
\end{widetext}
where $f=da$ is the field strength for the dynamical gauge field, and $R$ is the Riemann curvature tensor. The first term comes from the $\Theta_{SU(2)}=\pi$ response of the fermion topological band. The second and third terms are the $U(1)$ and gravitational theta-terms induced by the fermions. The gauge field strength $f$ satisfies the cocycle condition
\beq \label{eq: FTP-constraint}
\int \left(\frac{f}{\pi}+w_2^{TM}+w_2^{SO(3)}\right)=0\hspace{5pt}({\rm mod}\hspace{2pt}2),
\eeq
where $w_2^{TM}$ is the second Stiefel-Whitney class of the tangent bundle on $Y^4$, $w_2^{SO(3)}$ is the second Stiefel-Whitney class of the $SO(3)$ gauge bundle (physically it measures the $\mathbb{Z}_2$-valued $SO(3)$ monopole number and serves as an obstruction to lifting the gauge bundle to an $SU(2)$ one), and the integration is taken on arbitrary 2-cycles on $Y^4$. The Maxwell term for $f$ is suppressed in the above equation for simplicity. Physically this cocycle condition simply represents the fact that charge-$1$ objects under $a_{\mu}$ must carry spin-$1/2$ of the global $SO(3)$ symmetry and must also be a fermion. When $w_2^{TM}$ is trivial, this requires an $SO(3)$ monopole to be accompanied by a half $U(1)$ magnetic-charge, a conclusion we have drawn previously in less formal terms.

To show that Eq.~\eqref{Lthetapiover2} is time-reversal invariant, we only need to show that $2\tilde{S}_{\Theta=\pi/2}$ is trivial (mod $2\pi$). This was shown explicitly in Ref.~\onlinecite{dqcpdual} (Sec.~VII A therein). This also provides an explicit example, in which $\Theta=\pi$ is trivial in the sense that $\tilde{S}_{\Theta=\pi}=2\tilde{S}_{\Theta=\pi/2}$ is trivial.

Similar result can also be obtained in the magnetic picture (with an inverted Maxwell coupling $e^2$). Integrating out the bosonic ($M$) degrees of freedom gives
\beq
\tilde{S}'_{\Theta=\pi/2}=\frac{\pi}{2}\left(\frac{1}{16\pi^2}\int{\rm tr}F\wedge F-\frac{1}{2\pi^2}\int \tilde{f}\wedge \tilde{f} \right),
\eeq
where the inverted sign of the $U(1)$ theta-term and the absence of the gravitational term is simply reflecting the fact that for a bosonic integer quantum hall state in two dimensions with $U(2)=U(1)\times SU(2)/\mathbb{Z}_2$ symmetry, the spin and charge hall conductance are opposite in sign and the net thermal hall conductance is zero\cite{Lu2012,Senthil2012}. The cocycle condition for the dual field strength is now
\beq
\int \left(\frac{\tilde{f}}{\pi}+w_2^{SO(3)}\right)=0\hspace{5pt}({\rm mod}\hspace{2pt}2).
\eeq

Following an argument similar to that in Ref.~\onlinecite{dqcpdual} (Sec.~VII A therein), one can show that $\tilde{S}'_{\Theta=\pi}=2\tilde{S}'_{\Theta=\pi/2}$ is trivial (mod $2\pi$). Therefore the effective theory in the magnetic picture is also time-reversal invariant.

\subsection{Surface states}
Perhaps the most striking property of a topological insulator is the presence of protected surface states. It is natural then to ask about the physics at the surface of the Fractional Topological Paramagnets. Specifically we consider an interface between the vacuum and a material in a Fractional Topological Paramagnet phase.   The gauged SPT point of view then makes it natural that both  $E_{bT\frac{1}{2}}M_{f\frac{1}{2}}$ and $E_{f\frac{1}{2}}M_{b\frac{1}{2}}$ have protected states at such an interface.

Protected surface states for $U(1)$ quantum spin liquids with time reversal were described in Ref. \onlinecite{Wang2016}. As discussed there, in states where both $E$ and $M$ are non-trivial (i.e not simply a boson transforming trivially under the global symmetry) the surface to the vacuum necessarily has protected states. Of the 15 families of $U(1)$ quantum spin liquids with $SO(3) \times \mc{T}$, only $E_{bT\frac{1}{2}}M_{f\frac{1}{2}}$ and $E_{f\frac{1}{2}}M_{b\frac{1}{2}}$ therefore necessarily have protected surface states.  In both these cases the parent SPTs (either in the $E$ or $M$ points of view) are such that the surface exhibits the phenomenon of Symmetry Enforced Gaplessness, {\i.e}, there is no symmetry preserving gapped surface even with topological order. Symmetry preserving surfaces are necessarily  gapless.
For the Fractional Topological Paramagnets  a gapless surface state is readily described from the fermion point of view. Both states then have 2 gapless surface Dirac cones (one for each spin) that is coupled to the bulk $U(1)$ gauge field.  Time reversal acts differently on the surface Dirac fermions in the two states (the time reversal is inherited from that on the bulk fermionic quasiparticle).


\section{Combining $U(1)$ quantum spin liquids and bosonic SPTs under symmetry $SO(3)\times\mc{T}$} \label{sec:QSL+SPT}

We have thus far described the distinct possible realizations of symmetry for the bulk excitations of  $U(1)$ quantum spin liquids with time reversal and $SO(3)$ spin rotational symmetries. However, strictly speaking, this is not the complete classification of such spin liquids. We can in principle obtain distinct spin liquids with the same symmetry fractionalization patterns by simply combining spin liquids with SPT states protected by the global $SO(3) \times \mc{T}$ symmetry.  This was demonstrated for time reversal invariant $U(1)$ spin liquids in Ref. \onlinecite{Wang2016}. Further it was shown that not all SPTs remain non-trivial when combined with a spin liquid. In other words some SPTs can ``dissolve" into some spin liquids without leading to a distinct state. Determining the distinct spin liquids that result when SPTs are combined with spin liquids is a delicate but unavoidable task that is part of any classification of symmetry enriched spin liquids. In this section we undertake this task for the $SO(3) \times \mc{T}$ symmetric $U(1)$ QSLs of primary interest in this paper. We will show that each of the 15 families of such $U(1)$ spin liquids described so far is further refined to give a total of 168 distinct phases. We expect that this is the complete classification of $U(1)$ QSLs enriched with $SO(3) \times \mc{T}$ symmetry.

Bosonic SPTs with symmetry $SO(3)\times\mc{T}$ are classified by $\mathbb{Z}_2^4$. The four root states all admit surface $Z_2$ topological order $\{1,e,m,\epsilon\}$, with different assignments of fractional quantum numbers to the anyons $e,m,\epsilon$ (notice that $e,m$ here denote the anyons in the $2d$ surface topological order, which are not to be confused with $E,M$ in earlier sections denoting electric and magnetic charges in the $3d$ bulk $U(1)$ gauge theory). These surface $Z_2$ topological orders realize symmetries in a way that is impossible for a purely two dimensional system (see Table \ref{table:bSPTs}. More details can be found in Appendix \ref{app:bSPTsurface}).   The surface topological order  provides  a non-perturbative  characterization of these SPTs; we therefore label the SPTs themselves by their surface $Z_2$ topological orders. The four root states generate in total 16 distinct SPTs, and each can be viewed as a combination of some of the root states. For example, if $efmf$ and $e\half m\half$ are taken as two root states, weakly coupling them produces a new SPT denoted by $efmf\oplus e\half m\half$. In this example the notation of the state can be simplified because a surface phase transition can be induced such that the bound state of the $\epsilon$'s from the surface $efmf$ and $e\half m\half$ is condensed. This condensation will not change the bulk property, but the surface now has $Z_2$ topological order $ef\half mf\half$, where both the $e$ and $m$ are spin-1/2 fermions. So for simplicity $efmf\oplus e\half m\half$ can be denoted by $ef\half mf\half$.

\begin{table}
\begin{tabular}{|c|c|c|c|c|c|}
\hline
&$T^2_e$&$T^2_m$&$S_e$&$S_m$&comments\\
\hline
$eTmT$&-1&-1&1&1&\\
\hline
$efmf$&1&1&1&1&$e$ and $m$ are fermions\\
\hline
$e\half m\half$&1&1&$\half$&$\half$&\\
\hline
$e\half mT$&1&-1&$\half$&1&\\
\hline
\end{tabular}
\caption{Properties of the surface $Z_2$ topological orders of the four root states of bosonic SPTs with symmetry $SO(3)\times\mc{T}$.} \label{table:bSPTs}
\end{table}

Below in Sec. \ref{subsec: SPT trivialization} we use the same strategy as in Ref. \onlinecite{Wang2016} to determine if these nontrivial SPTs are trivial or still nontrivial in the presence of the excitations in the quantum spin liquids. Then in Sec. \ref{subsec:QSL+SPT}, we apply these results to obtain the enriched classification of $U(1)$ quantum spin liquids combined with SPTs.

\subsection{SPTs in the presence of nontrivial excitations} \label{subsec: SPT trivialization}

Table \ref{table:bSPTs2} and table \ref{table:bSPTs3} show whether the nontrivial SPTs are trivial or nontrivial in the presence of fractional excitations with all possible statistics and relevant quantum numbers. Below we explain the reasons for the entries of these tables. The notations that will be used below are defined in the captions of these tables.

\begin{table}
\begin{tabular}{|c|c|c|c|c|c|c|}
\hline
&$C^2$&$\tilde C^2$&$C^2T$&$C^2\half$&$\tilde C^2\half$&$C^2T\half$\\
\hline
$eTmT$&$\times$&$\times$&$\surd$&$\times$&$\times$&$\times$\\
\hline
$efmf$&$\times$&$\times$&$\times$&$\times$&$\times$&$\times$\\
\hline
$e\half m\half$&$\times$&$\times$&$\times$&$\times$&$\times$&$\times$\\
\hline
$e\half mT$&$\times$&$\times$&$\times$&$\surd$&$\times$&$\times$\\
\hline
$efT mfT$&$\times$&$\times$&$\times$&$\times$&$\times$&$\times$\\
\hline
$eT\half mT\half$&$\times$&$\times$&$\times$&$\times$&$\times$&$\times$\\
\hline
$eT\half mT$&$\times$&$\times$&$\times$&$\times$&$\times$&$\surd$\\
\hline
$ef\half mf\half$&$\times$&$\times$&$\times$&$\times$&$\times$&$\times$\\
\hline
$efmf\oplus e\half mT$&$\times$&$\times$&$\times$&$\times$&$\times$&$\times$\\
\hline
$e\half mT\half$&$\times$&$\times$&$\times$&$\times$&$\times$&$\times$\\
\hline
$efT\half mfT\half$&$\times$&$\times$&$\times$&$\times$&$\times$&$\times$\\
\hline
$efT mfT\oplus e\half mT$&$\times$&$\times$&$\times$&$\times$&$\times$&$\times$\\
\hline
$eT\half mT\half\oplus e\half mT$&$\times$&$\times$&$\times$&$\times$&$\times$&$\times$\\
\hline
$ef\half mf\half\oplus e\half mT$&$\times$&$\times$&$\times$&$\times$&$\times$&$\times$\\
\hline
$ef\half mf\half\oplus eT\half mT$&$\times$&$\times$&$\times$&$\times$&$\times$&$\times$\\
\hline
\end{tabular}
\caption{Triviality of the root states of bosonic SPTs with symmetry $SO(3)\times\mc{T}$ in the presence of nontrivial bosonic excitaions. The rows represent the nontrivial SPT states, and the columns represent the quantum numbers of the bosonic excition. $C^2$ means the elementary boson carries electric charge $1$, and $\tilde C^2$ means it carries magnetic charge $1$. Notice electric (magnetic) charge is even (odd) under time reversal. $T$ means the elementary boson is a Kramers doublet, and $\half$ means it carries spin-$\half$. A cross (hook) means the topological order is anomalous (non-anomalous) in the presence of the excitation from the quantum spin liquid.} \label{table:bSPTs2}
\end{table}

\subsubsection*{SPTs with component $efmf$ always enrich the classification of the quantum spin liquids}

When time reversal symmetry is broken on its surface, $efmf$ has surface thermal Hall conductance $\kappa_{xy}=4\ ({\rm mod})\ 8$ in units of $\frac{\pi^2}{3}\frac{k_B^2}{h}T$.{\footnote{To characterize the SPT $efmf$ more formallly, one can consider its response to a change of the background metric. Then this SPT is characterized by a bulk gravitational response term given by $\frac{1}{24\pi}\int{\rm tr}R\wedge R$, where $R$ is the Riemann curvature tensor. In this formal language, because none of the $U(1)$ quantum spin liquids discussed has a gravitational response term that can cancel this one, this SPT cannot be ``absorbed" by any of these spin liquids.}} Thus it always enriches the classification of the quantum spin liquids\cite{VishwanathSenthil2013,Wang2016}.  The same is true for all SPTs that are obtained by combining $efmf$ and other root states. Besides $efmf$, these include $efTmfT$, $ef\half mf\half$, $efmf\oplus e\half mT$, $efT\half mfT\half$, $efTmfT\oplus e\half mT$, $ef\half mf\half\oplus e\half mT$ and $ef\half mf\half\oplus eT\half mT$.

\subsubsection*{SPTs with a component $Z_2$ topological order where both $e$ and $m$ carry spin-1/2 are anomalous in the presence of the nontrivial excitations}

These SPTs include $e\half m\half$, $eT\half mT\half$, $ef\half mf\half$, $e\half mT\half$, $efT\half mfT\half$, $eT\half mT\half\oplus e\half mT$, $ef\half mf\half\oplus e\half mT$ and $ef\half mf\half\oplus eT\half mT$. In this case, the $SO(3)$ $\Theta=2\pi$ (see Appendix \ref{app:bSPTsurface}).{\footnote{{More formally, this is characterized by $\frac{\Theta}{16\pi^2}\int{\rm tr}_{SO(3)}F\wedge F$, a response term to a background $SO(3)$ gauge field, where $F$ is the $SO(3)$ field strength and $\Theta=2\pi$ for these states.}}}  But as discussed in Sec. \ref{sec: FTP}, none of the spin liquids have $\Theta=2\pi$. So coupling these SPTs with spin liquids cannot change $\Theta$ from $2\pi$ to $0\ ({\rm mod}\ 4\pi)$, and all these surface states will remain anomalous even when coupled to a spin liquid. Below we provide more physical reasoning to demonstrate their anomalies in the presence of the excitations from the spin liquids.

To see their anomalies, we can first assume that such a state can exist in a purely two dimensional system. Then in the case where the nontrivial excitations carry quantum numbers $C^2$, $\tilde C^2$ or $C^2T$, we can tunnel an $SO(3)$ monopole through the system, which leaves a flux. This is a local process, but due to the spin-1/2 of $e$ and $m$, both $e$ and $m$ will see a $-1$ phase factor around the flux, no matter how far they are away from it. To cancel this nonlocal effect, an $\epsilon$ has to be generated in the process of tunneling the monopole. As shown in Appendix \ref{app:sigmaxy}, this process will not induce any polarization charge or spin because of the symmetry of the system. Therefore, this local process generates a {\it single neutral and spinless} fermion in the system, which is impossible and shows these states are still anomalous even in the presence of the nontrivial excitations.

When the quantum numbers of the nontrivial excitations are $C^2\half$, $\tilde C^2\half$ or $C^2T\half$, tunneling an $SO(3)$ monopole is no longer a local process, but tunneling a bound state of an $SO(3)$ monopole and half of a $U(1)$ monopole is still local. Again, this process will generate a single neutral and spinless fermion in the system, which is impossible and shows the anomalies of these states in the presence of these nontrivial excitations.

\subsubsection*{$eTmT$ is anomalous in the presence of non-Kramers bosons}

It is known that the anomaly of $eTmT$ only comes from time reversal symmetry, so the presence of bosons with quantum numbers $C^2$, $\tilde C^2$, $C^2\half$ or $\tilde C^2\half$ will not remove the anomaly.

\subsubsection*{$e\frac{1}{2}mT$ and $eT\frac{1}{2}mT$ are anomalous in the presence of nontrivial excitations with quantum numbers $C^2$, $\tilde C^2$, $C^2T$ and $\tilde C^2\frac{1}{2}$}

It turns out that $e\half mT$ and $eT\half mT$ are also still anomalous in the presence of nontrivial excitations with quantum numbers $C^2$, $\tilde C^2$, $C^2T$ or $\tilde C^2\half$. To see this, for the cases where excitations carry quantum numbers $C^2$, $\tilde C^2$ or $C^2T$, we can again tunnel an $SO(3)$ monopole through the system. This will leave a flux such that $e$ and $\epsilon$ see a $-1$ phase factor no matter how far they are away from it. To cancel this nonlocal effect, an $m$ will have to be generated in the process. As argued in Appendix \ref{app:sigmaxy}, this process cannot induce any polarization charge or spin. Because the $SO(3)$ flux background is invariant under time reversal, such a local process generates a Kramers doublet that carries no other nontrivial quantum number. But there are no such local degrees of freedom in these cases, so this is impossible. Thus these $Z_2$ topological orders are still anomalous.

If the excitations carry quantum number $\tilde C^2\half$, one can tunnel a bound state of an $SO(3)$ monopole and half of a $U(1)$ monopole. Similar argument shows an $m$ needs to be produced in the process. Again, because both $SO(3)$ and $U(1)$ commute with $\mc{T}$, the flux background left on the system is time reversal invariant. This again shows that a local process generates a Kramers doublet with no other nontrivial quantum number and thus it is impossible.

\subsubsection*{$(eTmT,bC^2T)$, $(e\half mT,bC^2\half)$, $(eT\half mT,bC^2T\half)$, $(e\frac{1}{2}mT,fC^2T\frac{1}{2})$, $(eTmT,fC^2)$ and $(eT\frac{1}{2}mT,fC^2\frac{1}{2})$ are non-anomalous}

Denote $eTmT$ in the presence of bosons with quantum number $C^2T$ by $(eTmT,bC^2T)$. It turns out this is non-anomalous. \cite{Wang2016} To see this, we can attach a boson with quantum number $C^2T$ to the $e$ particle, then $eTmT$ will be relabelled as $eC^2mT$. This is a non-anomalous state. To construct it, one can first construct $eC^2T$, which is non-anomalous because the topological order can be confined by condensing $m$ without breaking any symmetry. Then putting the $\epsilon$ into a quantum spin Hall state makes it $eC^2mT$. \cite{Ran2008,Qi2008}

Similarly, with parallel notations, $(e\half mT,bC^2\half)$, $(eT\half mT,bC^2T\half)$, $(e\frac{1}{2}mT,fC^2T\frac{1}{2})$, $(eTmT,fC^2)$ and $(eT\frac{1}{2}mT,fC^2\frac{1}{2})$ are also non-anomalous.

\subsubsection*{Other entries in table \ref{table:bSPTs2} and \ref{table:bSPTs3} are anomalous}

For other entries in Table \ref{table:bSPTs2} and Table \ref{table:bSPTs3}, the arguments utilized above do not apply. However, they are still expected to be anomalous. Below we sketch the logic to show this, and more details can be found in Appendix \ref{app:non-edgeability}.

Suppose any of these $Z_2$ topological orders is non-anomalous, that is, it can be realized in a purely two dimensional system, it must allow a physical edge, i.e. a boundary that separates this state and the trivial vacuum. It is believed that the $K$-matrix formalism can describe all two dimensional Abelian topological orders, and in particular, $K$-matrix theory naturally allows a physical edge.\cite{Wen2004Book} So if no $K$-matrix description of a $Z_2$ topological order exists, it should not be edgeable, i.e. it is anomalous. We note the $K$-matrix formalism has already been applied to check edgeability or to classify SPTs and symmetry-enriched topological orders in the literature.\cite{Wang2013,Levin2012,Lu2012,LuVishwanath2013}

Indeed, in Appendix \ref{app:non-edgeability} we will show all other entries are not edgeable. This implies they are still anomalous.

\begin{table}
\begin{tabular}{|c|c|c|c|c|c|c|}
\hline
&$C^2$&$\tilde C^2$&$C^2T$&$C^2\half$&$\tilde C^2\half$&$C^2T\half$\\
\hline
$eTmT$&$\surd$&$\times$&$\times$&$\times$&$\times$&$\times$\\
\hline
$efmf$&$\times$&$\times$&$\times$&$\times$&$\times$&$\times$\\
\hline
$e\half m\half$&$\times$&$\times$&$\times$&$\times$&$\times$&$\times$\\
\hline
$e\half mT$&$\times$&$\times$&$\times$&$\times$&$\times$&$\surd$\\
\hline
$efT mfT$&$\times$&$\times$&$\times$&$\times$&$\times$&$\times$\\
\hline
$eT\half mT\half$&$\times$&$\times$&$\times$&$\times$&$\times$&$\times$\\
\hline
$eT\half mT$&$\times$&$\times$&$\times$&$\surd$&$\times$&$\times$\\
\hline
$ef\half mf\half$&$\times$&$\times$&$\times$&$\times$&$\times$&$\times$\\
\hline
$efmf\oplus e\half mT$&$\times$&$\times$&$\times$&$\times$&$\times$&$\times$\\
\hline
$e\half mT\half$&$\times$&$\times$&$\times$&$\times$&$\times$&$\times$\\
\hline
$efT\half mfT\half$&$\times$&$\times$&$\times$&$\times$&$\times$&$\times$\\
\hline
$efT mfT\oplus e\half mT$&$\times$&$\times$&$\times$&$\times$&$\times$&$\times$\\
\hline
$eT\half mT\half\oplus e\half mT$&$\times$&$\times$&$\times$&$\times$&$\times$&$\times$\\
\hline
$ef\half mf\half\oplus e\half mT$&$\times$&$\times$&$\times$&$\times$&$\times$&$\times$\\
\hline
$ef\half mf\half\oplus eT\half mT$&$\times$&$\times$&$\times$&$\times$&$\times$&$\times$\\
\hline
\end{tabular}
\caption{Trivialness of the root states of bosonic SPTs with symmetry $SO(3)\times\mc{T}$ in the presence of nontrivial fermionic excitaions. The rows represent the nontrivial SPT states, and the columns represent the quantum numbers of the fermionic excition. $C^2$ means the elementary fermion carries electric charge $1$, and $\tilde C^2$ means it carries magnetic charge $1$. Notice electric (magnetic) charge is even (odd) under time reversal. $T$ means the elementary fermion is a Kramers doublet, and $\half$ means it carries spin-$\half$. A cross (hook) means the topological order is anomalous (non-anomalous) in the presence of the excitation from the quantum spin liquid.} \label{table:bSPTs3}
\end{table}

\subsection{Enriched classification of quantum spin liquids combined with SPTs} \label{subsec:QSL+SPT}

In the previous subsection, we have shown when the nontrivial bosonic SPTs are in an environment with some nontrivial particles, which ones are still nontrivial and which ones become trivial. In most cases, the nontrivial SPTs remain nontrivial. Then each of the quantum spin liquids can become $2^4=16$ distinct ones after being weakly coupled with the bosonic SPTs. In the presence of the excitations in the quantum spin liquids, the cases where nontrivial SPTs become trivial are when $eTmT$ coupled with $E_{bT}M_b$, $E_{bT}M_f$ or $E_fM_b$, when $e\half mT$ coupled with $E_{b\half}M_b$ or $E_{fT\half}M_b$ and when $eT\half mT$ coupled with $E_{bT\half}M_b$, $E_{bT\half}M_{f\half}$, $E_{f\half}M_b$ or $E_{f\half}M_{b\half}$. For these quantum spin liquids, each can become $2^3=8$ distinct ones after weakly coupled with the bosonic SPTs. All these SPT-enriched quantum spin liquids are different from each other. Therefore, when weakly coupled with bosonic SPTs with time reversal and $SO(3)$ spin rotational symmetries, there are in total $6\times 16+9\times 8=168$ distinct $U(1)$ quantum spin liquids.

\section{A general framework to classify symmetry enriched $U(1)$ quantum spin liquids} \label{sec: general framework}

The above discussion on the classification of $SO(3)\times\mc{T}$ symmetric $U(1)$ quantum spin liquids provides a good example. In this section we describe a general framework to classify symmetry enriched $U(1)$ quantum spin liquids. It involves three steps: enumerating all putative states, examining the anomalies of these states, and coupling these states to 3D bosonic SPTs with the same symmetry. This framework is physics-based. After discussing this framework, we will briefly discuss a supplementary formal approach to classify such states, which can be potentially more useful for thinking about these problems more abstractly.

\subsection{Enumerate putative states}

We begin with the first step: enumerating all putative states. As discussed earlier, different symmetry enriched $U(1)$ quantum spin liquids are distinguished by the properties of their excitations, and to determine the phase, we need to specify the statistics and symmetry quantum numbers of the excitations.

We start with the simpler case where the symmetry $G$ is unitary and connected, that is, all elements in the symmetry group are unitary and they can all be continuously connected to the identity element. In this case, the symmetry cannot exchange the type of the fractional excitations. Also, one can tune $\theta$ such that the charge-monopole lattice is of the $\theta=0$-type, and both $E$ and $M$ are bosons (this is shown more explicitly in the examples in Sec. \ref{subsec: SO(N)}). To fully determine the properties of the excitations, we just need to specify the symmetry quantum numbers of $E$ and $M$. More precisely, we need to assign (projective) representations to $E$ and $M$, which are classified by the second group cohomology $H^2(G,U(1))$. While doing this, we also need to keep in mind that $E$ and $M$ are equivalent in this case, so that, for example, $E_{b\half}M_b$ and $E_bM_{b\half}$ are the same $SO(3)$ symmetric phase. When this is done, all putative states will be listed.

Next we go to the more complicated case where the symmetry is $G\times\mc{T}$ (or, more generally, $G\rtimes\mc{T}$), with $G$ a connected unitary group. Again, the elements in $G$ will not change the type of fractional excitations. However, time reversal will necessarily change some types of fractional excitations, and we will always take the convention that the emergent electric (magnetic) field is even (odd) under time reversal. Then there are two types of charge-monopole lattice, with $\theta=0$ and $\theta=\pi$, respectively.

Consider the states at $\theta=0$ first. Then the $U(1)$ quantum spin liquids are classified by the statistics and symmetry quantum numbers carried by $E$ and $M$. As for statistics, the only constraint at this point is just that $E$ and $M$ cannot both be fermions. Below we discuss the symmetry quantum numbers, or in other words, projective representations.

Let us start from the case with $\theta=0$. Here we need to distinguish two types of projective representations: the electric (standard) one and the magnetic (twisted) one. The electric projective representations are applicable to $E$, and they are classified by $\mc{H}^2(G\times\mc{T},U_T(1))$, where $G\times\mc{T}$ acts on the $U(1)$ coefficient by taking the complex conjugate if the group element is anti-unitary. This is the standard classification of the projective representations of a group with anti-unitary elements. However, another type of projective representations apply to $M$, which are classified by another group cohomology $\mc{H}^2(G\times\mc{T},U_T^M(1))$ (see Appendix \ref{app: projective representations} for more details). This group cohomology differs from the standard one in the group action on the $U(1)$ factors, and this difference comes from the convention that the magnetic (electric) field is odd (even) under time reversal. After assigning statistics and symmetry quantum numbers to $E$ and $M$ as above, all putative $G\times\mc{T}$ symmetric $U(1)$ quantum spin liquids with $\theta=0$ will be listed.

As for states with $\theta=\pi$, the properties of all excitations are determined by the properties of the $(\frac{1}{2},\pm1)$ dyons, which must be bosons. So to list all putative states, we only need to assign symmetry quantum numbers to these two dyons. As discussed in Appendix \ref{app: projective representations}, these symmetry quantum numbers are given by the dyonic (mixed) projective representations, which are classified by another group cohomology $H^2(G\times\mc{T},U^D(1)\times U^D(1))$. After assigning symmetry quantum numbers to the $\left(\half,\pm1\right)$ dyons, all putative $G\times\mc{T}$ symmetric $U(1)$ quantum spin liquids with $\theta=\pi$ will be listed.

If the symmetry group is $G$ or $G\times\mc{T}$, where $G$ is unitary but not connected, the elements in $G$ can also permute fractional excitations, as we  will see in examples below. The putative states in this more complicated scenario can be listed in a similar manner as above: one has to fix the shape of the charge-monopole lattice, specify the statistics of the relevant excitations, and specify the symmetry quantum numbers of the relevant excitations. The first two steps are identical as the previous cases, but the classification of symmetry quantum numbers will be more complicated in this case, and in this paper we do not attempt to give a mathematical framework for this step, although it can be done in a case-by-case manner.

\subsection{Examine the anomalies} \label{subsec: general-anomaly-detection}

After enumerating all putative symmetry enriched $U(1)$ quantum spin liquids, we need to examine which ones are anomalous. A general way of doing this is to consider whether the corresponding SPT of this spin liquid can exist. Denote the symmetry of the $U(1)$ quantum spin liquid by $G$, which is supposed to be a completely general on-site symmetry in this subsection (it can contain anti-unitary elements, and its unitary elements do not need to be continuously connected with identity). By the corresponding SPT of a spin liquid, we mean an SPT protected by a $U(1)$ central extension of $G$,{\footnote{In the more standard terminology, this $U(1)$ central extension of $G$ is the projective symmetry group of $G$, which was first introduced in Ref. \onlinecite{Wen2002}.}} which becomes this spin liquid once this $U(1)$ symmetry is gauged. Clearly, by definition, as long as this SPT can exist, the spin liquid state must be anomaly-free. If this SPT is intrinsically inconsistent, then the corresponding spin liquid state must be anomalous. To see this, suppose this SPT is problematic but the corresponding spin liquid is anomaly-free, then we will show this leads to a contradiction, because of a systematic method to ungauge the gauge theory and generate the corresponding SPT.

\begin{figure}[h!]
  \centering
  \includegraphics[width=0.48\textwidth]{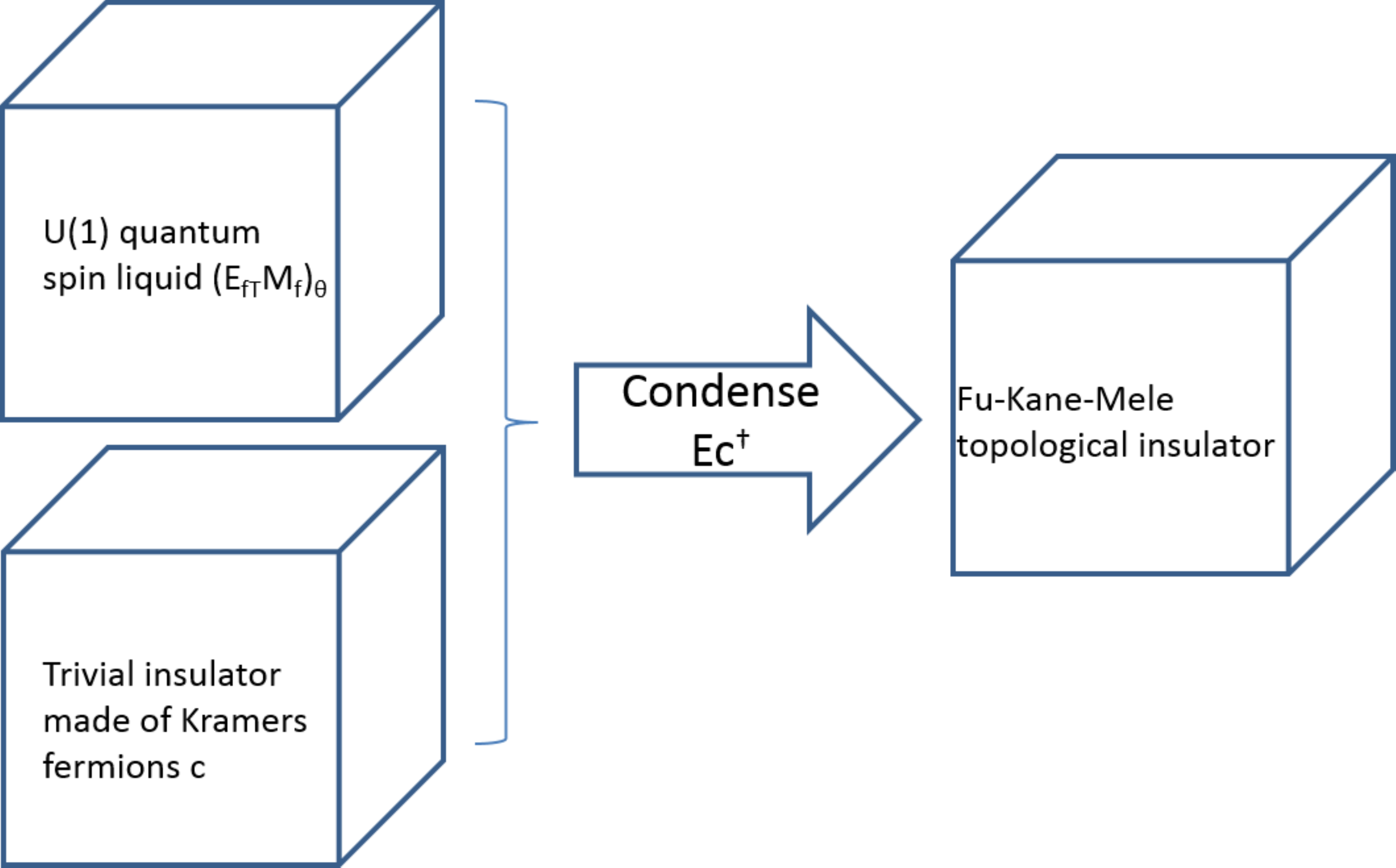}\\
  \caption{There is a systematic ungauging procedure that takes a symmetry enriched $U(1)$ quantum spin liquid to its corresponding SPT. Consider the time reversal symmetric $U(1)$ quantum spin liquid $(E_{fT}M_f)_\theta$ (the upper left system), and we will try to get its corresponding SPT from the perspective of the electric charge. To do this, we first introduce an auxiliary trivial time reversal symmetric insulator made of fermions that are Kramers doublets, where these fermions are denoted by $c$ (the lower left system). Next we condense the bound state of $E$, the electric charge of the $U(1)$ spin liquid, and $c^\dag$, the holes in the auxiliary trivial insulator. This bound state is a boson and a Kramers singlet, so this condensation will preserve the time reversal symmetry. The dynamical $U(1)$ gauge field in the $U(1)$ gauge theory will be confined, and the resulting state is precisely the Fu-Kane-Mele topological insulator,\cite{FuKaneMele2007} which, viewed from the perspective of the electric charge, is the corresponding SPT of $(E_{fT}M_f)_\theta$ (the right system).}\label{fig: ungauge}
\end{figure}

More precisely, suppose this spin liquid can be realized, one can bring in a trivial insulator made of particles that have the same properties (same statistics and quantum numbers) as the particles that make up this corresponding SPT, and condense the bound state between the particle in this trivial insulator and the electric charge or magnetic monopole of the spin liquid. This is a systematic method to ungauge the gauge theory: it will confine the dynamical $U(1)$ gauge field, and the resulting state will be precisely the corresponding SPT of the spin liquid state (an example is shown in Figure \ref{fig: ungauge}).\cite{Metlitski2016} This leads to a contradiction to the original assumption that such SPT is problematic. Therefore, a sufficient and necessary condition for a symmetry enriched $U(1)$ quantum spin liquid to be anomaly-free is that its corresponding SPT is consistent.

How do we check whether the corresponding SPT is consistent? One way is to consider whether it has a consistent surface state. This condition  - known as ``edgeability" - was defined in Ref. \onlinecite{Wang2013}. Assuming such SPT is consistent, one can first condense certain charges on the surface of this SPT and get a surface superfluid. Then one can try to condense certain vortices to restore the symmetry on the surface. If the symmetric surface state is consistent (but possibly anomalous), one can build up the three dimensional bulk SPT (for example, by a layer construction or a Walker-Wang type construction). If the putative symmetric surface state is inconsistent, then this SPT is inconsistent, because it has an invalid edge state.

\begin{figure}[h!]
  \centering
  \includegraphics[width=0.4\textwidth]{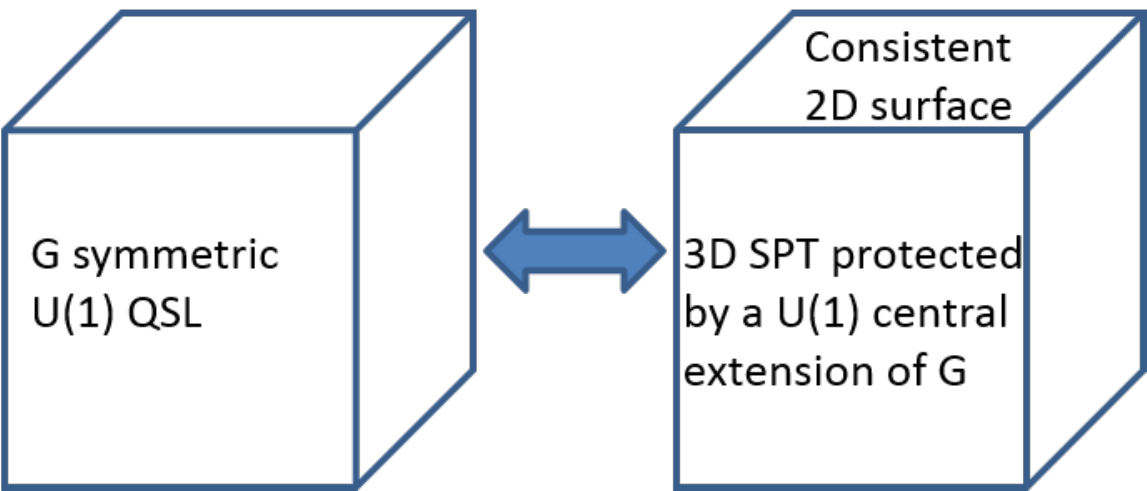}\\
  \caption{That the $G$ symmetric $U(1)$ quantum spin liquid is anomaly-free is equivalent to that it has a corresponding SPT, which is in turn equivalent to that this SPT can have a consistent (but possibly anomalous) 2D surface state.}\label{fig: QSL-SPT}
\end{figure}

In summary, a systematic physical way to examine whether a putative symmetry enriched $U(1)$ quantum spin liquid is anomalous is to check whether its corresponding SPT can have a legitimate surface state. If so, this spin liquid state is non-anomalous. Otherwise, it is anomalous. These relations is sketched in Figure \ref{fig: QSL-SPT}.

This method of anomaly detection applies to any symmetry enriched $U(1)$ quantum spin liquids, but for some particular cases, there are more physical ways of doing it by focusing on the spin liquid state itself, instead of its corresponding SPT. For example, we have used the $SO(3)$ monopole to detect the anomaly of some putative $SO(3)$ symmetric spin liquid states in Sec. \ref{sec: anomalies}. However, for some more subtle cases, the anomalies are examined by considering the corresponding SPTs. Some examples are given in Sec. \ref{sec: Z2Z2T}, where $Z_2\times\mc{T}$ symmetric states are discussed.

\subsection{Couple the spin liquids with SPTs}

The above two steps classify symmetry enriched $U(1)$ quantum spin liquids in terms of the properties of the bulk excitations. To complete the classification of the symmetry enriched $U(1)$ quantum spin liquids, one has to consider coupling these spin liquids and 3D bosonic SPTs with the same symmetry. In general, when an SPT is coupled with a $U(1)$ spin liquid, the result is still a $U(1)$ spin liquid with the bulk fractional excitations carrying the same symmetry fractionalization pattern, but the new state can have a different type of surface compared to the original one, due to the nontrivial surface of the SPT. Therefore, one has to check if this SPT can be ``absorbed" by the $U(1)$ spin liquid. Physically, this amounts to checking if the nontrivial surface of the SPT remains nontrivial if it is coupled with the bulk excitations in the spin liquid. Examples of such excercises are given in Sec. \ref{sec:QSL+SPT}.

\subsection{A formal framework}

We would like to close this section by briefly discussing a more formal approach to classify symmetry enriched $U(1)$ quantum spin liquids. In this formal approach, the problem amounts to classifying the action, or more precisely, the universal part of the partition function, of the $U(1)$ gauge theories. To encode the information about symmetries, in this action the $U(1)$ gauge field should be coupled to a background gauge field corresponding to the symmetry and a background spacetime metric.  If the global symmetry includes time reversal the equivalent of coupling a background gauge field is to place the theory on an unorientable space-time manifold.

Note that we are considering spin liquids that arise in a UV system made out of bosons. To impose this restriction directly in the low energy continuum theory we demand that the low energy theory can be consistently formulated on an arbitrary non-spin space-time manifold.  On an orientable manifold, this is achieved by requiring that the emergent gauge field be either an ordinary $U(1)$ gauge field (when the emergent electric charge $E$ is a boson) or that it is a Spin$_c$ connection\footnote{A Spin$_c$ connection differs from an ordinary $U(1)$ gauge field through a modification of its flux quantization condition: the curvature $F$ of a Spin$_c$ connection satisfies $\int \frac{F}{2\pi} = \int \frac{w_2^{TM}}{2}~~~(mod ~Z)$ on oriented 2-cycles where $w_2^{TM}$ is the second Stieffel-Whitney class of the tangle bundle of the manifold.  For more detail  see Refs. \onlinecite{Metlitski2015,Seiberg2016} and references therein.} (when $E$ is a fermion). On an unorientable manifold, there is a  generalization of a Spin$_c$ connection known as Pin$_{c \pm}$ connections - the $\pm$ sign correspond to the two possibilities that $E$ is non-Kramers or Kramers under time reversal (more detail is in Ref. \onlinecite{Metlitski2015}).  We will not make an explicit distinction in the schematic discussion below between these different kinds of $U(1)$ connections.

Denote the $U(1)$ gauge field by $a$, the gauge field corresponding to the global symmetry by $b$, and the background metric by $g$. In general, the action can be written in a form
\beq
\begin{split}
S[a;b,g]
=S_{\rm U(1)}[a]+S_{\rm SPT}[b,g]+S_{\rm mixed}[a;b,g]
\end{split}
\eeq
The first term contains the Maxwell action and the $\theta$ term of a $U(1)$ gauge field, and it is present in general for a $U(1)$ quantum spin liquid and are independent of symmetries. The third term, $S_{\rm SPT}[b,g]$, depends only on $b$ and $g$. This term physically describes 3D bosonic SPTs with the same symmetry as the $U(1)$ spin liquid, and adding it into the action means coupling a $U(1)$ spin liquid and a bosonic SPT with the same symmetry. As discussed before, this will potentially change the system into a different $U(1)$ spin liquid. In order to see if such an SPT can be ``absorbed" into a $U(1)$ spin liquid, one needs to check if the universal part of the partition function will change due to the presence of this term. The last term, $S_{\rm mixed}[a;b,g]$, only involves terms that couple $a$ with $b$ and/or $g$. This term encodes the information about symmetry fractionalization on the bulk excitations.

In general, such an action is constrained by gauge invariance. In addition, certain constraints on these fields may apply analogous to the modification of the flux quantization condition for Spin$_c$ connections when $E$ is a fermion. For example, for fractional topological paramagnets, there is a constraint on such fields given by (\ref{eq: FTP-constraint}). To classify symmetry enriched $U(1)$ quantum spin liquids, one can first write down all possible such actions and then classify the resulting universal part of the partition function. We leave this for future work.

\section{$U(1)$ quantum spin liquids enriched by $Z_2\times\mc{T}$ symmetry} \label{sec: Z2Z2T}

In this section we apply the above general framework to classify $U(1)$ quantum spin liquids enriched by $Z_2\times\mc{T}$ symmetry. This symmetry can be relevant for experimental candidates of quantum spin liquids made of non-Kramers quantum spins, {\it i.e.} for example, two-level systems made of $m_z=\pm 1$ states of a spin-1 atom, where time reversal flips $S_z$ and $Z_2$ acts as a $\pi$ spin rotation around the $x$ axis. Below we will first list all putative states, including the anomalous ones. Then we will examine the anomalies of these states. We will leave the problem of coupling these spin liquids with SPTs for future work.

It turns out there are two types of $Z_2$ actions that deserve separate discussions. In the first type, the $Z_2$ symmetry does not change one type of fractional excitation to another. More precisely, the electric charge and magnetic monopole will both retain their characters under this type of $Z_2$ action. In the second type, the $Z_2$ symmetry changes the fractional excitations. In particular, it can change the electric charge into the anti-electric charge, and at the same time change the magnetic monopole into the anti-magnetic monopole. This type of $Z_2$ action is physically a charge conjugation. One may wonder whether it is possible to change an electric charge into a magnetic monopole, but Ref. \onlinecite{KravecMcGreevy2013} pointed out this is impossible in a strictly 3D system.

Below we will discuss these two types of $Z_2$ actions in turn.

\subsection{$Z_2$ not acting as a charge conjugation} \label{subsec: Z2Z2T-simple}

We start from the case where $Z_2$ does not act as a charge conjugation, that is, it does not change a type of fractional excitation to another type.

We will begin with the simpler case that has $\theta=0$. In this case, to classify the quantum numbers of the electric charge, it is appropriate to look at the projective representations of $Z_2\times\mc{T}$, which are classified by $\mathbb{Z}_2^2$, where the nontrivial projective representations can be viewed as being a Kramers doublet under the original time reversal and/or under a new anti-unitary symmetry $\mc{T}'$, whose generator is the product of the generator of $Z_2$ and the generator of $\mc{T}$. Although it is not meaningful to talk about whether the magnetic monopole is a Kramers singlet or doublet under $\mc{T}$ or $\mc{T}'$, there are still two types of quantum numbers of the magnetic monopole under $Z_2\times\mc{T}$: on the monopole the $Z_2$ and $\mc{T}$ can either commute or anti-commute.{\footnote{If $\mc{T}$ and $Z_2$ commute (anti-commute), $\mc{T}'$ defined above will also commute (anti-commute) with $Z_2$.}} This relation between $Z_2$ and $\mc{T}$ is gauge invariant for the monopole, but not gauge invariant for the charge.

Therefore, we can make a list of putative $U(1)$ quantum spin liquids with this type of symmetry, and there are $3\times 2^2\times 2=24$ of them, as listed in Table \ref{table: main-text-Z2Z2T-simple Z2-theta=0-non-anomalous} and Table \ref{table: main-text-Z2Z2T-simple Z2-theta=0-anomalous}. It turns out the 15 states in Table \ref{table: main-text-Z2Z2T-simple Z2-theta=0-non-anomalous} are anomaly-free, and the 9 states in Table \ref{table: main-text-Z2Z2T-simple Z2-theta=0-anomalous} are anomalous, which can be grouped into three anomaly classes. We will give the construction of the non-anomalous states in Appendix \ref{app: Z2Z2T}. Later in Sec. \ref{subsec: Z2Z2T-anomaly-strategy}, we will give the strategy to show the anomalies of the states in Table \ref{table: main-text-Z2Z2T-simple Z2-theta=0-anomalous}, and we will finish the arguments for this anomaly-detection in Appendix \ref{app: Z2Z2T}.

\begin{table}
\center
\begin{tabular}{|c|c|c|c|}
\hline
& $T^2_E$ & $T'^2_E$ & $[\mc{T},Z_2]_M$\\
\hline
$E_bM_b$ & 1 & 1 & +\\
\hline
$E_{bT}M_b$ & $-1$ & $1$ & +\\
\hline
$E_{bT'}M_b$ & $1$ & $-1$ & +\\
\hline
$E_{bTT'}M_b$ & $-1$ & $-1$ & +\\
\hline
$E_bM_{b-}$ & $1$ & $1$ & $-$\\
\hline
$E_fM_b$ & $1$ & $1$ & +\\
\hline
$E_{fT}M_b$ & $-1$ & $1$ & +\\
\hline
$E_{fT'}M_b$ & $1$ & $-1$ & +\\
\hline
$E_{fTT'}M_b$ & $1$ & $-1$ & +\\
\hline
$E_{fTT'}M_{b-}$ & $-1$ & $-1$ & $-$\\
\hline
$E_bM_f$ & $1$ & $1$ & $+$\\
\hline
$E_{bT}M_f$ & $-1$ & $1$ & $+$\\
\hline
$E_{bT'}M_f$ & $1$ & $-1$ & $+$\\
\hline
$E_{bTT'}M_f$ & $-1$ & $-1$ & $+$\\
\hline
$E_bM_{f-}$ & $1$ & $1$ & $-$\\
\hline
\end{tabular}
\caption{List of non-anomalous $Z_2\times\mc{T}$ symmetric $U(1)$ quantum spin liquids that have $\theta=0$ and have $Z_2$ not acting as a charge conjugation. All these states are anomaly-free. $T_E^2=1$ ($T^2_E=-1$) represents the case where $E$ is a Kramers singlet (doublet) under $\mc{T}$. $T'^2_E=1$ ($T'^2_E=-1$) represents the case where $E$ is a Kramers singlet (doublet) under $\mc{T}'$. $[\mc{T},Z_2]_M=+$ ($[\mc{T},Z_2]_M=-$) represents the case where $Z_2$ and $\mc{T}$ commute (anti-commute) on $M$.} \label{table: main-text-Z2Z2T-simple Z2-theta=0-non-anomalous}
\end{table}

\begin{table}
\center
\begin{tabular}{|c|c|c|c|c|}
\hline
& $T^2_E$ & $T'^2_E$ & $[\mc{T},Z_2]_M$ & anomaly class\\
\hline
$E_{bT'}M_{f-}$ & $1$ & $-1$ & $-$ & class a\\
\hline
$E_{fT}M_{b-}$ & $-1$ & $1$ & $-$ & class a\\
\hline
$E_{bT'}M_{b-}$ & $1$ & $-1$ & $-$ & class a\\
\hline
$E_{fT'}M_{b-}$ & $1$ & $-1$ & $-$ & class b\\
\hline
$E_{bT}M_{f-}$ & $-1$ & $1$ & $-$ & class b\\
\hline
$E_{bT}M_{b-}$ & $-1$ & $1$ & $-$ & class b\\
\hline
$E_{bTT'}M_{f-}$ & $-1$ & $-1$ & $-$ & class c\\
\hline
$E_{bTT'}M_{b-}$ & $-1$ & $-1$ & $-$ & class c\\
\hline
$E_fM_{b-}$ & $1$ & $1$ & $-$ & class c\\
\hline
\end{tabular}
\caption{List of anomalous $Z_2\times\mc{T}$ symmetric $U(1)$ quantum spin liquids that have $\theta=0$ and have $Z_2$ not acting as a charge conjugation. All these states are anomaly-free. $T_E^2=1$ ($T^2_E=-1$) represents the case where $E$ is a Kramers singlet (doublet) under $\mc{T}$. $T'^2_E=1$ ($T'^2_E=-1$) represents the case where $E$ is a Kramers singlet (doublet) under $\mc{T}'$. $[\mc{T},Z_2]_M=+$ ($[\mc{T},Z_2]_M=-$) represents the case where $Z_2$ and $\mc{T}$ commute (anti-commute) on $M$. The last column indicates the anomaly classes.} \label{table: main-text-Z2Z2T-simple Z2-theta=0-anomalous}
\end{table}

Before moving to the case with $\theta=\pi$, we note that one point deserves immediate clarification. That is, one may wonder, for example, whether $E_{bT}M_b$ and $E_{bT'}M_b$ are truly distinct, since they are related to each other by relabelling $\mc{T}\leftrightarrow\mc{T}'$ and $E\leftrightarrow M$. At the first glance, these two states indeed seem to have identical physical properties when examined on their own. However, once the definitions of $\mc{T}$ and $\mc{T}'$ are fixed, these states are distinct. One physical way to see this is to consider the two states at the same time, clearly without breaking either $\mc{T}$ or $\mc{T}'$, one state cannot be connected to another without encountering a phase transition. Therefore, all these 24 states are truly distinct.

Now we turn to the states with $\theta=\pi$. In this case, the quantum number of the $\left(\half,1\right)$ dyon determines the quantum numbers of all other dyons. However, the $\left(\half,1\right)$ dyon does not have any projective representation of the $Z_2\times\mc{T}$ symmetry, so there is only one state: $(E_{fTT'}M_f)_\theta$, as described in Table \ref{table:Z2Z2T-simple Z2-theta=pi}. The electric charge has to be Kramers doublet under both $\mc{T}$ and $\mc{T}'$, because it is a bound state of the $\left(\half,1\right)$ and $\left(\half,-1\right)$ dyons, which have $\pi$ mutual braiding and are exchanged under both $\mc{T}$ and $\mc{T}'$. Naively, the $M$ particle (the $(0,2)$ dyon in this context) can either have $Z_2$ and $\mc{T}$ commuting or anti-commuting. But it turns out the latter possibility can be ruled out, as shown in Appendix \ref{app: Z2Z2T-simple}. This $(E_{fTT'}M_{f})_\theta$ state can be viewed as a descendant of the $SO(3)\times\mc{T}$ symmetric $(E_{fT}M_f)_\theta$, so it is anomaly-free.

\begin{table} [h!]
\begin{tabular}{|c|c|c|c|}
\hline
& $T^2_E$ & $T'^2_E$ & $[\mc{T},Z_2]_M$\\
\hline
$(E_{fTT'}M_{f})_\theta$ & $-1$ & $-1$ & $+$\\
\hline
\end{tabular}
\caption{The $Z_2\times\mc{T}$ symmetric $U(1)$ quantum spin liquids that have $\theta=\pi$ and have $Z_2$ not acting as a charge conjugation. This state is anomaly-free. $T_E^2=1$ ($T^2_E=-1$) represents the case where $E$ is a Kramers singlet (doublet) under $\mc{T}$. $T'^2_E=1$ ($T'^2_E=-1$) represents the case where $E$ is a Kramers singlet (doublet) under $\mc{T}'$. $[\mc{T},Z_2]_M=+$ ($[\mc{T},Z_2]_M=-$) represents the case where $Z_2$ and $\mc{T}$ commute (anti-commute) on $M$.} \label{table:Z2Z2T-simple Z2-theta=pi}
\end{table}

In summary, if $Z_2$ does not change one type of fractional excitation into another type, there are 16 distinct anomaly-free $Z_2\times\mc{T}$ symmetric $U(1)$ quantum spin liquids.

\subsection{$Z_2$ acting as a charge conjugation}

Now we turn to the more complicated case where the $Z_2$ symmetry acts as a charge conjugation. Let us first pause to lay out the principle of organizing these states. Let us focus on the case with $\theta=0$ for the moment. In this case, it is meaningful to discuss whether $E$ is a Kramers doublet under the original time reversal $\mc{T}$, and whether $M$ is a Kramers doublet under $\mc{T}'$. Also, notice now it is also meaningful to ask whether $Z_2$ squares to $+1$ or $-1$ for both $E$ and $M$ (see Appendix~\ref{app: remarks on TR} for more details). We will use $(\cdots)_-$ to indicate that $Z_2$ acts as a charge conjugation, and a subscript $Z$ to represent that certain excitation has $Z_2$ squaring to $-1$. For example, $(E_{fT}M_{bT'Z})_-$ means $Z_2$ flips both the electric charge and magnetic charge, and $E$ is a fermionic Kramers doublet under $\mc{T}$, while $M$ is a boson where $Z_2$ squares to $-1$, and $M$ is also a Kramers doublet under $\mc{T}'$.

With this notation, we can list all $3\times 2^2\times 2^2=48$ possible distinct states with $\theta=0$ and $Z_2$ acting as a charge conjugation, and they are shown in Table \ref{table: main-text-Z2Z2T-conjugation Z2-theta=0-nonanomalous} and Table \ref{table:Z2Z2T-conjugation Z2-theta=0-anomalous}.

\begin{table} [h!]
\begin{tabular}{|c|c|c|c|c|}
\hline
& $T^2_E$ & $Z_{E}^2$ & $T'^2_M$ & $Z_M^2$\\
\hline
$(E_bM_b)_-$ & $1$ & $1$ & $1$ & $1$\\
\hline
$(E_{bZ}M_b)_-$ & $1$ & $-1$ & $1$ & $1$\\
\hline
$(E_{bT}M_b)_-$ & $-1$ & $1$ & $1$ & $1$\\
\hline
$(E_{bTZ}M_b)_-$ & $-1$ & $-1$ & $1$ & $1$\\
\hline
$(E_bM_{bZ})_-$ & $1$ & $1$ & $1$ & $-1$\\
\hline
$(E_bM_{bT'})_-$ & $1$ & $1$ & $-1$ & $1$\\
\hline
$(E_bM_{bT'Z})_-$ & $1$ & $1$ & $-1$ & $-1$\\
\hline
$(E_{f}M_b)_-$ & $1$ & $1$ & $1$ & $1$\\
\hline
$(E_{fZ}M_b)_-$ & $1$ & $-1$ & $1$ & $1$\\
\hline
$(E_{fT}M_b)_-$ & $-1$ & $1$ & $1$ & $1$\\
\hline
$(E_{fTZ}M_b)_-$ & $-1$ & $-1$ & $1$ & $1$\\
\hline
$(E_bM_f)_-$ & $1$ & $1$ & $1$ & $1$\\
\hline
$(E_bM_{fZ})_-$ & $1$ & $1$ & $1$ & $-1$\\
\hline
$(E_bM_{fT'})_-$ & $1$ & $1$ & $-1$ & $1$\\
\hline
$(E_bM_{fT'Z})_-$ & $1$ & $1$ & $-1$ & $-1$\\
\hline
$(E_{fT}M_{bT'})_-$ & $-1$ & $1$ & $-1$ & $1$\\
\hline
$(E_{bT}M_{fT'})_-$ & $-1$ & $1$ & $-1$ & $1$\\
\hline
$(E_{fZ}M_{bT'Z})_-$ & $1$ & $-1$ & $-1$ & $-1$\\
\hline
$(E_{bTZ}M_{fZ})_-$ & $-1$ & $-1$ & $1$ & $-1$\\
\hline
$(E_{fTZ}M_{bZ})_-$ & $-1$ & $-1$ & $1$ & $-1$\\
\hline
$(E_{bZ}M_{fT'Z})_-$ & $1$ & $-1$ & $-1$ & $-1$\\
\hline
\end{tabular}
\caption{List of anomaly-free $Z_2\times\mc{T}$ symmetric $U(1)$ quantum spin liquids that have $\theta=0$ and have $Z_2$ acting as a charge conjugation. $T_E^2=1$ ($T^2_E=-1$) represents the case where $E$ is a Kramers singlet (doublet) under $\mc{T}$. $T'^2_M=1$ ($T'^2_M=-1$) represents the case where $M$ is a Kramers singlet (doublet) under $\mc{T}'$. $Z_{E,M}^2$ represents the result of acting the charge conjugation twice on $E$ and $M$, respectively.} \label{table: main-text-Z2Z2T-conjugation Z2-theta=0-nonanomalous}
\end{table}

\begin{table} [h!]
\begin{tabular}{|c|c|c|c|c|c|}
\hline
& $T^2_E$ & $Z_{E}^2$ & $T'^2_M$ & $Z_M^2$ & anomaly class\\
\hline
$(E_{bZ}M_{bZ})_-$ & $1$ & $-1$ & $1$ & $-1$ & class 1\\
\hline
$(E_{bTZ}M_{bT'Z})_-$ & $-1$ & $-1$ & $-1$ & $-1$ & class 1\\
\hline
$(E_{fT}M_{bZ})_-$ & $-1$ & $1$ & $1$ & $-1$ & class 1\\
\hline
$(E_{bZ}M_{fT'})_-$ & $1$ & $-1$ & $-1$ & $1$ & class 1\\
\hline
$(E_{fT}M_{bT'Z})_-$ & $-1$ & $1$ & $-1$ & $-1$ & class 1\\
\hline
$(E_{bTZ}M_{fT'})_-$ & $-1$ & $-1$ & $-1$ & $1$ & class 1\\
\hline
$(E_{bTZ}M_{bZ})_-$ & $-1$ & $-1$ & $1$ & $-1$ & class 2\\
\hline
$(E_{f}M_{bZ})_-$ & $1$ & $1$ & $1$ & $-1$ & class 2\\
\hline
$(E_{bTZ}M_f)_-$ & $-1$ & $-1$ & $1$ & $1$ & class 2\\
\hline
$(E_{bZ}M_{bT'Z})_-$ & $1$ & $-1$ & $-1$ & $-1$ & class 3\\
\hline
$(E_{bZ}M_f)_-$ & $1$ & $-1$ & $1$ & $1$ & class 3\\
\hline
$(E_fM_{bT'Z})_-$ & $1$ & $1$ & $-1$ & $-1$ & class 3\\
\hline
$(E_{bZ}M_{bT'})_-$ & $1$ & $-1$ & $-1$ & $1$ & class 4\\
\hline
$(E_{fTZ}M_{bT'})_-$ & $-1$ & $-1$ & $-1$ & $1$ & class 4\\
\hline
$(E_{fTZ}M_{bT'Z})_-$ & $-1$ & $-1$ & $-1$ & $-1$ & class 4\\
\hline
$(E_{bT}M_{bT'Z})_-$ & $-1$ & $1$ & $-1$ & $-1$ & class 4\\
\hline
$(E_{bT}M_{fZ})_-$ & $-1$ & $1$ & $1$ & $-1$ & class 4\\
\hline
$(E_{bZ}M_{fZ})_-$ & $1$ & $-1$ & $1$ & $-1$ & class 4\\
\hline
$(E_{bT}M_{bZ})_-$ & $-1$ & $1$ & $1$ & $-1$ & class 5\\
\hline
$(E_{bT}M_{fT'Z})_-$ & $-1$ & $1$ & $-1$ & $-1$ & class 5\\
\hline
$(E_{bTZ}M_{fT'Z})_-$ & $-1$ & $-1$ & $-1$ & $-1$ & class 5\\
\hline
$(E_{bTZ}M_{bT'})_-$ & $-1$ & $-1$ & $-1$ & $1$ & class 5\\
\hline
$(E_{fZ}M_{bT'})_-$ & $1$ & $-1$ & $-1$ & $1$ & class 5\\
\hline
$(E_{fZ}M_{bZ})_-$ & $1$ & $-1$ & $1$ & $-1$ & class 5\\
\hline
$(E_{bT}M_{bT'})_-$ & $-1$ & $1$ & $-1$ & $1$ & class 6\\
\hline
$(E_{bT}M_f)_-$ & $-1$ & $1$ & $1$ & $1$ & class 6\\
\hline
$(E_fM_{bT'})_-$ & $1$ & $1$ & $-1$ & $1$ & class 6\\
\hline
\end{tabular}
\caption{List of anomalous $Z_2\times\mc{T}$ symmetric $U(1)$ quantum spin liquids that have $\theta=0$ and have $Z_2$ acting as a charge conjugation at $\theta=\pi$. $T_E^2=1$ ($T^2_E=-1$) represents the case where $E$ is a Kramers singlet (doublet) under $\mc{T}$. $T'^2_M=1$ ($T'^2_M=-1$) represents the case where $M$ is a Kramers singlet (doublet) under $\mc{T}'$. $Z_{E,M}^2$ represents the result of acting the charge conjugation twice on $E$ and $M$, respectively. The last column lists the anomaly classes.} \label{table:Z2Z2T-conjugation Z2-theta=0-anomalous}
\end{table}

Similarly, for states with $\theta=\pi$ and $Z_2$ acting as a charge conjugation, there are only two states: $(E_{fT}M_{fT'})_{\theta-}$ and $(E_{fT}M_{fT'})_{\theta-Z}$. In both states, $Z_2$ takes the $(\frac{1}{2},1)$ dyon into the $(-\frac{1}{2},-1)$ dyon. Because time reversal takes $(\frac{1}{2},1)$ into $(\frac{1}{2},-1)$, then we know $\mc{T}'$ takes $(\frac{1}{2},1)$ to $(-\frac{1}{2},1)$. This implies that $M$, the bound state of $(\frac{1}{2},1)$ and $(-\frac{1}{2},1)$, is a Kramers doublet under $\mc{T}'$. \cite{Wang2014, Wang2016} The difference in these two states is that $Z_2$ squares to $+1$ ($-1$) on the $\left(\frac{1}{2},1\right)$ dyon in the former (latter). In fact, the former state is just the time reversal symmetric $(E_{fT}M_f)_\theta$ further equipped with a charge conjugation symmetry, so it must be anomaly-free.

\begin{table} [h!]
\begin{tabular}{|c|c|c|c|c|c|c|}
\hline
& $T^2_E$ & $Z_{E}^2$ & $T'^2_M$ & $Z_M^2$ & $Z_D^2$ & comments\\
\hline
$(E_{fT}M_{fT'})_{\theta-}$ & $-1$ & $1$ & $-1$ & $1$ & $1$ &\\
\hline
$(E_{fT}M_{fT'})_{\theta-Z}$ & $-1$ & $1$ & $-1$ & $1$ & $-1$ & anomalous, class 1\\
\hline
\end{tabular}
\caption{List of $Z_2\times\mc{T}$ symmetric $U(1)$ quantum spin liquids that have $\theta=\pi$ and have $Z_2$ acting as a charge conjugation. $T_E^2=1$ ($T^2_E=-1$) represents the case where $E$ is a Kramers singlet (doublet) under $\mc{T}$. $T'^2_M=1$ ($T'^2_M=-1$) represents the case where $M$ is a Kramers singlet (doublet) under $\mc{T}'$. $Z_{E,M,D}^2$ represents the result of acting the charge conjugation twice on $E$, $M$ and the $\left(\half,1\right)$ dyon, respectively.} \label{table:Z2Z2T-conjugation Z2-theta=pi}
\end{table}

So without examining anomalies, there are in total 50 possible distinct $Z_2\times\mc{T}$ symmetric $U(1)$ quantum spin liquids where $Z_2$ acts as a charge conjugation. It turns out that, together with the anomaly-free $(E_{fT}M_{fT'})_{\theta-}$, 22 of these states are free of anomaly. The other 28 states are all anomalous, and there are 6 anomaly classes. The strategy to show the anomalies will be given in Sec. \ref{subsec: Z2Z2T-anomaly-strategy}, and the arguments for this anomaly-detection will be completed in Appendix \ref{app: Z2Z2T}.

Therefore, combined with the 16 states where $Z_2$ does not permute any excitation, there are in total 38 distinct non-anomalous $Z_2\times\mc{T}$ symmetric $U(1)$ spin liquid states, and they can be found in Table \ref{table: main-text-Z2Z2T-simple Z2-theta=0-non-anomalous}, Table \ref{table:Z2Z2T-simple Z2-theta=pi}, Table \ref{table: main-text-Z2Z2T-conjugation Z2-theta=0-nonanomalous} and Table \ref{table:Z2Z2T-conjugation Z2-theta=pi}.

We note that models that discuss $Z_2\times\mc{T}$ symmetric $U(1)$ quantum spin liquids have been proposed in the literature, and a prototypical set includes but is not limited to Ref. \onlinecite{SavaryBalents2011, SavaryBalents2016, RoechnerBalentsSchmidt2016}. In these models, the $Z_2\times\mc{T}$ symmetric $U(1)$ quantum spin liquid states are $(E_bM_b)_-$.

\subsection{Strategy of anomaly-detection} \label{subsec: Z2Z2T-anomaly-strategy}

In this subsection we lay out the strategy to show the anomaly of the other 36 states. It turns out to be easier to first show that $(E_{bZ}M_{bZ})_-$ is anomalous with the $Z_2$ symmetry (independent of time reversal), and this will be done later in this section. This immediately implies that $(E_fM_{bZ})_-$ and $(E_{bZ}M_f)_-$ are also anomalous with $Z_2$ symmetry, because these states can be related to $(E_{bZ}M_{bZ})_-$ by tuning $\theta$ by $2\pi$. It also immediately implies that $(E_{bTZ}M_{bZ})_-$, $(E_{bZ}M_{bT'Z})_-$, $(E_{bTZ}M_{bT'Z})_-$, $(E_{fT}M_{bZ})_-$, $(E_{bZ}M_{fT'})_-$ are anomalous with $Z_2\times\mc{T}$ symmetry, because breaking $\mc{T}$ will make them one of $(E_{bZ}M_{bZ})_-$, $(E_fM_{bZ})_-$ and $(E_{bZ}M_f)_-$. Furthermore, this means $(E_{fT}M_{fT'})_{\theta-Z}$ is anomalous, because even if the time reversal symmetry is broken, this state is smoothly connected to the anomalous $(E_{bZ}M_{bZ})_-$.

Next, using a generalization of the method for showing the anomaly of $(E_{bZ}M_{bZ})_-$, we show $(E_{bT}M_{bT'})_-$ and $E_{bTT'}M_{b-}$ are anomalous with $Z_2\times\mc{T}$ symmetry in Appendix \ref{app: Z2Z2T}. It turns out this is enough to show the remaining states are all anomalous. More precisely,

\begin{itemize}
\item[1.]

showing that $(E_{bT}M_{bT'})_-$ is anomalous is sufficient to show that the other entries in Table \ref{table:Z2Z2T-conjugation Z2-theta=0-anomalous} are anomalous.

\item[2.]

showing that $E_{bTT'}M_{b-}$ is anomalous is sufficient to show that the rest entries in Table \ref{table: main-text-Z2Z2T-simple Z2-theta=0-anomalous} are anomalous.

\end{itemize}

To see the first claim, let us consider $(E_{bT}M_{bZ})_-$ and $(E_{bZ}M_{bT'})_-$. These two states must be simultaneously anomalous or anomaly-free, because they are related to each other by the relabelling $\mc{T}\leftrightarrow\mc{T}'$ and $E\leftrightarrow M$. Suppose they are anomaly-free, then by combining them with the states that will be constructed in Appendix \ref{app: Z2Z2T-free-fermion}, we will get $(E_{bT}M_{bT'})_-$, which is in contradiction with that $(E_{bT}M_{bT'})_-$ is anomalous. This means if $(E_{bT}M_{bT'})_-$ is anomalous, then $(E_{bT}M_{bZ})_-$ and $(E_{bZ}M_{bT'})_-$ will also be anomalous. Combining these three anomalous states with the anomaly-free states constructed in Appendix \ref{app: Z2Z2T-free-fermion}, one can show all other entries in Table \ref{table:Z2Z2T-conjugation Z2-theta=0-anomalous} are also anomalous.

To see the second claim, consider $E_{fT}M_{b-}$ and $E_{fT'}M_{b-}$. These two states must also simultaneously be anomalous or anomaly-free, because they are related to each other by the relabelling $\mc{T}\leftrightarrow\mc{T}'$ and $E\leftrightarrow M$. Suppose they are anomaly-free. By combining them and the anomaly-free states constructed in Appendix \ref{app: Z2Z2T-free-fermion}, we can get $E_{bTT'}M_{b-}$, which contradicts that $E_{bTT'}M_{b-}$ is anomalous. This means if $E_{bTT'}M_{b-}$ is anomalous, then $E_{fT}M_{b-}$ and $E_{fT'}M_{b-}$ must also be anomalous. Combining these anomalous states with the anomaly-free states constructed in Appendix \ref{app: Z2Z2T-free-fermion}, one can show all other entries in Table \ref{table: main-text-Z2Z2T-simple Z2-theta=0-anomalous} are also anomalous.

\subsubsection*{Anomaly of $(E_{bZ}M_{bZ})_-$}

In the spirit of Sec. \ref{subsec: general-anomaly-detection}, here we will show that $(E_{bZ}M_{bZ})_-$ is anomalous with a $Z_2$ charge-conjugation symmetry (independent of time reversal), by showing its corresponding SPT has an inconsistent surface.

To show the anomaly of $(E_{bZ}M_{bZ})_-$, we will consider from the perspective of $E_{bZ}$, and suppose there is an SPT made of $E_{bZ}$ that after gauging becomes $(E_{bZ}M_{bZ})_-$. On the surface of this SPT, we first condense the bound state of two $E_{bZ}$, and this makes the surface a superfluid with a $Z_4$ symmetry. There will be various vortices, and the $4\pi$ vortex is the minimal trivial boson. So we can then condense the $4\pi$ vortices to restore the $U(1)$ symmetry, and this gives a symmetric gapped surface state where the $U(1)$ charge of the excitations is quantized in units of $1/2$.

The particle contents of the surface can be written as $\{1,M_{bZ},X,N_I\}\times\{1,E_{bZ}\}$, where $M_{bZ}$ is the remnant of the strength-1 monopole, $X$ carries half charge under $U(1)$, and $N_I$'s are neutral. In general, there can be many flavors of $N_I$, but only the case with a single $X$ needs to be considered, because other $X$'s can be related to a single one by attaching certain $N_I$. Notice in this notation the inverses and bound states of these excitations are understood to be implicitly displayed. We would like to check whether such a surface is consistent. That is, it has consistent braiding, fusion and symmetry transformation rules.

As for braiding, we know $E_{bZ}$ is local, $M_{bZ}$ and $X$ have mutual $\pi$ statistics, and $M_{bZ}$ has no mutual statistics with $N_I$. $X$ and $N_I$ can have complicated braiding though, and it can even be non-Abelian.

As for fusion, we have
\beq \label{eq:fusion-1}
M_{bZ}\times M_{bZ}=1
\eeq
\beq \label{eq:fusion-2}
X\times X=E_{bZ}+ E_{bZ}N_1+ E_{bZ}N_2M_{bZ}
\eeq
and
\beq \label{eq:fusion-3}
N_I\times N_J=N_k+ M_{bZ}N_k
\eeq
(\ref{eq:fusion-1}) comes from that this surface is obtained by condensing $M_{bZ}^2$, and (\ref{eq:fusion-2}) and (\ref{eq:fusion-3}) are obtained under the constraint due to charge conservation. Notice in (\ref{eq:fusion-2}) the fusion product cannot be $E_{bZ}M_{bZ}$, because this will be inconsistent with the general condition that a particle and its anti-particle should have the same topological spin. Also, all potential fusion multiplicities are suppressed, and they turn out to be unimportant for our discussion.

Now if we are willing to break the $U(1)$ symmetry on the surface by condensing $E_{bZ}M_{bZ}$,{\footnote{This is a boson in this surface topological order.}} $X$ will be confined and $N_I$'s will remain, and we will be left with $\{1,M_{bZ},N_I\}$ that has a $Z_4$ symmetry. Notice that $\{1,M_{bZ},N_I\}$ is closed under fusion and braiding, and it is known that in three dimensions there is no bosonic SPT protected by $Z_4$ symmetry. This means $N_I$ can be further confined (without breaking the $Z_4$ symmetry), and we are left with $\{1,M_{bZ}\}$. In other words, $N_I$'s can be viewed as emergent particles of a system made of $M_{bZ}$ in the presence of the $Z_4$ symmetry but in the absence of the $U(1)$ symmetry. However, because neither $M_{bZ}$ nor $N_I$ carries a $U(1)$ charge, even in the presence of $U(1)$ symmetry $N_I$ can still be viewed as emergent particles of a system made of $M_{bZ}$.

So we can get rid of $N_I$ and be left with $\{1,M_{bZ},X\}\times\{1,E_{bZ}\}$. Now the fusion of $X$ must be
\beq
X\times X=E_{bZ}
\eeq
The only possible consistent topological order of this state is a $Z_2$ topological order (or its twisted version, the double-semion theory).

Let us turn to symmetry assignment, and we will particularly consider how charge-conjugation acts on $X$. Notice when defining the charge-conjugation action on $X$, there is an ambiguity due to our freedom to multiply it by a gauge transformation. But because this topological order is a $Z_2$ gauge theory, the action of the global charge-conjugation symmetry twice should have an unambiguous result on $X$. So in order to be consistent with the above fusion rule, $X$ must go to $\pm iX$ upon acting with charge-conjugation twice. Below we show this is impossible.

Suppose the action of charge-conjugation on $X$ is implemented by a generic matrix $C$
\beq
X_i\rightarrow C_{ij}X_j^\dag,
\quad
X_i^\dag\rightarrow C_{ij}^*X_j
\eeq
Notice the indices label different components of $X$ that differ by some local operations. This implies that acting charge-conjugation twice on $X$ gives
\beq
X_i\rightarrow (CC^*)_{ij}X_j
\eeq

Now consider the operator $X_iM_{ij}X_j$ with an arbitrary matrix $M$, which is a charge-1 boson, so the charge-conjugation acting on it twice gives $-1$. This requires
\beq
(CC^*)^TM(CC^*)=-M
\eeq
Because $M$ is arbitrary, this is possible only if $CC^*=\pm i$, which confirms the previous statement that $X\rightarrow\pm iX$ upon acted by $Z_2$ twice. However, no matrix $C$ can possibly satisfy $CC^*=\pm i$. To see this, suppose $CC^*=\pm i$, then $(CC^*)^2=-1$. On the other hand, $C^*C=\mp i$ and $(CC^*)^2=CC^*CC^*=C(C^*C)C^*=1$, which contradicts with the previous result.

The above contradiction shows that there cannot even be any $X$. So the surface is just $\{1,E_{bZ},M_{bZ}\}$, where everything is local. This means that there is a charge neutral excitation that has $Z_2$ squaring to $-1$, which contradicts the original assumption. Therefore, this SPT cannot exist, and furthermore, $(E_{bZ}M_{bZ})_-$ is anomalous with a $Z_2$ charge-conjugation symmetry. Notice from the above argument we see the anomaly of $(E_{bZ}M_{bZ})_-$ is independent of time reversal.

Note this argument can be easily modified to show that $E_{b\frac{1}{2}}M_{b\frac{1}{2}}$ is anomalous with $SO(3)$ symmetry, by changing every $Z_2$ symmetry by $SO(3)$ symmetry, and changing every excitation with charge-conjugation squaring to $-1$ by an excitation with spin-1/2. This is of course consistent with our conclusion from Sec. \ref{sec: anomalies}, where we have used the $SO(3)$ monopole to show the anomaly.

\subsection{Anomaly classes} \label{subsec: Z2Z2T-anomaly-class}

Before finishing this section, we make some brief remarks on the anomaly classes of these anomalous spin liquid states. Here by an anomaly class, we mean a group of anomalous states which can be turned into each other by coupling it with a state that is anomaly-free. Because the anomalous spin liquid states can in principle be realized on the surface of some $4+1$-d $Z_2\times\mc{T}$ symmetric bosonic SPTs, analysing the anomaly classes of these 37 anomalous spin liquid states gives some information on the properties of these SPTs.

We first discuss the anomalous spin liquid states with $Z_2$ acting as a charge conjugation. It is straightforward to check that within each of the following 6 groups of anomalous states, all states have the same anomaly:
\begin{itemize}

\item[1.] $(E_{bZ}M_{bZ})_-$, $(E_{bTZ}M_{bT'Z})_-$, $(E_{fT}M_{bZ})_-$, $(E_{bZ}M_{fT'})_-$, $(E_{fT}M_{bT'Z})_-$, $(E_{bTZ}M_{fT'})_-$, $(E_{fT}M_{fT'})_{\theta-Z}$.

\item[2.] $(E_{bTZ}M_{bZ})_-$, $(E_fM_{bZ})_-$, $(E_{bTZ}M_f)_-$.

\item[3.] $(E_{bZ}M_{bT'Z})_-$, $(E_{bZ}M_f)_-$, $(E_fM_{bT'Z})_-$.

\item[4.] $(E_{bZ}M_{bT'})_-$, $(E_{fTZ}M_{bT'})_-$, $(E_{fTZ}M_{bT'Z})_-$, $(E_{bT}M_{bT'Z})_-$, $(E_{bT}M_{fZ})_-$, $(E_{bZ}M_{fZ})_-$.

\item[5.] $(E_{bT}M_{bZ})_-$, $(E_{bT}M_{fT'Z})_-$, $(E_{bTZ}M_{fT'Z})_-$, $(E_{bTZ}M_{bT'})_-$, $(E_{fZ}M_{bT'})_-$, $(E_{fZ}M_{bZ})_-$.

\item[6.] $(E_{bT}M_{bT'})_-$, $(E_{f}M_{bT'})_-$, $(E_{bT}M_f)_-$.

\end{itemize}

It is clear that combining group 1 and group 4 results in group 3, combining group 1 and group 5 results in group 2, and combining group 4 and group 5 results in group 6. This implies that the 4D bosonic SPTs with $Z_2\times\mc{T}$ symmetry at least have a classification of $\mathbb{Z}_2^3$. Group cohomology gives precisely the same classification,\cite{Chen2013} and our results suggest that the surface states of these SPTs can be the above anomalous $U(1)$ gauge theories.

Notice there is another SPT that is beyond group cohomology, and that SPT is protected purely by $Z_2$ symmetry.\cite{Wen2014,Bi2015a} One physical realization of the bulk of this SPT is to consider a decorated domain wall construction, where on each $Z_2$ domain wall we place an $efmf$ state. \cite{Wen2014} The surface properties of this beyond-group-cohomology SPT is unclear at this point, and it may be interesting to work it out.

Taking all these together, we propose that the complete classification of 4D bosonic SPTs with $Z_2\times\mc{T}$ symmetry is $\mathbb{Z}_2^4$. This agrees with the classification of 4D bosonic SPTs with $Z_2\times Z_2^P$ symmetry, where $Z_2^P$ is a reflection symmetry that results in a trivial action when acted twice.\cite{Cheng2017}

Next we discuss the anomalous spin liquid states with $Z_2$ not acting as a charge conjugation, whose anomaly classes can be organized as
\begin{itemize}

\item[a.]

$E_{bT'}M_{f-}$, $E_{fT}M_{b-}$, $E_{bT'}M_{b-}$.

\item[b.]

$E_{bT}M_{f-}$, $E_{fT'}M_{b-}$, $E_{bT}M_{b-}$.

\item[c.]

$E_{bTT'}M_{b-}$, $E_{bTT'}M_{f-}$, $E_fM_{b-}$.

\end{itemize}

First notice all these states are anomalous with the full $Z_2\times\mc{T}$ symmetry. If $\mc{T}$ is broken, all these states should be non-anomalous $Z_2$ symmetric states. Second, class a and class b differ by the relabelling $\mc{T}\leftrightarrow\mc{T}'$ and $E\leftrightarrow M$, and class c can be obtained by combining states in class a and class b. Notice before that class 2 and class 3 differ by this relabelling, so do class 4 and class 5, and that class 6 can be obtained by combining class 2 and class 3, or by combining class 4 and class 5. This suggests states in class c here are of the same anomaly class as states in class 6 as above. We do not attempt to completely settle down the relation among these anomaly classes in this paper.

\section{$U(1)$ quantum spin liquids enriched by some other symmetries}\label{sec: other symmeties}

In the spirit of the general framework in Sec. \ref{sec: general framework}, in this section we briefly discuss $U(1)$ quantum spin liquids with some other symmetries. Part of the motivation comes from the existing lattice models that realize $U(1)$ quantum spin liquids with $O(2)\times\mc{T}=(U(1)\rtimes Z_2)\times\mc{T}$ symmetry.\cite{Motrunich2002, HermeleFisherBalents2004, Motrunich2005,Levin2006} In all these models, the improper $Z_2$ rotation of the $O(2)$ symmetry acts as a charge conjugation. Ref. \onlinecite{Motrunich2002, HermeleFisherBalents2004, Motrunich2005} studied a couple of different lattice models that realize $(E_{b}M_{b\frac{1}{2}})_-$, where $\theta=0$ and both $E$ and $M$ are bosons, and $M$ carries half charge under the $U(1)$ subgroup of $O(2)$. Ref. \onlinecite{Levin2006} constructed two other models of $O(2)\times\mc{T}$ symmetric $U(1)$ quantum spin liquids, where one of them has a bosonic monopole and the other has a fermionic monopole. These two states are $(E_bM_b)_-$ and $(E_bM_f)_-$, respectively. Notice, for simplicity, in this section we will not consider the refined classification that considers coupling these spin liquids with SPTs.

\subsection{$SO(N)\times\mc{T}$ symmetry} \label{subsec: SO(N)Z2T}

We can generalize our classification of $SO(3)\times\mc{T}$ symmetric $U(1)$ quantum spin liquids to $SO(N)\times\mc{T}$ symmetric $U(1)$ quantum spin liquids, with the integer $N>3$. The projective representations of $SO(N)\times\mc{T}$ have the same classification as those of $SO(3)\times\mc{T}$: there is a spinor representation of $SO(N)$ and a Kramers doublet representation of time reversal. Therefore, the enumeration of states with this symmetry goes in a parallel way as those with $SO(3)\times\mc{T}$ symmetry, and all non-anomalous states with $SO(3)\times\mc{T}$ can be generalized to their $SO(N)\times\mc{T}$ analogs. Furthermore, $\pi_1(SO(N))=Z_2$ and the monopole structure of an $SO(N)$ gauge field is similar to that of an $SO(3)$ gauge field, so the generalization of the anomalous states in the $SO(3)\times\mc{T}$ case will still be anomalous. Therefore, we conclude that with $SO(N)\times\mc{T}$ symmetry there will also be 15 distinct non-anomalous $U(1)$ quantum spin liquids, and they have similar properties as those with only $SO(3)\times\mc{T}$ symmetry in terms of the bulk fractional excitations.

For the special case of $SO(2)\times\mc{T}\cong U(1)\times\mc{T}$, its projective representations on $E$ are classified by $\mathbb{Z}_2^2$. One of the nontrivial root projective representation corresponds to Kramers doublet of time reversal, while the other corresponds to half-charge of $SO(2)$, which is protected by time reversal here. Therefore, although $SO(2)$ has no projective representation on its own, the projective representations of $SO(2)\times\mc{T}$ on $E$ can still be viewed as descendants of those of $SO(3)\times\mc{T}$.

If $\theta=0$, there is no projective representation on $M$. So there are $3\times 2^2=12$ putative states with $\theta=0$, which can all be viewed as descendant states of $SO(3)\times\mc{T}$ symmetric states when the symmetry is reduced to $SO(2)\times\mc{T}$. Notice some distinct $SO(3)\times\mc{T}$ symmetric states have the same $SO(2)\times\mc{T}$ symmetric descendant, because there is no fractional quantum number on monopoles anymore. The descendants of the 15 anomaly-free states of course remain anomaly-free. By inspecting the anomaly classes of anomalous $SO(3)\times\mc{T}$ symmetric states listed in Sec. \ref{sec: anomalies}, we see in each anomaly class there is at least one state that has a trivial monopole when the symmetry is reduced to $U(1)\times\mc{T}$, which means all these anomalous states will become anomaly-free. So all these $12$ $SO(2)\times\mc{T}$ symmetric states with $\theta=0$ are anomaly-free.

If $\theta=\pi$, the properties of the $\left(\half,\pm1\right)$ dyons will determine the phase, which does not have any projective representation of this symmetry. So it contributes one anomaly-free state, which can be viewed as a descendant of the $SO(3)\times\mc{T}$ symmetric $(E_{fT}M_f)_\theta$ state when the symmetry is broken down to $U(1)\times\mc{T}$.

Therefore, there are in total $13$ distinct anomaly-free $U(1)$ quantum spin liquids with $U(1)\times\mc{T}$ symmetry.

\subsection{$SO(N)$ symmetry} \label{subsec: SO(N)}

In the following few subsections we will consider the case in the absence of time reversal symmetry. In this subsection we start by discussing $U(1)$ quantum spin liquids with only $SO(3)$ spin rotation symmetry, which can be realized in systems with a spin chirality term in the Hamiltonian.

As for the elementary excitations, we focus on $(1,0)$ and $(q_e,1)$, with $q_e$ a real number. It is not hard to see there must be bosons for some $q_e$, and we will consider such a bosonic $(q_e, 1)$ excitation. Due to the absence of time reversal symmetry, $\theta$ in (\ref{eq: gauge theory}) can be tuned continuously, so that the $(q_e,1)$ particle chosen above can be tuned to $(0,1)$ by Witten effect. That is to say, now we fix $(0,1)$ to be a boson. Similarly, we can make $(1,0)$ also a boson. Therefore, the statistics of the elementary excitations is irrelevant here due to the absence of time reversal symmetry.

Next we turn to their quantum numbers under symmetry. Due to the absence of time reversal symmetry, the distinction between electric charge and magnetic monopole also becomes irrelevant. By enumerating the quantum numbers of $E=(1,0)$ and $M=(0,1)$ under $SO(3)$, we can have $E_bM_b$, $E_{b\half}M_b$ and $E_{b\half}M_{b\half}$, and these exhaust all (including anomalous) $SO(3)$ symmetric $U(1)$ spin liquids. $E_bM_b$ and $E_{b\half}M_b$ are clearly not anomalous, and their description as a gauged SPT is similar to their cousins with $SO(3)\times\mc{T}$ symmetry. Also, as argued in Sec. \ref{sec: anomalies}, $E_{b\half}M_{b\half}$ is anomalous even with only $SO(3)$ symmetry.

Therefore, with only $SO(3)$ symmetry, there are only two (anomaly-free) distinct possible symmetry realizations in the $U(1)$ quantum spin liquids: $E_bM_b$ and $E_{b\half}M_b$. This concludes the classification of $U(1)$ quantum spin liquids with $SO(3)$ symmetry.

Similar reasoning as above can be applied to $U(1)$ quantum spin liquids with $SO(N)$ symmetry with $N>3$. First, all these $SO(N)$ groups have one nontrivial projective representation, the spinor representation. Also, as mentioned above, the monopole properties of an $SO(N)$ gauge field is similar to that of an $SO(3)$ gauge field. Therefore, the arguments for the enumeration of all the states, construction of the non-anomalous states and examination of the anomalous states are parallel to that of $SO(3)$, and gives only 2 distinct $SO(N)$ symmetric $U(1)$ quantum spin liquids.

For the special case of $SO(2)\cong U(1)$, because of the absence of any nontrivial projective representation of this symmetry, there will be only a single type of symmetric $U(1)$ quantum spin liquid.

\subsection{$Z_2$ symmetry}

For the case with $Z_2$ symmetry, as above, all states can be symmetrically tuned so that it has the $\theta=0$-type of charge-monopole lattice with both $E$ and $M$ bosonic. Also notice there is no projective representation of $Z_2$, so nontrivial states must have $Z_2$ acting as charge conjugation. Then there can be $E_bM_b$, $(E_bM_b)_-$, $(E_{bZ}M_b)_-$ and $(E_{bZ}M_{bZ})_-$. The first three states can clearly be realized, but as shown before, $(E_{bZ}M_{bZ})_-$ is anomalous with a $Z_2$ symmetry. So there are 3 distinct $Z_2$ symmetric $U(1)$ quantum spin liquids: $E_bM_b$, $(E_bM_b)_-$ and $(E_{bZ}M_b)_-$.

\subsection{$O(2)$ symmetry}

Similar considerations can be applied to the case with $O(2)$ symmetry. There will still be two spin liquid states where the improper $Z_2$ component does not act as a charge conjugation, and these are the descendants of $E_bM_b$ and $E_{b\frac{1}{2}}M_b$ with $SO(3)\times\mc{T}$ symmetry. As shown in Sec. \ref{sec: anomalies}, the descendant of $E_{b\frac{1}{2}}M_{b\frac{1}{2}}$ is still anomalous with $O(2)$ symmetry. Again, for a complete classification, states where $Z_2$ acts as a charge conjugation need to be taken into account. Unlike the case with $O(2)\times\mc{T}$ symmetry, the fractional excitations always have integer charges under the $U(1)$ subgroup of $O(2)$, if the improper $Z_2$ component acts as a charge conjugation. So these states include $(E_bM_b)_-$, $(E_{bZ}M_b)_-$, $(E_{bZ}M_{bZ})_-$, and the first two are anomaly-free, while the last one is anomalous. Therefore, there are in total 4 distinct non-anomalous $O(2)$ symmetric $U(1)$ quantum spin liquids: $E_bM_b$, $E_{b\frac{1}{2}}M_b$, $(E_{b}M_b)_-$ and $(E_{bZ}M_b)_-$.


\section{Discussion} \label{sec:discussion}

In this paper we have classified and characterized 3D symmetry enriched $U(1)$ quantum spin liquids. One of our focuses is on such spin liquids with time reversal and $SO(3)$ spin rotational symmetries. 26 states were enumerated based on the properties of the bulk fractional excitations, among which only 15 can be realized in 3D lattice systems. We explain in details how to view these quantum spin liquids as gauged version of some SPTs. The other 11 are shown to be anomalous, i.e. they cannot be realized in a 3D bosonic system with these symmetries. In Appendix \ref{app: 4d surface}, they are constructed on the surface of some 4D bosonic short-range entangled states.

The anomalies of the anomalous states become clear when the properties of the $SO(3)$ monopoles are examined. Although checking the topological defects of the gauge field that corresponds to certain symmetry has been widely applied to detect anomalies, to the best of our knowledge, the properties of $SO(3)$ monopoles have not been investigated in previous studies. We expect it to be helpful in studying other problems that involves $SO(3)$ symmetries.

When combined with bosonic SPTs with time reversal and $SO(3)$ spin rotational symmetry, we find a further refined classification which shows there are 168 different $U(1)$ quantum spin liquids.

After warming up with the example of $SO(3)\times\mc{T}$ symmetric $U(1)$ quantum spin liquids, we have described a general framework to classify such spin liquid states with a general symmetry. This approach is again physics-based, and it has the advantage of providing us with  intuition both on the classification and the physical characterization. However,  it is not always easy to implement this framework,  and so it is desirable to find a simpler systematic way to do the classification.  The field theoretic formal approach discussed in Sec. \ref{sec: general framework} may be potentially helpful.  Another possibly helpful formal approach is to generalize the categorical theory that is used to study $2d$ SETs to $U(1)$ quantum spin liquids.\cite{Barkeshli2014,Lan2016} This may be  possible because in both cases the excitations are all particle like, although there are infinitely many types of fractional excitations in a $U(1)$ quantum spin liquid.

In the spirit of this general framework, we have also discussed $U(1)$ quantum spin liquids with some other symmetries, and found some very rich structures. In particular, we discussed $U(1)$  spin liquids with $Z_2\times\mc{T}$ symmetry in great detail. Based on the properties of the bulk fractional excitations, there are $38$ such $Z_2\times\mc{T}$ symmetric states that are free of anomaly. The anomalies of the other 37 such $Z_2\times\mc{T}$ symmetric states are detected based on the method in the general framework. The study of $Z_2\times\mc{T}$ symmetric $U(1)$ quantum spin liquids have some implications on some SPTs, as discussed in Appendix \ref{app: Z2Z2T}.

Besides looking for a simper systematic classification of these symmetry enriched $U(1)$ quantum spin liquids, the other most important open questions are of course which microscopic models and experimental systems realizing  these different symmetry enriched $U(1)$ quantum spin liquids,  and how to detect and distinguish them numerically or experimentally. One particular interesting theoretical aspect of this question is how lattice symmetries interplay with these quantum spin liquids. These are beyond the scope of the current paper and are worth further investigating in the future, and we note some recent progress in this aspect.\cite{Li2016,Chen2017,Chen2017a,Zou2017}

Another interesting theoretical challenge is to classify and characterize symmetry enriched gapped quantum spin liquids, some of which can be obtained by condensing some excitations in the $U(1)$ quantum spin liquids. One complication in this problem is that there usually exists loop-like excitations, whose properties are not completely understood to date. Because some of these gapped quantum spin liquids are descendants of the $U(1)$ quantum spin liquids, relating the properties of loop-like excitations in the former to the properties of the particle-like excitations in the latter may shed light on this problem.

\section{Acknowledgement}

We acknowledge fruitful discussions with Maissam Barkeshli, Meng Cheng, Meng Guo, Max Metlistki,  Yang Qi, Chenjie Wang and Boyu Zhang.   T. S. was supported
by  NSF grant DMR-1608505, and partially through a Simons Investigator Award from the Simons Foundation.   L. Z. acknowledges support by the 2016 Boulder Summer School for Condensed
Matter and Materials Physics - where a part of this work was done - through NSF grant  DMR-13001648.   C. W. was supported by the Harvard Society of Fellows.

\begin{appendix}

\section{Some remarks on time reversal and charge-conjugation symmetry} \label{app: remarks on TR}

In this appendix we discuss some general properties of time reversal and charge-conjugation symmetry. In particular, we would like to check the values of $\mathcal{T}^2$ and $\mathcal{C}^2$ on a fractionalized excitation. By abuse of notation, we will denote this excitation by $E$ in general, which includes but does not limit to the case where $E$ is the electric charge of a $U(1)$ quantum spin liquid. In general, $E$ can be a multicomponent object, and let us denote the $i$th component as $E_i$.

We first consider time-reversal symmetry $\mathcal{T}$. We assume the anti-unitary time reversal symmetry acts on $E_i$ as
\beq \label{eq: generic-TR}
\mathcal{T}: E_i\rightarrow U_{ij}E_j
\eeq
with $U$ a generic matrix. That is, time reversal changes $E$ to something that differs from it only by a local operator. We also assume $E_i^\dag M_{ij}E_j$ is always a local object, for any matrix $M$. Notice in the above notation, some components of $E$ can be a bound state of the fractionalized excitation $E$ and some local particles.

Straightforward algebra indicates that acting time reversal twice, the original $E$ becomes
\beq
E_i\rightarrow (U^*U)_{ij}E_j
\eeq
and the local object becomes
\beq
E_i^\dag M_{ij}E_j\rightarrow E_i^\dag((U^*U)^\dag M(U^*U))_{ij}E_j
\eeq

Suppose the system is made of Kramers singlets. That is, $E_i^\dag M_{ij}E_j$ is invariant upon acted by time reversal twice for any $M$, which is equivalent to that
\beq
(U^*U)^\dag M(U^*U)=M
\eeq
for any matrix $M$. This is possible only if $U^*U=e^{i\phi}I$, where $I$ is the identity matrix. Because $(U^*U)^2=e^{2i\phi}I$ and $U^*UU^*U=U^*(UU^*)U=I$. This implies $e^{i\phi}=\pm 1$. That is, $T^2$ acting on such fractionalized excitations must give $\pm 1$.\cite{Wang2016}

The above argument also shows even if there is microscopic Kramers doublet in the system, as long as time reversal takes the form (\ref{eq: generic-TR}) and $T^2$ is well-defined, $T^2$ can only be $\pm 1$. However, if there are microscopic Kramers doublets in the system, it is possible to have $T^4=-1$. Notice in the case where $T^4=-1$, $T^2$ does not have to be well-defined.

Now $T^4=-1$ (or $T^2=\pm i$) can happen if time reversal changes the relevant excitation by a nonlocal operation, and a typical example for this case is $(eCmC)_T\epsilon$ mentioned in the main text, where under time reversal $e$ and $m$ are exchanged. It can also happen if time reversal changes this excitation by a local operation. In this case, time reversal has to attach a local Kramers doublet to the relevant nontrivial excitation. To see this, without loss of generality, let us assume the time reversal partner of $E$ is $F$, i.e., $E\rightarrow F$ under time reversal. To have $T^2=\pm i$ for $E$, we need $F\rightarrow \pm iE$ under time reversal. These also imply $T^2=\mp i$ for $F$, so $E$ and $F$ defer by a local Kramers doublet.

We now consider charge-conjugation symmetry $\mathcal{C}$. In general it is a unitary symmetry acting on $E_i$ as
\beq
\mathcal{C}: E_i\to V_{ij}E_j^{\dagger},
\eeq
where $V$ is a matrix. Taking the hermitian conjugate of the above equation, we have
\beq
\mathcal{C}: E^{\dagger}_i\to V^*_{ij}E_j.
\eeq
So $\mathcal{C}$ acting twice should give
\beq
\mathcal{C}^2: E_i\to (VV^*)_{ij}E_j.
\eeq
Again we require the local operator $E^{\dagger}_iM_ijE_j$ to have $\mathcal{C}^2=1$, for any matrix $M$. Following the same logic as we did for time-reversal symmetry, we conclude that $\mathcal{C}^2=\pm 1$ on the fractionalized excitation $E$.

Also notice that the value of $\mathcal{C}^2$ is invariant under a $U(1)$ gauge transform $U_{\theta}$, namely $(U_{\theta}\mathcal{C})^2=\mathcal{C}^2$. This means that the value $\mathcal{C}^2=\pm1$ is a physically meaningful quantity.

The simple discussion here on $\mathcal{C}^2$ should be enough for our purpose. In the context of gapped topological orders (for example in $\mathbb{Z}_{2N}$ gauge theories), the mathematically more precise meaning of $\mathcal{C}^2$ on fractionalized excitations has been discussed in Ref.~\cite{Barkeshli2014}.

\section{An SPT: $eCm\half$} \label{app: eCmhalf}

In this appendix we describe a 3D SPT, $eCm\half$, under symmetry $U(1)\times SO(3)$, $U(1)\times\mc{T}\times SO(3)$ or $(U(1)\rtimes\mc{T})\times SO(3)$. The defining property of this SPT is that it can have a symmetric surface $Z_2$ topological order with excitations $\{1,e,m,\epsilon\}$, where $e$ carries charge-1/2 under $U(1)$ and $m$ carries spin-1/2 under $SO(3)$. We begin by giving field theoretic descriptions of this surface state with the above symmetries.

If the symmetry is simply $U(1)\times SO(3)$, a field theory for this surface can be described by the following Lagrangian:
\beq \label{eq: NCCP1}
\begin{split}
&\mc{L}=\sum_{s=\pm}|(\partial_\mu-ia_\mu)z_s|^2+V(|z|^2)\\
&\qquad\qquad\qquad
+\frac{1}{4e^2}(\epsilon^{\mu\nu\lambda}\partial_\nu a_\lambda)^2 -\frac{1}{2\pi}Ada
\end{split}
\eeq
where $Ada$ is a shorthand for $\epsilon^{\mu\nu\lambda}A_\mu\partial_\nu a_\lambda$. $z_s$ is a two-component complex field that transforms as a doublet under $SO(3)$, $a_\mu$ is a non-compact $U(1)$ gauge field, and $A$ is a background $U(1)$ gauge field corresponding to the global $U(1)$ symmetry.

Condensing a bound state of two $z$'s in the singlet channel (i.e. letting $\la z_1\partial_x z_2-z_2\partial_xz_1\ra\neq 0$ for example) gives us the above surface topological order, where the uncondensed single $z$ will be identified as $m$ that carries spin-1/2. After this condensation, the flux of $a$ is quantized in unit of $\pi$, and the last term in the above Lagrangian implies that this $\pi$-flux carries charge-1/2 under the global $U(1)$. This $\pi$-flux can be identified as $e$. The resulting state is precisely $eCm\half$.

If the symmetry is $(U(1)\rtimes\mc{T})\times SO(3)$, the surface theory of $eCm\half$ can still be described by the field theory given by (\ref{eq: NCCP1}), but now the spin operator is represented as $\vec S\sim z^\dag\vec\sigma i\partial_tz$ and under time reversal
\beq
z\rightarrow z^\dag,
\
\vec a\rightarrow\vec a,
\
a_0\rightarrow -a_0
\eeq
In order to obtain $eCm\half$, we need to make $\la z_1\partial_x z_2-z_2\partial_xz_1\ra=\la z_1^\dag\partial_x z_2^\dag-z_2^\dag\partial_xz_1^\dag\ra\neq 0$.

If the symmetry is $U(1)\times\mc{T}\times SO(3)$, the surface state of $eCm\half$ can be described by a field theory similar to (\ref{eq: NCCP1}):
\beq
\begin{split}
&\mc{L}=\sum_{s,\alpha=\pm}|(\partial_\mu-ia_\mu)z_{s\alpha}|^2+V(|z|^2)\\
&\qquad\qquad\qquad
+\frac{1}{4e^2}(\epsilon^{\mu\nu\lambda}\partial_\nu a_\lambda)^2 -\frac{1}{2\pi}Ada
\end{split}
\eeq
Notice now each component of $z_s$ contains two components, $z_{s\alpha}$ with $\alpha=\pm$, and the generators of spin rotations become $\vec S=\frac{1}{2}z^\dag_{s\alpha}\vec\tau_{ss'}z_{s'\alpha}$. Under time reversal
\beq
z_{s\alpha}\rightarrow(\sigma_2)_{\alpha\alpha'}(\tau_2)_{ss'}z_{s'\alpha'},
\
\vec a\rightarrow-\vec a,
\
a_0\rightarrow a_0
\eeq
where $\sigma$ and $\tau$ are the standard Pauli matrices. The reason to give more components to $z$ is to make it not a Kramers doublet. To get $eCm\half$, we can also condense the bound state of two $z$'s in the singlet channel, that is, let $\la z^T\sigma_1\tau_2\partial_xz\ra\neq 0$. Similar argument as above implies the resulting state is $eCm\half$.

To see that this state as a strictly 2D system is anomalous, consider tunneling a $U(1)$ monopole through this 2D system. This will leave a $2\pi$ flux on the system. For such a local process, no excitations far away should be able to tell the existence of this $2\pi$-flux. But $e$ carries half charge under $U(1)$, it will pick up a nontrivial phase factor upon circling around this $2\pi$-flux, regardless how far it locates away from it. To cancel this phase factor, an $m$ needs to be present at the $2\pi$-flux. Because $m$ carries spin-1/2, this then implies tunneling a monopole leaves a spin-1/2 on the surface. This is not possible for a strictly 2D system with symmetry $U(1)\times SO(3)$. Notice that time reversal symmetry is not involved in the anomaly, so this surface is still anomalous even if the symmetry is $U(1)\times SO(3)$.

To visualize this SPT, the simplest way is to do a layer construction similar to that used in Ref. \onlinecite{Wang2013}. Because similar method will be used in Appendix \ref{app: 4d surface} to construct $4+1$-d systems whose surfaces realize the anomalous quantum spin liquids, we do not explicitly display it for $eCm\half$ here.

Notice when the symmetry is $(U(1)\rtimes\mc{T})\times SO(3)$, to realize $eCm\half$, we have assumed that the microscopic bosons are Kramers singlet. Below we argue that for microscopic Kramers doublet charged bosons, $eCm\half$ cannot be realized. This fact is important, because otherwise gauging $eCm\half$ in such a system would lead to $E_{bT}M_{b\half}$, which is argued to be anomalous in Sec. \ref{sec: anomalies}.

Suppose $eCm\half$ can be realized in a system made of Kramers doublet charged bosons. Fusing two $e$'s gives a charge-1 local particle, which must be a Kramers doublet by assumption. Given the excitation content of this theory and that time reversal keeps the $U(1)$ charge, the time reversal action on $e$ can always be represented as
\beq
e_i\rightarrow U_{ij}e_j
\eeq
Now notice $e_iM_{ij}e_j$ is a local charge-1 operator for any matrix $M$, so this operator must be a Kramers doublet, which implies that
\beq
(U^*U)^TM(U^*U)=-M
\eeq
This is possible only if $U^*U=\pm i$. As shown in Appendix \ref{app: remarks on TR}, no matrix $U$ can have this property. This implies that $eCm\frac{1}{2}$ cannot be realized in a system made of Kramers doublet charged bosons.

\section{Anomalous spin liquids as surface states of some $4+1$-d systems} \label{app: 4d surface}

In this appendix we will show the anomalous spin liquids can be obtained on the surface of some $4+1$-d systems. The simplest way to construct these $4+1$-d surface states is the following layer construction, which has been widely used to construct topological states\cite{Wang2013,Jian2014,Kravec2015,ChenHermele2016}.

\begin{figure}
  \centering
  \includegraphics[width=0.4\textwidth]{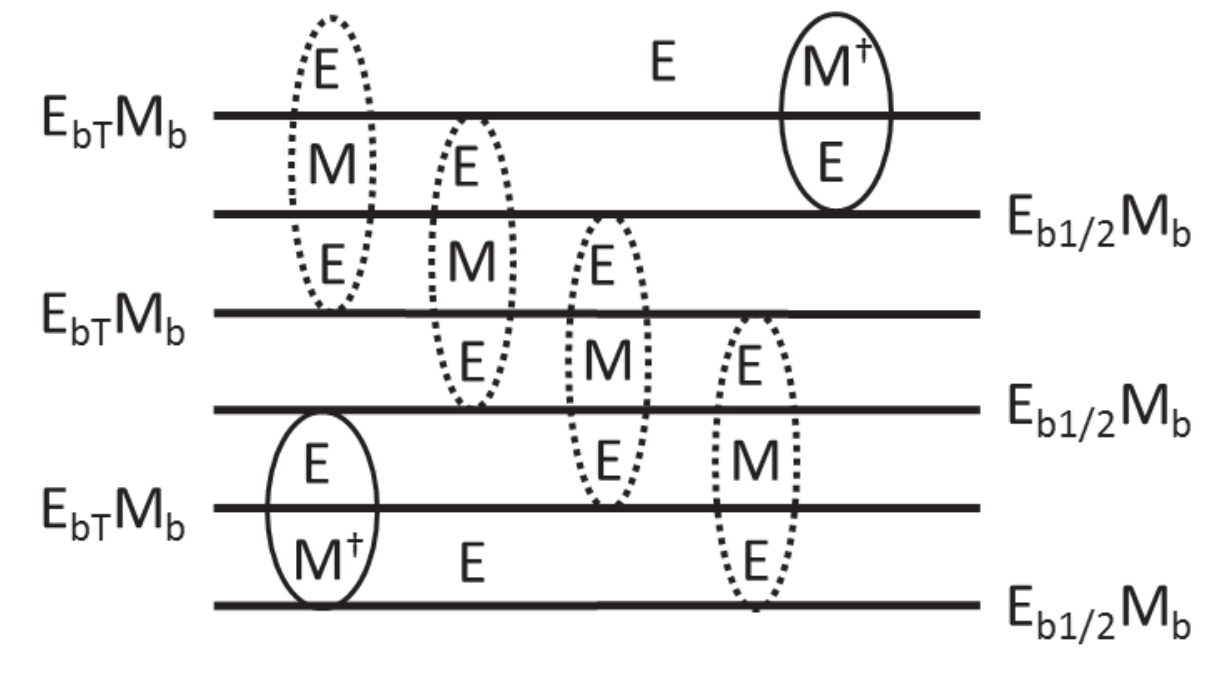}\\
  \caption{Layer construction of the $4+1$-d system whose surface realize an anomalous spin liquid.}\label{fig: layer construction}
\end{figure}

For example, to construct a $4+1$-d system whose surface can realize $E_{bT}M_{b\half}$, one can start by stacking alternating layers of non-anomalous spin liquids $E_{bT}M_b$ and $E_{b\half}M_b$ (see Fig. \ref{fig: layer construction}). Then on the $i$th, $i+1$th and $i+2$th layers, one can condense the bound state of $E_i$, $M_{i+1}$ and $E_{i+2}$ with the subscript indicating the layer index. This bound state, $B_i=E_iM_{i+1}E_{i+2}$, is a trivial boson, and $B_i$'s with different $i$'s commute, so they can be simultaneously condensed without breaking any symmetry. After this condensation, the gauge field in the $4+1$-d bulk will be confined (or Higgsed) and this bulk becomes short-range entangled, but on the surface some nontrivial excitations still survive. These survivors are $E_1$ and $M^\dag_1E_2$ on the top surface, and $E_N$ and $E_{N-1}M^\dag_N$ on the bottom surface. Now the top (bottom) surface realizes $E_{bT}M_{b\half}$, and $E_1$ ($E_{N-1}M^\dag_N$) and $M^\dag_1E_2$ ($E_N$) can be viewed as the electric charge and magnetic monopole, respectively. The $4+1$-d system constructed here is an SPT under symmetry $SO(3)\times\mathcal{T}$ because its surface, $E_{bT}M_{b\half}$, is anomalous.

To obtain $4+1$-d systems whose surface realize all other anomalous spin liquids, one only needs to replace each layer by the appropriate non-anomalous spin liquid and condense the proper bound states.

For some spin liquids, there is a more isotropic construction of the corresponding $4+1$-d systems by using a non-linear Sigma model (NLSM) with appropriate topological terms and anisotropies, similar to that used in Ref. \onlinecite{Bi2015}. For example, to construct the corresponding $4+1$-d bulk of $E_{bT\half}M_{b\half}$, consider a $4+1$-d $O(6)$ NLSM with a theta-term at $\theta=2\pi$. Its surface theory is a $3+1$-d six-component NLSM with a Wess-Zumino-Witten (WZW) term at level-1, with Lagrangian
\beq
\begin{split}
&\mc{L}=\frac{1}{g}(\partial_\mu n_a)^2\\
&\quad
+\frac{2\pi i}{\Omega_5}\int_0^1du\epsilon^{abcdef}n_a\partial_un_b\partial_xn_c \partial_yn_d\partial_zn_e\partial_\tau n_f
\end{split}
\eeq
with $\Omega_5$ the surface area of a five-dimensional unit sphere. The six-component vector transforms under time reversal as
\beq
n_{1,2,3}\rightarrow -n_{1,2,3}
\quad
n_{4,5,6}\rightarrow n_{4,5,6}
\eeq
This theory is invariant under $O(6)\times\mc{T}$.

To see how $E_{bT\half}M_{b\half}$ can be accessed by the above theory, let us first add some $SO(3)\times SO(3)$ anisotropy, such that the first (second) three components transform as a vector under the first (second) $SO(3)$. Consider the weak coupling limit of the theory where both the first and the second three components are ordered. Now disorder the second three components by proliferating its hedgehog defects. In this way, the second three components themselves form a trivial state that preserves the second $SO(3)$ symmetry. Due to the WZW term, the hedgehog defects of the first three components carry spin-1/2 under the second $SO(3)$, and it will be identified as the magnetic monopole of $E_{bT\half}M_{b\half}$ later.

Now disorder the first three components by proliferating the spin wave excitations while keeping its hedgehog defects gapped. In this way we will get a $U(1)$ spin liquid. One way to see this is to write the first three components in the $CP^1$ representation,
\beq \label{eq: CP1}
n_a=z_\alpha^\dag\sigma_a^{\alpha\beta}z_\beta,
\quad
{\rm for\ }
a=1,2,3
\eeq
where $z$ is a two-component complex spinon field with $|z_1|^2+|z_2|^2=1$, and $\sigma$'s are the standard Pauli matrices. Under the previous defined time reversal and the first $SO(3)$, the spinon field transforms as a Kramers doublet and $SU(2)$ doublet. This spinon will be identified as the electric charge of $E_{bT\half}M_{b\half}$ later.

The Lagrangian that only involves the first three components can now be written as the following gauge theory
\beq
\mc{L}'=|(\partial_\mu-ia_\mu)z|^2+V(|z|^2)+\frac{1}{4e^2}(\epsilon^{\mu\nu\lambda} \partial_\nu a_\lambda)^2
\eeq
where $a_\mu$ is an emergent $U(1)$ gauge field due to the $U(1)$ gauge redundancy in (\ref{eq: CP1}), i.e., $n_a$ is invariant when $z\rightarrow ze^{i\theta}$ for any real $\theta$. The ordered state of the first three components corresponds to the Higgs phase of the $U(1)$ gauge theory, and proliferating spin wave excitations corresponds to making spinons gapped and give rise to a $U(1)$ spin liquid, where the gapped spinons are the electric charge. The monopole of this $U(1)$ spin liquid, which is the source of magnetic flux, should be identified as the un-proliferated hedgehog defect, which is the source of the skyrmions.

Finally, adding a weak anisotropy to collapse the $SO(3)\times SO(3)$ symmetry to its diagonal $SO(3)$ subgroup, we get $E_{bT\half}M_{b\half}$ with $SO(3)\times\mc{T}$ symmetry, where the electric charge is a Kramers doublet and $SU(2)$ doublet, and the magnetic monopole is an $SU(2)$ doublet.

The construction above gives a $4+1$-d system whose surface realizes $E_{bT\half}M_{b\half}$. If time reversal is ignored, the above construction gives the $4+1$-d system whose surface realizes $E_{b\half}M_{b\half}$. The $4+1$-d system whose surface can realize $E_{bT}M_{b\half}$ can be obtained similarly, where time reversal acts in the same way as before, while under $SO(3)$ only the last three components transform as a vector and the first three components do not transform.

Notice in the construction based on NLSMs, even if $SO(3)$ is broken to $Z_2\times Z_2$, all components still transform nontrivially under this symmetry. Then it is believed that the constructed $4+1$-d states are still nontrivial SPTs. This motivates us to conjecture that even if the symmetry is broken to $Z_2\times Z_2\times\mc{T}$, the descendants of the anomalous states remain anomalous because they still live on the surface of some SPTs.

\section{Classification of some SPTs} \label{app: SPTs}

In this appendix we classify some SPTs which, once gauged, can become some of the $U(1)$ quantum spin liquids studied in the main text.

\subsection{Bosonic SPT with symmetry $(U(1)\rtimes\mc{T})\times SO(3)$}

We start with bosonic SPT with symmetry $(U(1)\rtimes\mc{T})\times SO(3)$, where the microscopic boson is a Kramers singlet. Without $SO(3)$ symmetry, the classification of this SPT is well established. They are classified by $\mathbb{Z}_2^3$, where the three root states are $eCmC$, $eTmT$ and $efmf$.\cite{VishwanathSenthil2013} With $SO(3)$ symmetry, Appendix \ref{app: eCmhalf} shows there is another root state: $eCm\half$. In fact, there two more root states: $e\half mT$ and $e\half m\half$. That these two are nontrivial SPTs can be inferred from the classification of bosonic SPT with symmetry $U(1)\times\mc{T}$. Indeed, once $SO(3)$ is broken to $U(1)$, $e\half mT$ and $e\half m\half$ become $eCmT$ and $eCmC$, respectively.

Therefore, we propose the classification of these SPTs is $\mathbb{Z}_2^6$. Notice that among the six root states, only two of them need protection from the $U(1)$ symmetry: $eCmC$ and $eCm\half$.

\subsection{Bosonic SPT with symmetry $U(1)\times\mc{T}\times SO(3)$}

If the symmetry is $U(1)\times\mc{T}\times SO(3)$, the understanding of bosonic SPTs with symmetry $U(1)\times\mc{T}$ implies there is one more root state: $eCmT$. This state is protected by both $U(1)$ and time reversal.

Therefore, we propose the classification of these SPTs is $\mathbb{Z}_2^7$. The properties of the surface $Z_2$ topologically ordered states of the root states are summarized in Table \ref{table:SPTs}.
\begin{table}
\begin{tabular}{|c|c|c|c|c|c|c|c|}
\hline
&$q_e$&$q_m$&$T^2_e$&$T^2_m$&$S_e$&$S_m$&comments\\
\hline
$eCmC$&$\half$&$\half$&1&1&1&1&\\
\hline
$eTmT$&0&0&-1&-1&1&1&\\
\hline
$efmf$&0&0&1&1&1&1&e and m are fermions\\
\hline
$eCm\half$&$\half$&0&1&1&1&$\half$&\\
\hline
$e\half m\half$&0&0&1&1&$\half$&$\half$&\\
\hline
$e\half mT$&0&0&1&-1&$\half$&1&\\
\hline
$eCmT$&0&0&1&-1&1&1&\\
\hline
\end{tabular}
\caption{Surface $Z_2$ topological ordered states of SPTs under symmetry $U(1)\times\mc{T}\times SO(3)$. The topological sectors are denoted by $\{1,e,m,\epsilon\}$. $q_e$ and $q_m$ represents the charge of $e$ and $m$ under $U(1)$, $T^2_e$ and $T^2_m$ represents the Kramersness of $e$ and $m$ under time reversal, and $S_e$ and $S_m$ represents the spin of $e$ and $m$, respectively. If the symmetry is $(U(1)\rtimes\mc{T})\times SO(3)$, $eCmT$ will be absent and all other six root states remain.} \label{table:SPTs}
\end{table}

\subsection{SPT with symmetry $\left((U(1)\times SU(2))/Z_2\right)\times\mc{T}$ of fermions}

For fermionic SPT with symmetry $\left((U(1)\times SU(2))/Z_2\right)\times{T}$, free fermion band theory gives a $\mathbb{Z}$ classification, and each state can be labelled by an integer $k$, which is basically the number of pairs of massless Dirac fermions on the surface.

The root state can have a surface state with two massless Dirac fermions with the following Hamiltonian
\beq \label{eq: two-Dirac}
H=\psi^\dag(-i\partial_x\sigma_x-i\partial_y\sigma_z)\otimes\tau_0\psi
\eeq
where $\sigma$ and $\tau$ are the standard Pauli matrices with $\sigma_0=\tau_0=I$, and $\sigma$ acts on the internal indices of the Dirac fermions and $\tau$ acts on the spin indices. Under $U(1)$,
\beq
\psi\rightarrow\psi e^{i\theta}
\eeq
Under $SU(2)$,
\beq
\psi\rightarrow\sigma_0\otimes U\psi
\eeq
with $U$ an $SU(2)$ matrix. And under time reversal
\beq
\psi\rightarrow i\sigma_y\otimes\tau_0\psi^\dag
\eeq
Notice the inverse of this root state, i.e. the state that can trivialize the root state when coupled together, can have the same surface Hamiltonian as this root state except that time reversal acts as $\psi\rightarrow-i\sigma_2\otimes\tau_0\psi^\dag$. This means the state labelled by $k$ and that labelled by $-k$ are identical after gauging the $U(1)$, because the aforementioned sign difference in the time reversal action can be eliminated by a $U(1)$ gauge transformation. \cite{Wang2016}

When the $U(1)$ symmetry is gauged, the monopole of the corresponding $U(1)$ gauge field is a Kramers doublet that carries spin-1/2.\cite{WangSenthil2014,Borokhov2002} Because the method that leads to this result will be used extensively later, it is helpful to review it here.

Since the surface is described by two free Dirac fermions, which is a conformal field theory, one can use state-operator correspondence to study the properties of monopoles, by imagining putting the surface on a sphere with $2\pi$ flux threading out. Guaranteed by the index theorem, each Dirac fermion will contribute a zero mode in the background of the $2\pi$ flux, in our case denoted by $f_1$ and $f_2$, respectively. We also denote the flux background with both zero modes empty by $|0\ra$. Because the time reversal symmetry flips the $U(1)$ charge here, the physical gauge invariant states must have one of the zero modes being occupied. That is, it should be $f_1^\dag|0\ra$ or $f_2^\dag|0\ra$, which are bosonic. In light of state-operator correspondence, these two states correspond to two different charge-neutral monopole operators, denoted by $M_1$ and $M_2$, respectively. Also, $|0\ra$ corresponds to the operator of the $(-1,1)$ dyon, and $f_1^\dag f_2^\dag|0\ra$ corresponds to the operator of $(1,1)$ dyon. Then the quantum numbers of the monopole can be read off from the properties of these states.

For example, for the surface theory described above, because the two Dirac fermions transform as spin-1/2 under $SU(2)$, the monopoles $M_1\sim f_1^\dag|0\ra$ and $M_2\sim f_2^\dag|0\ra$ also transform as spin-1/2. As for time reversal, these two states transform as
\beq
\begin{split}
&M_1\sim f_1^\dag|0\ra\rightarrow f_1f_1^\dag f_2^\dag|0\ra=f_2^\dag|0\ra\sim M_2\\
&M_2\sim f_2^\dag|0\ra\rightarrow f_2f_1^\dag f_2^\dag|0\ra=-f_1^\dag|0\ra\sim -M_1
\end{split}
\eeq
This means the monopoles are Kramers doublet under time reversal. Therefore, after gauging the $U(1)$ symmetry, this state becomes $E_{bT\half}M_{f\half}$.

Now we turn to the classification of such fermionic SPTs. Upon adding interactions, the free fermion classification collapses to $\mathbb{Z}_4$.\cite{WangSenthil2014} It can be shown that the state with 8 massless Dirac fermions on the surface is trivial, and the state with 4 massless Dirac fermions on the surface is equivalent to $eTmT\half$, a bosonic SPT with symmetry $SO(3)\times\mc{T}$. There can be interacting SPTs beyond band theory, which can be viewed as bosonic SPTs with symmetry $SO(3)\times\mc{T}$. They are classified by $\mathbb{Z}_2^4$. One of the root states of these bosonic SPTs coincide with a free fermion SPT that can have 4 massless Dirac fermions on the surface, so we propose the complete classification is $\mathbb{Z}_4\times\mathbb{Z}_2^3$.

\subsection{SPT with symmetry $((U(1)\rtimes\mc{T})\times SU(2))/Z_2$ of Kramers singlet fermions}

Consider fermionic SPT with symmetry $((U(1)\rtimes\mc{T})\times SU(2))/Z_2$ and assume $T^2=1$ for these fermions. Free fermion band theory gives a $\mathbb{Z}_2$ classification. The root state can have a surface state with two massless Dirac fermions, described by the same Hamiltonian as (\ref{eq: two-Dirac}), with the only difference that under time reversal
\beq
\psi\rightarrow\sigma_y\otimes\tau_y\psi
\eeq

When the $U(1)$ symmetry of these Dirac fermions is gauged, the monopole of the corresponding $U(1)$ gauge field carries spin-1/2. \cite{WangSenthil2016} This can also be seen by using the method of state-operator correspondence reviewed above. Notice in this case the time reversal symmetry does not flip the $U(1)$ charge, so it is convenient to momentarily suppose equipping the system with a further charge conjugation symmetry. We will determine the properties of the monopoles in the presence of this further symmetry first, and then break this symmetry. Because the properties of the monopoles are described by some discrete data, breaking this symmetry will not change any of them.

Again, each Dirac fermion will contribute a zero mode to the $2\pi$ flux background, and the neutral bosonic monopoles correspond to the two states with one zero mode occupied: $M_{1,2}\sim f_{1,2}^\dag|0\ra$. Because the two Dirac fermions carry spin-1/2 under $SU(2)$, as above, the monopoles must also carry spin-1/2. It is not meaningful to discuss whether monopoles are Kramers doublet or not under time reversal, so this finishes determining the properties of the monopoles. From this discussion, we see that after gauging the $U(1)$ symmetry this state becomes $E_{f\half}M_{b\half}$.

Although it is not meaningful to discuss whether monopoles are Kramers doublet or not under time reversal, it is interesting and helpful to determine how monopoles transform under time reversal. To be consistent with that time reversal and $SU(2)$ commute on the monopole, under time reversal the monopole operators must transform as $M_{1,2}\rightarrow M_{1,2}^\dag$ (with a possible phase ambiguity).

How do we understand this time reversal action on monopoles from the point of view of state-operator correspondence? This is a little tricky because in this case the $2\pi$ flux will be turned into a $-2\pi$ flux under time reversal, which also has two zero modes, denoted by $\tilde f_{1,2}$, where $\tilde f_{1,2}$ is contributed by $\psi_{1,2}$, respectively. In particular, denote $|\tilde 0\ra$ as the state with $-2\pi$ flux background and neither zero mode occupied. This is the time reversal partner of $|0\ra$, so it corresponds to the $(-1,-1)$ dyon. Similarly, $\tilde f_1^\dag\tilde f_2^\dag|\tilde 0\ra$ is the time reversal partner of $f_1^\dag f_2^\dag|0\ra$, so it corresponds to the $(1,-1)$ dyon.

Under time reversal,
\beq
\begin{split}
&f_1^\dag|0\ra\rightarrow\tilde f_2^\dag|\tilde 0\ra\\
&f_2^\dag|0\ra\rightarrow-\tilde f_1^\dag|\tilde 0\ra
\end{split}
\eeq
where an unimportant $U(1)$ phase factor has been suppressed. For this to be compatible with that $M_{1,2}\rightarrow M_{1,2}^\dag$ under time reversal, we must identify (with an unimportant phase factor)
\beq
M_1^\dag\sim\tilde f_2^\dag|\tilde 0\ra,
\quad
M_2^\dag\sim-\tilde f_1^\dag|\tilde 0\ra
\eeq
To the best of our knowledge, this identification of the Hermitian conjugate of the monopoles in the context of state-operator correspondence has not been given before, and it will be used later. We remark that this identification is true as long as the long-distance conformal field theory is described by two massless Dirac fermions, and it should be independent of the microscopic symmetries of the system, although we obtained it by considering a system with a particular symmetry.

Now we return to the classification of these fermionic SPTs. Upon adding interaction, the nontrivial state is stable. There are also SPTs beyond band theory with root state that can be viewed as bosonic SPTs with symmetry $SO(3)\times\mc{T}$, and they can be classified by $\mathbb{Z}_2^4$. One of the four root states becomes trivial in the presence of fermions with this symmetry (see Table \ref{table:bSPTs3}), therefore, we propose that the total classification is $\mathbb{Z}_2^4$.

\section{Relations between the classification of $SO(3)\times\mc{T}$ symmetric $U(1)$ quantum spin liquids and the classification of some relevant SPTs} \label{app: QSLandSPTs}

In the main text $SO(3)\times\mc{T}$ symmetric $U(1)$ quantum spin liquids can be classified into 15 phases, as summarized in Table \ref{table:theta=0} and Table \ref{table:theta=pi}. In particular, how they can be viewed as gauged SPTs are also discussed. It is helpful to understand the relation between the classification of the $U(1)$ quantum spin liquids and the classification of the relevant SPTs. Below we give some examples.

From the point of view of $E$, $E_bM_b$, $E_bM_f$, $E_bM_{b\half}$ and $E_bM_{f\half}$ can all be regarded as the gauged insulators of Kramers singlet bosons with symmetry $(U(1)\rtimes\mc{T})\times SO(3)$. In Appendix \ref{app: SPTs} we propose that the bosonic SPTs under this symmetry are classified by $\mathbb{Z}_2^6$, where only two of the six root states requires protection from the $U(1)$ symmetry. It is not hard to see, after gauging this $U(1)$ symmetry, the $\mathbb{Z}_2^2$ subset of SPTs coming from these two root states become precisely the above four quantum spin liquids.

With the same symmetry as above, if the bosons are Kramers doublet, Appendix \ref{app: eCmhalf} shows that only one of the two root states survives. Gauging the trivial insulator and the nontrivial SPT from the other root state leads to $E_{bT}M_b$ and $E_{bT}M_f$, respectively.

From the point of view of $M$, $E_bM_b$, $E_{bT}M_b$, $E_{b\half}M_b$, $E_{bT\half}M_b$, $E_fM_b$, $E_{fT}M_b$, $E_{f\half}M_b$ and $E_{fT\half}M_b$ can be thought of as the gauged bosonic insulators with symmetry $U(1)\times\mc{T}\times SO(3)$. In Appendix \ref{app: SPTs} we propose that the bosonic SPTs under this symmetry are classified by $\mathbb{Z}_2^7$, where only three of the root states requires protection from the $U(1)$ symmetry. Gauging the $\mathbb{Z}_2^3$ subset of the SPTs generated by these three root states gives precisely the above eight quantum spin liquids.

From the point of view of $M$, both $E_bM_{f\half}$ and $E_{bT\half}M_{f\half}$ can be viewed as a gauged topological superconductor of fermions with symmetry $\left((U(1)\times SU(2))/Z_2\right)\times\mc{T}$. In Appendix \ref{app: SPTs} we propose that the topological superconductors with this symmetry are classified by $\mathbb{Z}_4\times\mathbb{Z}_2^3$, where the first $\mathbb{Z}_4$ factor represents those can be realized with free fermions, and the last $\mathbb{Z}_2^3$ factor corresponds to interacting topological superconductors beyond band theory. For states that can be realized by band theory, the nontrivial topological superconductors can have $2k$ massless Dirac fermions on the surface, where $k=0,1,2,3\ ({\rm mod}\ 4)$. Gauging states with even $k$ leads to $E_bM_{f\half}$ (up to a bosonic SPT $eTmT\half$) and gauging states with odd $k$ leads to $E_{bT\half}M_{f\half}$. For states beyond band theory, gauging them results in the same quantum spin liquid as their corresponding state within band theory up to a bosonic SPT with symmetry $SO(3)\times\mc{T}$.

From the point of view of $E$, both $E_{f\half}M_b$ and $E_{f\half}M_{b\half}$ can be viewed as gauged topological insulators of fermions with symmetry $((U(1)\rtimes\mc{T})\times SU(2))/Z_2$, where the microscopic fermions are Kramers singlets. In Appendix \ref{app: SPTs} we propose that these topological insulators are classified by $\mathbb{Z}_2^5$, where the first $\mathbb{Z}_2$ factor corresponds to those realizable by free fermions, and the nontrivial state can have $2$ massless Dirac fermions on the surface. Gauging the trivial state leads to $E_{f\half}M_b$ and gauging the nontrivial state leads to $E_{f\half}M_{b\half}$. Gauing the states beyond band theory gives the same quantum spin liquids as their corresponding free fermion cousins up to a bosonic SPT under symmetry $SO(3)\times\mc{T}$.

The above examples show that gauging different SPTs may results in the same quantum spin liquid, and no one single class of SPTs will lead to all quantum spin liquids after gauging, so the classification of these quantum spin liquids does not form a simple group structure, while the classification of SPTs does. As mentioned in the introduction, viewing a single quantum spin liquid as two different gauged SPTs leads to some helpful dualities on the surface theories of these SPTs, which can be inferred from the above discussion.

Also, if two different quantum spin liquids can be viewed as two different SPTs with the same microscopic constitutes and symmetry, the quantum phase transition between them can also be viewed as the gauged version of the quantum phase transition between the two corresponding SPTs. For example, the quantum phase transition between $E_{f\half}M_b$ and $E_{f\half}M_{b\half}$ can be viewed as the gauged version of the quantum phase transition between the trivial and nontrivial fermionic insulators with symmetry $((U(1)\rtimes\mc{T})\times SU(2))/Z_2$. We will not go into the details in this paper.

\section{Bosonic SPTs with $SO(3)\times\mc{T}$ symmetry} \label{app:bSPTsurface}

In this appendix we discuss bosonic SPTs with symmetry $SO(3)\times\mc{T}$, with the assumption that the microscopic degrees of freedom are non-Kramers bosons with spin-1. Group cohomology gives classification $\mathbb{Z}_2^3$,\cite{Chen2013} but it misses one root state\cite{VishwanathSenthil2013}, and the complete classification should be $\mathbb{Z}_2^4$. The four root states all have anomalous surface $Z_2$ topological orders, where symmetries are realized in a way impossible in a purely two dimensional system (see Table \ref{table:bSPTs}). Among the four root states, $eTmT$ and $efmf$ are protected by time reversal alone. Below we review the anomalies of $e\half m\half$ and $e\half mT$.

Suppose $e\half m\half$ is realizable in a purely two dimensional system, then tunneling an $SO(3)$ monopole through it leaves a $\pi$-flux seen by both $e$ and $m$. Because such a local process should not have nonlocal observable effect, an $\epsilon$, the fermionic bound state of $e$ and $m$, must be trapped at this $\pi$-flux. Due to time reversal symmetry, there is no polarization spin in this process and this flux is just a fermion. Therefore, a local process generates a fermion in this 2D system, which is impossible. Notice this anomaly is just the $SO(3)$ version of the anomaly of $eCmC$ under symmetry $U(1)\times\mc{T}$. In fact, when the symmetry $SO(3)\times\mc{T}$ is broken down to $U(1)\times\mc{T}$, the descendant state of $e\half m\half$ is precisely $eCmC$, which has a $U(1)$ theta angle to be $2\pi$. This also implies the $SO(3)$ $\Theta=2\pi$ for $e\frac{1}{2}m\frac{1}{2}$.

As for $e\half mT$, tunneling an $SO(3)$ monopole leaves a $\pi$-flux seen by $e$ and $\epsilon$, so this process must trap an $m$. Because $SO(3)$ commutes with $\mc{T}$, the $SO(3)$ flux is invariant under time reversal, and a Kramers doublet is generated by this local process. This contradicts the assumption that there is no local Kramers doublet particles. Again, this anomaly is the $SO(3)$ version of the anomaly of $eCmT$ under symmetry $U(1)\times\mc{T}$, and the descendant state of $e\half mT$ is just $eCmT$ when the symmetry $SO(3)\times\mc{T}$ is broken down to $U(1)\times\mc{T}$.

\section{Constraints on the Hall conductances due to time reversal and spin rotational symmetries} \label{app:sigmaxy}

Suppose in addition to a $U(1)_c$ charge conservation symmetry, a two dimensional gapped system also has time reversal symmetry and spin rotational symmetry. One can also consider the spin quantum Hall conductance, $\sigma_{xy}^s$, which relates the spin current due to a gradient Zeeman field.\cite{Senthil1999} This can be formally viewed as the response of the system to a probe gauge field, $A_s$, which corresponds to the $S_z$ rotation symmetry, $U(1)_s$. There can also be quantum spin Hall conductance, $\sigma_{xy}^{cs}$, which relates the spin current and the electric field of the gauge field $A_c$, the gauge field corresponding to symmetry $U(1)_c$.\cite{Kane2005} This appendix discusses the constraints on these Hall conductances due to time reversal and spin rotational symmetries. The results are useful in determining what polarization charge or spin will be generated when flux is inserted in the system, or equivalently, when a monopole tunnels through the system.

To this end, we first reorganize the charge conservation and $S_z$ rotation symmetries in terms of two other $U(1)$ symmetries, denoted by $U(1)_\uparrow$ and $U(1)_\downarrow$. These two $U(1)$ symmetries can be viewed as separate conservations of spin-up and spin-down particles. The corresponding gauge fields and charges of these two symmetries are related to $A_c$ and $A_s$ by
\beq \label{eq: reorganizingU(1)}
\begin{split}
&A_\uparrow=A_c+A_s,
\qquad
A_\downarrow=A_c-A_s\\
&Q_\uparrow=\frac{Q_c+Q_s}{2},
\qquad
Q_\downarrow=\frac{Q_c-Q_s}{2}
\end{split}
\eeq
Notice under the charge quantization condition that $Q_1$ and $Q_2$ can independently take any integers, $Q_c$ and $Q_s$ have to either be both even or both odd.

Now we can discuss the Hall conductances due to coupling to $A_\uparrow$ and $A_\downarrow$. The general Hall response theory reads
\beq
\mc{L}=\frac{1}{4\pi}(\sigma_{xy}^\uparrow A_\uparrow dA_\uparrow+\sigma_{xy}^\downarrow A_\downarrow dA_\downarrow+2\sigma_{xy}^{\uparrow\downarrow}A_\uparrow dA_\downarrow)
\eeq
where $AdB$ is a shorthand for $\epsilon_{\mu\nu\lambda}A_\mu\partial_\nu B_\lambda$. Using (\ref{eq: reorganizingU(1)}), we get
\beq \label{eq: sigmas}
\begin{split}
&\sigma_{xy}^c=\sigma_{xy}^\uparrow+\sigma_{xy}^\downarrow+2\sigma_{xy}^{\uparrow \downarrow}\\
&\sigma_{xy}^s=\sigma_{xy}^\uparrow+\sigma_{xy}^\downarrow-2\sigma_{xy}^{\uparrow \downarrow}\\
&\sigma_{xy}^{cs}=\sigma_{xy}^\uparrow-\sigma_{xy}^\downarrow
\end{split}
\eeq

Clearly any element in the spin rotational symmetry that takes spin-up to spin-down (such as rotation around $x$-axis by $\pi$) requires $\sigma_{xy}^\uparrow=\sigma_{xy}^\downarrow$. Below we study the constraints from time reversal symmetry. Notice that the $U(1)_s$ charge is always odd under time reversal, but the $U(1)_c$ charge can either be time reversal even or odd, and we discuss these two cases separately.

We start from the case where the $U(1)_c$ charge is even under time reversal, which means
\beq
A_{1,2}^0\rightarrow A_{2,1}^0,
\qquad
\vec A_{\uparrow,\downarrow}\rightarrow -\vec A_{\downarrow,\uparrow}
\eeq
For the response theory to be time reversal symmetric, we need
\beq
\sigma_{xy}^\uparrow=-\sigma_{xy}^\downarrow,
\quad
\sigma_{xy}^{\uparrow\downarrow}=0
\eeq

For the case where the $U(1)_c$ is odd under time reversal, time reversal transformation takes
\beq
A_{\uparrow,\downarrow}^0\rightarrow -A_{\uparrow,\downarrow}^0,
\qquad
\vec A_{\uparrow,\downarrow}\rightarrow \vec A_{\uparrow,\downarrow}
\eeq
For the response theory to be time reversal symmetric, we need
\beq
\sigma_{xy}^\uparrow=\sigma_{xy}^\downarrow=\sigma_{xy}^{\uparrow\downarrow}=0
\eeq

From these constraints and (\ref{eq: sigmas}) one can easily obtain the constraints on $\sigma_{xy}^c$, $\sigma_{xy}^s$ and $\sigma_{xy}^{cs}$. We note all these constraints can also be obtained simply by applying Laughlin's flux insertion argument.

We notice that all these Hall conductance vanishes if the system has both time reversal symmetry and spin rotational symmetry that contains at least $O(2)\simeq U(1)_s\rtimes Z_2$, where $U(1)_s$ is a rotational symmetry around one axis and $Z_2$ is the $\pi$-rotation around another axis perpendicular to the previous one. This implies that inserting flux or tunneling a monopole through such two dimensional systems will not lead to any polarization charge or spin.

\section{Non-edgeability of some $Z_2$ topological orders in the presence of nontrivial particles} \label{app:non-edgeability}

In the Sec. \ref{subsec: SPT trivialization} we claimed some $Z_2$ topological orders are not edgeable even in the presence of some nontrivial particles, i.e. they do not allow for a physical edge separating it and the trivial vacuum. In this appendix, we will justify this claim by showing that these $Z_2$ topological orders allow no $K$-matrix theory to describe them.

\subsection{Brief review of the $K$-matrix theory}

We begin with a brief review of some general aspects of the $K$-matrix theory. For more details, see Ref. \onlinecite{Wen2004Book} and Ref. \onlinecite{Levin2012,Lu2012,LuVishwanath2013}.

The Lagrangian of a $K$-matrix theory of a system that couples to an external $U(1)$ gauge field $A_c$ is given by
\beq \label{eq:K-matrix}
\mc{L}=\frac{K_{IJ}}{4\pi}a_Ida_J-\frac{q_c^I}{2\pi}A_cda_I
\eeq
with $K$ a symmetric invertible matrix with all entries integers, and $q_c$ a vector with all entries integers.

An excitation of this theory can be represented by an integral excitation vector, $l$. The charge of this excitation under the external gauge field $A_c$ is
\beq
l^TK^{-1}q_c
\eeq
And this excitation has self-statistics angle
\beq
\pi l^TK^{-1}l
\eeq
For two excitations represented by excitation vectors $l_1$ and $l_2$, respectively, the mutual braiding angle between them is
\beq
2\pi l_1^TK^{-1}l_2
\eeq

A simple example is that
\beq
K=
\left(
\begin{array}{cc}
&2\\
2&
\end{array}
\right)
\eeq
which represents $Z_2$ topological order, where $e$ can be taken to be represented by excitation vector $(1,0)^T$ and $m$ can be taken to be represented by $(0,1)^T$.

The 2+1-d bulk theory (\ref{eq:K-matrix}) allows the following boundary theory:
\beq
\mc{L}=\frac{1}{4\pi}(K_{IJ}\partial_t\phi_I\partial_x\phi_J -V_{IJ}\partial_x\phi_I\partial_x\phi_J)
\eeq
where $\phi_I$ are bosonic fields such that $e^{il_I\phi_I}$ is the annihilation operator of excitation $l$ on the boundary. These bosonic fields satisfy Kac-Moody algebra
\beq
[\phi_I(x),\partial_y\phi_J(y)]=2\pi i(K^{-1})_{IJ}\delta(x-y)
\eeq
$V_{IJ}$ is called the velocity matrix that gives the velocities of these bosonic fields.

The above summarizes the topological properties of the $K$-matrix theory, (\ref{eq:K-matrix}). Below we review the symmetry actions on this theory.

In general symmetries act on the gauge fields $a_I$ as a matrix. For example, we denote the time reversal action as
\beq
a_I\rightarrow T_{IJ}a_J
\eeq
with $T$ an integral matrix. Notice the above equation only gives the transformation of the spatial components of the gauge fields, and the temporal components should have a minus sign in front due to the anti-unitary nature of time reversal symmetry.

It is important to notice the above does not fully specify the symmetry action, and to that end, one needs to specify how the boundary bosonic fields transform. In general, they transform as
\beq
\phi_I\rightarrow T_{IJ}\phi_J+t_I
\eeq
with $t_I$ a real vector. \cite{Levin2012}

To make the bulk theory (\ref{eq:K-matrix}) invariant under anti-unitary time reversal symmetry, we need
\beq \label{eq:TRonK}
K\rightarrow T^TKT=-K
\eeq
If the $U(1)$ charge is even under time reversal, we further require
\beq \label{eq:TRonqce}
q_c\rightarrow T^Tq_c=q_c
\eeq
while if the $U(1)$ charge is odd under time reversal we require
\beq \label{eq:TRonqco}
q_c\rightarrow T^Tq_c=-q_c
\eeq

\subsection{Non-edgeability of some $Z_2$ topological orders in the presence of nontrivial particles}

Now by showing some $Z_2$ topological orders even in the presence of nontrivial particles do not allow a $K$-matrix theory description, we show their non-edgeability because $K$-matrix theories are supposed to capture all two dimensional Abelian states.

Here we list the $Z_2$ topological orders of interests. We denote $eTmT$ in the presence of bosons with quantum numbers $C^2T\half$ by $(eTmT,bC^2T\half)$, and $eTmT$ in the presence of fermions with quantum number $\tilde C^2$ by $(eTmT, f\tilde C^2)$. Besides these two, we will also consider $(e\half mT, bC^2T\half)$, $(eT\half mT,bC^2\half)$, $(eTmT, fC^2T)$, $(eTmT, fC^2\half)$, $(eTmT, f\tilde C^2\half)$, $(eTmT, fC^2T\half)$, $(e\half mT, fC^2\half)$ and $(eT\half mT,fC^2T\half)$, with similar notations as before.

We immediately have two main difficulties in showing their non-edgeability. First, in some cases the nontrivial particles carry spin-1/2 and we need to incorporate $SU(2)$ symmetry in the $K$-matrix theory, but continuous non-Abelian symmetries are usually not manifest in a $K$-matrix theory and dealing with them directly is generally difficult. To resolve this difficulty, we will instead just show that the descendants of the relevant states are still not edgeable when the $SU(2)$ symmetry is broken down to $U(1)$, which is sufficient to show the original states are non-edgeable with the full $SU(2)$ symmetry. To distinguish this $U(1)$ from the original charge $U(1)$, we will denote the charge $U(1)$ by $U(1)_c$, and this $U(1)$ by $U(1)_s$. Unit charge under $U(1)_s$ will be denoted by $C'^2$. Therefore, for example, we will consider $(eTmT,fC^2C'^2)$ instead of $(eTmT,fC^2\half)$. Now (\ref{eq:K-matrix}) needs to be modified to include the coupling to $A_s$, the external gauge field corresponding to $U(1)_s$
\beq
\mc{L}=\frac{K_{IJ}}{4\pi}a_Ida_J-\frac{q_c^I}{2\pi}A_cda_I -\frac{q_s^I}{2\pi}A_sda_I
\eeq
Notice the charge of $U(1)_s$ is always odd under time reversal, so time reversal symmetry requires that
\beq \label{eq:TRonqs}
q_s\rightarrow T^Tq_s=-q_s
\eeq

The second difficulty is that in general the state that we are interested in may be described by a $K$-matrix with a large dimension, but dealing with a large-dimensional $K$-matrix is daunting. However, the following observation suggests we actually only need to consider a $2\times 2$ $K$-matrix.

Notice all these $Z_2$ topological orders come from $eTmT$, $e\half mT$ and $eT\frac{1}{2}mT$. For both cases, the nontrivial topological quasiparticles only need a single component to describe them. This is because whenever there are two components of them, one can condense some bound states of them that are singlets under all symmetries. This will not change the topological order or the symmetry of the system, but only one component will be left over.\cite{Wang2013} More concretely, this means to describe the putative $Z_2$ topological orders that we are interested in, we should always be able to write the $K$-matrix as
\beq
K=2\sigma_x\oplus L
\eeq
where $L$ is an invertible symmetric integral matrix that can be large in dimension, and $L$ describes only local excitations. For bosons, $L$ can be written as $\sigma_x\oplus\sigma_x\oplus\cdots\oplus\sigma_x$, while for fermions, $L$ can be written as $\sigma_z\oplus\sigma_z\oplus\sigma_x\oplus\cdots$. In this form, any excitation with an excitation vector of the form $(1,0,\cdots)^T$ can be identified as $e$, and all excitations with an excitation vector of the form $(0,1,\cdots)^T$ can be viewed as $m$, where the ``$\cdots$" can be nonzero. At this moment, the nontrivial topological quasiparticle that we are after, for example, the Kramers doublet $e$ particle in $eTmT$, can still be represented by $(1,0,\cdots)^T$ with ``$\cdots$" nonzero. But we can always bind proper local excitations to this excitation so that the excitation vector becomes $(1,0,0,0,\cdots)^T$ with ``$\cdots$" all zeros.

The argument above shows that, in order to show the non-edgeability of those $Z_2$ topological orders, it is sufficient to show that no $2\times 2$ $K$-matrix can describe the topological quasiparticles with the corresponding quantum numbers, up to binding local excitations. Let us demonstrate this via the following example. In $(eTmT, bC^2TC'^2)$, $eTmT$ can be relabelled as, for example, $eC^2C'^2mT$. The above statement means that, in order to show that $(eTmT, bC^2TC'^2)$ is not edgeable, it is sufficient to show that none of $eTmT$, $eC^2C'^2mT$ and all other states related to these by binding a local excitation made up of $bC^2TC'^2$ can be realized by a $2\times 2$ $K$-matrix.

Because time reversal should not convert a local excitation into a nonlocal one, we expect that the matrix $T$ can be written in the following form
\beq
T=
\left(
\begin{array}{cc}
T_0&T_1\\
0&T_2
\end{array}
\right)
\eeq
where $T_0$ is a $2\times 2$ integral matrix. Plugging this form of $K$ and $T$ into (\ref{eq:TRonK}), we see time reversal symmetry requires that
\beq
T_0^T\sigma_xT_0=-\sigma_x
\eeq
It is easy to show the only solutions are $T_0=\pm\sigma_z$ or $T_0=\pm\epsilon$, where $\epsilon=i\sigma_y$.

Notice in all the cases we consider, the quantum numbers of $e$ and $m$ are always nontrivial. If $T_0=\pm\epsilon$, then $T_0^2=-1$. This does not allow Kramers doublet structure, and it also does not allow nonzero $q_c$ and $q_s$ that satisfy (\ref{eq:TRonqce}) or (\ref{eq:TRonqco}) and (\ref{eq:TRonqs}). So this choice of $T_0$ can never work.

So we can focus on the case with $T_0=\pm\sigma_z$. Without loss of generality, we take $T_0=-\sigma_z$. Notice now $e$, represented by the excitation vector $(1,0)^T$, is always a Kramers singlet independent of $t$. If the second entry of $t$ is $\pi/2$, $m$, represented by excitation vector $(0,1)^T$, is a Kramers doublet.

With this choice of $T_0$, in order to satisfy (\ref{eq:TRonqce}) or (\ref{eq:TRonqco}) and (\ref{eq:TRonqs}), $q_c$ and $q_s$ can only be taken as
\beq
q_c=(0,q_1)^T
\qquad
q_s=(q_2,0)^T
\eeq
when the charge under $A_c$ is even under time reversal, or
\beq
q_c=(q_1,0)^T
\qquad
q_s=(q_2,0)^T
\eeq
when the charge under $A_c$ is odd under time reversal. In the first case, $e$ carries charge $q_1/2$ under $A_c$ and zero charge under $A_s$. In the second case, $e$ carries zero charge under both $A_c$ and $A_s$.

\subsubsection*{$eTmT$ is not edgeable in the presence of $bC^2TC'^2$}

The above discussions immediately imply that $eTmT$ is not edgeable in the presence of $bC^2TC'^2$. This is because $e$ cannot be a Kramers doublet, an odd number of $bC^2TC'^2$s have to be attached to it to cancel its Kramersness. Then $e$ carries nonzero charge under $A_s$, which is in contradiction with $e$ always carrying zero charge under $A_s$. So $(eTmT,bC^2T\half)$ is not edgable.

\subsubsection*{$eC'^2 mT$ is not edgeable in the presence of $bC^2TC'^2$}

Here $e$ is not a Kramers doublet but it carries nonzero charge under $A_s$. To cancel this charge, an odd number of $bC^2TC'^2$s have to be attached to $e$, which makes it a Kramers doublet. This is again impossible as argued above. One can also try to switch the label between $e$ and $m$, then it becomes $eTmC'^2$. The argument above implies this is inconsistent even in the presence of $bC^2TC'^2$. So $(e\half mT,eC^2T\half)$ is not edgeable.

\subsubsection*{$eTC'^2mT$ is not edgeable in the presence of $bC^2C'^2$}

Here no matter how many $bC^2C'^2$'s are attached, the $Z_2$ topological order always has both $e$ and $m$ being Kramers doublet. This cannot be realized. So $(eT\half mT,bC^2\half)$ is not edgeable.

\subsubsection*{$eTmT$ is not edgeable in the presence of $f\tilde C^2$ or $f\tilde C^2C'^2$}

Here $e$ is a Kramers doublet, so an odd number of $f\tilde C^2$s or $f\tilde C^2C'^2$ need to be attached to it, which makes it become $\epsilon$ and the new $e$ carry $\tilde C^2$. This is in contradiction with $e$ carrying zero charge under $A_c$ in this case. So $(eTmT,fC^2)$ and $(eTmT,fC^2\half)$ are not edgeable.

\subsubsection*{$eTmT$ and $eTC'^2mT$ are not edgeable in the presence of $fC^2T$ or $fC^2TC'^2$}

Here $e$ and $m$ are Kramers doublets, and attaching any number of $fC^2T$ or $fC^2TC'^2$ always leaves both $e$ and $m$ Kramers doublets, so $(eTmT,fC^2T)$, $(eTmT,fC^2T\half)$ and $(eT\half mT,fC^2T\half)$ are not edgeable.

\subsubsection*{$eTmT$ and $eC'^2mT$ are not edgeable in the presence of $fC^2C'^2$}

Here an odd number of $fC^2C'^2$s need to be attached to $e$, which converts it to $\epsilon$ and make the new $e$ carry nonzero charge under $A_c$ and $A_s$. But $e$ cannot carry nonzero charge under both $A_c$ and $A_s$. So $(eTmT,fC^2\half)$ and $(e\frac{1}{2}mT,fC^2\frac{1}{2})$ are not edgeable.

In summary, none of the $Z_2$ topological orders is edgeable in the presence of the relevant nontrivial excitations. This implies they are all still anomalous.

\section{Projective representations: the electric (standard), the magnetic (twisted) and the dyonic (mixed) ones} \label{app: projective representations}

In this appendix we discuss in detail various projective representations of a symmetry group $G$: the electric (standard), the magnetic (twisted) and the dyonic (mixed) ones. We always assume this group $G$ can contain time reversal, but besides, for simplicity, all other elements form a connected unitary group. That is, these elements are all unitary and they can all be continuously connected to the identity element. We will see although all these projective representations are classified by some group cohomologies, but different cases are classified by different group cohomologies.

\subsection{Electric (standard) projective representations}

We begin with the familiar case of standard projective representations. Although our results will be identical to the ones in textbooks, we will use a different formulation that is more appropriate for our purposes and easier to generalize to twisted projective representations.

Suppose there is a symmetry $G$, which can in principle contain anti-unitary element. If all elements of $G$ only change an excitation by a local operation, then it is appropriate to discuss the standard projective representations of $G$ on this excitation.

A prototypical example of this case is that the relevant excitation is the electric charge $E$ of a $U(1)$ quantum spin liquid. In general, the action of an element $g\in G$ on $E$ can be written as
\beq
E_i\rightarrow U(g)_{ij}E_j
\eeq
Here different components of $E_i$ differ from each other by a local operation, and $U(g)$ is a matrix representation of $g$.

Because $E_i^*A_{ij}E_j$ is a local operator for any matrix $A$, this operator is supposed to transform in the linear representation of $G$. For $g_1,g_2\in G$, acting $g_1$ followed by $g_2$ on this operator gives
\beq
\begin{split}
&E^\dag\left(U(g_1)^{s(g_2)}U(g_2)\right)^\dag\cdot A^{s(g_2g_1)}\\
&\qquad\qquad\qquad\qquad
\cdot \left(U(g_1)^{s(g_2)}U(g_2)\right)E
\end{split}
\eeq
where now $E$ represents a column vector with components $E_i$, and, for an arbitrary matrix $M$, $M^{s(g)}=M$ if $g$ is unitary, while $M^{s(g)}=M^*$ if $g$ is anti-unitary. For the special case where $M$ is just a phase factor, $s(g)=1$ ($s(g)=-1$) if $g$ is unitary (anti-unitary).

The above result should be identical to the one obtained by acting $g_2g_1$ on this local operator directly:
\beq
E^\dag\cdot U^\dag(g_2g_1)\cdot A^{s(g_2g_1)}\cdot U(g_2g_1)\cdot E
\eeq
For these two results to be identical for an arbitrary matrix $A$, we must have
\beq
U(g_1)^{s(g_2)}U(g_2)=\omega(g_2,g_1)U(g_2g_1)
\eeq
where $\omega(g_2,g_1)$ is a phase factor.

The above equation can be written in the following equivalent way:
\beq
U(g_2g_1)=\omega(g_2,g_1)^{-1}U(g_1)^{s(g_2)}U(g_2)
\eeq
For any $g_1,g_2,g_3\in G$, applying this equation to $U(g_1g_2g_3)$ yields
\beq \nonumber
\begin{split}
&U(g_1g_2g_3)\\
=&\omega(g_1,g_2g_3)^{-1}U(g_2g_3)^{s(g_1)}U(g_1)\\
=&\omega(g_1,g_2g_3)^{-1}\omega(g_2,g_3)^{-s(g_1)} U(g_3)^{s(g_1g_2)}U(g_2)^{s(g_1)}U(g_1)\\
=&\omega(g_1g_2,g_3)^{-1}U(g_3)^{s(g_1g_2)}U(g_1g_2)\\
=&\omega(g_1g_2,g_3)^{-1}\omega(g_1,g_2)^{-1}U(g_3)^{s(g_1g_2)} U(g_2)^{s(g_1)}U(g_1)
\end{split}
\eeq
This implies the following associativity condition for the phase factor $\omega$'s:
\beq \label{eq: E-associative}
\omega(g_1g_2,g_3)\omega(g_1,g_2) =\omega(g_1,g_2g_3)\omega(g_2,g_3)^{s(g_1)}
\eeq

There is a gauge freedom for the phase factor $\omega$'s. To see this, notice the symmetry action of $g\in G$ on $E$ can be modified by a gauge transformation:
\beq
E_i\rightarrow \lambda(g)U(g)_{ij}E_j\equiv \tilde U(g)_{ij}E_j
\eeq
where $\lambda(g)$ is a $U(1)$ phase factor. The action of $g\in G$ on any local operator will be the same, which means $\tilde U(g)$ is an equally good representation of $g$. Under this transformation, it is straightforward to check that
\beq
\tilde U(g_2)^{s(g_1)}\tilde U(g_1)=\tilde\omega(g_1,g_2)\tilde U(g_1g_2)
\eeq
where
\beq \label{eq: factor-system-gauge-freedom}
\tilde\omega(g_1,g_2)=\omega(g_1,g_2)\cdot \frac{\lambda(g_1)\lambda(g_2)^{s(g_1)}}{\lambda(g_1g_2)}
\eeq

The factor systems $\omega$ and $\tilde\omega$ related in this way should be regarded to be in the same class, because they give rise to identical results in any local operator. It is straightforward to check that the relation (\ref{eq: factor-system-gauge-freedom}) is an equivalence relation, that is, it is reflexive, symmetric and transitive. Furthermore, it is clear if $\omega_1$ and $\omega_2$ are the two classes of factor systems corresponding to representations $U_1(g)$ and $U_2(g)$, respectively, $\omega_1\cdot\omega_2$ will be the factor system of the representation $U_1(g)\cdot U_2(g)$. This defines a multiplication operation among the classes of factor systems. With this multiplication, the classes of factor systems form an Abelian group, where the trivial element is the class of factor systems of a linear representation. In fact, this Abelian group form a structure of group cohomology $H^2(G,U_T(1))$.\cite{Chen2013} In this group cohomology, the $n$-cochains $\omega_n(g_1,g_2,\cdots,g_n)$ take value as a phase factor, the 1-coboundary operation is defined as
\beq
d_1\omega_1(g_1,g_2)
=\frac{\omega_1(g_1)\omega_1(g_2)^{s(g_1)}}{\omega_1(g_1g_2)}
\eeq
and the 2-coboundary operation is defined as
\beq
d_2\omega_2(g_1,g_2,g_3)=\frac{\omega_2(g_2,g_3)^{s(g_1)}\omega_2(g_1,g_2g_3)} {\omega_2(g_1g_2,g_3)\omega_2(g_1,g_2)}
\eeq

It is straightforward to check that $d_2d_1=1$. Also, the solutions to the associativity condition (\ref{eq: E-associative}) are 2-cocycles, and different solutions are identified up to a 1-coboundary. Therefore, the classes of factor systems, or the (standard) projective representations, are indeed classified by this cohomology. Below we will see the twisted and mixed projective representations are also classified by some group cohomologies, which are however different from this $H^2(G,U_T(1))$.

\subsection{Magnetic (twisted) projective representations}

Next we turn to twisted projective representations. A prototypical example where it is appropriate to consider twisted projective representations is, when the symmetry includes time reversal, to consider the fractional quantum numbers on the magnetic monopole, $M$.

Suppose $g\in G$ is unitary, its action on $M$ can be represented as
\beq
M_i\rightarrow U(g)_{ij}M_j
\eeq
Suppose $g\in G$ is anti-unitary, its action on $M$ can be represented as
\beq
M_i\rightarrow U(g)_{ij}M_j^*
\eeq

Again, because $M_i^*A_{ij}M_j$ is a local operator for any matrix $A$, it is supposed to transform in the linear representation of $G$. Similar to the analysis in the previous case, this implies, for $g_1,g_2\in G$,
\beq
U(g_1)^{s(g_2)}U(g_2)^{s(g_1)}=\omega(g_2,g_1)U(g_2g_1)
\eeq
where $\omega(g_2,g_1)$ is a phase factor.

Similar as standard projective representations, these phase factors need to satisfy an associativity condition:
\beq
\begin{split}
&\omega(g_1g_2,g_3)\omega(g_1,g_2)^{s(g_3)}\\
&\qquad\qquad
= \omega(g_1,g_2g_3)\omega(g_2,g_3)^{s(g_1)}
\end{split}
\eeq
Further, there is also a gauge freedom that leads to the following equivalence relation
\beq
\begin{split}
\omega(g_1,g_2)\sim&\tilde\omega(g_1,g_2)\\
=&\omega(g_1,g_2)\cdot\frac{\lambda(g_1)^{s(g_2)}\lambda(g_2)^{s(g_1)}} {\lambda(g_1g_2)}
\end{split}
\eeq
where $\lambda$'s are phase factors.

Just as in the case of the standard projective representations, the classes of factor systems of a twisted projective representation also form an Abelian group, whose multiplication, trivial element, and inverse element are defined in parallel as in the case of the standard projective representation. This group is also described by a cohomology, denoted by $H^2(G,U_T^M(1))$. This cohomology is different from the previous one, $H^2(G,U_T(1))$, in the coboundary operations. In this cohomology, the $n$-cochains $\omega_n(g_1,g_2,\cdots,g_n)$ still take values as phase factors, the 1-coboundary operation is defined as
\beq
d_1\omega_1(g_1,g_2)=\frac{\omega_1(g_1)^{s(g_2)}\omega_1(g_2)^{s(g_1)}} {\omega_1(g_1g_2)}
\eeq
and the 2-coboundary operation is defined as
\beq
d_2\omega_2(g_1,g_2,g_3)=\frac{\omega_2(g_1,g_2g_3)\omega_2(g_2,g_3)^{s(g_1)}} {\omega_2(g_1g_2,g_3)\omega_2(g_1,g_2)^{s(g_3)}}
\eeq

It is straightforward to check $d_2d_1=1$. Again, the solutions of the associativity condition are 2-cocycles, and they are identified up to a 1-coboundary. Therefore, twisted projective representations are classified by $H^2(G,U_T^M(1))$.

Interestingly, for the cases with $G=1$, $G=Z_2$ and $G=U(1)$, $H^2\left(G\times\mc{T},U_T^M(1)\right)=H^2\left(G\times Z_2,U(1)\right)$, where the latter is the standard group cohomology with $G\times Z_2$ acting trivially on the $U(1)$ coefficient. It will be interesting to know if this relation is always true.

\subsection{Dyonic (mixed) projective representations}

In the case of a $G$ symmetric $U(1)$ quantum spin liquid at $\theta=\pi$, the property of the phase is determined by the $\left(\half,1\right)$ and $\left(\half,-1\right)$ dyons. Denote these two dyons by $D^{(+)}$ and $D^{(-)}$, respectively. The symmetry quantum numbers of these two dyons are given by the dyonic (mixed) projective representations.

Again, assume the only part of the symmetry that can change the type of fractional excitations is time reversal, the action of $g\in G$ on $D^{(+)}$ and $D^{(-)}$ can be written as
\beq
D^{(+)}_i\rightarrow U_+(g)_{ij}D^{(+)},
\quad
D^{(-)}_i\rightarrow U_-(g)_{ij}D^{(-)}
\eeq
if $g$ is unitary, and
\beq
D^{(+)}_i\rightarrow U_+(g)_{ij}D^{(-)},
\quad
D^{(-)}_i\rightarrow U_-(g)_{ij}D^{(+)}
\eeq
if $g$ is anti-unitary.

Now using that $D_i^{(+)*}A_{ij}D_j^{(+)}$ and $D_i^{(-)*}A_{ij}D_j^{(-)}$ are local operators for any matrix $A$, we get
\beq
U_i^{s(g_2)}(g_1)U_{{s(g_1)\cdot i}}(g_2)=\omega_i(g_2,g_1)U_i(g_2g_1)
\eeq
where $i=\pm$, and $\omega_i(g_2,g_1)$ is a phase factor.

In this case the associativity condition becomes
\beq
\begin{split}
&\omega_i(g_1,g_2g_3)\omega_i(g_2,g_3)^{s(g_1)}\\
&\qquad\qquad
=\omega_i(g_1g_2,g_3)\omega_{{s(g_3)\cdot i}}(g_1,g_2)
\end{split}
\eeq
And the equivalence relation becomes
\beq
\begin{split}
\omega_i(g_1,g_2)\sim&\tilde\omega_i(g_1,g_2)\\
=&\omega_i(g_1,g_2)\cdot\frac{\lambda_{{s(g_2)\cdot i}}(g_1)\lambda_i(g_2)^{s(g_1)}} {\lambda_i(g_1g_2)}
\end{split}
\eeq

Similar as the twisted projective representations, the mixed projective representations also form an Abelian group and are also classified by a group cohomology, denoted by $H^2\left(G,U^D(1)\times U^D(1)\right)$. Here the $n$-cochains $\omega_{i,n}(g_1,g_2,\cdots,g_n)$ take values as a phase factor (for $i=\pm$ separately), the 1-coboundary operation is defined as
\beq
d_1\omega_{i,1}(g_1,g_2)=\frac{\omega_{{s(g_2)\cdot i},1}(g_1)\omega_{i,1}(g_2)^{s(g_1)}} {\omega_{i,1}(g_1g_2)}
\eeq
and the 2-coboundary operation is defined as
\beq
d_2\omega_{i,2}(g_1,g_2,g_3)=\frac{\omega_{i,2}(g_1,g_2g_3)\omega_{i,2}(g_2,g_3)^{s(g_1)}} {\omega_{i,2}(g_1g_2,g_3)\omega_{{s(g_3)\cdot i},2}(g_1,g_2)}
\eeq
It is straightforward to check $d_2d_1=1$. Clearly, the solutions of the associativity condition are 2-cocycles, and different solutions are identified up to a 1-coboundary. So the mixed projective representations are classified by $H^2(G,U^D(1)\times U^D(1))$.

\section{Examine the anomalies of $Z_2\times\mc{T}$ symmetric $U(1)$ quantum spin liquids with $\theta=0$} \label{app: Z2Z2T}

In this appendix we will give more details of the anomaly-detection of the 72 different putative $U(1)$ quantum spin liquids with $Z_2\times\mc{T}$ symmetry that have $\theta=0$. Among these states, $Z_2$ does not act as a charge conjugation for $24$ of them and acts as a charge conjugation for the other $48$. These 72 states are all listed in Sec. \ref{sec: Z2Z2T}, and they are copied in Table \ref{table:Z2Z2T-simple Z2-theta=0-non-anomalous}, Table \ref{table:Z2Z2T-simple Z2-theta=0-anomalous}, Table \ref{table: app-Z2Z2T-conjugation Z2-theta=0-nonanomalous} and Table \ref{table: app-Z2Z2T-conjugation Z2-theta=0-anomalous} for convenience.

\begin{table}
\center
\begin{tabular}{|c|c|c|c|}
\hline
& $T^2_E$ & $T'^2_E$ & $[\mc{T},Z_2]_M$\\
\hline
$E_bM_b$ & 1 & 1 & +\\
\hline
$E_{bT}M_b$ & $-1$ & $1$ & +\\
\hline
$E_{bT'}M_b$ & $1$ & $-1$ & +\\
\hline
$E_{bTT'}M_b$ & $-1$ & $-1$ & +\\
\hline
$E_bM_{b-}$ & $1$ & $1$ & $-$\\
\hline
$E_fM_b$ & $1$ & $1$ & +\\
\hline
$E_{fT}M_b$ & $-1$ & $1$ & +\\
\hline
$E_{fT'}M_b$ & $1$ & $-1$ & +\\
\hline
$E_{fTT'}M_b$ & $1$ & $-1$ & +\\
\hline
$E_{fTT'}M_{b-}$ & $-1$ & $-1$ & $-$\\
\hline
$E_bM_f$ & $1$ & $1$ & $+$\\
\hline
$E_{bT}M_f$ & $-1$ & $1$ & $+$\\
\hline
$E_{bT'}M_f$ & $1$ & $-1$ & $+$\\
\hline
$E_{bTT'}M_f$ & $-1$ & $-1$ & $+$\\
\hline
$E_bM_{f-}$ & $1$ & $1$ & $-$\\
\hline
\end{tabular}
\caption{List of non-anomalous $Z_2\times\mc{T}$ symmetric $U(1)$ quantum spin liquids that have $\theta=0$ and have $Z_2$ not acting as a charge conjugation. All these states are anomaly-free. $T_E^2=1$ ($T^2_E=-1$) represents the case where $E$ is a Kramers singlet (doublet) under $\mc{T}$. $T'^2_E=1$ ($T'^2_E=-1$) represents the case where $E$ is a Kramers singlet (doublet) under $\mc{T}'$. $[\mc{T},Z_2]_M=+$ ($[\mc{T},Z_2]_M=-$) represents the case where $Z_2$ and $\mc{T}$ commute (anti-commute) on $M$.} \label{table:Z2Z2T-simple Z2-theta=0-non-anomalous}
\end{table}

\begin{table}
\center
\begin{tabular}{|c|c|c|c|c|}
\hline
& $T^2_E$ & $T'^2_E$ & $[\mc{T},Z_2]_M$ & anomaly class\\
\hline
$E_{bT'}M_{f-}$ & $1$ & $-1$ & $-$ & class a\\
\hline
$E_{fT}M_{b-}$ & $-1$ & $1$ & $-$ & class a\\
\hline
$E_{bT'}M_{b-}$ & $1$ & $-1$ & $-$ & class a\\
\hline
$E_{fT'}M_{b-}$ & $1$ & $-1$ & $-$ & class b\\
\hline
$E_{bT}M_{f-}$ & $-1$ & $1$ & $-$ & class b\\
\hline
$E_{bT}M_{b-}$ & $-1$ & $1$ & $-$ & class b\\
\hline
$E_{bTT'}M_{f-}$ & $-1$ & $-1$ & $-$ & class c\\
\hline
$E_{bTT'}M_{b-}$ & $-1$ & $-1$ & $-$ & class c\\
\hline
$E_fM_{b-}$ & $1$ & $1$ & $-$ & class c\\
\hline
\end{tabular}
\caption{List of anomalous $Z_2\times\mc{T}$ symmetric $U(1)$ quantum spin liquids that have $\theta=0$ and have $Z_2$ not acting as a charge conjugation. All these states are anomaly-free. $T_E^2=1$ ($T^2_E=-1$) represents the case where $E$ is a Kramers singlet (doublet) under $\mc{T}$. $T'^2_E=1$ ($T'^2_E=-1$) represents the case where $E$ is a Kramers singlet (doublet) under $\mc{T}'$. $[\mc{T},Z_2]_M=+$ ($[\mc{T},Z_2]_M=-$) represents the case where $Z_2$ and $\mc{T}$ commute (anti-commute) on $M$. The last column indicates the anomaly classes.} \label{table:Z2Z2T-simple Z2-theta=0-anomalous}
\end{table}

\begin{table} [h!]
\begin{tabular}{|c|c|c|c|c|}
\hline
& $T^2_E$ & $Z_{E}^2$ & $T'^2_M$ & $Z_M^2$\\
\hline
$(E_bM_b)_-$ & $1$ & $1$ & $1$ & $1$\\
\hline
$(E_{bZ}M_b)_-$ & $1$ & $-1$ & $1$ & $1$\\
\hline
$(E_{bT}M_b)_-$ & $-1$ & $1$ & $1$ & $1$\\
\hline
$(E_{bTZ}M_b)_-$ & $-1$ & $-1$ & $1$ & $1$\\
\hline
$(E_bM_{bZ})_-$ & $1$ & $1$ & $1$ & $-1$\\
\hline
$(E_bM_{bT'})_-$ & $1$ & $1$ & $-1$ & $1$\\
\hline
$(E_bM_{bT'Z})_-$ & $1$ & $1$ & $-1$ & $-1$\\
\hline
$(E_{f}M_b)_-$ & $1$ & $1$ & $1$ & $1$\\
\hline
$(E_{fZ}M_b)_-$ & $1$ & $-1$ & $1$ & $1$\\
\hline
$(E_{fT}M_b)_-$ & $-1$ & $1$ & $1$ & $1$\\
\hline
$(E_{fTZ}M_b)_-$ & $-1$ & $-1$ & $1$ & $1$\\
\hline
$(E_bM_f)_-$ & $1$ & $1$ & $1$ & $1$\\
\hline
$(E_bM_{fZ})_-$ & $1$ & $1$ & $1$ & $-1$\\
\hline
$(E_bM_{fT'})_-$ & $1$ & $1$ & $-1$ & $1$\\
\hline
$(E_bM_{fT'Z})_-$ & $1$ & $1$ & $-1$ & $-1$\\
\hline
$(E_{fT}M_{bT'})_-$ & $-1$ & $1$ & $-1$ & $1$\\
\hline
$(E_{bT}M_{fT'})_-$ & $-1$ & $1$ & $-1$ & $1$\\
\hline
$(E_{fZ}M_{bT'Z})_-$ & $1$ & $-1$ & $-1$ & $-1$\\
\hline
$(E_{bTZ}M_{fZ})_-$ & $-1$ & $-1$ & $1$ & $-1$\\
\hline
$(E_{fTZ}M_{bZ})_-$ & $-1$ & $-1$ & $1$ & $-1$\\
\hline
$(E_{bZ}M_{fT'Z})_-$ & $1$ & $-1$ & $-1$ & $-1$\\
\hline
\end{tabular}
\caption{List of anomaly-free $Z_2\times\mc{T}$ symmetric $U(1)$ quantum spin liquids that have $\theta=0$ and have $Z_2$ acting as a charge conjugation. $T_E^2=1$ ($T^2_E=-1$) represents the case where $E$ is a Kramers singlet (doublet) under $\mc{T}$. $T'^2_M=1$ ($T'^2_M=-1$) represents the case where $M$ is a Kramers singlet (doublet) under $\mc{T}'$. $Z_{E,M}^2$ represents the result of acting the charge conjugation twice on $E$ and $M$, respectively.} \label{table: app-Z2Z2T-conjugation Z2-theta=0-nonanomalous}
\end{table}

\begin{table} [h!]
\begin{tabular}{|c|c|c|c|c|c|}
\hline
& $T^2_E$ & $Z_{E}^2$ & $T'^2_M$ & $Z_M^2$ & anomaly class\\
\hline
$(E_{bZ}M_{bZ})_-$ & $1$ & $-1$ & $1$ & $-1$ & class 1\\
\hline
$(E_{bTZ}M_{bT'Z})_-$ & $-1$ & $-1$ & $-1$ & $-1$ & class 1\\
\hline
$(E_{fT}M_{bZ})_-$ & $-1$ & $1$ & $1$ & $-1$ & class 1\\
\hline
$(E_{bZ}M_{fT'})_-$ & $1$ & $-1$ & $-1$ & $1$ & class 1\\
\hline
$(E_{fT}M_{bT'Z})_-$ & $-1$ & $1$ & $-1$ & $-1$ & class 1\\
\hline
$(E_{bTZ}M_{fT'})_-$ & $-1$ & $-1$ & $-1$ & $1$ & class 1\\
\hline
$(E_{bTZ}M_{bZ})_-$ & $-1$ & $-1$ & $1$ & $-1$ & class 2\\
\hline
$(E_{f}M_{bZ})_-$ & $1$ & $1$ & $1$ & $-1$ & class 2\\
\hline
$(E_{bTZ}M_f)_-$ & $-1$ & $-1$ & $1$ & $1$ & class 2\\
\hline
$(E_{bZ}M_{bT'Z})_-$ & $1$ & $-1$ & $-1$ & $-1$ & class 3\\
\hline
$(E_{bZ}M_f)_-$ & $1$ & $-1$ & $1$ & $1$ & class 3\\
\hline
$(E_fM_{bT'Z})_-$ & $1$ & $1$ & $-1$ & $-1$ & class 3\\
\hline
$(E_{bZ}M_{bT'})_-$ & $1$ & $-1$ & $-1$ & $1$ & class 4\\
\hline
$(E_{fTZ}M_{bT'})_-$ & $-1$ & $-1$ & $-1$ & $1$ & class 4\\
\hline
$(E_{fTZ}M_{bT'Z})_-$ & $-1$ & $-1$ & $-1$ & $-1$ & class 4\\
\hline
$(E_{bT}M_{bT'Z})_-$ & $-1$ & $1$ & $-1$ & $-1$ & class 4\\
\hline
$(E_{bT}M_{fZ})_-$ & $-1$ & $1$ & $1$ & $-1$ & class 4\\
\hline
$(E_{bZ}M_{fZ})_-$ & $1$ & $-1$ & $1$ & $-1$ & class 4\\
\hline
$(E_{bT}M_{bZ})_-$ & $-1$ & $1$ & $1$ & $-1$ & class 5\\
\hline
$(E_{bT}M_{fT'Z})_-$ & $-1$ & $1$ & $-1$ & $-1$ & class 5\\
\hline
$(E_{bTZ}M_{fT'Z})_-$ & $-1$ & $-1$ & $-1$ & $-1$ & class 5\\
\hline
$(E_{bTZ}M_{bT'})_-$ & $-1$ & $-1$ & $-1$ & $1$ & class 5\\
\hline
$(E_{fZ}M_{bT'})_-$ & $1$ & $-1$ & $-1$ & $1$ & class 5\\
\hline
$(E_{fZ}M_{bZ})_-$ & $1$ & $-1$ & $1$ & $-1$ & class 5\\
\hline
$(E_{bT}M_{bT'})_-$ & $-1$ & $1$ & $-1$ & $1$ & class 6\\
\hline
$(E_{bT}M_f)_-$ & $-1$ & $1$ & $1$ & $1$ & class 6\\
\hline
$(E_fM_{bT'})_-$ & $1$ & $1$ & $-1$ & $1$ & class 6\\
\hline
\end{tabular}
\caption{List of anomalous $Z_2\times\mc{T}$ symmetric $U(1)$ quantum spin liquids that have $\theta=0$ and have $Z_2$ acting as a charge conjugation at $\theta=\pi$. $T_E^2=1$ ($T^2_E=-1$) represents the case where $E$ is a Kramers singlet (doublet) under $\mc{T}$. $T'^2_M=1$ ($T'^2_M=-1$) represents the case where $M$ is a Kramers singlet (doublet) under $\mc{T}'$. $Z_{E,M}^2$ represents the result of acting the charge conjugation twice on $E$ and $M$, respectively. The last column lists the anomaly classes.} \label{table: app-Z2Z2T-conjugation Z2-theta=0-anomalous}
\end{table}

We will first show that the 15 states in Table \ref{table:Z2Z2T-simple Z2-theta=0-non-anomalous} and the 21 states in Table \ref{table: app-Z2Z2T-conjugation Z2-theta=0-nonanomalous} are anomaly-free, and give their constructions. Then we will show that the 9 states in Table \ref{table:Z2Z2T-simple Z2-theta=0-anomalous} and the 27 states in Table \ref{table: app-Z2Z2T-conjugation Z2-theta=0-anomalous} are anomalous.

Among the 36 anomaly-free states mentioned above, 26 of them have at least one of $E$ and $M$ being a trivial boson. These states clearly do not suffer from any anomaly, and they can be viewed as some gauged trivial insulators. The other 10 states,
\beq \label{eq: nontrivial-Z2Z2T}
\begin{split}
&E_{bT}M_f, E_{bT'}M_f, E_{bTT'}M_f, E_{fTT'}M_{b-},\\
&(E_{fT}M_{bT'})_-, (E_{bT}M_{fT'})_-, (E_{fZ}M_{bT'Z})_-,\\ &(E_{bTZ}M_{fZ})_-, (E_{fTZ}M_{bZ})_-, (E_{bZ}M_{fT'Z})_-,
\end{split}
\eeq
can be viewed as gauged free-fermion SPTs, which will be constructed below. To show that all other states are anomalous, as discussed in Sec. \ref{subsec: Z2Z2T-anomaly-strategy}, it is sufficient to show that $(E_{bT}M_{bT'})_-$ and $E_{bTT'}M_{b-}$ are anomalous.

The rest of this appendix is organized as follows. In Appendix \ref{app: Z2Z2T-free-fermion} we will construct the relevant free-fermion SPTs, which after gauging give rise to states in (\ref{eq: nontrivial-Z2Z2T}). Then we will show that $(E_{bT}M_{bT'})_-$ is anomalous in Appendix \ref{app: (EbTMbT')-}, and that $E_{bTT'}M_{b-}$ is anomalous in Appendix \ref{app: EbTT'Mb-}.

\subsection{Constructions of the relevant free-fermion SPTs} \label{app: Z2Z2T-free-fermion}

This subsection gives the construction of the free-fermion SPTs corresponding to states in (\ref{eq: nontrivial-Z2Z2T}). All these free-fermion topological insulators have two Dirac cones on the surface, and the surface Hamiltonian can be written as
\beq
H=\sum_{i=1}^2\psi_i^\dag(-i\partial_x\sigma_x-i\partial_y\sigma_z)\psi_i
\eeq
The differences among these states are in the symmetry assignments. Denote $\psi=(\psi_1,\psi_2)^T$, in all cases there is a $U(1)$ symmetry under which $\psi\rightarrow e^{i\theta}\psi$. We will also assign time reversal and $Z_2$ symmetries to these states, such that there is no symmetry-allowed fermion bilinear term that can open a gap on the surface. Then we will show the bosonic monopoles of these topological insulators have the desired nontrivial properties, using the method in Ref. \onlinecite{Borokhov2002} (reviewed in Appendix \ref{app: SPTs}).

Let us start with the case where $Z_2$ does not act as a charge conjugation, and give the construction of free-fermion SPTs corresponding to $E_{bT}M_f$ and $E_{fTT'}M_{b-}$. For the corresponding SPT of $E_{bT}M_f$, let the symmetries be assigned as
\beq
\begin{split}
&\mc{T}:\ \psi\rightarrow \sigma_y\psi^\dag\\
&Z_2:\ \psi\rightarrow \tau_y\psi\\
&\mc{T}':\ \psi\rightarrow\sigma_y\tau_y\psi^\dag
\end{split}
\eeq

Clearly the action of $\mc{T}$ and $Z_2$ commute on the fermion, so after gauging the fermion will become the $M_f$. Now we check the symmetry quantum number of the $E$, which is the monopole of $\psi$. Using state-operator correspondence, this is equivalent to checking the properties of the two zero-energy state in the presence of a $2\pi$ flux background with one of the two zero modes being occupied. Denote these zero modes by $f_1$ and $f_2$, which are related to $\psi_1$ and $\psi_2$, respectively. And denote the state with a $2\pi$ flux background and none of the zero modes being occupied by $|0\ra$. Because $\mc{T}$ will flip the charge but keep the flux, under $\mc{T}$,
\beq
\begin{split}
&f_1^\dag|0\ra\rightarrow f_1f_1^\dag f_2^\dag|0\ra=f_2^\dag|0\ra\\
&f_2^\dag|0\ra\rightarrow f_2f_1^\dag f_2^\dag|0\ra=-f_1^\dag|0\ra
\end{split}
\eeq
Notice the above transformations can be modified by an unimportant phase factor. This means $E$ will be a Kramers doublet under $\mc{T}$. In fact, here the particle-hole-like $\mc{T}$ is enough to protect the Dirac cones, and it is shown this is sufficient to show that $E$ is a Kramers doublet under $\mc{T}$.\cite{WangSenthil2014} Under $\mc{T}'$,
\beq
\begin{split}
&f_1^\dag|0\ra\rightarrow f_2f_1^\dag f_2^\dag|0\ra=-f_1^\dag|0\ra\\
&f_2^\dag|0\ra\rightarrow-f_1f_1^\dag f_2^\dag|0\ra=-f_2^\dag|0\ra
\end{split}
\eeq
This means $E$ will be a Kramers singlet under $\mc{T}'$. This is consistent with that $\mc{T}'$ is not enough to protect the Dirac cones. Therefore, after gauging this state indeed becomes $E_{bT}M_f$.

From this, $E_{bT'}M_f$ can be constructed similarly, and $E_{bTT'}M_f$ can be obtained by combining $E_{bT}M_f$ and $E_{bT'}M_f$. To obtain $E_{fTT'}M_{b-}$, let the symmetries be assigned as
\beq
\begin{split}
&\mc{T}:\ \psi\rightarrow\sigma_y\psi\\
&Z_2:\ \psi\rightarrow\tau_x\psi
\end{split}
\eeq
One can show that this state becomes $E_{fTT'}M_{b-}$ after gauging by using state-operator correspondence, but an alternative point of view can be obtained by considering this state as a descendant of the corresponding SPT of the $SO(3)\times\mc{T}$ symmetric $E_{f\half}M_{b\half}$, which has been described in details in Appendix \ref{app: SPTs}. To see it, denote the three generators of $SO(3)$ by $S_x$, $S_y$ and $S_z$, and denote the generator of $\mc{T}$ by $t$. Now break the $SO(3)\times\mc{T}$ to $Z_2\times\mc{\tilde T}$, where the $Z_2$ is generated by $\exp\left(iS_x\pi\right)$, and $\mc{\tilde T}$ is generated by $\exp\left(iS_y\pi\right)\cdot t$. It is straightforward to check that the descendant state is $E_{f\tilde T\tilde T'}M_{b-}$

Now we turn to states with $Z_2$ acting as a charge conjugation. Let us start with the example corresponding to $(E_{fTZ}M_{bZ})_-$. This state is actually the gauged version of the free-fermion topological insulator in class CII, which has been discussed in Ref. \onlinecite{Potter2016} (but using a different notation as here). The time reversal and charge conjugation symmetries are assigned as
\beq
\begin{split}
&\mc{T}:\ \psi\rightarrow\sigma_y\psi\\
&Z_2:\ \psi\rightarrow\tau_y\psi^\dag
\end{split}
\eeq
Now let us first examine the $\mc{T}'$ action on the two states corresponding to monopoles, whose action on $\psi$ is
\beq
\mc{T}':\ \psi\rightarrow\sigma_y\tau_y\psi^\dag
\eeq
This is the same $\mc{T}'$ action as in the corresponding SPT of $E_{bT}M_f$, so $f_1^\dag|0\ra$ and $f_2^\dag|0\ra$ correspond to Kramers singlets under $\mc{T}'$.

Next let us examine the $Z_2$ action on $f_1^\dag|0\ra$ and $f_2^\dag|0\ra$. Notice the $2\pi$ flux background is converted into a $-2\pi$ flux background, which also has two zero modes. Using the method in Ref. \onlinecite{Potter2016}, we argue that for such systems with two symmetry-protected Dirac cones, the value of charge-conjugation squared on the neutral monopole is the same as the value of charge-conjugation squared on the Dirac fermions. The simplest way to see this is to notice that these states have $\theta=2\pi$. For the state corresponding to a trivial insulator, which has $\theta=0$, the monopole has trivial quantum number, that is, the value of charge-conjugation squared is $1$. Then one can tune $\theta$ by $2\pi$ to get a state corresponding to the topological insulator. In intermediate process of tuning $\theta$, the time reversal symmetry is generically broken. But the existence of such nontrivial topological insulator implies at the end the system will have time reversal symmetry when $\theta$ becomes $2\pi$. On the other hand, this process will not break the charge conjugation symmetry. Then according to the Witten effect,\cite{WITTEN1979} the original $(1,1)$ dyon now becomes the $(0,1)$ monopole, and this new monopole has the value of charge-conjugation squared to be $-1$.

Ref. \onlinecite{Potter2016} obtained this result by considering tuning $\theta$ of the $U(1)$ gauge theory. It can actually also be obtained directly by using state-operator correspondence. Recall it has been shown in Appendix \ref{app: SPTs} that $M_1^\dag\sim f_2^\dag|\tilde 0\ra$ and $M_2^\dag\sim -f_1^\dag|\tilde 0\ra$, where $|\tilde 0\ra$ is the state with $-2\pi$ flux background and neither zero mode being occupied. Under charge conjugation, both charge and flux will be occupied. So under a convention of phase factors we can choose $|0\ra\rightarrow f_1^\dag f_2^\dag|\tilde 0\ra$ under charge conjugation, then the monopole operators transform as
\beq
\begin{split}
&M_1\sim f_1^\dag|0\ra\rightarrow\tilde f_2\tilde f_1^\dag \tilde f_2^\dag|\tilde 0\ra=-\tilde f_1^\dag|\tilde 0\ra\sim M_2^\dag\\
&M_2\sim f_2^\dag|0\ra\rightarrow -\tilde f_1\tilde f_1^\dag \tilde f_2^\dag|\tilde 0\ra=-\tilde f_2^\dag|\tilde 0\ra\sim -M_1^\dag
\end{split}
\eeq
Again, unimportant $U(1)$ phase factors have been suppressed. The above transformation shows that the value of charge-conjugation squared is indeed $-1$ on the monopoles. Therefore, after gauging this state becomes $(E_{fTZ}M_{bZ})_-$.

Next we turn to the free-fermion SPT corresponding to $(E_{fT}M_{bT'})_-$, where the assignments of the time reversal and charge conjugation symmetries are
\beq
\begin{split}
&\mc{T}:\ \psi\rightarrow\sigma_y\psi\\
&Z_2:\ \psi\rightarrow\psi^\dag
\end{split}
\eeq
Now we check the whether these monopoles are Kramers doublets under $\mc{T}'$, whose action on $\psi$ is
\beq
\mc{T}':\ \psi\rightarrow\sigma_y\psi^\dag
\eeq
This is the same as the $\mc{T}$ action in the SPT corresponding to $E_{bT}M_f$, so here the monopole must be a Kramers doublet under $\mc{T}'$.

As for the value of charge-conjugation squared on the monopole,
\beq
\begin{split}
&f_1^\dag|0\ra\rightarrow f_1\tilde f_1^\dag\tilde f_2^\dag|\tilde 0\ra\sim\tilde f_2^\dag|\tilde 0\ra\sim M_1^\dag\\
&f_2^\dag|0\ra\rightarrow f_2\tilde f_1^\dag\tilde f_2^\dag|\tilde 0\ra\sim-\tilde f_1^\dag|\tilde 0\ra\sim M_2^\dag
\end{split}
\eeq
so the value of charge-conjugation squared is $1$ for the monopoles. Therefore, after gauging this state becomes $(E_{fT}M_{bT'})_-$.

Finally, for the free-fermion SPT corresponding to $(E_{fZ}M_{bT'Z})_-$, the assignments of the time reversal and charge conjugation symmetries are
\beq
\begin{split}
&\mc{T}:\ \psi\rightarrow\sigma_y\tau_y\psi\\
&Z_2:\ \psi\rightarrow\tau_y\psi^\dag
\end{split}
\eeq
This state has the same $Z_2$ action as the one giving rise to $(E_{fTZ}M_{bZ})_-$, and the same $\mc{T}'$ action as the one giving rise to $(E_{fT}M_{bT'})_-$. In light of the previous discussion, the monopole should have charge-conjugation squared to be $-1$ and be a Kramers doublet under $\mc{T}'$. Therefore, after gauging this state becomes $(E_{fZ}M_{bT'Z})_-$.

By now the constructions of free-fermion SPTs corresponding to states in (\ref{eq: nontrivial-Z2Z2T}) are given. Before we finish this subsection, we make some remarks on free-fermion topological insulators with time reversal and a unitary $Z_2$ symmetry, which may or may not act as a charge conjugation. In all these free-fermion topological insulators, the surface will always have an even number of Dirac cones in order to have $\theta=0$. If it has 4 Dirac cones on the surface, the corresponding $U(1)$ quantum spin liquid in general has a trivial monopole. In order to get a $U(1)$ quantum spin liquid with nontrivial monopole, the corresponding free-fermion SPT should have only 2 surface Dirac cones. We have actually exhausted all possible free-fermion topological insulators with two surface Dirac cones, and they only give the 10 $U(1)$ quantum spin liquids in (\ref{eq: nontrivial-Z2Z2T}) that have nontrivial monopole. On the other hand, if a nontrivial fermionic topological insulator is equivalent to a bosonic SPT, the monopole must also be trivial.\cite{Wang2014,WangSenthil2014} Thus, if a state with fermionic charge and nontrivial monopole is anomaly-free and distinct from the above three (such as $(E_fM_{bT'})_-$), it implies the existence of an intrinsically interacting fermionic SPT, which is a nontrivial fermionic SPT that cannot be realized by free-fermions and is not equivalent to a bosonic SPT.\cite{Wang2016b,Cheng2017a} These SPTs are very interesting, but in the discussion below we will argue that no other spin liquid state with fermionic charge and nontrivial monopole is anomaly-free, which means no such intrinsically interacting fermionic SPT can be found with $U(1)$, time reversal and $Z_2$ symmetries (even if the fermions are allowed to transform projectively under these symmetries).

\subsection{Anomaly of $(E_{bT}M_{bT'})_-$} \label{app: (EbTMbT')-}

In this subsection, by using the same logic as before, we will examine the anomaly of $(E_{bT}M_{bT'})_-$. That is, we will consider the corresponding SPT from the perspective of $M_{bT'}$, and check whether it is possible to have a consistent surface topological order. However, unlike in the case of $(E_{bZ}M_{bZ})_-$, where we can reach the conclusion by quite general arguments, here we need to examine some rather detailed properties of the surface states.

Again, we will first condense the bound state of two $M_{bT'}$ on the surface, which reduces the surface symmetry to $\mc{T}'\times Z_2$. We would like to point out that there are two possibilities for the surface at this point: the surface can either be a simple superfluid, or the surface superfluid has to coexist with another anomalous topological order. The latter happens if the bulk is still a nontrivial SPT even if the bulk symmetry is broken down to $\mc{T}'\times Z_2$, in which case there must be another anomalous surface topological order of an SPT with $\mc{T}'\times Z_2$ symmetry, if this symmetry is to be preserved.

For the case of a simple superfluid surface, we can show there is inconsistency of the surface topological order. As for the more complicated case where the surface superfluid has to coexist with another anomalous topological order, we need the properties of the surfaces of 3D bosonic SPTs with $\mc{T}'\times Z_2$ symmetry, which, to the best of our knowledge, are lacking in the literature. So we will first discuss the classification of such SPTs, and then show that  there will still be some inconsistency even for the more complicated case. This leads us to concluding that there is no such corresponding SPTs that can become $(E_{bT}M_{bT'})_-$ after gauging, which means $(E_{bT}M_{bT'})_-$ is anomalous.

Before the detailed discussion on this problem, let us first collect a few useful tools that will be applied repeatedly below.

\begin{itemize}

\item[1.]

In a topological order, a particle always has the same topological spin as its anti-particle. That is, denote $a^{-1}$ as the anti-particle of $a$, then
\beq \label{eq: TO-general-1}
\theta_a=\theta_{a^{-1}}
\eeq

\item[2.]

Suppose $a$ and $b$ are two anyons in a topological order. Suppose $c$ is a possible fusion product of $a$ and $b$, that is, $a\times b=N_{ab}^cc+\cdots$ with $N_{ab}^c$ the fusion multiplicity. Then $(R^{ab}_c)^2=\frac{\theta_c}{\theta_a\theta_b}$, where $(R^{ab}_c)^2$ is the mutual braiding between $a$ and $b$ when their fusion product is fixed to be $c$, and $\theta_{a}$ is the topological spin of $a$. In the case of Abelian topological order, the mutual braiding between $a$ and $b$ can be simply denoted as $\theta_{a,b}$, and the above formula becomes
\beq \label{eq: TO-general-2}
\theta_{a,b}=\frac{\theta_c}{\theta_a\theta_b}
\eeq

\item[3.]

Braiding and fusion commute in a topological order. For example, in an Abelian topological order,
\beq \label{eq: TO-general-3}
\theta_{ab,c}=\theta_{a,c}\theta_{b,c}
\eeq

\item[4.]

If the time reversal partner of an anyon $a$ is $b$, and $c=a\times b$ is the bound state of $a$ and $b$, then the Kramers parity of $c$ is determined by
\beq \label{eq: TO-general-4}
T_c^2=\theta_{c}
\eeq

\end{itemize}

\subsubsection*{Simple superfluid}

We start our discussion with the case of a simple superfluid surface. This superfluid has vortices with vorticity quantized in units of $\pi$, and the minimal trivial vortex is the $4\pi$ vortex. We will condense these $4\pi$ vortices to restore the full symmetry of the surface and get a symmetric gapped surface topological order, where the $U(1)$ charge is quantized in units of 1/2. The $\pi$ vortices and the $2\pi$ vortices will remain gapped, and we will denote the $\pi$ vortex by $v$ and the $2\pi$ vortex by $E_{bT}$.

The vortex condensation above will in general generate a charge-1/2 boson, which we denote by $X$. Physically, $X$ is the $2\pi$ vortex of the $4\pi$ vortices. This $X$ should be an Abelian boson. To see this, let us go to the energy scale below which we can consider only the $4\pi$ vortices that are to be condensed. Limiting ourselves below this energy scale should not change the topological data of $X$. Although the $\pi$ and $2\pi$ vortices are nontrivial, below this energy scale they do not play any role. Then because the $4\pi$ vortices are trivial bosons, $X$, the $2\pi$ vortex of the $4\pi$ vortices, is expected to be a simple Abelian boson.

The bound state of two $X$'s can be combined with $M_{bT'}$ to generate a charge-neutral bosonic excitation, which will be denoted by $N$. The bound state of two $N$'s have trivial braiding with all other excitations, so this should be viewed as a local excitation. Therefore, the particle contents of the surface theory can be written as
\beq
\{1,X,N,X^{-1},v,E_{bT},v^{-1}\}\times\{1,M_{bT'}\}
\eeq
The various bound states of these excitations are understood to be implicitly displaced. Also, $X^{-1}$ represents the excitation that can fuse with $X$ into the trivial vacuum, $1$, which does not carry any quantum number, and it should be distinguished from $XM_{bT'}^\dag$.

Below we will determine the braiding, fusion and symmetry assignments of these excitations. Without loss of generality, we will always take $v$ to be neutral, because its charge can always be cancelled by binding it with certain amount of $X$ and $M_{bT'}$.

We start with braiding. For self-braiding, the only uncertain part is about $v$: it can either be Abelian or non-Abelian. Now we turn to mutual braiding. The mutual braiding within the charge sector (built up with $X$ and $N$) is always trivial. For the vortex sector (built up with $v$ and $E_{bT}$), the braiding between $v$ and $E_{bT}$ is trivial because $v$ is neutral and $E_{bT}$ is the remnant of the $2\pi$ vortex, and the braiding between $v$ and $v^{-1}$ is to be determined.

The mutual braiding between the charge sector and the vortex sector can be determined in the following way. Because condensing $X$ will make the surface back into the simple superfluid, we can view $X$ as something that is condensed in the superfluid phase. From the Meissner effect we know the vortices come with certain fluxes in the superfluid phase, and this combined object of vortices and fluxes should be local with respect to the $X$ condensate. Therefore, the mutual braiding between the vortices themselves with the $X$ condensate is the {\it conjugate} of the charge-flux Aharonov-Bohm phase. This tells us
\beq \label{eq: mutual-braiding-(EbTMbT')-}
\theta_{X,v}=-i,
\
\theta_{X,E_{bT}}=
\theta_{N,v}=-1,
\
\theta_{N,E_{bT}}=1
\eeq
Notice the third relation comes from the identification $N=X^2 M_{bZ}^\dag$ and that $X^2$ is condensed in the superfluid ($N$ is not condensed in the superfluid, so we cannot say $\theta_{v,N}=1$ because $N$ is neutral). The mutual braiding listed here will be used repeatedly below.

Now we turn to fusion. Most fusion rules can be determined by the charge and vorticity assignment:
\beq \label{eq: fusion-(EbTMbT')-simple}
X\times X=NM_{bZ},
\
N\times N=1,
\
E_{bT}\times E_{bT}=1
\eeq

However, there is some flexibility for $v$. For example, even if the $v$ is Abelian, we can have either $v\times v=E_{bT}$ or $v\times v=E_{bT}\times N$. Of course when $v$ is non-Abelian, we must have $v\times v=E_{bT}+E_{bT}N$ (with potential fusion multiplicities suppressed). Because $N$ is a boson that is local with respect to $E_{bT}$, $N$ must have trivial braiding with $v$ in this non-Abelian case, otherwise $v$ and its anti-particle would have opposite topological spins, which violates (\ref{eq: TO-general-1}) and is thus disallowed. However, $\theta_{v,N}=-1$. This implies $v$ cannot be non-Abelian. {\footnote{In contrast, if the local particle is a fermion, $v$ can still be non-Abelian.}} Furthermore, for the same reason, the fusion rule for $v$ has to be
\beq
v\times v=E_{bT}
\eeq


Finally we discuss the symmetry assignment. The $U(1)$ charges of these excitations are clear: $X$ carries half charge, $M_{bT'}$ carries unit charge, and other excitations are neutral. The assignment of the $\mc{T}$ and $Z_2$ symmetries is constrained by some general rules. First, $\mc{T}$ should conjugate the topological spins of the excitations, and $Z_2$ should keep their topological spins. Second, the behavior of charge and vorticity under various symmetries is fixed. For example, because the charge flips and the vorticity does not change under $\mc{T}$, $\mc{T}$ will take $v$ to either $v$ or $vN$. Because of the fusion rule, $v\times v=E_{bT}$, $v$ cannot become $v$ under time reversal. That is, $v$ should go to $vN$.

Putting all these constraints together, there is actually not too much freedom for this topological order. One choice is the $Z_4$ gauge theory listed in Table \ref{table: symmetry-(EbTMbT')-simple},{\footnote{To fit into the usual notation of a $Z_4$ gauge theory, one can take the $Z_4$ charge to be $X$, and take the $Z_4$ flux to be $Xv$.}} and the only thing that one can modify on top of this state is to change the values of $Z_2^2$ for $X$ and $v$, and the value of $T'^2$ for $X$. Notice in all these cases, $N$ is a Kramers doublet under $\mc{T}'$.

\begin{table}[h!]
\centering
\begin{tabular}{|c|c|c|c|c|}
\hline
& $X$ & $N$ & $v$ & $E_{bT}$\\
\hline
$U(1)$ & $\frac{1}{2}$ & 0 & 0 & 0\\
\hline
$\mc{T}$ &$X^{-1}$ & $N$ & $vN$ & $E_{bT}$\\
\hline
$T^2$ & &$-1$& & $-1$\\
\hline
$Z_2$ & $X^{-1}$ & $N$ & $v^{-1}$ & $E_{bT}$\\
\hline
$Z_2^2$ & $\pm 1$ & $1$ & $\pm 1$ & $1$\\
\hline
$\mc{T}'$ & $X$ & $N$ & $v^{-1}N$ & $E_{bT}$\\
\hline
$T'^2$ & $\pm 1$ & $-1$ &  & $\pm 1$\\
\hline
\end{tabular}
\caption{Symmetry assignments of the surface topological order from the simple superfluid surface of the corresponding SPT of $(E_{bT}M_{bT'})_-$. The first row lists all nontrivial excitations, from which the symmetry assignments on all their bound states can be inferred. The second row lists the charges of these excitations under $U(1)$. The third row lists the time reversal partners of these excitations. The fourth row lists the values of $T^2$ of these excitations, with empty entries representing that $T^2$ is not well-defined. The fifth row lists the $Z_2$ partners of these excitations. The sixth row lists the values of $Z_2^2$ of these excitations. The seventh row lists the $\mc{T}'$ partners of these excitations. And the last row lists the values of $T'^2$ of these excitations.} \label{table: symmetry-(EbTMbT')-simple}
\end{table}

The above theory is actually inconsistent. To see this, notice that since the $\mc{T}$ partner of $v$ is $vN$ and $\theta_{v,N}=-1$, $v$ must be a semion or anti-semion, so that $\mc{T}$ can conjugate the topological spin of $v$. This means the bound state of $v$ and its $\mc{T}'$ partner, $v^{-1}N$, is a boson, so this bound state should be a Kramers singlet under $\mc{T}'$ according to (\ref{eq: TO-general-4}). However, as discussed above, this bound state is $v\times v^{-1}N=N=X^2M_{bT'}^\dag$, which is a Kramers doublet under $\mc{T}'$. This contradiction shows that the simple superfluid surface is inconsistent.

\subsubsection*{Superfluid coexisting with another anomalous topological order}

Now we turn to the case where the surface superfluid has to exist with another anomalous topological order. As discussed earlier, this happens if the bulk remains to be a nontrivial SPT when the bulk symmetry is also reduced to $\mc{T}'\times Z_2$. We will call such SPTs the reduced bulk SPTs. To complete the discussion, we need the properties of 3D bosonic SPTs with this symmetry, which will be discussed below.\\

{\it\quad\quad\quad\underline{3D bosonic SPTs with $\mc{T}'\times Z_2$ symmetry}}\\

Notice there exists local Kramers doublet under $\mc{T}'$, so more precisely, the symmetry group of the surface superfluid should be denoted by $Z_4^{T'}\times Z_2$. From group cohomology, the classification of such SPTs is $\mathbb{Z}_2^3$, and there should still be another SPT whose surface is $efmf$, and this SPT is beyond group cohomology. So we propose that the complete classifications of such SPTs is $\mathbb{Z}_2^4$. This proposal is further supported by the classification of 3D bosonic SPTs with $Z_4^P\times Z_2$ symmetry, where $Z_4^P$ is a reflection symmetry that results in a trivial action when acted four times. SPTs with $Z_2\times Z_4^T$ are believed to have the same classification of SPTs with $Z_2\times Z_4^P$, where the latter are classified by $\mathbb{Z}_2^4$.\cite{Cheng2017}

What are the surface topological orders of the other three root states? We show that they can all be $Z_2$ topological orders, and they are denoted by $(eT'_imT'_i)_{T'}$, $eZmZ$ and $eT'_imZ$. Below we explain the properties of these states.

The first state, $(eT'_imT'_i)_{T'}$, is protected by $\mc{T}'$ alone, and in this state $e$ and $m$ are exchanged under $\mc{T}'$. Furthermore, $\mc{T}'$ acting on $e$ or $m$ four times gives $-1$ (the meaning of $T'_i$). The action of $Z_2$ is trivial on both $e$ and $m$. In fact, this state is the descendant of $(eCmC)_{T'}\epsilon$ when $e^4$ is condensed without breaking time reversal, and $(eCmC)_{T'}\epsilon$ is a surface state of the bosonic topological insulator made of Kramers bosons.

To justify that this is a legitimate surface state of an SPT protected by $\mc{T}'$, we need to show this descendant state is still a nontrivial SPT. That is, the bosonic topological insulator made of Kramers bosons is still a nontrivial SPT when double charge is condensed without breaking time reversal. This can be seen by checking the time reversal domain wall of this state. Consider breaking $\mc{T}'$ in two opposite ways in the two sides of a 2D domain wall, while keeping a unitary $Z_2$ symmetry intact through the entire system (this unitary $Z_2$ is just the symmetry generated by acting the generator of $\mc{T}'$ twice). Notice before the double charges are condensed, the time reversal domain wall of this bosonic topological insulator is the elementary bosonic integer quantum Hall state,\cite{Senthil2012} because it has $\sigma_{xy}=2e^2/h$.\cite{VishwanathSenthil2013} When the double charge is condensed, this bosonic integer quantum Hall state becomes the Levin-Gu state.\cite{Levin2012a,Xu2013} That is to say, the time reversal domain wall of this descendant state is a Levin-Gu state. But this cannot happen unless the original $\mc{T}'$ symmetric system is a nontrivial SPT. {\footnote{More precisely, this is because the Levin-Gu state is the root state of 2D $Z_2$ SPTs, which means there is no 2D $Z_2$ symmetric short-range entangled bosonic state that becomes the Levin-Gu state when it is stacked with its time reversal partner.}}

The above discussion establishes that there is a 3D bosonic SPT protected by $\mc{T}'$, and its surface can be $(eT'_imT'_i)_{T'}$. For our purposes, it will be useful to think about what state this SPT becomes if the symmetry is enhanced to the full $((U(1)\rtimes Z_2)\times \mc{T})/Z_2=(O(2)\times\mc{T})/Z_2$ symmetry of $M_{bT'}$, which can be obtained by considering what the surface topological order becomes when the symmetry is enhanced. Because the $\pi$ rotation of the $U(1)$ is locked with acting $\mc{T}'$ twice, $e$ and $m$ should carry half charge under $U(1)$. The entire symmetry assignment of this surface state is shown in Table \ref{table: symmetry-(eT'imT'i)T'}.

\begin{table}[h!]
\centering
\begin{tabular}{|c|c|c|c|}
\hline
& $e$ & $m$ & $\epsilon\equiv em^\dag$\\
\hline
$U(1)$ & $\frac{1}{2}$ & $\frac{1}{2}$ & 0\\
\hline
$\mc{T}$ &$m^\dag$ & $e^\dag$ & $\epsilon$\\
\hline
$T^2$ & & & $-1$\\
\hline
$Z_2$ & $e^\dag$ & $m^\dag$ & $\epsilon$\\
\hline
$Z_2^2$ & $1$ & $1$ & $1$\\
\hline
$\mc{T}'$ & $m$ & $e$ & $\epsilon$\\
\hline
$T'^2$ & $\pm i$ & $\pm i$ & $1$\\
\hline
\end{tabular}
\caption{Symmetry assignments of the surface topological order of the corresponding SPT of $(E_{fT}M_{bT'})_-$. The first row lists all nontrivial excitations, from which the symmetry assignments on all their bound states can be inferred. The second row lists the charges of these excitations under $U(1)$. The third row lists the time reversal partners of these excitations. The fourth row lists the values of $T^2$ of these excitations, with empty entries representing that $T^2$ is not well-defined. The fifth row lists the $Z_2$ partners of these excitations. The sixth row lists the values of $Z_2^2$ of these excitations. The seventh row lists the $\mc{T}'$ partners of these excitations. And the last row lists the values of $T'^2$ of these excitations, with $\pm i$ standing for that $T'^4=-1$ on the excitation.} \label{table: symmetry-(eT'imT'i)T'}
\end{table}

It is straightforward to check that this state can be the surface topological order of the corresponding SPT of $(E_{fT}M_{bT'})_-$ (viewed from the perspective of $M_{bT'}$). This observation implies that if the reduced bulk SPT is $(eT'_imT'_i)_{T'}$, we can reduce the surface state into a simple superfluid by coupling the original SPT to the corresponding SPT of $(E_{fT}M_{bT'})_-$. Because coupling $(E_{bT}M_{bT'})_-$ and $(E_{fT}M_{bT'})_-$ should result in $(E_fM_{bT'})_-$, if we can show in the scenario of a simple superfluid surface, no SPT made of $M_{bT'}$ can become $(E_fM_{bT'})_-$ after gauging, it is sufficient to show no SPT with $(eT'_imT'_i)_{T'}$ reduced bulk SPT can become $(E_{bT}M_{bT'})_-$ after gauging.

Next we turn to explaining the properties of $eZmZ$. In this surface $Z_2$ topological order, the $Z_2$ symmetry acting on $e$ or $m$ twice gives a $-1$ phase factor, and $\mc{T}'$ acts trivially on $e$ and $m$. Again, we need to show that the bulk with this surface topological order is a nontrivial SPT with $\mc{T}'\times Z_2$ symmetry, or equivalently, that $eZmZ$ is anomalous with $\mc{T}'\times Z_2$ symmetry.

The way to understand the anomaly of $eZmZ$ is to relate it to $eCmC$, the surface state of a nontrivial SPT with $\mc{T}'\times U(1)$ symmetry. Notice this symmetry is not $(U(1)\rtimes\mc{T})/Z_2$, the symmetry of charged Kramers bosons. In particular, the $\pi$ rotation of the $U(1)$ here is not locked with acting time reversal twice, as in the latter symmetry. There is a bosonic topological insulator at $\theta=2\pi$ with $\mc{T}'\times U(1)$ symmetry, and this is independent of the presence of local bosons that are Kramers doublets under $\mc{T}'$. The surface state of this bosonic topological insulator is $eCmC$, which means a $Z_2$ topological order with both $e$ and $m$ carrying half charge under $U(1)$, and $\mc{T}'$ acts trivially on $e$ and $m$. Again, the time reversal domain wall of this bosonic topological insulator will have the character of an elementary bosonic integer quantum Hall state. Breaking the $U(1)$ symmetry down to $Z_2$ results in the $eZmZ$ state, and as before, the time reversal domain wall of this descendant 3D state will be a Levin-Gu state, which is disallowed unless the 3D bulk is a nontrivial SPT. This shows that $eZmZ$ is still anomalous with $\mc{T}'\times Z_2$ symmetry, and we also see $\mc{T}'$ plays a role in protecting this state, even though it appears to act on $e$ and $m$ trivially.{\footnote{We point out that in order to establish that $eZmZ$ is anomalous, it is actually important that $\mc{T}'$ acts trivially on both $e$ and $m$, so that this $eZmZ$ state can be viewed as the descendant of $eCmC$ after the $U(1)$ is broken down to $Z_2$.}}

Again, it will be useful for us to understand what this state becomes when the symmetry is enhanced to the full $(O(2)\times\mc{T})/Z_2$ symmetry of $M_{bT'}$. Because only the $Z_2$ acts nontrivially in this state, when the symmetry is enhanced, $U(1)$ and $\mc{T}'$ should still act trivially. This means there is a corresponding SPT of $(E_bM_{bT'})_-$ that becomes $eZmZ$ when the symmetry is broken down to $\mc{T}'\times Z_2$.  Therefore, when the reduced bulk SPT is $eZmZ$, we can always cancel the anomaly of $eZmZ$ by coupling the original putative SPT to this corresponding SPT of $(E_bM_{bT'})_-$.

Lastly, we turn to discuss $eT'_imZ$, which is a $Z_2$ topological order with $e$ having $T'^4=-1$ and $m$ having $Z_2$ squaring to $-1$. As argued in Appendix \ref{app: remarks on TR}, in order for $e$ to have $T'^4=-1$, $\mc{T}'$ should attach $M_{bT'}$, a local Kramers doublet under $\mc{T}'$, to $e$. This surface state is anomalous because when two $Z_2$ fluxes are inserted in the system, both $m$ and $\epsilon$ will see a nontrivial phase factor when moving around it. But this is a local process, so this $-1$ phase factor must be compensated by something nucleated in the $Z_2$ flux when it is inserted, and this nucleated object must be $e$. That is to say, a local process produces an object with $T^4=-1$, which is absent in the system by assumption. Therefore, this state is anomalous.

What does $eT'_imZ$ become when the symmetry is enhanced to the full $(O(2)\times\mc{T})/Z_2$ symmetry? It turns out the symmetry cannot be enhanced to the full $(O(2)\times\mc{T})/Z_2$ symmetry. In other words, there is no bosonic SPT with $(O(2)\times\mc{T})/Z_2$ symmetry whose descendant can be $eT'_imZ$. However, we still know that, because the $\pi$ rotation of the $U(1)$ is locked with acting $\mc{T}'$ twice, when the $4\pi$ vortices are condensed and the full symmetry is recovered on the surface, $e$ should carry half charge of $U(1)$, and $m$ carries integer charge but has $Z_2$ squaring to $-1$.

For completeness, we also mention that if the reduced bulk SPT is $efmf$, when the full symmetry is recovered, none of $e$ and $m$ get nontrivial action under the symmetry, and its anomaly can be cancelled by just coupling it with another $efmf$ state with the full symmetry, which after gauging becomes $(E_bM_{bT'})_-$.

The above discussion implies that, in the scenario where the surface superfluid has to coexist with another anomalous topological order, in order to show there is no SPT made of $M_{bT'}$ that can become $(E_{bT}M_{bT'})_-$ after gauging, it is sufficient to show:

\begin{itemize}

\item[1.] In the scenario of a simple superfluid surface, there is no SPT made of $M_{bT'}$ that can become $(E_fM_{bT'})_-$.

\item[2.] In the scenario of $eT'_imZ$ reduced bulk SPT, there is no SPT made of $M_{bT'}$ that can become $(E_{bT}M_{bT'})_-$.

\item[3.] In the scenario of $eT'_imZ$ reduced bulk SPT, there is no SPT made of $M_{bT'}$ that can become $(E_fM_{bT'})_-$.

\end{itemize}

Below we will show these three statements in turn.\\

{\it\quad\quad\quad\underline{Simple superfluid surface for $(E_fM_{bT'})_-$}}\\

We start from the first statement. The similar argument as before implies that the surface topological order in this case will be a $Z_4$ topological order:
\beq
\{1,X,N,X^{-1},v,E_f,v^{-1}\}\times\{1,M_{bT'}\}
\eeq
with the symbols standing for parallel excitations as before, and a difference is that here $E_f$ is a fermion with no symmetry fractionalization in terms of $\mc{T}$ and $Z_2$.

In this case, most of the topological data (braiding and fusion) will be the same as in the simple superfluid for $(E_{bT}M_{bT'})_-$. The only difference is that, because $E_f$ is a fermion, the fusion product of two $v$'s should be $E_fN$, otherwise the antiparticle of $v$ will have a different topological spin from $v$. That is,
\beq
v\times v=E_fN
\eeq
Because $E_fN$ is still a fermion, $v$ has topological spin $\theta_v=\exp\left(i\left(\frac{\pi}{4}+\frac{\pi n}{2}\right)\right)$ with $n$ an integer. In order for $\mc{T}$ to conjugate $\theta_v$, the $\mc{T}$ partner of $v$ has to be $vX$ or $vX^{-1}$. This then changes the charge of $v$, which is not legitimate.

This establishes the first statement: in the scenario of a simple superfluid surface, there is no SPT made of $M_{bT'}$ that can become $(E_fM_{bT'})_-$.\\

{\it\quad\quad\underline{$eT'_imZ$ reduced bulk SPT for $(E_{bT}M_{bT'})_-$}}\\

Now we turn to the second statement, where the surface superfluid has to coexist with the anomalous topological order, $eT'_imZ$. In the superfluid phase of the surface, the vortices and anyons of this anomalous topological order are distinct. One of their distinctions is that the vorticies carry (logarithmically) expensive energy cost, while the anyons have a finite energy gap. The fusion rules of $e$ and $m$ in the superfluid phase is that
\beq \label{eq: fusion-(EbTMbT')-complicated}
e\times e=M_{bT'},\ m\times m=1
\eeq
These seemingly innocuous fusion rules actually deserve further clarification. Purely in terms of topological sectors, there is no difference in $M_{bT'}$ and $1$ in the right hand sides of these fusion rules, because they can be turned into each other by binding an $M_{bT'}^\dag$, a local excitation. However, in terms of symmetry quantum numbers, $M_{bT'}$ and $1$ are of course different. It makes sense to talk about fusing two $e$ particles or two $m$ particles that have the same symmetry quantum number (with the difference due to attaching an $M_{bZ}$ resolved), and the above equations should be interpreted as the fusion rules of fusing two {\it identical} $e$'s or $m$'s. This distinction is important when we try to determine the fusion rules of $e$ and $m$ after the surface gets into the symmetric topologically ordered phase, where the right hand sides of these fusion rules can potentially be modified by multiplying something condensed in the superfluid phase.

When $4\pi$ vortices are condensed and the full symmetry is restored, $e$ will remain deconfined because it will carry half charge under $U(1)$ ($m$ will of course also remain deconfined because it is neutral). Then the surface topological order can be written as
\beq
\{1,X,N,X^{-1},v,E_{bT},v^{-1},e,m\}\times\{1,M_{bT'}\}
\eeq
where the symbols have parallel meaning as in the case of a simple superfluid. As before, the various bound states of these excitations are understood to be implicitly displayed.

Again, $X$ will be an Abelian boson that carries half charge, and $N$ will be a boson that is a Kramers doublet under $\mc{T}'$. Because condensing $X$ makes the surface back to the superfluid coexisting with $eT'_imZ$, the superfluid phase can again be viewed as a condensate of $X$. This has two important consequences.

First, the mutual braiding in (\ref{eq: mutual-braiding-(EbTMbT')-}) still holds, and $\{e,m\}$ will have trivial braiding with $\{X,N,X^{-1}\}$. Then $N^2$ is trivial, and (\ref{eq: fusion-(EbTMbT')-simple}) still holds. However, the mutual braiding between $v$ and $\{e,m\}$ is undetermined at this point. We only know, because $\theta_{E_{bT},m}=1$, we should have $\theta_{v,m}=\pm1$.

Second, no condensate in the superfluid can be multiplied to the right hand sides of the fusion rules in (\ref{eq: fusion-(EbTMbT')-complicated}) without violating charge conservation, which means that (\ref{eq: fusion-(EbTMbT')-complicated}) is also the right fusion rules for $e$ and $m$ in the topologically ordered phase.

Now let us determine the fusion rule of two $v$'s. Just from charge conservation and vorticity conservation, there seem to be many possible fusion products of two $v$'s:
\beq
\begin{split}
&E_{bT},E_{bT}N,E_{bT}m,E_{bT}mN,\\
&E_{bT}eX^{-1},E_{bT}eX^{-1}N,E_{bT}\epsilon X^{-1},E_{bT}\epsilon X^{-1}N
\end{split}
\eeq
However, based on some topological arguments, in the following we can rule out all of them but $E_{bT}m$ and $E_{bT}mN$.

To see this, first notice that for each Abelian anyon, it has a unique braiding phase factor with all fusion products of two $v$'s. Because of this, some of the above cannot simultaneously be the fusion products. For example, because $E_{bT}$ and $E_{bT}m$ have different braiding phase factors with $e$, they cannot both be the fusion products of two $v$'s. Using this, we see the fusion products can be one of the four possible pairs
\beq
\begin{split}
&\{E_{bT},E_{bT}N\},
\
\{E_{bT}m,E_{bT}mN\},\\
&\{E_{bT}eX^{-1},E_{bT}eX^{-1}N\},
\
\{E_{bT}\epsilon X^{-1},E_{bT}\epsilon X^{-1}N\}
\end{split}
\eeq

However, because $N$ is a boson with $\theta_{v,N}=-1$, within each pair at most one of them can be the fusion product of two $v$'s, otherwise the anti-particle of $v$ will have opposite topological spin of $v$, and (\ref{eq: TO-general-1}) is violated. This means $v$ has to be Abelian again, and the entire topological order is Abelian. Together with $\theta_{E_{bT},v}=1$, this also implies $E_{bT}N$ cannot be a fusion product of two $v$'s. Furthermore, because $\theta_{X,v}=-i$, the last four excitations cannot be fusion products of two $v$'s, otherwise $v$ would have an antiparticle that has a different topological spin from itself.

In fact, $E_{bT}$ cannot be a fusion product of two $v$'s, either. This is because $\theta_{E_{bT},e}=-1$, which means $\theta_{v,e}=\pm i$ and $\theta_{v,e^2}=-1$, if $v\times v=E_{bT}$. However, $e^2$ is local, so $\theta_{v,e^2}=1$. This contradiction implies that $E_{bT}$ cannot be the fusion product of two $v$'s.

So we are finally left with two possibilities:
\beq
v\times v=E_{bT}m
\eeq
or
\beq
v\times v=E_{bT}mN
\eeq
In the first possibility, $\theta_{v,m}=1$, while $\theta_{v,m}=-1$ in the second possibility.

Now we discuss the symmetry assignment. Recall that in the superfluid phase, the {\it topological sectors} of $e$ and $m$ are transformed as
\beq
\mc{T}':\quad
e\rightarrow eM_{bT'},
\quad
m\rightarrow m
\eeq
under $\mc{T}'$, and
\beq
Z_2:\quad
e\rightarrow e,
\quad
m\rightarrow m
\eeq
or
\beq
Z_2:\quad
e\rightarrow eM_{bT'},
\quad
m\rightarrow m
\eeq
under $Z_2$. Notice that the above expressions only imply the action on the topological sectors, and in this case there are two possibilities for the $Z_2$ transformation.

In the symmetric topologically ordered phase, the right hand sides of these transformation rules can be multiplied by something condensed in the superfluid phase. Also, according to the values of $T'^2$ of $e$ and $m$ and the fact that the $\pi$ rotation of $U(1)$ is locked with acting $\mc{T}'$ twice, $e$ carries half charge and $m$ carries no charge under $U(1)$ in the topologically ordered phase. In order for $\mc{T}'$ to maintain the $U(1)$ charge, and for $Z_2$ to flip the $U(1)$ charge, the unique choice for the transformation rules of $e$ and $m$ in the topologically ordered phase are
\beq \label{eq: em-T'-eTi'mZ}
e\rightarrow eN,
\quad
m\rightarrow m
\eeq
under $\mc{T}'$, and
\beq \label{eq: em-Z2-eTi'mZ-1}
e\rightarrow eM_{bT'}^\dag N,
\quad
m\rightarrow m
\eeq
or
\beq \label{eq: em-Z2-eTi'mZ-2}
e\rightarrow eM_{bT'}^\dag,
\quad
m\rightarrow m
\eeq
under $Z_2$, corresponding to the two possible $Z_2$ transformations in the superfluid phase, respectively.

To further determine the symmetry assignments, it is crucial to determine the symmetry actions on $v$. Let us consider the $Z_2$ action on $v$ first, and we will begin with the case where $v\times v=E_{bT}m$ and $e\rightarrow eM_{bT'}^\dag N$ and $m\rightarrow m$ under $Z_2$. Notice in this case $\theta_{v,m}=1$. As the $Z_2$ flips both the charge and the vorticity, the options for the $Z_2$ partner of $v$ are:
\beq
\begin{split}
&v^{-1}, v^{-1}N, v^{-1}m, v^{-1}mN,\\
&v^{-1}eX^{-1}, v^{-1}eX^{-1}N,
v^{-1}\epsilon X^{-1}, v^{-1}\epsilon X^{-1}N
\end{split}
\eeq
All of them except $v^{-1}$ and $v^{-1}m$ can be ruled out, because in those cases the $Z_2$ action cannot keep the topological spin of $v$ invariant. If the $Z_2$ partner of $v$ is $v^{-1}$, then under the $Z_2$ transformation $\theta_{e,v}$ becomes
\beq
\begin{split}
&\theta_{eNM_{bT'}^\dag,v^{-1}}\\
=&\theta_{eNM_{bT'}^\dag,v^3}
=\theta_{e,v^{3}}\theta_{N,v^{3}}\\ =&\theta_{e,v}\theta_{e,v^2}\theta_{N,v}\theta_{N,v^2}=\theta_{e,v}\theta_{N,v} =-\theta_{e,v}
\end{split}
\eeq
which is disallowed. In the above we have used (\ref{eq: TO-general-2}) and (\ref{eq: TO-general-3}). The above discussion implies that the $Z_2$ partner of $v$ can only be $v^{-1}m=vE_{bT}$.

Just from that $\mc{T}$ flips the charge but keeps the vorticity, the options for the $\mc{T}$ partner of $v$ are:
\beq
\begin{split}
&v,vN,vm,vmN,\\
&veX^{-1},veX^{-1}N,
v\epsilon X^{-1},v\epsilon X^{-1}N
\end{split}
\eeq
The last four can be ruled out due to some topological reasons. For example, suppose $v$ becomes $veX^{-1}$ under $\mc{T}$. In order for $\mc{T}$ to conjugate the topological spin of $v$, using the (\ref{eq: mutual-braiding-(EbTMbT')-}) and $\theta_{e,v}=\pm 1$, the topological spin of $v$ must be $\theta_v=\exp\left(i\left(\pm\frac{\pi}{4}+n\pi\right)\right)$, with $n$ an integer. Then the bound state  of two $v$'s must be a fermion, contradicting to the fusion rule $v\times v=E_{bT}m$. This means $veX^{-1}$ cannot be the $\mc{T}$ partner of $v$. Similar arguments show that $veX^{-1}N$, $v\epsilon X^{-1}$ and $v\epsilon X^{-1}N$ cannot be the $\mc{T}$ partner of $v$, either.

If the $\mc{T}$ partner of $v$ is $vm$, then $v$ is either a boson or a fermion, and $\theta_{vvm}=1$. But $\theta_{vvm}$ should be locked to the Kramers parity of $vvm=E_{bT}$,  which is $-1$. This contradiction implies that $vm$ cannot be the $\mc{T}$ partner in this case. The time reversal partner of $v$ can also not be $vNm$, because in this case under time reversal $\theta_{v,e}$ becomes
\beq
\theta_{vNm,eM_{bT'}^\dag}=-\theta_{v,e}\neq \theta_{v,e}^*
\eeq
which is disallowed. This is actually another reason why $vm$ cannot be the time reversal partner of $v$, because $vm$ cannot conjugate $\theta_{v,e}$, either. If the time reversal partner of $v$ is $v$, then $v$ has a well defined Kramers parity, but its $Z_2$ partner, $v^{-1}m=vE_{bT}$, has an opposite Kramers parity. This is disallowed, otherwise $Z_2$ and time reversal cannot commute for the spin liquid. To see this, suppose $v$ transforms under time reversal as
\beq
v_i\rightarrow T_{ij}v_j
\eeq
and under $Z_2$ as
\beq
v_i\rightarrow C_{ij}\tilde v_j
\eeq
Its $Z_2$ partner, $\tilde v$, transforms under time reversal as
\beq
\tilde v_i\rightarrow \tilde T_{ij}\tilde v_j
\eeq
Because $v_i^*M_{ij}v_j$ is a local operator that has no charge or vorticity, its Kramers parity should be $1$ and $Z_2$ should commute with time reversal on this operator, for any matrix $M$. This requires $T^*T=\pm 1$, $\tilde T^*\tilde T=\pm 1$, and $TC=e^{i\phi}C^*\tilde T$, with $\phi$ a phase. Taking all these together, we get
\beq
T^*T=C\tilde T^*\tilde TC^{-1}=\tilde T^*\tilde T
\eeq
That is, $v$ and $\tilde v$ should have the same Kramers parity. Notice to claim that for $v_i^*M_{ij}v_j$ the Kramers parity is $1$ and $Z_2$ commutes with $\mc{T}$, it is important that this operator is not only local, but also carries no charge or vorticity, otherwise by a gauge transformation its Kramers parity and the commutation relation can be changed.

So the time reversal partner can only be $vN$. Notice in this case $v$ must be a semion or anti-semion, because $\theta_{v,N}=-1$ and time reversal conjugates the topological spin of $v$. Then $\theta_{vvN}=1$, which means the Kramers parity of $vvN=E_{bT}mN$ is $1$, so $N$ has to be a Kramers doublet.

The above discussion implies that if $v\times v=E_{bT}m$ and the $Z_2$ action on $e$ and $m$ is given by (\ref{eq: em-Z2-eTi'mZ-1}), the $Z_2$ partner of $v$ can only be $vE_{bT}$, and the $\mc{T}$ partner of $v$ can only be $vN$. Using similar arguments, one can actually check that if $v\times v=E_{bT}mN$ and the $Z_2$ action on $e$ and $m$ is given by (\ref{eq: em-Z2-eTi'mZ-1}), the $Z_2$ partner of $v$ can still only be $vE_{bT}$, and the $\mc{T}$ partner of $v$ can only be $vN$. In both cases, the entire symmetry assignments are largely determined, as shown in Table \ref{table: symmetry-partner-(EbTMbT')-complicated-1-1}.

This surface state is actually problematic. To see this, notice under $\mc{T}$ the topological sector of $eX^{-1}v$ is invariant, so $eX^{-1}v$ has a well defined Kramers parity. However, under the $Z_2$ this topological sector becomes $eX^{-1}vE_{bT}$, which carries an opposite Kramers parity. As discussed above, this is disallowed, otherwise $\mc{T}$ and $Z_2$ cannot commute for the spin liquid.

\begin{table}[h!]
\centering
\begin{tabular}{|c|c|c|c|c|c|c|}
\hline
& $X$ & $N$ & $v$ & $E_{bT}$ & $e$ & $m$\\
\hline
$U(1)$ & $\frac{1}{2}$ & 0 & 0 & 0 & $\frac{1}{2}$ & $0$\\
\hline
$\mc{T}$ &$X^{-1}$ & $N$ & $vN$ & $E_{bT}$ & $eM_{bT'}^\dag$ & $m$ \\
\hline
$Z_2$ & $X^{-1}$ & $N$ & $vE_{bT}$ & $E_{bT}$ & $eNM_{bT'}^\dag$ & $m$\\
\hline
\end{tabular}
\caption{Symmetry assignments of the surface topological order if the $Z_2$ action on $e$ and $m$ is given by (\ref{eq: em-Z2-eTi'mZ-1}), for both the case with $v\times v=E_{bT}m$ and the case with $v\times v=E_{bT}mN$. The first row lists all nontrivial excitations, from which the symmetry assignments on all their bound states can be inferred. The second row lists the charges of these excitations under $U(1)$. The third row lists the time reversal partners of these excitations. And the fourth row lists the $Z_2$ partners of these excitations.} \label{table: symmetry-partner-(EbTMbT')-complicated-1-1}
\end{table}

The above discussion implies that if the $Z_2$ action on $e$ and $m$ is given by (\ref{eq: em-Z2-eTi'mZ-1}), the surface state of this SPT is problematic. Now we are only left with the case where the $Z_2$ action on $e$ and $m$ is given by (\ref{eq: em-Z2-eTi'mZ-2}). In this case, one can use similar method to constrain the rest of the symmetry assignments, and the resulting symmetry assignment is shown in Table \ref{table: symmetry-partner-(EbTMbT')-complicated-1-2}. There is also a problem of this topological order: the $Z_2$ partner of $v$ is $v^{-1}$, so it has a well defined value for charge-conjugation squared. However, its $\mc{T}$ partner has an opposite value of charge-conjugation squared, because the values of charge-conjugation squared for $N$ and $m$ are $1$ and $-1$, respectively. This contradicts the fact that $Z_2$ and $\mc{T}$ should commute for the spin liquid.

\begin{table}[h!]
\centering
\begin{tabular}{|c|c|c|c|c|c|c|}
\hline
& $X$ & $N$ & $v$ & $E_{bT}$ & $e$ & $m$\\
\hline
$U(1)$ & $\frac{1}{2}$ & 0 & 0 & 0 & $\frac{1}{2}$ & $0$\\
\hline
$\mc{T}$ &$X^{-1}$ & $N$ & $vm/vmN$ & $E_{bT}$ & $eM_{bT'}^\dag N$ & $m$ \\
\hline
$Z_2$ & $X^{-1}$ & $N$ & $v^{-1}$ & $E_{bT}$ & $eM_{bT'}^\dag$ & $m$\\
\hline
\end{tabular}
\caption{Symmetry assignments of the surface topological order if the $Z_2$ action on $e$ and $m$ is given by (\ref{eq: em-Z2-eTi'mZ-2}). The first row lists all nontrivial excitations, from which the symmetry assignments on all their bound states can be inferred. The second row lists the charges of these excitations under $U(1)$. The third row lists the time reversal partners of these excitations. And the fourth row lists the $Z_2$ partners of these excitations.} \label{table: symmetry-partner-(EbTMbT')-complicated-1-2}
\end{table}

To see this, suppose $\mc{T}$ acts as
\beq
v_i\rightarrow T_{ij}\tilde v_j
\eeq
and $Z_2$ acts as
\beq
v_i\rightarrow C_{ij}v_j^*,
\quad
\tilde v_i\rightarrow\tilde C_{ij}\tilde v_j^*
\eeq
where $T$, $C$ and $\tilde C$ are three unitary matrices (in fact, being invertible is enough for the following argument).

For an arbitrary matrix $M$, because $v_i^*M_{ij}v_j$ is a {\it neutral} local operator that carries {\it no vorticity}, $Z_2$ and time reversal should commute on it. Demanding the results of acting time reversal first and $Z_2$ later and the results of acting $Z_2$ first and time reversal later to be the same gives
\beq
(T\tilde C)^\dag M^*(T\tilde C)=(C^*T^*)^\dag M^*C^*T^*
\eeq
For this equation to be true for arbitrary $M$, we must have
\beq
T\tilde C=e^{i\phi}C^*T^*
\eeq
with $\phi$ a phase. Or equivalently,
\beq
\tilde C=e^{i\phi}T^{-1}C^*T^*
\eeq

Taking the complex conjugation on both sides yields $\tilde C^*=e^{-i\phi}(T^{-1})^*CT$. So
\beq \label{eq: same-C^2}
\tilde C\tilde C^*=T^{-1}C^*T^*(T^{-1})^*CT=T^{-1}C^*CT
\eeq
Notice because $v_i^*M_{ij}v_j$ is local and neutral, $Z_2$ squares to $1$ for it because the system is made of $M_{bZ}$. Combined with the discussion in Appendix \ref{app: remarks on TR}, the above equation implies that $\tilde C\tilde C^*=C^*C$. That is, $v$ and $\tilde v$ should have the same charge-conjugation squared.

Taking all these arguments together, the second statement is established: in the scenario of $eT'_imZ$ reduced bulk SPT, there is no SPT made of $M_{bT'}$ that can become $(E_{bT}M_{bT'})_-$.\\

{\it\quad\quad\underline{$eT'_imZ$ reduced bulk SPT for $(E_fM_{bT'})_-$}}\\

Finally, we turn to the last statement. Using similar arguments as before, the surface topological order can be written as
\beq
\{1,X,N,X^{-1},v,E_f,v^{-1},e,m\}\times\{1,M_{bT'}\}
\eeq
with the similar symbols representing parallel excitations as before.

As before, in this topological order, the fusion rules for $e$ and $m$ are still given by (\ref{eq: fusion-(EbTMbT')-complicated}). The symmetry assignments for $e$ and $m$ are such that $e$ carries half charge under $U(1)$, while $m$ carries no charge, and the other symmetry assignments for $e$ and $m$ are given by (\ref{eq: em-T'-eTi'mZ}) and (\ref{eq: em-Z2-eTi'mZ-1}), or (\ref{eq: em-T'-eTi'mZ}) and (\ref{eq: em-Z2-eTi'mZ-2}).

In this case, most of the topological data will be the same as the case with $eT'_imZ$ reduced bulk SPT for $(E_{bT}M_{bT'})_-$. The only difference is in the fusion product of two $v$'s. Modifying the arguments before while keeping in mind that now $E_f$ is a fermion, we find two possible fusion rules for two $v$'s
\beq
v\times v=E_fm
\eeq
or
\beq
v\times v=E_fNm
\eeq
In the first possibility, $\theta_{v,m}=-1$, while $\theta_{v,m}=1$ in the second possibility. In both cases the right hand side of the fusion rules are fermions, so the topological spin of $v$ must be $\theta_{v}=\exp\left(i\left(\frac{\pi}{4}+\frac{n\pi}{2}\right)\right)$, with $n$ an integer.

In this case, in order for $\mc{T}$ to keep the vorticity of $v$ and conjugate the topological spin of $v$, the $\mc{T}$ partner of $v$ can be one of the following:
\beq
veX^{-1},\ veX^{-1}N,\ v\epsilon X^{-1},\ v\epsilon X^{-1}N
\eeq
Which one of these can conjugate the topological spin of $v$ depends on the values of $\theta_v=\exp\left(i\left(\frac{\pi}{4}+\frac{n\pi}{2}\right)\right)$, $\theta_{e,v}=\pm 1$ and $\theta_{\epsilon,v}=\pm 1$. However, no matter which one of the above four excitations is the $\mc{T}$ partner of $v$, $\theta_{v,m}$ becomes $-\theta_{v,m}\neq \theta_{v,m}^*$. This means
there is no consistent symmetry assignment for this topological order.

This establishes the third statement: in the scenario of $eT'_imZ$ reduced bulk SPT, there is no SPT made of $M_{bT'}$ that can become $(E_fM_{bT'})_-$.

Taking all the arguments above together, we have established that $(E_{bT}M_{bT}')_-$ is anomalous with $Z_2\times\mc{T}$ symmetry. Notice unlike $(E_{bZ}M_{bZ})_-$, which is anomalous even if there is only the $Z_2$ symmetry, here both the $\mc{T}$ and $Z_2$ symmetries are responsible for the anomaly.

\subsection{Anomaly of $E_{bTT'}M_{b-}$} \label{app: EbTT'Mb-}

In this subsection we show the anomaly of $E_{bTT'}M_{b-}$, by showing that no SPT made of $M_{b-}$ will become $E_{bTT'}M_{b-}$ after gauging. As a reminder, in this case $Z_2$ does not act as a charge-conjugation, and it anticommutes with $\mc{T}$ on $M_{b-}$. As before, we will first condense double charge on the surface to get a surface superfluid, whose minimal trivial vortex is the $4\pi$ vortex. We will then proliferate these $4\pi$ vortices to restore the full symmetry and get a surface topological order. Again, there are two scenarios for the surface superfluid: it can either be a simple superfluid, or this superfluid has to coexist with another anomalous topological order. We will discuss these cases in turn.

\subsubsection*{Simple superfluid}

We begin with the case of a simple superfluid. Using similar argument as before, we see the symmetric surface topological order obtained by condensing $4\pi$ vortices can be written as
\beq
\{1,X,N,X^{-1},v,E_{bTT'},v^{-1}\}\times\{1,M_{b-}\}
\eeq
The symbols stand for parallel meanings as before, while now the $2\pi$ vortex is $E_{bTT'}$, a Kramers doublet under both $\mc{T}$ and $\mc{T}'$.

Similar arguments as before show that the braiding and fusion are similar to the surface state obtained from the simple superfluid for the corresponding SPT of $(E_{bT}M_{bT'})_-$, and the symmetry assignment is shown in Table \ref{table: symmetry-EbTT'Mb-simple}. In particular, in order for $E_{bTT'}=v\times v$ to be a Kramers doublet under both $\mc{T}$ and $\mc{T}'$, both of the $\mc{T}$ and $\mc{T}'$ partners of $v$ should be $vN$. Notice $Xv$ will become $XvM_{b-}^\dag$ under $\mc{T}$, and it is invariant under $Z_2$. Below we show this is disallowed.

\begin{table}[h!]
\centering
\begin{tabular}{|c|c|c|c|c|}
\hline
& $X$ & $N$ & $v$ & $E_{bTT'}$\\
\hline
$U(1)$ & $\frac{1}{2}$ & 0 & 0 & 0\\
\hline
$\mc{T}$ &$X^{-1}$ & $N$ & $vN$ & $E_{bTT'}$\\
\hline
$Z_2$ & $X$ & $N$ & $v$ & $E_{bTT'}$\\
\hline
$\mc{T}'$ & $X^{-1}$ & $N$ & $vN$ & $E_{bTT'}$\\
\hline
\end{tabular}
\caption{Symmetry assignments of the surface topological order from the simple superfluid surface of the corresponding SPT of $E_{bTT'}M_{b-}$. The first row lists all nontrivial excitations, from which the symmetry assignments on all their bound states can be inferred. The second row lists the charges of these excitations under $U(1)$. The third row lists the $\mc{T}$ partners of these excitations. The fourth row lists the $Z_2$ partners of these excitations. The fifth row lists the $\mc{T}'$ partners of these excitations.} \label{table: symmetry-EbTT'Mb-simple}
\end{table}

For notational simplicity, denote $Xv$ and $XvM_{b-}^\dag$ by $x$ and $y$, respectively. Suppose the $\mc{T}$ action is
\beq
\mc{T}:
x_i\rightarrow T_{ij}y_j,\
y_i\rightarrow\tilde T_{ij}x_j
\eeq
and the $Z_2$ action is
\beq
Z_2:
x_i\rightarrow Z_{ij}x_j,\
y_i\rightarrow\tilde Z_{ij}y_j
\eeq
with $T$, $\tilde T$, $Z$ and $\tilde Z$ four invertible matrices.

Because, for any matrix $M$, $x_i^*M_{ij}x_j$ is local and neutral, $Z_2$ and $\mc{T}$ should commute on this operator. This gives a condition
\beq
T\tilde Z=e^{i\phi_1}Z^*T
\eeq
Because, for any matrix $M$, $x_i^*M_{ij}y_j$ is local and carries charge-1, $Z_2$ and $\mc{T}$ should anti-commute on this operator, because the system is made of $M_{b-}$. This, together with the above condition, yields
\beq
\tilde TZ=e^{i\phi_2}\tilde Z^*\tilde T
\eeq
with $e^{i(\phi_1-\phi_2)}=-1$.

Combining these two equations, we get
\beq
Z=e^{i\phi_1}T^*\tilde Z^*T^{*-1}=e^{i\phi_2}\tilde T^{-1}\tilde Z^*\tilde T
\eeq
Using that $e^{i(\phi_1-\phi_2)}=-1$, the above equation yields
\beq
\tilde Z^*=-\tilde TT^*\tilde Z^*T^{*-1}\tilde T^{-1}
\eeq

Now notice $y_i^*M_{ij}y_j$ should have Kramers parity $1$ for any matrix $M$, this implies that $\tilde T^*T=e^{i\phi_3}$, or $\tilde TT^*=e^{-i\phi_3}$, where $\phi_3$ is another phase. Plugging this into the above equation yields
\beq
\tilde Z^*=-\tilde Z^*
\eeq
so $\tilde Z=0$. This is disallowed.

The above argument shows that in the scenario of a simple superfluid, there is no SPT made of $M_{b-}$ that can become $E_{bTT'}M_{b-}$ after gauging.

\subsubsection*{Superfluid coexisting with another anomalous topological
order}

Next we turn to the more complicated case where the superfluid has to coexist with another anomalous topological order. Again, this happens when the bulk remains as a nontrivial SPT when the bulk symmetry is also reduced. For this purpose, let us first clarify what the reduced symmetry is. The reduced symmetry group has a $Z_2^T$ subgroup, with an anti-unitary generator $t$ that satisfies $t^2=1$. It also has a $Z_2$ subgroup, with a unitary generator $g$ that satisfies $g^2=1$. However, $gt+tg=0$. Denote $t'=gt$, then $t'^2=gtgt=-1$, so $t'$ generates a $Z_4^{T'}$ symmetry. Notice $t'$ gets inverted when conjugated with both $t$ and $g$, that is, $gt'g=ggtg=tg=-gt=-t'$ and $tt't=tgtt=tg=-gt=-t'$. So we will denote this group by $D_4^T\equiv Z_4^{T'}\rtimes Z_2=Z_4^{T'}\rtimes Z_2^T$. Now it is also easy to get the full symmetry, which also has $U(1)$ charge conservation. In the gauge where $t^2$ and $g^2$ are fixed to be identity, because each charge-1 boson has $Z_2$ and $\mc{T}$ anti-commuting, the $\pi$ rotation of the $U(1)$ is locked with $t'^2$, and the full symmetry group can be written as $(U(1)\times D_4^T)/Z_2$.\\

{\it\quad\quad\quad\underline{3D bosonic SPTs with $D_4^T$ symmetry}}\\

Now let us discuss 3D bosonic SPTs with $D_4^T$ symmetry. By group cohomology, these SPTs are classified by $\mathbb{Z}_2^3$. There should be another root state, $efmf$, which is beyond group cohomology. So the full classification of these SPTs is expected to be $\mathbb{Z}_2^4$. Among the other root states, $eTmT$ should be one of them. The $U(1)$ symmetry can be added to $efmf$ and $eTmT$ with a trivial action, so reduced bulk SPTs with such anomalies can be easily cancelled, and we will ignore these two states from now on. We propose two other root states: $(eTT'_imT'_i)_{ZT'}\epsilon{TT'}$ and $(eT'_imT'_i)_{ZT'}$.

The root state $(eTT'_imT'_i)_{ZT'}\epsilon{TT'}$ is actually the descendant of the corresponding SPT of $E_{fTT'}M_{b-}$ (viewed from the perspective of $M_{b-}$), when the $U(1)$ symmetry is broken to its $Z_2$ subgroup generated by its $\pi$ rotation. The surface state of this corresponding SPT of $E_{fTT'}M_{b-}$ can be a $Z_2$ topological order, with symmetries assigned as in Table \ref{table: EfTT'Mb-}. These symmetry assignments can be derived from the corresponding SPT of $E_{fTT'}M_{b-}$ viewed from the perspective of $E_{fTT'}$, which is described in Appendix \ref{app: Z2Z2T-free-fermion}. Notice the fusion rules are $e\times e=m\times m=M_{b-}$.

\begin{table}[h!]
\centering
\begin{tabular}{|c|c|c|c|}
\hline
& $e$ & $m$ & $\epsilon=em^\dag$\\
\hline
$U(1)$ & $\frac{1}{2}$ & $\half$ & 0\\
\hline
$\mc{T}$ &$e^\dag$ & $m^\dag$ & $\epsilon$\\
\hline
$T^2$ & $-1$ &$1$& $-1$\\
\hline
$Z_2$ & $m$ & $e$ & $\epsilon$\\
\hline
$Z_2^2$ & & & $1$\\
\hline
$\mc{T}'$ & $m^\dag$ & $e^\dag$ & $\epsilon$\\
\hline
$T'^2$ & $\pm i$ & $\pm i$ & $-1$\\
\hline
\end{tabular}
\caption{Symmetry assignments of the $Z_2$ surface topological order of the bosonic SPT made of $M_{b-}$ that will become $E_{fTT'}M_{b-}$ after gauging. The first row lists all nontrivial excitations, from which the symmetry assignments on all their bound states can be inferred. The second row lists the charges of these excitations under $U(1)$. The third row lists the time reversal partners of these excitations. The fourth row lists the values of $T^2$ of these excitations, with empty entries representing that $T^2$ is not well-defined. The fifth row lists the $Z_2$ partners of these excitations. The sixth row lists the values of $Z_2^2$ of these excitations. The seventh row lists the $\mc{T}'$ partners of these excitations. And the last row lists the values of $T'^2$ of these excitations, with $\pm i$ standing for that $T'^4=-1$.} \label{table: EfTT'Mb-}
\end{table}

When the $U(1)$ symmetry is reduced to its $Z_2$ subgroup generated by its $\pi$ rotation, the symmetry becomes $D_4^T$. The resulting state is still anomalous, which can be seen by checking its $Z_4^{T'}$ domain wall. Consider breaking the $Z_4^{T'}$ symmetry in two different ways on the two sides of a 2D domain wall, while keeping a $Z_2$ subgroup generated by $t'^2$ intact across the entire system. Then this domain wall has a $Z_2\times Z_2$ symmetry, and by relating it to the corresponding SPT of $E_{fTT'}M_{b-}$, we see this time reversal domain wall is a Levin-Gu state (protected by the $Z_2$ generated by $t'^2$). The existence of this decorated domain wall shows the descendant state is still a nontrivial SPT with the $D_4^T$ symmetry.

Next we turn to $(eT'_imT'_i)_{ZT'}$, whose symmetry assignments are shown in Table \ref{table: (eT'imT'iZ)-TT'epsilonTZ}. Or more precisely, the symmetry assignments are
\beq
\begin{split}
&\mc{T}:\ e\rightarrow ie,\ m\rightarrow m\\
&Z_2: e\rightarrow m,\ m\rightarrow e
\end{split}
\eeq
Notice the fusion rules are $e\times e=m\times m=M_{b-}$.

\begin{table}[h!]
\centering
\begin{tabular}{|c|c|c|c|}
\hline
& $e$ & $m$ & $\epsilon=em^\dag$\\
\hline
$\mc{T}$ &$e$ & $m$ & $\epsilon$\\
\hline
$T^2$ & $1$ & $1$ & $1$\\
\hline
$Z_2$ & $m$ & $e$ & $\epsilon$\\
\hline
$Z_2^2$ & & & $1$\\
\hline
$\mc{T}'$ & $m$ & $e$ & $\epsilon$\\
\hline
$T'^2$ & $\pm i$ & $\pm i$ & $1$\\
\hline
\end{tabular}
\caption{Symmetry assignments of the $Z_2$ surface topological order $(eT'_imT'_iZ)_{TT'}\epsilon{TZ}$. The first row lists all nontrivial excitations, from which the symmetry assignments on all their bound states can be inferred. The second row lists the $\mc{T}$ partners of these excitations. The third row lists the values of $T^2$ of these excitations. The fourth row lists the $Z_2$ partners of these excitations. The fifth row lists the values of $Z_2^2$ of these excitations. The sixth row lists the $\mc{T}'$ partners of these excitations. And the last row lists the values of $T'^2$ of these excitations, with $\pm i$ standing for that $T'^4=-1$.} \label{table: (eT'imT'iZ)-TT'epsilonTZ}
\end{table}

To show that this state is anomalous, consider breaking the $D_4^{T}$ symmetry to $Z_2\times Z_2^T$, where the first $Z_2$ is generated by $t'^2$, and the $Z_2^T$ is generated by $t$. Notice in this case the $Z_2$ generated by $g$ is broken. This can be done, for example, by considering that $\mc{T}$ acts on the local boson as
\beq
M_1\rightarrow M_2,\
M_2\rightarrow M_1
\eeq
and $Z_2$ acts on the local boson as
\beq
M_1\rightarrow M_1,\
M_2\rightarrow-M_2
\eeq
Giving $M_1M_2$ a nonzero expectation value will break the $D_4^T$ to $Z_2\times Z_2^T$ in the above way.

With this reduced symmetry, this state becomes $eZmZ$, where the symbol $Z$ stands for half charge under the $Z_2$ (generated by $t'^2$). This reduced state is anomalous. To see this, define $\tilde t=t'^2\cdot t$, which generates another anti-unitary symmetry, $\tilde T$. The state $eZmZ$ can then be relabelled as $e\tilde T m\tilde T$, so it is anomalous. Enhancing the symmetry back to $D_4^T$ will not add Kramers doublet boson under $\tilde T$ to the system, so it will not remove this anomaly. This also shows that $(eT'_imT'_i)_{ZT'}$ is distinct from the previous $(eTT'_imT'_i)_{ZT'}\epsilon TT'$, because when the symmetry is reduced to $Z_2\times\mc{T}$ in the above way, the latter becomes $eZTmZ$, which is a distinct anomalous state.

For later purpose, let us now consider the $U(1)$ charge of $e$ and $m$ when the $4\pi$ vortices are condensed and the full symmetry is restored. Again, because the $\pi$ rotation of the $U(1)$ is locked with $t'^2$, $e$ and $m$ should carry half charge in the topological order.

The above discussion implies that, in the scenario where the surface superfluid has to coexist with another anomalous topological order, in order to show there is no SPT made of $M_{b-}$ that can become $E_{bTT'}M_{b-}$ after gauging, it is sufficient to show:

\begin{itemize}

\item[1.] In the scenario of a simple superfluid surface, there is no SPT made of $M_{b-}$ that can become $E_fM_{b-}$.

\item[2.] In the scenario of $(eT'_imT'_i)_{ZT'}$ reduced bulk SPT, there is no SPT made of $M_{b-}$ that can become $E_{bTT'}M_{b-}$.

\item[3.] In the scenario of $(eT'_imT'_i)_{ZT'}$ reduced bulk SPT, there is no SPT made of $M_{b-}$ that can become $E_fM_{b-}$.

\end{itemize}

Below we will show these three statements in turn.\\

{\it\quad\quad\quad\underline{Simple superfluid surface for $E_fM_{b-}$}}\\

We start with the first statement. Using similar arguments as before, the surface topological order in this case can be written as
\beq
\{1,X,N,X^{-1},v,E_f,v^{-1}\}\times\{1,M_{b-}\}
\eeq
The symbols stand for parallel meanings as in the previous cases, and now the $2\pi$ vortex, $E_f$, is a fermion and is a Kramers singlet under both $\mc{T}$ and $\mc{T}'$.

In this case, most of the topological data will be the same as in the case with simple superfluid surface for $E_{bTT'}M_{b-}$. The only difference is in the fusion rule of $v$. The similar arguments as before implies that the fusion rule of two $v$'s in this case is
\beq
v\times v=E_fN
\eeq
otherwise the antiparticle of $v$ will have an opposite topological spin as itself. Because the right hand side is a fermion, the topological spin of $v$ will be $\theta_v=\exp\left(i\left(\frac{\pi}{4}+\frac{n\pi}{2}\right)\right)$, with $n$ an integer.

Now let us consider the $\mc{T}$ partner of $v$. As $\mc{T}$ should keep the vorticity and flip the charge, there are two possible $\mc{T}$ partners of $v$: $v$ and $vN$. Because $\theta_{v}=\exp\left(i\left(\frac{\pi}{4}+\frac{n\pi}{2}\right)\right)$ and $\theta_{v,N}=-1$, neither of these options will conjugate $\theta_v$ under $\mc{T}$. So this is inconsistent.


This establishes the first statement: in the scenario of a simple superfluid surface, there is no SPT made of $M_{b-}$ that can become $E_fM_{b-}$.\\

{\it\quad\quad\underline{$(eT'_imT'_i)_{ZT'}$ reduced bulk SPT for $E_{bTT'}M_{b-}$}}\\

Next we turn to the second statement. In this case, similar arguments as before show the surface topological order can be written as
\beq
\{1,X,N,X^{-1},v,E_{bTT'},v^{-1},e,m\}\times\{1,M_{b-}\}
\eeq
with the symbols standing for parallel meanings as before.

Recall that both $e$ and $m$ will carry half charge under $U(1)$. By charge conservation and vorticity conservation, the possible fusion products of two $v$'s are
\beq
\begin{split}
&E_{bTT'}, E_{bTT'}N, E_{bTT'}\epsilon, E_{bTT'}\epsilon N\\
&E_{bTT'}eX^{-1}, E_{bTT'}eX^{-1}N,\\
&E_{bTT'}mX^{-1}, E_{bTT'}mX^{-1}N
\end{split}
\eeq

Similar arguments as before imply that there are two possible fusion rules for $v$:
\beq
v\times v=E_{bTT'}\epsilon
\eeq
or
\beq
v\times v=E_{bTT'}\epsilon N
\eeq
In both cases, the bound state of two $v$'s is fermionic, so the topological spin of $v$ should be $\theta_v=\exp\left(i\left(\frac{\pi}{4}+\frac{n\pi}{2}\right)\right)$, with $n$ an integer. In order for $\mc{T}$ to conjugate $\theta_v$, the $\mc{T}$ partner of $v$ should be one of
\beq
veX^{-1}, veX^{-1}N, vmX^{-1}, vmX^{-1}N
\eeq

From the symmetry actions on $e$ and $m$ in the superfluid phase, by multiplying something condensed in the superfluid to be consistent with charge conservation, we get the symmetry actions on $e$ and $m$ in the topologically ordered phase:
\beq
\begin{split}
&\mc{T}:\ e\rightarrow eM_{b-}^\dag N,\
m\rightarrow mM_{b-}^\dag N\\
&Z_2:\ e\rightarrow m,\
m\rightarrow e
\end{split}
\eeq
Combining this with that $\theta_{v,N}=-1$, we see no matter which one of the four excitations is the $\mc{T}$ partner of $v$, $\theta_{\epsilon,v}$ cannot be conjugated by $\mc{T}$, which is disallowed.

This establishes the second statement: in the scenario of $(eT'_imT'_i)_{ZT'}$ reduced bulk SPT, there is no SPT made of $M_{b-}$ that can become $E_{bTT'}M_{b-}$.\\

{\it\quad\quad\underline{$(eT'_imT'_i)_{ZT'}$ reduced bulk SPT for $E_fM_{b-}$}}\\

Finally we come to the last statement. Similar arguments as before imply the topological order can be written as
\beq
\{1,X,N,X^{-1},v,E_f,v^{-1},e,m\}\times\{1,M_{b-}\}
\eeq
with symbols standing for parallel meanings as before. Notice the $2\pi$ vortex, $E_f$, is a fermion, and it is a Kramers singlet under both $\mc{T}$ and $\mc{T}'$.

Most of the topological data can be easily determined using similar arguments as before:
\beq
\begin{split}
&\theta_{X,v}=-i, \theta_{X,E_{bTT'}}=-1, \theta_{v,e}=\pm 1, \theta_{v,m}=\pm 1,\\
&X\times X=NM_{b-}, N\times N=1, E_{bTT'}\times E_{bTT'}=1
\end{split}
\eeq
There are two possible fusion rules for $v$:
\beq
v\times v=E_f\epsilon
\eeq
or
\beq
v\times v=E_f\epsilon N
\eeq
In both cases, the bound state of two $v$'s is a boson, so $v$ is a boson, fermion, semion or anti-semion.

Knowing that $\mc{T}$ should flip the charge and keep the vorticity, the $\mc{T}$ partner of $v$ can be one of
\beq
\begin{split}
&v,vN,v\epsilon,v\epsilon N,\\
&veX^{-1},veX^{-1}N,vmX^{-1},vmX^{-1}N
\end{split}
\eeq

Because $v$ is a boson, fermion, semion, or anti-semion, and $\theta_{X,v}=-i$, $\theta_{v,e}=\pm 1$ and $\theta_{v,m}=\pm 1$, the last four options can be ruled out, because in those cases $\mc{T}$ will not conjugate $\theta_v$. Because the $\mc{T}$ partner of $e$ is $eM_{b-}^\dag N$, in order to conjugate $\theta_{e,v}$, the $\mc{T}$ partner of $v$ cannot be $v$ or $vN$. That is, the $\mc{T}$ partner of $v$ is either $v\epsilon$ or $v\epsilon N$.

If $v\times v=E_f\epsilon$ and the $\mc{T}$ partner of $v$ is $v\epsilon$, then $v\times v\epsilon=E_f$. Because $v$ and $v\epsilon$ are $\mc{T}$ partners and $E_f$ is a fermion, $E_f$ must be a Kramers doublet under $\mc{T}$, which contradicts the original assumption. The same reasoning rules out the possibility that $v\times v=E_f\epsilon N$ and the $\mc{T}$ partner of $v$ is $v\epsilon N$.

If $v\times v=E_f\epsilon$ and the $\mc{T}$ partner of $v$ is $v\epsilon N$, then $v\times v\epsilon N=E_fN$. This means $N$ is a Kramers doublet under $\mc{T}$. On the other hand, the $\mc{T}$ partner of $e$ is $eM_{b-}^\dag N$, so $N$ can be viewed as the bound state of $e$ and its $\mc{T}$ partner. This implies that $N$ is a Kramers singlet under $\mc{T}$, which leads to a contradiction. The same reasoning also rules out the possibility that $v\times v=E_f\epsilon N$ and the $\mc{T}$ partner of $v$ is $v\epsilon$.

Putting all these analysis together, we have established the last statement: in the scenario of $(eT'_imT'_i)_{ZT'}$ reduced bulk SPT, there is no SPT made of $M_{b-}$ that can become $E_fM_{b-}$.

Therefore, we have shown that no SPT made of $M_{b-}$ can become $E_{bTT'}M_{b-}$ after gauging, which means $E_{bTT'}M_{b-}$ is anomalous with $Z_2\times\mc{T}$ symmetry.

As discussed before, we have already shown all states in Table \ref{table:Z2Z2T-simple Z2-theta=0-anomalous} and Table \ref{table:Z2Z2T-conjugation Z2-theta=0-anomalous} are anomalous.

\section{$Z_2\times\mc{T}$ symmetric $U(1)$ quantum spin liquids with $\theta=\pi$ and $Z_2$ not acting as a charge conjugation} \label{app: Z2Z2T-simple}

In this appendix, we discuss $Z_2\times\mc{T}$ symmetric $U(1)$ quantum spin liquids with $\theta=\pi$ and $Z_2$ not acting as a charge conjugation.

As discussed in Sec. \ref{subsec: Z2Z2T-simple}, in this case the quantum numbers of the $\left(\half,1\right)$ dyon determine the phase. Since there is no nontrivial projective representation of the $Z_2\times\mc{T}$ symmetry on the $\left(\half,1\right)$ dyon, we expect only one state: $(E_{fTT'}M_f)_\theta$. In this state, the electric charge must be a Kramers doublet under both $\mc{T}$ and $\mc{T}'$, and $M$ (the $(0,2)$ dyon in this context) has $Z_2$ and $\mc{T}$ commuting with each other. One may wonder whether it is possible to have $Z_2$ and $\mc{T}$ anti-commuting with each other in this case. Below we show this is not possible.

Denote the $\left(\half,1\right)$ dyon by $D^{(+)}$, and its time reversal partner, the $\left(\half,-1\right)$ dyon, by $D^{(-)}$. Notice $M$ is a bound state of $D^{(+)}$ and $D^{(-)\dag}$. Suppose the $Z_2$ action on $D^{(+)}$ and $D^{(-)}$ is
\beq \label{eq: Z2-action}
Z_2: D^{(+)}_i\rightarrow Z_{ij}D^{(+)}_j,
\quad
D^{(-)}_i\rightarrow \tilde Z_{ij}D^{(-)}_j
\eeq
and the $\mc{T}$ action on $D^{(+)}$ and $D^{(-)}$ is
\beq \label{eq: T-action}
\mc{T}: D^{(+)}_i\rightarrow T_{ij}D^{(-)}_j,
\quad
D^{(-)}_i\rightarrow \tilde T_{ij}D^{(+)}_j
\eeq

For any matrix $M$, the operator $D_i^{(+)\dag}M_{ij}D_j^{(+)}$ is local, so the actions of $Z_2$ and $\mc{T}$ should commute on it. This gives the condition
\beq
(Z^*T)^\dag M^*(Z^*T)=(T\tilde Z)^\dag M^*(T\tilde Z)
\eeq
In order for this equation to be satisfied by any matrix $M$, we need to have
\beq
Z^*T=e^{i\phi_1}T\tilde Z
\eeq
or, equivalently,
\beq \label{eq: constraint-1}
Z=e^{-i\phi_1}T^*\tilde Z^*(T^*)^{-1}
\eeq

Consider the local operator $D_i^{(-)\dag}M_{ij}D_j^{(-)}$ in a similar way, we get the condition
\beq
\tilde Z^*\tilde T=e^{i\phi_2}\tilde T Z
\eeq
or, equivalently,
\beq \label{eq: constraint-2}
Z=e^{-i\phi_2}\tilde T^{-1}\tilde Z^*\tilde T
\eeq

On the other hand, acting time reversal twice on the local operator $D^{(+)*}_iM_{ij}D_j^{(+)}$ should results in a trivial action, which implies that
\beq
T^*\tilde T=e^{i\phi_3}
\eeq
or, equivalently,
\beq
\tilde T=e^{i\phi_3}(T^*)^{-1}
\eeq

Combining this equation and (\ref{eq: constraint-2}) yields
\beq
Z=e^{-i\phi_2}T^*\tilde Z^*(T^*)^{-1}
\eeq
Comparing this equation and (\ref{eq: constraint-1}) yields
\beq \label{eq: constraint-3}
e^{i\phi_1}=e^{i\phi_2}
\eeq

Now consider the $(0,2)$ dyon, which is represented as $D^{(+)}_iM_{ij}D^{(-)\dag}_j$. Acting on this operator by $Z_2$ and $\mc{T}$ with different orders using (\ref{eq: Z2-action}) and (\ref{eq: T-action}), and using the constraints (\ref{eq: constraint-1}), (\ref{eq: constraint-2}) and (\ref{eq: constraint-3}), we see that the $Z_2$ and $\mc{T}$ commute on $D^{(+)}_iM_{ij}D^{(-)\dag}_j$, which proves the assertion stated at the beginning of this appendix.

The above argument can also be applied to $O(2)\times\mc{T}$ symmetric $U(1)$ quantum spin liquids discussed in Sec. \ref{subsec: O(2)Z2T}. If such a spin liquid has $\theta=\pi$ and the improper $Z_2$ component not acting as a charge conjugation, then on $M$ the actions of $Z_2$ and $\mc{T}$ should commute.

\section{$U(1)$ quantum spin liquids with some other symmetries} \label{app: other symmetries}

In this appendix we briefly discuss $U(1)$ quantum spin liquids with some other symmetries. The classifications of these symmetry enriched $U(1)$ quantum spin liquids are quite complicated, which we leave for future work. In this appendix we only lay out the principle of enumerating the putative states and make some comments, but we will not attempt to finish the procedure of examining the anomalies.

\subsection{$O(2)\times\mc{T}$ symmetry} \label{subsec: O(2)Z2T}

First we consider the case where the $SO(3)\times\mc{T}$ symmetry is broken down to $O(2)\times\mc{T}\cong(U(1)\rtimes Z_2)\times\mc{T}$. Physically, here $U(1)$ can represent spin rotations around one axis, while the $Z_2$ transformation is a $\pi$ spin rotation around another axis perpendicular to this one. This case is rather complicated, and we do not attempt to complete the anomaly detection and determine the final classification. Instead, we will just give a way to systematically list all putative (possibly anomalous) states and make some comments.

The structure of projective representations of $O(2)\times\mc{T}$ is rich. On the electric charge, it is classified by $\mathbb{Z}_2^3$, and the three root projective representations physically correspond to having half charge under the $U(1)$ subgroup, being a Kramers doublet under $\mc{T}$ and being a Kramers doublet under $\mc{T}'$ (the anti-unitary symmetry whose generator is the product of the generators of $Z_2$ and $\mc{T}$). If $\theta=0$, then on the magnetic monopole the projective representations are classified by $\mathbb{Z}_2^2$, which physically correspond to having half charge under the $U(1)$ subgroup and having the discrete $Z_2$ anti-commuting with $\mc{T}$. So at $\theta=0$, if the discrete $Z_2$ symmetry does not act as a charge conjugation, there are $3\times 2^3\times 2^2=96$ putative states. We will not write down the long list of all these states, since it is straightforward and not particularly illuminating. Notice some of these are descendant states of an $SO(3)\times\mc{T}$ symmetric state. Because the anomaly argument for the $SO(3)\times\mc{T}$ symmetric states should also apply to these $O(2)\times\mc{T}$ symmetric states, their anomalies can be determined immediately.

At $\theta=\pi$, there are only two putative states, $(E_{fTT'}M_f)_\theta$ and $(E_{fTT'}M_f)_{\theta\half}$, which are the descendants of $(E_{fT}M_{f})_\theta$ and $(E_{fT}M_{f})_{\theta\half}$ with $SO(3)\times\mc{T}$ symmetry, respectively. In the former state the $\left(\half,1\right)$ dyon carries integer $U(1)$ charge, while in the latter it carries half charge. As discussed in Sec. \ref{sec: anomalies}, the former state is anomaly-free while the latter is anomalous even with $O(2)$ symmetry. Here the actions of $Z_2$ and $\mc{T}$ commute on $M$ (the $(0,2)$ dyon in this context), and it is shown in Appendix \ref{app: Z2Z2T-simple} that, in this case, it is impossible to have the actions of $Z_2$ and $\mc{T}$ anti-commuting on $M$.

In the case where the discrete $Z_2$ symmetry acts as a charge conjugation, the possible fractional quantum numbers on the electric charge have a structure of $\mathbb{Z}_2^3$: having half charge under the $U(1)$ subgroup, being a Kramers doublet under $\mc{T}$, and having charge conjugation squaring to $-1$. If $\theta=0$, the possible fractional quantum numbers on the magnetic monopole also have a structure of $\mathbb{Z}_2^2$: being a Kramers doublet under $\mc{T}'$, and having charge conjugation squaring to $-1$. So there are $3\times 2^3\times 2^2=96$ such putative states. At $\theta=\pi$, there are two putative states: $(E_{fT}M_{fT'})_{\theta-}$ and $(E_{fT}M_{fT'})_{\theta-Z}$. In the former, the $\left(\half,1\right)$ dyon carries a linear representation of the symmetry, while this dyon has charge conjugation squaring to $-1$ for the latter. This former state is anomaly-free, and it can be obtained by equipping its $Z_2\times\mc{T}$ symmetric cousin with a further $U(1)$ symmetry. The latter is anomalous, because even if the symmetry is broken to $Z_2\times\mc{T}$ it is still anomalous.

We finish this subsection by briefly commenting on a few models that realize $O(2)\times\mc{T}$ symmetric $U(1)$ quantum spin liquids. Ref. \onlinecite{Motrunich2002, HermeleFisherBalents2004, Motrunich2005} studied a couple of different lattice models that realize three dimensional $U(1)$ quantum spin liquid phases with $O(2)\times\mc{T}$ symmetry, and the particular phase realized in these works is $(E_{b}M_{b\frac{1}{2}})_-$. Ref. \onlinecite{Levin2006} constructed two models of $O(2)\times\mc{T}$ symmetric $U(1)$ quantum spin liquids, where one of them has a bosonic monopole and the other has a fermionic monopole. These two states are $(E_bM_b)_-$ and $(E_bM_f)_-$, respectively.

\subsection{$Z_2\times Z_2\times\mc{T}$ symmetry}

Now consider the case where the $SO(3)\times\mc{T}$ symmetry is broken down to $Z_2\times Z_2\times\mc{T}$, where these two $Z_2$'s can represent $\pi$ spin rotations around two perpendicular axes. It is known that the projective representations (on the electric charge) of $Z_2\times Z_2\times\mc{T}$ symmetry are classified by $\mathbb{Z}_2^4$, where two of the four $\mathbb{Z}_2$'s are descendants of the projective representations of $SO(3)\times\mc{T}$, and the other two $\mathbb{Z}_2$'s come from the nontrivial interplay between $Z_2\times Z_2$ and time reversal. More precisely, each of the two $Z_2$'s together with time reversal can form a new anti-unitary symmetry, and the other two root nontrivial projective representations can be viewed as having Kramers doublets under such new anti-unitary symmetries. If $\theta=0$, on the magnetic monopole the projective representations are classified by $\mathbb{Z}_2^3$, and the nontrivial root projective representations physically correspond to the two $Z_2$'s and $\mc{T}$ anti-commuting.

If $\theta=\pi$, the phase is determined by the $\left(\half,\pm1\right)$ dyons, which has one nontrivial projective representation that corresponds to that the two $Z_2$ symmetries anti-commute. The state with the two $Z_2$ symmetries commuting is a descendant of the $SO(3)\times\mc{T}$ symmetric $(E_{fT}M_f)_\theta$, so it is still anomaly-free. The state with the two $Z_2$ symmetries anti-commuting is a descendant of the $SO(3)\times\mc{T}$ symmetric $(E_{fT}M_f)_{\theta\half}$, which we conjecture is still anomalous with $Z_2\times Z_2\times\mc{T}$ symmetry.

The descendants of the 15 non-anomalous spin liquid states with $SO(3)\times\mc{T}$ still remain non-anomalous and distinct, and we conjecture all the anomalous states remain anomalous even if the symmetry is broken to $Z_2\times Z_2\times\mc{T}$. This is of course just a partial classification, because, on the one hand, states labelled by the other projective representations of $Z_2\times Z_2\times\mc{T}$ should be taken into account, and on the other hand, states where $Z_2\times Z_2$ can act as charge conjugation should also be considered. This is not attempted in this paper.

\subsection{$Z_2\times Z_2$ symmetry}

Parallel considerations as in Sec. \ref{subsec: SO(N)} can be applied to the case with $Z_2\times Z_2$ symmetry. Here the projective representations are only classified by $\mathbb{Z}_2$, and the nontrivial projective representation is the descendant of the projective representation of $SO(3)$. Therefore, the descendants of $E_bM_b$ and $E_{b\frac{1}{2}}M_b$ will remain distinct and non-anomalous when the symmetry is broken from $SO(3)$ to $Z_2\times Z_2$. We conjecture that the descendant of $E_{b\frac{1}{2}}M_{b\frac{1}{2}}$ remains anomalous. Again, to have a complete classification, states where $Z_2\times Z_2$ permutes fractional excitations should be taken into account. We do not attempt it here.

\end{appendix}

\bibliography{SEU(1)QSL}

\begin{thebibliography}{86}%
\makeatletter
\providecommand \@ifxundefined [1]{%
 \@ifx{#1\undefined}
}%
\providecommand \@ifnum [1]{%
 \ifnum #1\expandafter \@firstoftwo
 \else \expandafter \@secondoftwo
 \fi
}%
\providecommand \@ifx [1]{%
 \ifx #1\expandafter \@firstoftwo
 \else \expandafter \@secondoftwo
 \fi
}%
\providecommand \natexlab [1]{#1}%
\providecommand \enquote  [1]{``#1''}%
\providecommand \bibnamefont  [1]{#1}%
\providecommand \bibfnamefont [1]{#1}%
\providecommand \citenamefont [1]{#1}%
\providecommand \href@noop [0]{\@secondoftwo}%
\providecommand \href [0]{\begingroup \@sanitize@url \@href}%
\providecommand \@href[1]{\@@startlink{#1}\@@href}%
\providecommand \@@href[1]{\endgroup#1\@@endlink}%
\providecommand \@sanitize@url [0]{\catcode `\\12\catcode `\$12\catcode
  `\&12\catcode `\#12\catcode `\^12\catcode `\_12\catcode `\%12\relax}%
\providecommand \@@startlink[1]{}%
\providecommand \@@endlink[0]{}%
\providecommand \url  [0]{\begingroup\@sanitize@url \@url }%
\providecommand \@url [1]{\endgroup\@href {#1}{\urlprefix }}%
\providecommand \urlprefix  [0]{URL }%
\providecommand \Eprint [0]{\href }%
\providecommand \doibase [0]{http://dx.doi.org/}%
\providecommand \selectlanguage [0]{\@gobble}%
\providecommand \bibinfo  [0]{\@secondoftwo}%
\providecommand \bibfield  [0]{\@secondoftwo}%
\providecommand \translation [1]{[#1]}%
\providecommand \BibitemOpen [0]{}%
\providecommand \bibitemStop [0]{}%
\providecommand \bibitemNoStop [0]{.\EOS\space}%
\providecommand \EOS [0]{\spacefactor3000\relax}%
\providecommand \BibitemShut  [1]{\csname bibitem#1\endcsname}%
\let\auto@bib@innerbib\@empty
\bibitem [{\citenamefont {Levin}\ and\ \citenamefont
  {Stern}(2012)}]{Levin2012}%
  \BibitemOpen
  \bibfield  {author} {\bibinfo {author} {\bibfnamefont {M.}~\bibnamefont
  {Levin}}\ and\ \bibinfo {author} {\bibfnamefont {A.}~\bibnamefont {Stern}},\
  }\href {\doibase 10.1103/PhysRevB.86.115131} {\bibfield  {journal} {\bibinfo
  {journal} {Phys. Rev. B}\ }\textbf {\bibinfo {volume} {86}},\ \bibinfo
  {pages} {115131} (\bibinfo {year} {2012})}\BibitemShut {NoStop}%
\bibitem [{\citenamefont {Neupert}\ \emph {et~al.}(2011)\citenamefont
  {Neupert}, \citenamefont {Santos}, \citenamefont {Ryu}, \citenamefont
  {Chamon},\ and\ \citenamefont {Mudry}}]{NeupertSantosRyuEtAl2011}%
  \BibitemOpen
  \bibfield  {author} {\bibinfo {author} {\bibfnamefont {T.}~\bibnamefont
  {Neupert}}, \bibinfo {author} {\bibfnamefont {L.}~\bibnamefont {Santos}},
  \bibinfo {author} {\bibfnamefont {S.}~\bibnamefont {Ryu}}, \bibinfo {author}
  {\bibfnamefont {C.}~\bibnamefont {Chamon}}, \ and\ \bibinfo {author}
  {\bibfnamefont {C.}~\bibnamefont {Mudry}},\ }\href {\doibase
  10.1103/PhysRevB.84.165107} {\bibfield  {journal} {\bibinfo  {journal} {Phys.
  Rev. B}\ }\textbf {\bibinfo {volume} {84,}},\ \bibinfo {pages} {165107}
  (\bibinfo {year} {2011})},\ \Eprint {http://arxiv.org/abs/1106.3989}
  {1106.3989} \BibitemShut {NoStop}%
\bibitem [{\citenamefont {Santos}\ \emph {et~al.}(2011)\citenamefont {Santos},
  \citenamefont {Neupert}, \citenamefont {Ryu}, \citenamefont {Chamon},\ and\
  \citenamefont {Mudry}}]{SantosNeupertRyuEtAl2011}%
  \BibitemOpen
  \bibfield  {author} {\bibinfo {author} {\bibfnamefont {L.}~\bibnamefont
  {Santos}}, \bibinfo {author} {\bibfnamefont {T.}~\bibnamefont {Neupert}},
  \bibinfo {author} {\bibfnamefont {S.}~\bibnamefont {Ryu}}, \bibinfo {author}
  {\bibfnamefont {C.}~\bibnamefont {Chamon}}, \ and\ \bibinfo {author}
  {\bibfnamefont {C.}~\bibnamefont {Mudry}},\ }\href {\doibase
  10.1103/PhysRevB.84.165138} {\bibfield  {journal} {\bibinfo  {journal} {Phys.
  Rev. B}\ }\textbf {\bibinfo {volume} {84,}},\ \bibinfo {pages} {165138}
  (\bibinfo {year} {2011})},\ \Eprint {http://arxiv.org/abs/1108.2440}
  {1108.2440} \BibitemShut {NoStop}%
\bibitem [{\citenamefont {Essin}\ and\ \citenamefont
  {Hermele}(2013)}]{EssinHermele2012}%
  \BibitemOpen
  \bibfield  {author} {\bibinfo {author} {\bibfnamefont {A.~M.}\ \bibnamefont
  {Essin}}\ and\ \bibinfo {author} {\bibfnamefont {M.}~\bibnamefont
  {Hermele}},\ }\href {\doibase 10.1103/PhysRevB.87.104406} {\bibfield
  {journal} {\bibinfo  {journal} {Phys. Rev. B}\ }\textbf {\bibinfo {volume}
  {87,}},\ \bibinfo {pages} {104406} (\bibinfo {year} {2013})},\ \Eprint
  {http://arxiv.org/abs/1212.0593} {1212.0593} \BibitemShut {NoStop}%
\bibitem [{\citenamefont {Mesaros}\ and\ \citenamefont
  {Ran}(2013)}]{MesarosRan2012}%
  \BibitemOpen
  \bibfield  {author} {\bibinfo {author} {\bibfnamefont {A.}~\bibnamefont
  {Mesaros}}\ and\ \bibinfo {author} {\bibfnamefont {Y.}~\bibnamefont {Ran}},\
  }\href {\doibase 10.1103/PhysRevB.87.155115} {\bibfield  {journal} {\bibinfo
  {journal} {Phys. Rev. B}\ }\textbf {\bibinfo {volume} {87,}},\ \bibinfo
  {pages} {155115} (\bibinfo {year} {2013})},\ \Eprint
  {http://arxiv.org/abs/1212.0835} {1212.0835} \BibitemShut {NoStop}%
\bibitem [{\citenamefont {Hung}\ and\ \citenamefont {Wan}(2013)}]{HungWan2013}%
  \BibitemOpen
  \bibfield  {author} {\bibinfo {author} {\bibfnamefont {L.-Y.}\ \bibnamefont
  {Hung}}\ and\ \bibinfo {author} {\bibfnamefont {Y.}~\bibnamefont {Wan}},\
  }\href {\doibase 10.1103/PhysRevB.87.195103} {\bibfield  {journal} {\bibinfo
  {journal} {Phys. Rev. B}\ }\textbf {\bibinfo {volume} {87,}},\ \bibinfo
  {pages} {195103} (\bibinfo {year} {2013})},\ \Eprint
  {http://arxiv.org/abs/1302.2951} {1302.2951} \BibitemShut {NoStop}%
\bibitem [{\citenamefont {Lu}\ and\ \citenamefont
  {Vishwanath}(2016)}]{LuVishwanath2013}%
  \BibitemOpen
  \bibfield  {author} {\bibinfo {author} {\bibfnamefont {Y.-M.}\ \bibnamefont
  {Lu}}\ and\ \bibinfo {author} {\bibfnamefont {A.}~\bibnamefont
  {Vishwanath}},\ }\href {\doibase 10.1103/PhysRevB.93.155121} {\bibfield
  {journal} {\bibinfo  {journal} {Phys. Rev. B}\ }\textbf {\bibinfo {volume}
  {93,}},\ \bibinfo {pages} {155121} (\bibinfo {year} {2016})},\ \Eprint
  {http://arxiv.org/abs/1302.2634} {1302.2634} \BibitemShut {NoStop}%
\bibitem [{\citenamefont {Wang}\ and\ \citenamefont
  {Senthil}(2013)}]{Wang2013}%
  \BibitemOpen
  \bibfield  {author} {\bibinfo {author} {\bibfnamefont {C.}~\bibnamefont
  {Wang}}\ and\ \bibinfo {author} {\bibfnamefont {T.}~\bibnamefont {Senthil}},\
  }\href {\doibase 10.1103/PhysRevB.87.235122} {\bibfield  {journal} {\bibinfo
  {journal} {Phys. Rev. B}\ }\textbf {\bibinfo {volume} {87}},\ \bibinfo
  {pages} {235122} (\bibinfo {year} {2013})}\BibitemShut {NoStop}%
\bibitem [{\citenamefont {Barkeshli}\ \emph {et~al.}(2014)\citenamefont
  {Barkeshli}, \citenamefont {Bonderson}, \citenamefont {Cheng},\ and\
  \citenamefont {Wang}}]{Barkeshli2014}%
  \BibitemOpen
  \bibfield  {author} {\bibinfo {author} {\bibfnamefont {M.}~\bibnamefont
  {Barkeshli}}, \bibinfo {author} {\bibfnamefont {P.}~\bibnamefont
  {Bonderson}}, \bibinfo {author} {\bibfnamefont {M.}~\bibnamefont {Cheng}}, \
  and\ \bibinfo {author} {\bibfnamefont {Z.}~\bibnamefont {Wang}},\ }\href@noop
  {} {\  (\bibinfo {year} {2014})},\ \Eprint {http://arxiv.org/abs/1410.4540}
  {1410.4540} \BibitemShut {NoStop}%
\bibitem [{\citenamefont {Tarantino}\ \emph {et~al.}(2015)\citenamefont
  {Tarantino}, \citenamefont {Lindner},\ and\ \citenamefont
  {Fidkowski}}]{Tarantino2015}%
  \BibitemOpen
  \bibfield  {author} {\bibinfo {author} {\bibfnamefont {N.}~\bibnamefont
  {Tarantino}}, \bibinfo {author} {\bibfnamefont {N.~H.}\ \bibnamefont
  {Lindner}}, \ and\ \bibinfo {author} {\bibfnamefont {L.}~\bibnamefont
  {Fidkowski}},\ }\href@noop {} {\bibfield  {journal} {\bibinfo  {journal} {New
  Journal of Physics, Volume}\ }\textbf {\bibinfo {volume} {18,}},\ \bibinfo
  {pages} {Issue3,035006} (\bibinfo {year} {2015})},\ \Eprint
  {http://arxiv.org/abs/1506.06754} {1506.06754} \BibitemShut {NoStop}%
\bibitem [{\citenamefont {Lan}\ \emph {et~al.}(2017)\citenamefont {Lan},
  \citenamefont {Kong},\ and\ \citenamefont {Wen}}]{Lan2016}%
  \BibitemOpen
  \bibfield  {author} {\bibinfo {author} {\bibfnamefont {T.}~\bibnamefont
  {Lan}}, \bibinfo {author} {\bibfnamefont {L.}~\bibnamefont {Kong}}, \ and\
  \bibinfo {author} {\bibfnamefont {X.-G.}\ \bibnamefont {Wen}},\ }\href@noop
  {} {\bibfield  {journal} {\bibinfo  {journal} {Phys. Rev. B}\ }\textbf
  {\bibinfo {volume} {95,}},\ \bibinfo {pages} {235140} (\bibinfo {year}
  {2017})},\ \Eprint {http://arxiv.org/abs/1602.05946} {1602.05946}
  \BibitemShut {NoStop}%
\bibitem [{\citenamefont {Xu}(2013)}]{Xu2013a}%
  \BibitemOpen
  \bibfield  {author} {\bibinfo {author} {\bibfnamefont {C.}~\bibnamefont
  {Xu}},\ }\href {\doibase 10.1103/PhysRevB.88.205137} {\bibfield  {journal}
  {\bibinfo  {journal} {Phys. Rev. B}\ }\textbf {\bibinfo {volume} {88,}},\
  \bibinfo {pages} {205137} (\bibinfo {year} {2013})},\ \Eprint
  {http://arxiv.org/abs/1307.8131} {1307.8131} \BibitemShut {NoStop}%
\bibitem [{\citenamefont {Chen}\ and\ \citenamefont
  {Hermele}(2016)}]{ChenHermele2016}%
  \BibitemOpen
  \bibfield  {author} {\bibinfo {author} {\bibfnamefont {X.}~\bibnamefont
  {Chen}}\ and\ \bibinfo {author} {\bibfnamefont {M.}~\bibnamefont {Hermele}},\
  }\href {\doibase 10.1103/PhysRevB.94.195120} {\bibfield  {journal} {\bibinfo
  {journal} {Phys. Rev. B}\ }\textbf {\bibinfo {volume} {94,}},\ \bibinfo
  {pages} {195120} (\bibinfo {year} {2016})},\ \Eprint
  {http://arxiv.org/abs/1602.00187} {1602.00187} \BibitemShut {NoStop}%
\bibitem [{\citenamefont {Ning}\ \emph {et~al.}(2016)\citenamefont {Ning},
  \citenamefont {Liu},\ and\ \citenamefont {Ye}}]{Ning2016}%
  \BibitemOpen
  \bibfield  {author} {\bibinfo {author} {\bibfnamefont {S.-Q.}\ \bibnamefont
  {Ning}}, \bibinfo {author} {\bibfnamefont {Z.-X.}\ \bibnamefont {Liu}}, \
  and\ \bibinfo {author} {\bibfnamefont {P.}~\bibnamefont {Ye}},\ }\href@noop
  {} {\bibfield  {journal} {\bibinfo  {journal} {Phys. Rev. B}\ }\textbf
  {\bibinfo {volume} {94,}},\ \bibinfo {pages} {245120} (\bibinfo {year}
  {2016})},\ \Eprint {http://arxiv.org/abs/1609.00985} {1609.00985}
  \BibitemShut {NoStop}%
\bibitem [{\citenamefont {Motrunich}\ and\ \citenamefont
  {Senthil}(2002)}]{Motrunich2002}%
  \BibitemOpen
  \bibfield  {author} {\bibinfo {author} {\bibfnamefont {O.~I.}\ \bibnamefont
  {Motrunich}}\ and\ \bibinfo {author} {\bibfnamefont {T.}~\bibnamefont
  {Senthil}},\ }\href {\doibase 10.1103/PhysRevLett.89.277004} {\bibfield
  {journal} {\bibinfo  {journal} {Phys. Rev. Lett.}\ }\textbf {\bibinfo
  {volume} {89}},\ \bibinfo {pages} {277004} (\bibinfo {year}
  {2002})}\BibitemShut {NoStop}%
\bibitem [{\citenamefont {Hermele}\ \emph
  {et~al.}(2004{\natexlab{a}})\citenamefont {Hermele}, \citenamefont {Fisher},\
  and\ \citenamefont {Balents}}]{HermeleFisherBalents2004}%
  \BibitemOpen
  \bibfield  {author} {\bibinfo {author} {\bibfnamefont {M.}~\bibnamefont
  {Hermele}}, \bibinfo {author} {\bibfnamefont {M.~P.~A.}\ \bibnamefont
  {Fisher}}, \ and\ \bibinfo {author} {\bibfnamefont {L.}~\bibnamefont
  {Balents}},\ }\href {\doibase 10.1103/PhysRevB.69.064404} {\bibfield
  {journal} {\bibinfo  {journal} {Phys. Rev. B}\ }\textbf {\bibinfo {volume}
  {69,}},\ \bibinfo {pages} {064404} (\bibinfo {year} {2004}{\natexlab{a}})},\
  \Eprint {http://arxiv.org/abs/cond-mat/0305401} {cond-mat/0305401}
  \BibitemShut {NoStop}%
\bibitem [{\citenamefont {Motrunich}\ and\ \citenamefont
  {Senthil}(2005)}]{Motrunich2005}%
  \BibitemOpen
  \bibfield  {author} {\bibinfo {author} {\bibfnamefont {O.~I.}\ \bibnamefont
  {Motrunich}}\ and\ \bibinfo {author} {\bibfnamefont {T.}~\bibnamefont
  {Senthil}},\ }\href {\doibase 10.1103/PhysRevB.71.125102} {\bibfield
  {journal} {\bibinfo  {journal} {Phys. Rev. B}\ }\textbf {\bibinfo {volume}
  {71}},\ \bibinfo {pages} {125102} (\bibinfo {year} {2005})}\BibitemShut
  {NoStop}%
\bibitem [{\citenamefont {Levin}\ and\ \citenamefont {Wen}(2005)}]{Levin2005}%
  \BibitemOpen
  \bibfield  {author} {\bibinfo {author} {\bibfnamefont {M.~A.}\ \bibnamefont
  {Levin}}\ and\ \bibinfo {author} {\bibfnamefont {X.-G.}\ \bibnamefont
  {Wen}},\ }\href {\doibase 10.1103/PhysRevB.71.045110} {\bibfield  {journal}
  {\bibinfo  {journal} {Phys. Rev. B}\ }\textbf {\bibinfo {volume} {71}},\
  \bibinfo {pages} {045110} (\bibinfo {year} {2005})}\BibitemShut {NoStop}%
\bibitem [{\citenamefont {Savary}\ and\ \citenamefont
  {Balents}(2012)}]{SavaryBalents2011}%
  \BibitemOpen
  \bibfield  {author} {\bibinfo {author} {\bibfnamefont {L.}~\bibnamefont
  {Savary}}\ and\ \bibinfo {author} {\bibfnamefont {L.}~\bibnamefont
  {Balents}},\ }\href {\doibase 10.1103/PhysRevLett.108.037202} {\bibfield
  {journal} {\bibinfo  {journal} {Phys. Rev. Lett.}\ }\textbf {\bibinfo
  {volume} {108,}},\ \bibinfo {pages} {037202} (\bibinfo {year} {2012})},\
  \Eprint {http://arxiv.org/abs/1110.2185} {1110.2185} \BibitemShut {NoStop}%
\bibitem [{\citenamefont {Savary}\ and\ \citenamefont
  {Balents}(2017)}]{SavaryBalents2016}%
  \BibitemOpen
  \bibfield  {author} {\bibinfo {author} {\bibfnamefont {L.}~\bibnamefont
  {Savary}}\ and\ \bibinfo {author} {\bibfnamefont {L.}~\bibnamefont
  {Balents}},\ }\href {\doibase 10.1103/PhysRevLett.118.087203} {\bibfield
  {journal} {\bibinfo  {journal} {Phys. Rev. Lett.}\ }\textbf {\bibinfo
  {volume} {118,}},\ \bibinfo {pages} {087203} (\bibinfo {year} {2017})},\
  \Eprint {http://arxiv.org/abs/1604.04630} {1604.04630} \BibitemShut {NoStop}%
\bibitem [{\citenamefont {Rochner}\ \emph {et~al.}(2016)\citenamefont
  {Rochner}, \citenamefont {Balents},\ and\ \citenamefont
  {Schmidt}}]{RoechnerBalentsSchmidt2016}%
  \BibitemOpen
  \bibfield  {author} {\bibinfo {author} {\bibfnamefont {J.}~\bibnamefont
  {Rochner}}, \bibinfo {author} {\bibfnamefont {L.}~\bibnamefont {Balents}}, \
  and\ \bibinfo {author} {\bibfnamefont {K.~P.}\ \bibnamefont {Schmidt}},\
  }\href {\doibase 10.1103/PhysRevB.94.201111} {\bibfield  {journal} {\bibinfo
  {journal} {Phys. Rev. B}\ }\textbf {\bibinfo {volume} {94,}},\ \bibinfo
  {pages} {201111} (\bibinfo {year} {2016})},\ \Eprint
  {http://arxiv.org/abs/1609.00484} {1609.00484} \BibitemShut {NoStop}%
\bibitem [{\citenamefont {Affleck}\ and\ \citenamefont
  {Marston}(1988)}]{Affleck1988}%
  \BibitemOpen
  \bibfield  {author} {\bibinfo {author} {\bibfnamefont {I.}~\bibnamefont
  {Affleck}}\ and\ \bibinfo {author} {\bibfnamefont {J.~B.}\ \bibnamefont
  {Marston}},\ }\href {\doibase 10.1103/PhysRevB.37.3774} {\bibfield  {journal}
  {\bibinfo  {journal} {Phys. Rev. B}\ }\textbf {\bibinfo {volume} {37}},\
  \bibinfo {pages} {3774} (\bibinfo {year} {1988})}\BibitemShut {NoStop}%
\bibitem [{\citenamefont {Lee}\ and\ \citenamefont
  {Nagaosa}(1992)}]{LeeNagaosa1992}%
  \BibitemOpen
  \bibfield  {author} {\bibinfo {author} {\bibfnamefont {P.~A.}\ \bibnamefont
  {Lee}}\ and\ \bibinfo {author} {\bibfnamefont {N.}~\bibnamefont {Nagaosa}},\
  }\href@noop {} {\bibfield  {journal} {\bibinfo  {journal} {Physical Review
  B}\ }\textbf {\bibinfo {volume} {46}},\ \bibinfo {pages} {5621} (\bibinfo
  {year} {1992})}\BibitemShut {NoStop}%
\bibitem [{\citenamefont {Wen}\ and\ \citenamefont {Lee}(1996)}]{WenLee1996}%
  \BibitemOpen
  \bibfield  {author} {\bibinfo {author} {\bibfnamefont {X.-G.}\ \bibnamefont
  {Wen}}\ and\ \bibinfo {author} {\bibfnamefont {P.~A.}\ \bibnamefont {Lee}},\
  }\href {\doibase 10.1103/PhysRevLett.76.503} {\bibfield  {journal} {\bibinfo
  {journal} {Phys. Rev. Lett.,}\ }\textbf {\bibinfo {volume} {76,}},\ \bibinfo
  {pages} {503} (\bibinfo {year} {1996})},\ \Eprint
  {http://arxiv.org/abs/cond-mat/9506065} {cond-mat/9506065} \BibitemShut
  {NoStop}%
\bibitem [{\citenamefont {Hermele}\ \emph
  {et~al.}(2004{\natexlab{b}})\citenamefont {Hermele}, \citenamefont {Senthil},
  \citenamefont {Fisher}, \citenamefont {Lee}, \citenamefont {Nagaosa},\ and\
  \citenamefont {Wen}}]{Hermele2004}%
  \BibitemOpen
  \bibfield  {author} {\bibinfo {author} {\bibfnamefont {M.}~\bibnamefont
  {Hermele}}, \bibinfo {author} {\bibfnamefont {T.}~\bibnamefont {Senthil}},
  \bibinfo {author} {\bibfnamefont {M.~P.~A.}\ \bibnamefont {Fisher}}, \bibinfo
  {author} {\bibfnamefont {P.~A.}\ \bibnamefont {Lee}}, \bibinfo {author}
  {\bibfnamefont {N.}~\bibnamefont {Nagaosa}}, \ and\ \bibinfo {author}
  {\bibfnamefont {X.-G.}\ \bibnamefont {Wen}},\ }\href {\doibase
  10.1103/PhysRevB.70.214437} {\bibfield  {journal} {\bibinfo  {journal} {Phys.
  Rev. B}\ }\textbf {\bibinfo {volume} {70}},\ \bibinfo {pages} {214437}
  (\bibinfo {year} {2004}{\natexlab{b}})}\BibitemShut {NoStop}%
\bibitem [{\citenamefont {Lee}\ and\ \citenamefont {Lee}(2005)}]{LeeLee2005}%
  \BibitemOpen
  \bibfield  {author} {\bibinfo {author} {\bibfnamefont {S.-S.}\ \bibnamefont
  {Lee}}\ and\ \bibinfo {author} {\bibfnamefont {P.~A.}\ \bibnamefont {Lee}},\
  }\href@noop {} {\bibfield  {journal} {\bibinfo  {journal} {Physical review
  letters}\ }\textbf {\bibinfo {volume} {95}},\ \bibinfo {pages} {036403}
  (\bibinfo {year} {2005})}\BibitemShut {NoStop}%
\bibitem [{\citenamefont {Lee}(2008)}]{Lee2008}%
  \BibitemOpen
  \bibfield  {author} {\bibinfo {author} {\bibfnamefont {S.-S.}\ \bibnamefont
  {Lee}},\ }\href@noop {} {\bibfield  {journal} {\bibinfo  {journal} {Phys.
  Rev. B}\ }\textbf {\bibinfo {volume} {78,}},\ \bibinfo {pages} {085129}
  (\bibinfo {year} {2008})},\ \Eprint {http://arxiv.org/abs/0804.3800}
  {0804.3800} \BibitemShut {NoStop}%
\bibitem [{\citenamefont {Senthil}(2008)}]{Senthil2008a}%
  \BibitemOpen
  \bibfield  {author} {\bibinfo {author} {\bibfnamefont {T.}~\bibnamefont
  {Senthil}},\ }\href@noop {} {\bibfield  {journal} {\bibinfo  {journal}
  {Physical Review B}\ }\textbf {\bibinfo {volume} {78}},\ \bibinfo {pages}
  {045109} (\bibinfo {year} {2008})}\BibitemShut {NoStop}%
\bibitem [{\citenamefont {Zou}\ and\ \citenamefont {Senthil}(2016)}]{Zou2016}%
  \BibitemOpen
  \bibfield  {author} {\bibinfo {author} {\bibfnamefont {L.}~\bibnamefont
  {Zou}}\ and\ \bibinfo {author} {\bibfnamefont {T.}~\bibnamefont {Senthil}},\
  }\href {\doibase 10.1103/PhysRevB.94.115113} {\bibfield  {journal} {\bibinfo
  {journal} {Phys. Rev. B}\ }\textbf {\bibinfo {volume} {94}},\ \bibinfo
  {pages} {115113} (\bibinfo {year} {2016})}\BibitemShut {NoStop}%
\bibitem [{\citenamefont {Wang}\ and\ \citenamefont
  {Senthil}(2016{\natexlab{a}})}]{Wang2016}%
  \BibitemOpen
  \bibfield  {author} {\bibinfo {author} {\bibfnamefont {C.}~\bibnamefont
  {Wang}}\ and\ \bibinfo {author} {\bibfnamefont {T.}~\bibnamefont {Senthil}},\
  }\href {\doibase 10.1103/PhysRevX.6.011034} {\bibfield  {journal} {\bibinfo
  {journal} {Phys. Rev. X}\ }\textbf {\bibinfo {volume} {6}},\ \bibinfo {pages}
  {011034} (\bibinfo {year} {2016}{\natexlab{a}})}\BibitemShut {NoStop}%
\bibitem [{\citenamefont {Metlitski}\ and\ \citenamefont
  {Vishwanath}(2016)}]{Metlitski2016}%
  \BibitemOpen
  \bibfield  {author} {\bibinfo {author} {\bibfnamefont {M.~A.}\ \bibnamefont
  {Metlitski}}\ and\ \bibinfo {author} {\bibfnamefont {A.}~\bibnamefont
  {Vishwanath}},\ }\href {\doibase 10.1103/PhysRevB.93.245151} {\bibfield
  {journal} {\bibinfo  {journal} {Phys. Rev. B}\ }\textbf {\bibinfo {volume}
  {93}},\ \bibinfo {pages} {245151} (\bibinfo {year} {2016})}\BibitemShut
  {NoStop}%
\bibitem [{\citenamefont {{Metlitski}}(2015)}]{Metlitski2015}%
  \BibitemOpen
  \bibfield  {author} {\bibinfo {author} {\bibfnamefont {M.~A.}\ \bibnamefont
  {{Metlitski}}},\ }\href@noop {} {\bibfield  {journal} {\bibinfo  {journal}
  {ArXiv e-prints}\ } (\bibinfo {year} {2015})},\ \Eprint
  {http://arxiv.org/abs/1510.05663} {arXiv:1510.05663 [hep-th]} \BibitemShut
  {NoStop}%
\bibitem [{\citenamefont {Xu}\ and\ \citenamefont {You}(2015)}]{qeddual}%
  \BibitemOpen
  \bibfield  {author} {\bibinfo {author} {\bibfnamefont {C.}~\bibnamefont
  {Xu}}\ and\ \bibinfo {author} {\bibfnamefont {Y.-Z.}\ \bibnamefont {You}},\
  }\href {\doibase 10.1103/PhysRevB.92.220416} {\bibfield  {journal} {\bibinfo
  {journal} {Phys. Rev. B}\ }\textbf {\bibinfo {volume} {92}},\ \bibinfo
  {pages} {220416} (\bibinfo {year} {2015})},\ \Eprint
  {http://arxiv.org/abs/cond-mat/0010440} {cond-mat/0010440} \BibitemShut
  {NoStop}%
\bibitem [{\citenamefont {Wang}\ and\ \citenamefont
  {Senthil}(2015)}]{Wang2015}%
  \BibitemOpen
  \bibfield  {author} {\bibinfo {author} {\bibfnamefont {C.}~\bibnamefont
  {Wang}}\ and\ \bibinfo {author} {\bibfnamefont {T.}~\bibnamefont {Senthil}},\
  }\href {\doibase 10.1103/PhysRevX.5.041031} {\bibfield  {journal} {\bibinfo
  {journal} {Phys. Rev. X}\ }\textbf {\bibinfo {volume} {5}},\ \bibinfo {pages}
  {041031} (\bibinfo {year} {2015})}\BibitemShut {NoStop}%
\bibitem [{\citenamefont {Mross}\ \emph {et~al.}(2016)\citenamefont {Mross},
  \citenamefont {Alicea},\ and\ \citenamefont {Motrunich}}]{dualdrMAM}%
  \BibitemOpen
  \bibfield  {author} {\bibinfo {author} {\bibfnamefont {D.~F.}\ \bibnamefont
  {Mross}}, \bibinfo {author} {\bibfnamefont {J.}~\bibnamefont {Alicea}}, \
  and\ \bibinfo {author} {\bibfnamefont {O.~I.}\ \bibnamefont {Motrunich}},\
  }\href {\doibase 10.1103/PhysRevLett.117.016802} {\bibfield  {journal}
  {\bibinfo  {journal} {Phys. Rev. Lett.}\ }\textbf {\bibinfo {volume} {117}},\
  \bibinfo {pages} {016802} (\bibinfo {year} {2016})}\BibitemShut {NoStop}%
\bibitem [{\citenamefont {Son}(2015)}]{sonphcfl}%
  \BibitemOpen
  \bibfield  {author} {\bibinfo {author} {\bibfnamefont {D.~T.}\ \bibnamefont
  {Son}},\ }\href {\doibase 10.1103/PhysRevX.5.031027} {\bibfield  {journal}
  {\bibinfo  {journal} {Phys. Rev. X}\ }\textbf {\bibinfo {volume} {5}},\
  \bibinfo {pages} {031027} (\bibinfo {year} {2015})}\BibitemShut {NoStop}%
\bibitem [{\citenamefont {Wang}\ and\ \citenamefont
  {Senthil}(2016{\natexlab{b}})}]{Wang2016a}%
  \BibitemOpen
  \bibfield  {author} {\bibinfo {author} {\bibfnamefont {C.}~\bibnamefont
  {Wang}}\ and\ \bibinfo {author} {\bibfnamefont {T.}~\bibnamefont {Senthil}},\
  }\href {\doibase 10.1103/PhysRevB.93.085110} {\bibfield  {journal} {\bibinfo
  {journal} {Phys. Rev. B}\ }\textbf {\bibinfo {volume} {93}},\ \bibinfo
  {pages} {085110} (\bibinfo {year} {2016}{\natexlab{b}})}\BibitemShut
  {NoStop}%
\bibitem [{\citenamefont {Geraedts}\ \emph {et~al.}(2016)\citenamefont
  {Geraedts}, \citenamefont {Zaletel}, \citenamefont {Mong}, \citenamefont
  {Metlitski}, \citenamefont {Vishwanath},\ and\ \citenamefont
  {Motrunich}}]{Geraedts197}%
  \BibitemOpen
  \bibfield  {author} {\bibinfo {author} {\bibfnamefont {S.~D.}\ \bibnamefont
  {Geraedts}}, \bibinfo {author} {\bibfnamefont {M.~P.}\ \bibnamefont
  {Zaletel}}, \bibinfo {author} {\bibfnamefont {R.~S.~K.}\ \bibnamefont
  {Mong}}, \bibinfo {author} {\bibfnamefont {M.~A.}\ \bibnamefont {Metlitski}},
  \bibinfo {author} {\bibfnamefont {A.}~\bibnamefont {Vishwanath}}, \ and\
  \bibinfo {author} {\bibfnamefont {O.~I.}\ \bibnamefont {Motrunich}},\ }\href
  {\doibase 10.1126/science.aad4302} {\bibfield  {journal} {\bibinfo  {journal}
  {Science}\ }\textbf {\bibinfo {volume} {352}},\ \bibinfo {pages} {197}
  (\bibinfo {year} {2016})}\BibitemShut {NoStop}%
\bibitem [{\citenamefont {Wang}\ and\ \citenamefont
  {Senthil}(2016{\natexlab{c}})}]{WangSenthil2016}%
  \BibitemOpen
  \bibfield  {author} {\bibinfo {author} {\bibfnamefont {C.}~\bibnamefont
  {Wang}}\ and\ \bibinfo {author} {\bibfnamefont {T.}~\bibnamefont {Senthil}},\
  }\href {\doibase 10.1103/PhysRevB.94.245107} {\bibfield  {journal} {\bibinfo
  {journal} {Phys. Rev. B}\ }\textbf {\bibinfo {volume} {94}},\ \bibinfo
  {pages} {245107} (\bibinfo {year} {2016}{\natexlab{c}})}\BibitemShut
  {NoStop}%
\bibitem [{\citenamefont {Wang}\ \emph
  {et~al.}(2017{\natexlab{a}})\citenamefont {Wang}, \citenamefont {Cooper},
  \citenamefont {Halperin},\ and\ \citenamefont
  {Stern}}]{WangCooperHalperinStern}%
  \BibitemOpen
  \bibfield  {author} {\bibinfo {author} {\bibfnamefont {C.}~\bibnamefont
  {Wang}}, \bibinfo {author} {\bibfnamefont {N.~R.}\ \bibnamefont {Cooper}},
  \bibinfo {author} {\bibfnamefont {B.~I.}\ \bibnamefont {Halperin}}, \ and\
  \bibinfo {author} {\bibfnamefont {A.}~\bibnamefont {Stern}},\ }\href@noop {}
  {\bibfield  {journal} {\bibinfo  {journal} {Phys. Rev. X}\ }\textbf {\bibinfo
  {volume} {7,}},\ \bibinfo {pages} {031029} (\bibinfo {year}
  {2017}{\natexlab{a}})},\ \Eprint {http://arxiv.org/abs/1701.00007}
  {1701.00007} \BibitemShut {NoStop}%
\bibitem [{\citenamefont {Hsin}\ and\ \citenamefont
  {Seiberg}(2016)}]{seiberg2}%
  \BibitemOpen
  \bibfield  {author} {\bibinfo {author} {\bibfnamefont {P.-S.}\ \bibnamefont
  {Hsin}}\ and\ \bibinfo {author} {\bibfnamefont {N.}~\bibnamefont {Seiberg}},\
  }\href {\doibase 10.1007/JHEP09(2016)095} {\bibfield  {journal} {\bibinfo
  {journal} {Journal of High Energy Physics}\ }\textbf {\bibinfo {volume}
  {2016}},\ \bibinfo {pages} {95} (\bibinfo {year} {2016})}\BibitemShut
  {NoStop}%
\bibitem [{\citenamefont {Wang}\ \emph
  {et~al.}(2017{\natexlab{b}})\citenamefont {Wang}, \citenamefont {Nahum},
  \citenamefont {Metlitski}, \citenamefont {Xu},\ and\ \citenamefont
  {Senthil}}]{dqcpdual}%
  \BibitemOpen
  \bibfield  {author} {\bibinfo {author} {\bibfnamefont {C.}~\bibnamefont
  {Wang}}, \bibinfo {author} {\bibfnamefont {A.}~\bibnamefont {Nahum}},
  \bibinfo {author} {\bibfnamefont {M.~A.}\ \bibnamefont {Metlitski}}, \bibinfo
  {author} {\bibfnamefont {C.}~\bibnamefont {Xu}}, \ and\ \bibinfo {author}
  {\bibfnamefont {T.}~\bibnamefont {Senthil}},\ }\href@noop {} {\bibfield
  {journal} {\bibinfo  {journal} {Phys. Rev. X}\ }\textbf {\bibinfo {volume}
  {7,}},\ \bibinfo {pages} {031051} (\bibinfo {year} {2017}{\natexlab{b}})},\
  \Eprint {http://arxiv.org/abs/1703.02426} {1703.02426} \BibitemShut {NoStop}%
\bibitem [{\citenamefont {Seiberg}\ \emph {et~al.}(2016)\citenamefont
  {Seiberg}, \citenamefont {Senthil}, \citenamefont {Wang},\ and\ \citenamefont
  {Witten}}]{Seiberg2016}%
  \BibitemOpen
  \bibfield  {author} {\bibinfo {author} {\bibfnamefont {N.}~\bibnamefont
  {Seiberg}}, \bibinfo {author} {\bibfnamefont {T.}~\bibnamefont {Senthil}},
  \bibinfo {author} {\bibfnamefont {C.}~\bibnamefont {Wang}}, \ and\ \bibinfo
  {author} {\bibfnamefont {E.}~\bibnamefont {Witten}},\ }\href@noop {}
  {\bibfield  {journal} {\bibinfo  {journal} {Annals of Physics}\ }\textbf
  {\bibinfo {volume} {374}},\ \bibinfo {pages} {395} (\bibinfo {year}
  {2016})},\ \Eprint {http://arxiv.org/abs/1606.01989} {1606.01989}
  \BibitemShut {NoStop}%
\bibitem [{\citenamefont {{Karch}}\ and\ \citenamefont
  {{Tong}}(2016)}]{karchtong}%
  \BibitemOpen
  \bibfield  {author} {\bibinfo {author} {\bibfnamefont {A.}~\bibnamefont
  {{Karch}}}\ and\ \bibinfo {author} {\bibfnamefont {D.}~\bibnamefont
  {{Tong}}},\ }\href {\doibase 10.1103/PhysRevX.6.031043} {\bibfield  {journal}
  {\bibinfo  {journal} {Physical Review X}\ }\textbf {\bibinfo {volume} {6}},\
  \bibinfo {eid} {031043} (\bibinfo {year} {2016})},\ \Eprint
  {http://arxiv.org/abs/1606.01893} {arXiv:1606.01893 [hep-th]} \BibitemShut
  {NoStop}%
\bibitem [{\citenamefont {Murugan}\ and\ \citenamefont
  {Nastase}(2017)}]{murugan}%
  \BibitemOpen
  \bibfield  {author} {\bibinfo {author} {\bibfnamefont {J.}~\bibnamefont
  {Murugan}}\ and\ \bibinfo {author} {\bibfnamefont {H.}~\bibnamefont
  {Nastase}},\ }\href {\doibase 10.1007/JHEP05(2017)159} {\bibfield  {journal}
  {\bibinfo  {journal} {Journal of High Energy Physics}\ }\textbf {\bibinfo
  {volume} {2017}},\ \bibinfo {pages} {159} (\bibinfo {year}
  {2017})}\BibitemShut {NoStop}%
\bibitem [{\citenamefont {{Kachru}}\ \emph {et~al.}(2016)\citenamefont
  {{Kachru}}, \citenamefont {{Mulligan}}, \citenamefont {{Torroba}},\ and\
  \citenamefont {{Wang}}}]{kachrubosonization}%
  \BibitemOpen
  \bibfield  {author} {\bibinfo {author} {\bibfnamefont {S.}~\bibnamefont
  {{Kachru}}}, \bibinfo {author} {\bibfnamefont {M.}~\bibnamefont
  {{Mulligan}}}, \bibinfo {author} {\bibfnamefont {G.}~\bibnamefont
  {{Torroba}}}, \ and\ \bibinfo {author} {\bibfnamefont {H.}~\bibnamefont
  {{Wang}}},\ }\href {\doibase 10.1103/PhysRevD.94.085009} {\bibfield
  {journal} {\bibinfo  {journal} {\prd}\ }\textbf {\bibinfo {volume} {94}},\
  \bibinfo {eid} {085009} (\bibinfo {year} {2016})},\ \Eprint
  {http://arxiv.org/abs/1608.05077} {arXiv:1608.05077 [hep-th]} \BibitemShut
  {NoStop}%
\bibitem [{\citenamefont {Vishwanath}\ and\ \citenamefont
  {Senthil}(2013)}]{VishwanathSenthil2013}%
  \BibitemOpen
  \bibfield  {author} {\bibinfo {author} {\bibfnamefont {A.}~\bibnamefont
  {Vishwanath}}\ and\ \bibinfo {author} {\bibfnamefont {T.}~\bibnamefont
  {Senthil}},\ }\href {\doibase 10.1103/PhysRevX.3.011016} {\bibfield
  {journal} {\bibinfo  {journal} {Phys. Rev. X}\ }\textbf {\bibinfo {volume}
  {3}},\ \bibinfo {pages} {011016} (\bibinfo {year} {2013})}\BibitemShut
  {NoStop}%
\bibitem [{\citenamefont {Metlitski}\ \emph {et~al.}(2013)\citenamefont
  {Metlitski}, \citenamefont {Kane},\ and\ \citenamefont
  {Fisher}}]{MetlitskiKaneFisher2013}%
  \BibitemOpen
  \bibfield  {author} {\bibinfo {author} {\bibfnamefont {M.~A.}\ \bibnamefont
  {Metlitski}}, \bibinfo {author} {\bibfnamefont {C.~L.}\ \bibnamefont {Kane}},
  \ and\ \bibinfo {author} {\bibfnamefont {M.~P.~A.}\ \bibnamefont {Fisher}},\
  }\href {\doibase 10.1103/PhysRevB.88.035131} {\bibfield  {journal} {\bibinfo
  {journal} {Phys. Rev. B}\ }\textbf {\bibinfo {volume} {88}},\ \bibinfo
  {pages} {035131} (\bibinfo {year} {2013})}\BibitemShut {NoStop}%
\bibitem [{\citenamefont {Wang}\ \emph {et~al.}(2014)\citenamefont {Wang},
  \citenamefont {Potter},\ and\ \citenamefont {Senthil}}]{Wang2014}%
  \BibitemOpen
  \bibfield  {author} {\bibinfo {author} {\bibfnamefont {C.}~\bibnamefont
  {Wang}}, \bibinfo {author} {\bibfnamefont {A.~C.}\ \bibnamefont {Potter}}, \
  and\ \bibinfo {author} {\bibfnamefont {T.}~\bibnamefont {Senthil}},\ }\href
  {\doibase 10.1126/science.1243326} {\bibfield  {journal} {\bibinfo  {journal}
  {Science}\ }\textbf {\bibinfo {volume} {343}},\ \bibinfo {pages} {629}
  (\bibinfo {year} {2014})}\BibitemShut {NoStop}%
\bibitem [{\citenamefont {Kravec}\ \emph {et~al.}(2015)\citenamefont {Kravec},
  \citenamefont {McGreevy},\ and\ \citenamefont {Swingle}}]{Kravec2015}%
  \BibitemOpen
  \bibfield  {author} {\bibinfo {author} {\bibfnamefont {S.}~\bibnamefont
  {Kravec}}, \bibinfo {author} {\bibfnamefont {J.}~\bibnamefont {McGreevy}}, \
  and\ \bibinfo {author} {\bibfnamefont {B.}~\bibnamefont {Swingle}},\ }\href
  {\doibase 10.1103/PhysRevD.92.085024} {\bibfield  {journal} {\bibinfo
  {journal} {Phys. Rev. D}\ }\textbf {\bibinfo {volume} {92}},\ \bibinfo
  {pages} {085024} (\bibinfo {year} {2015})}\BibitemShut {NoStop}%
\bibitem [{\citenamefont {Thorngren}(2015)}]{Thorngren2014}%
  \BibitemOpen
  \bibfield  {author} {\bibinfo {author} {\bibfnamefont {R.}~\bibnamefont
  {Thorngren}},\ }\href {\doibase 10.1007/JHEP02(2015)152} {\bibfield
  {journal} {\bibinfo  {journal} {Journal of High Energy Physics}\ }\textbf
  {\bibinfo {volume} {2015}},\ \bibinfo {pages} {152} (\bibinfo {year}
  {2015})}\BibitemShut {NoStop}%
\bibitem [{\citenamefont {Wu}\ and\ \citenamefont {Yang}(1975)}]{Wu1975}%
  \BibitemOpen
  \bibfield  {author} {\bibinfo {author} {\bibfnamefont {T.~T.}\ \bibnamefont
  {Wu}}\ and\ \bibinfo {author} {\bibfnamefont {C.~N.}\ \bibnamefont {Yang}},\
  }\href {\doibase 10.1103/physrevd.12.3845} {\bibfield  {journal} {\bibinfo
  {journal} {Phys. Rev. D}\ }\textbf {\bibinfo {volume} {12}},\ \bibinfo
  {pages} {3845} (\bibinfo {year} {1975})}\BibitemShut {NoStop}%
\bibitem [{\citenamefont {Goldhaber}(1976)}]{Goldhaber1976}%
  \BibitemOpen
  \bibfield  {author} {\bibinfo {author} {\bibfnamefont {A.~S.}\ \bibnamefont
  {Goldhaber}},\ }\href {\doibase 10.1103/physrevlett.36.1122} {\bibfield
  {journal} {\bibinfo  {journal} {Physical Review Letters}\ }\textbf {\bibinfo
  {volume} {36}},\ \bibinfo {pages} {1122} (\bibinfo {year}
  {1976})}\BibitemShut {NoStop}%
\bibitem [{\citenamefont {Chen}\ \emph {et~al.}(2013)\citenamefont {Chen},
  \citenamefont {Gu}, \citenamefont {Liu},\ and\ \citenamefont
  {Wen}}]{Chen2013}%
  \BibitemOpen
  \bibfield  {author} {\bibinfo {author} {\bibfnamefont {X.}~\bibnamefont
  {Chen}}, \bibinfo {author} {\bibfnamefont {Z.-C.}\ \bibnamefont {Gu}},
  \bibinfo {author} {\bibfnamefont {Z.-X.}\ \bibnamefont {Liu}}, \ and\
  \bibinfo {author} {\bibfnamefont {X.-G.}\ \bibnamefont {Wen}},\ }\href
  {\doibase 10.1103/PhysRevB.87.155114} {\bibfield  {journal} {\bibinfo
  {journal} {Phys. Rev. B}\ }\textbf {\bibinfo {volume} {87}},\ \bibinfo
  {pages} {155114} (\bibinfo {year} {2013})}\BibitemShut {NoStop}%
\bibitem [{\citenamefont {Senthil}\ and\ \citenamefont
  {Fisher}(2000)}]{Senthil2000}%
  \BibitemOpen
  \bibfield  {author} {\bibinfo {author} {\bibfnamefont {T.}~\bibnamefont
  {Senthil}}\ and\ \bibinfo {author} {\bibfnamefont {M.~P.~A.}\ \bibnamefont
  {Fisher}},\ }\href {\doibase 10.1103/physrevb.62.7850} {\bibfield  {journal}
  {\bibinfo  {journal} {Physical Review B}\ }\textbf {\bibinfo {volume} {62}},\
  \bibinfo {pages} {7850} (\bibinfo {year} {2000})}\BibitemShut {NoStop}%
\bibitem [{\citenamefont {Maciejko}\ \emph {et~al.}(2010)\citenamefont
  {Maciejko}, \citenamefont {Qi}, \citenamefont {Karch},\ and\ \citenamefont
  {Zhang}}]{fracTI1}%
  \BibitemOpen
  \bibfield  {author} {\bibinfo {author} {\bibfnamefont {J.}~\bibnamefont
  {Maciejko}}, \bibinfo {author} {\bibfnamefont {X.-L.}\ \bibnamefont {Qi}},
  \bibinfo {author} {\bibfnamefont {A.}~\bibnamefont {Karch}}, \ and\ \bibinfo
  {author} {\bibfnamefont {S.-C.}\ \bibnamefont {Zhang}},\ }\href {\doibase
  10.1103/PhysRevLett.105.246809} {\bibfield  {journal} {\bibinfo  {journal}
  {Phys. Rev. Lett.}\ }\textbf {\bibinfo {volume} {105}},\ \bibinfo {pages}
  {246809} (\bibinfo {year} {2010})}\BibitemShut {NoStop}%
\bibitem [{\citenamefont {{Swingle}}\ \emph {et~al.}(2011)\citenamefont
  {{Swingle}}, \citenamefont {{Barkeshli}}, \citenamefont {{McGreevy}},\ and\
  \citenamefont {{Senthil}}}]{fracTI2}%
  \BibitemOpen
  \bibfield  {author} {\bibinfo {author} {\bibfnamefont {B.}~\bibnamefont
  {{Swingle}}}, \bibinfo {author} {\bibfnamefont {M.}~\bibnamefont
  {{Barkeshli}}}, \bibinfo {author} {\bibfnamefont {J.}~\bibnamefont
  {{McGreevy}}}, \ and\ \bibinfo {author} {\bibfnamefont {T.}~\bibnamefont
  {{Senthil}}},\ }\href {\doibase 10.1103/PhysRevB.83.195139} {\bibfield
  {journal} {\bibinfo  {journal} {\prb}\ }\textbf {\bibinfo {volume} {83}},\
  \bibinfo {eid} {195139} (\bibinfo {year} {2011})},\ \Eprint
  {http://arxiv.org/abs/1005.1076} {arXiv:1005.1076 [cond-mat.str-el]}
  \BibitemShut {NoStop}%
\bibitem [{\citenamefont {Schnyder}\ \emph {et~al.}(2009)\citenamefont
  {Schnyder}, \citenamefont {Ryu},\ and\ \citenamefont
  {Ludwig}}]{SchnyderRyuLudwig}%
  \BibitemOpen
  \bibfield  {author} {\bibinfo {author} {\bibfnamefont {A.~P.}\ \bibnamefont
  {Schnyder}}, \bibinfo {author} {\bibfnamefont {S.}~\bibnamefont {Ryu}}, \
  and\ \bibinfo {author} {\bibfnamefont {A.~W.~W.}\ \bibnamefont {Ludwig}},\
  }\href {\doibase 10.1103/PhysRevLett.102.196804} {\bibfield  {journal}
  {\bibinfo  {journal} {Phys. Rev. Lett.}\ }\textbf {\bibinfo {volume} {102}},\
  \bibinfo {pages} {196804} (\bibinfo {year} {2009})}\BibitemShut {NoStop}%
\bibitem [{\citenamefont {Lu}\ and\ \citenamefont {Vishwanath}(2012)}]{Lu2012}%
  \BibitemOpen
  \bibfield  {author} {\bibinfo {author} {\bibfnamefont {Y.-M.}\ \bibnamefont
  {Lu}}\ and\ \bibinfo {author} {\bibfnamefont {A.}~\bibnamefont
  {Vishwanath}},\ }\href {\doibase 10.1103/PhysRevB.86.125119} {\bibfield
  {journal} {\bibinfo  {journal} {Phys. Rev. B}\ }\textbf {\bibinfo {volume}
  {86}},\ \bibinfo {pages} {125119} (\bibinfo {year} {2012})}\BibitemShut
  {NoStop}%
\bibitem [{\citenamefont {Senthil}\ and\ \citenamefont
  {Levin}(2013)}]{Senthil2012}%
  \BibitemOpen
  \bibfield  {author} {\bibinfo {author} {\bibfnamefont {T.}~\bibnamefont
  {Senthil}}\ and\ \bibinfo {author} {\bibfnamefont {M.}~\bibnamefont
  {Levin}},\ }\href {\doibase 10.1103/PhysRevLett.110.046801} {\bibfield
  {journal} {\bibinfo  {journal} {Phys. Rev. Lett.}\ }\textbf {\bibinfo
  {volume} {110}},\ \bibinfo {pages} {046801} (\bibinfo {year}
  {2013})}\BibitemShut {NoStop}%
\bibitem [{\citenamefont {Ran}\ \emph {et~al.}(2008)\citenamefont {Ran},
  \citenamefont {Vishwanath},\ and\ \citenamefont {Lee}}]{Ran2008}%
  \BibitemOpen
  \bibfield  {author} {\bibinfo {author} {\bibfnamefont {Y.}~\bibnamefont
  {Ran}}, \bibinfo {author} {\bibfnamefont {A.}~\bibnamefont {Vishwanath}}, \
  and\ \bibinfo {author} {\bibfnamefont {D.-H.}\ \bibnamefont {Lee}},\ }\href
  {\doibase 10.1103/PhysRevLett.101.086801} {\bibfield  {journal} {\bibinfo
  {journal} {Phys. Rev. Lett.}\ }\textbf {\bibinfo {volume} {101}},\ \bibinfo
  {pages} {086801} (\bibinfo {year} {2008})}\BibitemShut {NoStop}%
\bibitem [{\citenamefont {Qi}\ and\ \citenamefont {Zhang}(2008)}]{Qi2008}%
  \BibitemOpen
  \bibfield  {author} {\bibinfo {author} {\bibfnamefont {X.-L.}\ \bibnamefont
  {Qi}}\ and\ \bibinfo {author} {\bibfnamefont {S.-C.}\ \bibnamefont {Zhang}},\
  }\href {\doibase 10.1103/PhysRevLett.101.086802} {\bibfield  {journal}
  {\bibinfo  {journal} {Phys. Rev. Lett.}\ }\textbf {\bibinfo {volume} {101}},\
  \bibinfo {pages} {086802} (\bibinfo {year} {2008})}\BibitemShut {NoStop}%
\bibitem [{\citenamefont {Wen}(2004)}]{Wen2004Book}%
  \BibitemOpen
  \bibfield  {author} {\bibinfo {author} {\bibfnamefont {X.-G.}\ \bibnamefont
  {Wen}},\ }\href@noop {} {\emph {\bibinfo {title} {Quantum field theory of
  many-body systems}}}\ (\bibinfo  {publisher} {Oxford University Press},\
  \bibinfo {year} {2004})\ Chap.~\bibinfo {chapter} {7}\BibitemShut {NoStop}%
\bibitem [{\citenamefont {Wen}(2002)}]{Wen2002}%
  \BibitemOpen
  \bibfield  {author} {\bibinfo {author} {\bibfnamefont {X.-G.}\ \bibnamefont
  {Wen}},\ }\href@noop {} {\bibfield  {journal} {\bibinfo  {journal} {Phys.
  Rev., B}\ }\textbf {\bibinfo {volume} {65,}},\ \bibinfo {pages} {165113}
  (\bibinfo {year} {2002})},\ \Eprint {http://arxiv.org/abs/cond-mat/0107071}
  {cond-mat/0107071} \BibitemShut {NoStop}%
\bibitem [{\citenamefont {Fu}\ \emph {et~al.}(2007)\citenamefont {Fu},
  \citenamefont {Kane},\ and\ \citenamefont {Mele}}]{FuKaneMele2007}%
  \BibitemOpen
  \bibfield  {author} {\bibinfo {author} {\bibfnamefont {L.}~\bibnamefont
  {Fu}}, \bibinfo {author} {\bibfnamefont {C.~L.}\ \bibnamefont {Kane}}, \ and\
  \bibinfo {author} {\bibfnamefont {E.~J.}\ \bibnamefont {Mele}},\ }\href
  {\doibase 10.1103/PhysRevLett.98.106803} {\bibfield  {journal} {\bibinfo
  {journal} {Phys. Rev. Lett.}\ }\textbf {\bibinfo {volume} {98,}},\ \bibinfo
  {pages} {106803} (\bibinfo {year} {2007})},\ \Eprint
  {http://arxiv.org/abs/cond-mat/0607699} {cond-mat/0607699} \BibitemShut
  {NoStop}%
\bibitem [{\citenamefont {Kravec}\ and\ \citenamefont
  {McGreevy}(2013)}]{KravecMcGreevy2013}%
  \BibitemOpen
  \bibfield  {author} {\bibinfo {author} {\bibfnamefont {S.~M.}\ \bibnamefont
  {Kravec}}\ and\ \bibinfo {author} {\bibfnamefont {J.}~\bibnamefont
  {McGreevy}},\ }\href {\doibase 10.1103/PhysRevLett.111.161603} {\bibfield
  {journal} {\bibinfo  {journal} {Phys. Rev. Lett.}\ }\textbf {\bibinfo
  {volume} {111,}},\ \bibinfo {pages} {161603} (\bibinfo {year} {2013})},\
  \Eprint {http://arxiv.org/abs/1306.3992} {1306.3992} \BibitemShut {NoStop}%
\bibitem [{\citenamefont {Wen}(2015)}]{Wen2014}%
  \BibitemOpen
  \bibfield  {author} {\bibinfo {author} {\bibfnamefont {X.-G.}\ \bibnamefont
  {Wen}},\ }\href@noop {} {\bibfield  {journal} {\bibinfo  {journal} {Phys.
  Rev. B}\ }\textbf {\bibinfo {volume} {91,}},\ \bibinfo {pages} {205101}
  (\bibinfo {year} {2015})},\ \Eprint {http://arxiv.org/abs/1410.8477}
  {1410.8477} \BibitemShut {NoStop}%
\bibitem [{\citenamefont {Bi}\ and\ \citenamefont {Xu}(2015)}]{Bi2015a}%
  \BibitemOpen
  \bibfield  {author} {\bibinfo {author} {\bibfnamefont {Z.}~\bibnamefont
  {Bi}}\ and\ \bibinfo {author} {\bibfnamefont {C.}~\bibnamefont {Xu}},\
  }\href@noop {} {\bibfield  {journal} {\bibinfo  {journal} {Phys. Rev. B}\
  }\textbf {\bibinfo {volume} {91,}},\ \bibinfo {pages} {184404} (\bibinfo
  {year} {2015})},\ \Eprint {http://arxiv.org/abs/1501.02271} {1501.02271}
  \BibitemShut {NoStop}%
\bibitem [{\citenamefont {Cheng}(2017)}]{Cheng2017}%
  \BibitemOpen
  \bibfield  {author} {\bibinfo {author} {\bibfnamefont {M.}~\bibnamefont
  {Cheng}},\ }\href@noop {} {\bibfield  {journal} {\bibinfo  {journal} {privite
  communication}\ } (\bibinfo {year} {2017})}\BibitemShut {NoStop}%
\bibitem [{\citenamefont {Levin}\ and\ \citenamefont {Wen}(2006)}]{Levin2006}%
  \BibitemOpen
  \bibfield  {author} {\bibinfo {author} {\bibfnamefont {M.}~\bibnamefont
  {Levin}}\ and\ \bibinfo {author} {\bibfnamefont {X.-G.}\ \bibnamefont
  {Wen}},\ }\href {\doibase 10.1103/PhysRevB.73.035122} {\bibfield  {journal}
  {\bibinfo  {journal} {Phys. Rev. B}\ }\textbf {\bibinfo {volume} {73}},\
  \bibinfo {pages} {035122} (\bibinfo {year} {2006})}\BibitemShut {NoStop}%
\bibitem [{\citenamefont {Li}\ and\ \citenamefont {Chen}(2017)}]{Li2016}%
  \BibitemOpen
  \bibfield  {author} {\bibinfo {author} {\bibfnamefont {Y.~D.}\ \bibnamefont
  {Li}}\ and\ \bibinfo {author} {\bibfnamefont {G.}~\bibnamefont {Chen}},\
  }\href@noop {} {\bibfield  {journal} {\bibinfo  {journal} {Phys. Rev. B}\
  }\textbf {\bibinfo {volume} {95,}},\ \bibinfo {pages} {041106} (\bibinfo
  {year} {2017})},\ \Eprint {http://arxiv.org/abs/1607.02287} {1607.02287}
  \BibitemShut {NoStop}%
\bibitem [{\citenamefont {Chen}(2017{\natexlab{a}})}]{Chen2017}%
  \BibitemOpen
  \bibfield  {author} {\bibinfo {author} {\bibfnamefont {G.}~\bibnamefont
  {Chen}},\ }\href {\doibase 10.1103/PhysRevB.96.085136} {\bibfield  {journal}
  {\bibinfo  {journal} {Phys. Rev. B}\ }\textbf {\bibinfo {volume} {96}},\
  \bibinfo {pages} {085136} (\bibinfo {year} {2017}{\natexlab{a}})}\BibitemShut
  {NoStop}%
\bibitem [{\citenamefont {Chen}(2017{\natexlab{b}})}]{Chen2017a}%
  \BibitemOpen
  \bibfield  {author} {\bibinfo {author} {\bibfnamefont {G.}~\bibnamefont
  {Chen}},\ }\href {\doibase 10.1103/PhysRevB.96.195127} {\bibfield  {journal}
  {\bibinfo  {journal} {Phys. Rev. B}\ }\textbf {\bibinfo {volume} {96}},\
  \bibinfo {pages} {195127} (\bibinfo {year} {2017}{\natexlab{b}})}\BibitemShut
  {NoStop}%
\bibitem [{\citenamefont {Zou}(2017)}]{Zou2017}%
  \BibitemOpen
  \bibfield  {author} {\bibinfo {author} {\bibfnamefont {L.}~\bibnamefont
  {Zou}},\ }\href {\doibase 10.1103/PhysRevB.97.045130} {\bibfield  {journal}
  {\bibinfo  {journal} {Phys. Rev. B}\ }\textbf {\bibinfo {volume} {97,}},\
  \bibinfo {pages} {045130} (\bibinfo {year} {2017})},\ \Eprint
  {http://arxiv.org/abs/1711.03090} {1711.03090} \BibitemShut {NoStop}%
\bibitem [{\citenamefont {{Jian}}\ and\ \citenamefont {{Qi}}(2014)}]{Jian2014}%
  \BibitemOpen
  \bibfield  {author} {\bibinfo {author} {\bibfnamefont {C.-M.}\ \bibnamefont
  {{Jian}}}\ and\ \bibinfo {author} {\bibfnamefont {X.-L.}\ \bibnamefont
  {{Qi}}},\ }\href {\doibase 10.1103/PhysRevX.4.041043} {\bibfield  {journal}
  {\bibinfo  {journal} {Physical Review X}\ }\textbf {\bibinfo {volume} {4}},\
  \bibinfo {eid} {041043} (\bibinfo {year} {2014})},\ \Eprint
  {http://arxiv.org/abs/1405.6688} {arXiv:1405.6688 [cond-mat.str-el]}
  \BibitemShut {NoStop}%
\bibitem [{\citenamefont {Bi}\ \emph {et~al.}(2015)\citenamefont {Bi},
  \citenamefont {Slagle},\ and\ \citenamefont {Xu}}]{Bi2015}%
  \BibitemOpen
  \bibfield  {author} {\bibinfo {author} {\bibfnamefont {Z.}~\bibnamefont
  {Bi}}, \bibinfo {author} {\bibfnamefont {K.}~\bibnamefont {Slagle}}, \ and\
  \bibinfo {author} {\bibfnamefont {C.}~\bibnamefont {Xu}},\ }\href@noop {} {\
  (\bibinfo {year} {2015})},\ \Eprint {http://arxiv.org/abs/1504.04373}
  {1504.04373} \BibitemShut {NoStop}%
\bibitem [{\citenamefont {Wang}\ and\ \citenamefont
  {Senthil}(2014)}]{WangSenthil2014}%
  \BibitemOpen
  \bibfield  {author} {\bibinfo {author} {\bibfnamefont {C.}~\bibnamefont
  {Wang}}\ and\ \bibinfo {author} {\bibfnamefont {T.}~\bibnamefont {Senthil}},\
  }\href {\doibase 10.1103/PhysRevB.89.195124} {\bibfield  {journal} {\bibinfo
  {journal} {Phys. Rev. B}\ }\textbf {\bibinfo {volume} {89}},\ \bibinfo
  {pages} {195124} (\bibinfo {year} {2014})}\BibitemShut {NoStop}%
\bibitem [{\citenamefont {Borokhov}\ \emph {et~al.}(2002)\citenamefont
  {Borokhov}, \citenamefont {Kapustin},\ and\ \citenamefont
  {Wu}}]{Borokhov2002}%
  \BibitemOpen
  \bibfield  {author} {\bibinfo {author} {\bibfnamefont {V.}~\bibnamefont
  {Borokhov}}, \bibinfo {author} {\bibfnamefont {A.}~\bibnamefont {Kapustin}},
  \ and\ \bibinfo {author} {\bibfnamefont {X.}~\bibnamefont {Wu}},\ }\href@noop
  {} {\bibfield  {journal} {\bibinfo  {journal} {JHEP}\ }\textbf {\bibinfo
  {volume} {0211}},\ \bibinfo {pages} {049} (\bibinfo {year} {2002})},\ \Eprint
  {http://arxiv.org/abs/hep-th/0206054} {hep-th/0206054} \BibitemShut {NoStop}%
\bibitem [{\citenamefont {Senthil}\ \emph {et~al.}(1999)\citenamefont
  {Senthil}, \citenamefont {Marston},\ and\ \citenamefont
  {Fisher}}]{Senthil1999}%
  \BibitemOpen
  \bibfield  {author} {\bibinfo {author} {\bibfnamefont {T.}~\bibnamefont
  {Senthil}}, \bibinfo {author} {\bibfnamefont {J.~B.}\ \bibnamefont
  {Marston}}, \ and\ \bibinfo {author} {\bibfnamefont {M.~P.~A.}\ \bibnamefont
  {Fisher}},\ }\href {\doibase 10.1103/physrevb.60.4245} {\bibfield  {journal}
  {\bibinfo  {journal} {Physical Review B}\ }\textbf {\bibinfo {volume} {60}},\
  \bibinfo {pages} {4245} (\bibinfo {year} {1999})}\BibitemShut {NoStop}%
\bibitem [{\citenamefont {Kane}\ and\ \citenamefont {Mele}(2005)}]{Kane2005}%
  \BibitemOpen
  \bibfield  {author} {\bibinfo {author} {\bibfnamefont {C.~L.}\ \bibnamefont
  {Kane}}\ and\ \bibinfo {author} {\bibfnamefont {E.~J.}\ \bibnamefont
  {Mele}},\ }\href {\doibase 10.1103/PhysRevLett.95.226801} {\bibfield
  {journal} {\bibinfo  {journal} {Phys. Rev. Lett.}\ }\textbf {\bibinfo
  {volume} {95}},\ \bibinfo {pages} {226801} (\bibinfo {year}
  {2005})}\BibitemShut {NoStop}%
\bibitem [{\citenamefont {Potter}\ \emph {et~al.}(2016)\citenamefont {Potter},
  \citenamefont {Wang}, \citenamefont {Metlitski},\ and\ \citenamefont
  {Vishwanath}}]{Potter2016}%
  \BibitemOpen
  \bibfield  {author} {\bibinfo {author} {\bibfnamefont {A.~C.}\ \bibnamefont
  {Potter}}, \bibinfo {author} {\bibfnamefont {C.}~\bibnamefont {Wang}},
  \bibinfo {author} {\bibfnamefont {M.~A.}\ \bibnamefont {Metlitski}}, \ and\
  \bibinfo {author} {\bibfnamefont {A.}~\bibnamefont {Vishwanath}},\
  }\href@noop {} {\  (\bibinfo {year} {2016})},\ \Eprint
  {http://arxiv.org/abs/1609.08618} {1609.08618} \BibitemShut {NoStop}%
\bibitem [{\citenamefont {Witten}(1979)}]{WITTEN1979}%
  \BibitemOpen
  \bibfield  {author} {\bibinfo {author} {\bibfnamefont {E.}~\bibnamefont
  {Witten}},\ }\href {\doibase http://dx.doi.org/10.1016/0370-2693(79)90838-4}
  {\bibfield  {journal} {\bibinfo  {journal} {Physics Letters B}\ }\textbf
  {\bibinfo {volume} {86}},\ \bibinfo {pages} {283 } (\bibinfo {year}
  {1979})}\BibitemShut {NoStop}%
\bibitem [{\citenamefont {Wang}\ \emph
  {et~al.}(2017{\natexlab{c}})\citenamefont {Wang}, \citenamefont {Lin},\ and\
  \citenamefont {Gu}}]{Wang2016b}%
  \BibitemOpen
  \bibfield  {author} {\bibinfo {author} {\bibfnamefont {C.}~\bibnamefont
  {Wang}}, \bibinfo {author} {\bibfnamefont {C.-H.}\ \bibnamefont {Lin}}, \
  and\ \bibinfo {author} {\bibfnamefont {Z.-C.}\ \bibnamefont {Gu}},\
  }\href@noop {} {\bibfield  {journal} {\bibinfo  {journal} {Phys. Rev. B}\
  }\textbf {\bibinfo {volume} {95,}},\ \bibinfo {pages} {195147} (\bibinfo
  {year} {2017}{\natexlab{c}})},\ \Eprint {http://arxiv.org/abs/1610.08478}
  {1610.08478} \BibitemShut {NoStop}%
\bibitem [{\citenamefont {Cheng}\ \emph {et~al.}(2018)\citenamefont {Cheng},
  \citenamefont {Tantivasadakarn},\ and\ \citenamefont {Wang}}]{Cheng2017a}%
  \BibitemOpen
  \bibfield  {author} {\bibinfo {author} {\bibfnamefont {M.}~\bibnamefont
  {Cheng}}, \bibinfo {author} {\bibfnamefont {N.}~\bibnamefont
  {Tantivasadakarn}}, \ and\ \bibinfo {author} {\bibfnamefont {C.}~\bibnamefont
  {Wang}},\ }\href {\doibase 10.1103/PhysRevX.8.011054} {\bibfield  {journal}
  {\bibinfo  {journal} {Phys. Rev. X}\ }\textbf {\bibinfo {volume} {8}},\
  \bibinfo {pages} {011054} (\bibinfo {year} {2018})}\BibitemShut {NoStop}%
\bibitem [{\citenamefont {Levin}\ and\ \citenamefont {Gu}(2012)}]{Levin2012a}%
  \BibitemOpen
  \bibfield  {author} {\bibinfo {author} {\bibfnamefont {M.}~\bibnamefont
  {Levin}}\ and\ \bibinfo {author} {\bibfnamefont {Z.-C.}\ \bibnamefont {Gu}},\
  }\href@noop {} {\bibfield  {journal} {\bibinfo  {journal} {Phys. Rev. B}\
  }\textbf {\bibinfo {volume} {86,}},\ \bibinfo {pages} {115109} (\bibinfo
  {year} {2012})},\ \Eprint {http://arxiv.org/abs/1202.3120} {1202.3120}
  \BibitemShut {NoStop}%
\bibitem [{\citenamefont {Xu}\ and\ \citenamefont {Senthil}(2013)}]{Xu2013}%
  \BibitemOpen
  \bibfield  {author} {\bibinfo {author} {\bibfnamefont {C.}~\bibnamefont
  {Xu}}\ and\ \bibinfo {author} {\bibfnamefont {T.}~\bibnamefont {Senthil}},\
  }\href@noop {} {\bibfield  {journal} {\bibinfo  {journal} {Phys. Rev. B}\
  }\textbf {\bibinfo {volume} {87,}},\ \bibinfo {pages} {174412} (\bibinfo
  {year} {2013})},\ \Eprint {http://arxiv.org/abs/1301.6172} {1301.6172}
  \BibitemShut {NoStop}%
\end{thebibliography}%

\end{document}